\documentclass[economy,MS,openright,twoside]{dukedissertation}

\usepackage{amsmath, amssymb, amsfonts, amsthm}
\usepackage{graphicx}
\usepackage{cite}
\usepackage{color}
\usepackage{bm}
\usepackage{subfigure}
\usepackage{graphicx}
\usepackage{multirow}
\usepackage{setspace}
\usepackage{array}			 
\usepackage{listings}		 
\usepackage{mathtools}		 
\usepackage[automark]{scrpage2}
\usepackage{booktabs}
\usepackage{ctable}
							
\newcommand{\mc}[1]{\mathcal #1}
\newcommand{\ii}{\textrm {i}}
\newcommand{\ee}{\textrm {e}}
\newcommand{\ga}[1]{\gamma^{#1}}
\newcommand{\vga}{\vec{\gamma}}
\newcommand{\vpi}{\vec{\pi}}
\newcommand{\unit}{\mathbb{I}}
\newcommand{\de}[1]{\partial_{#1}}
\newcommand{\vnab}{\vec {\nabla}}
\newcommand{\br}[1]{\left( #1 \right)}
\newcommand{\obs}[1]{\left \langle #1 \right \rangle}
\newcommand{\qt}{\br{\vec q,t}}
\newcommand{\kt}{\br{\vec {q'},t}}
\newcommand{\brt}{\br{t}}
\newcommand{\brq}{\br{\vec q}}
\newcommand{\brk}{\br{\vec {q'}}}
\newcommand{\mqt}{\br{-\vec q,t}}
\newcommand{\brmq}{\br{-\vec q}}
\newcommand{\bru}{\br{u}}
\newcommand{\com}[1]{\left[ #1 \right]}
\newcommand{\acom}[1]{\left \{ #1 \right \}}
\newcommand{\chip}{\chi_0^+}
\newcommand{\chis}{\br{\chi_0^+}^*}
\newcommand{\bet}[1]{\left \lvert #1 \right \rvert}

\newcommand{\va}{\varepsilon}
\newcommand{\enorm}{\varepsilon_{\perp}}
\newcommand{\real}[1]{\operatorname{Re}\br{#1}} 
\newcommand{\imag}[1]{\operatorname{Im}\br{#1}} 
\newcommand{\qts}{\br{\vec q, t'}}
\newcommand{\ttau}{\br{\frac{t}{\tau}}}
\newcommand{\tln}[1]{\textrm{ln}\br{#1}}
\newcommand{\tcos}[1]{\textrm{cos}\br{#1}}
\newcommand{\tsin}[1]{\textrm{sin}\br{#1}}
\newcommand{\ttanh}[1]{\textrm{tanh}\br{#1}}

\newcommand{\tsinh}[1]{\textrm{sinh}\br{#1}}

\newcommand{\Tanh}{\ttanh{\frac{t}{\tau}}}

\newcommand{\IVac}{\int_{t_{Vac}}^t dt'}
\newcommand{\thetaT}{\theta \br{\vec q, t, t'}}
\newcommand{\qu}{\br{\vec q, u}}
\newcommand{\uua}[1]{u^{\alpha #1}}
\newcommand{\uu}{u^{\alpha}}
\newcommand{\muub}[1]{\br{1-u}^{\beta #1}}
\newcommand{\muu}{\br{1-u}^{\beta}}
\newcommand{\Hyper}[1]{ {}_2 \mathcal F_1 \br{#1}}

\newcommand{\be}[1]{\begin{equation} #1 \end{equation}}

\newcommand{\ma}[1]{\begin{pmatrix} #1 \end{pmatrix}}
\newcommand{\al}[1]{\begin{align} #1 \end{align}}

\newcommand{\wh}[1]{\textcolor{white}{#1}}

\definecolor{Gray}{gray}{0.9}
\definecolor{BrickRed}{cmyk}{0,0.89,0.94,0.28}
\definecolor{CadetBlue}{cmyk}{0.62,0.57,0.23,0}
\definecolor{Violet}{cmyk}{0.79,0.88,0,0}
\definecolor{ForestGreen}{cmyk}{0.91, 0, 0.88, 0.12}
\definecolor{Brown}{cmyk}{0, 0.81, 1, 0.60}

\definecolor{WildStrawberry}{cmyk}{0,0.96,0.39,0} 
\definecolor{CornflowerBlue}{cmyk}{0.65,0.13,0,0.4} 
\definecolor{CornflowerBlueDark}{cmyk}{0.65,0.13,0,0.6} 
\definecolor{CadetBlue}{cmyk}{0.62,0.57,0.23,0} 
\definecolor{Peach}{cmyk}{0,0.70,0.70,0.35} 
\definecolor{BurntOrange}{cmyk}{0,0.51,1,0} 
\definecolor{Red}{cmyk}{0,1,1,0} 
\definecolor{SkyBlue}{cmyk}{0.62,0,0.12,0.3} 

\definecolor{Blue}{cmyk}{1,1,0,0} 
\definecolor{Red}{cmyk}{0,1,1,0}
\definecolor{comment}{cmyk}{0,0,0,0}

\author{Christian Kohlf\"urst}
\title{Electron-Positron Pair Production in Structured Pulses of Electric Fields}
\supervisor{Dr. Reinhard Alkofer, Dr. Florian Hebenstreit}

\date{2012} 

\member{[Department Committee Member \#1]}
\member{[Department Committee Member \#2]}
\member{[External Committee Member]}

\usepackage[pdftex, pdfusetitle, plainpages=false, 
				letterpaper, bookmarks, bookmarksnumbered,
				colorlinks, linkcolor=black, citecolor=black,
	         filecolor=black, urlcolor=black]
				{hyperref}

\begin{document}

\maketitle

\abstract

\onehalfspacing

The so-called Schwinger effect, the creation of electron-positron pairs in strong electric fields, has been discussed for nearly eighty years even though an experimental verification of this effect seemed to be impossible: The generation of electric fields of the order of $10^{18}V/m$,
which are required to overcome the exponential suppression of the Schwinger effect, was not feasible. Nevertheless there has been substantial progress in the theoretical models and methods as well as a rapid development in LASER
and computer technology during the last decades. As a consequence, plans for high-intensity LASER facilities have been made and thus an experimental measurement of the Schwinger effect could become possible in the near future. \\

\onehalfspacing
In this master thesis the non-perturbative electron-positron pair production in time-dependent electric fields is investigated. 
The quantum kinetic formalism is employed in order to calculate the electron density for various field configurations. 
The corresponding set of first order, ordinary differential equations is analyzed and numerically solved. The focus of this study lies on the dynamically assisted 
Schwinger effect in pulsed electric fields with at least two different time scales.  
Furthermore, interference effects arising in setups with multiple pulses are examined and
first results for an optimization of the particle number yield by pulse-shaping are given.

\singlespacing


\tableofcontents 
\listoftables	
\listoffigures	

\acknowledgements

I am indebted to my many people, who supported me over the last years. \\

Foremost, I would like to express my gratitude to my advisor Prof. Reinhard Alkofer for giving me the opportunity to work on a, at least in my opinion, highly interesting research topic. For the support over the last year I further want to thank both Prof. Reinhard Alkofer and Dr. Florian Hebenstreit. 
In addition, a 'thank you' to Greg von Winckel and Mario Mitter in particular, who introduced me to optimal control theory. \\

Besides my advisors and collaborators, I would like to thank my colleagues and friends, where I want to mention a few people by name, Georg J\"ager, Alexander Hieden and Hans-Peter Schadler. Furthermore, I would like to express my gratitude to the whole SIC!QFT group. Particularly, I want to mention Matthias Blatnik. It was a pleasure to share an office with you. \\

My sincere thanks also go to Karin Dissauer and Richard Haider for a first proof-reading of this thesis. For the very comprehensive and detailed proof-reading I owe my deepest gratitude to Dr. Florian Hebenstreit. \\

Last but not least I would like to express my gratitude to my parents and my brother. This thesis would not have been possible without your permanent support.

%
%
%
\chapter{Introduction}

\section{Historical Overview}

One of the most important calculations of the 1930s regarding particle production was F. Sauters seminal investigation of an electron in a potential of the form $V=\nu x$ \cite{Sauter:1931zz}, corresponding to a static electric field. His computations within the framework of barrier scattering revealed that the transmission coefficient is exponentially suppressed and becomes sizable 
only at field strengths of the order of $10^{18}V/m$,  
\be{
P \approx \mathrm{exp} \br{-\pi \frac{m_e^2 c^3}{\hbar e \bet{E}}}.
}

A few years later W. Heisenberg and H. Euler \cite{springerlink:10.1007/BF01343663} published a seminal paper in which they investigated the non-linear effect of virtual electron-positron
pairs on the electromagnetic field as well as electron-positron pair production.
One of the results worth mentioning is the effective Lagrangian
\begin{multline}
\mathcal{L}_{eff} = \frac{1}{hc} \int_0^{\infty} \frac{d \eta}{\eta^3} \ee^{-\eta e E_{crit}} \\
  \times \left( \ii e^2 \eta^2 \br{\vec E \cdot \vec B} \frac{\br{\mathrm{cos}\br{\eta e\sqrt{\vec E^2 - \vec B^2 + 2\ii \br{\vec E \cdot \vec B}}} + c.c.}}{\br{\mathrm{cos}\br{\eta e\sqrt{\vec E^2 - \vec B^2 + 2\ii \br{\vec E \cdot \vec B}}} - c.c.}} 
 +1+\frac{e^2 \eta^2}{3} \br{\vec B^2 - \vec E^2} \right).
\end{multline}
The weak field expansion of this effective Lagrangian yields as leading order \\
 correction to the Maxwell Lagrangian \cite{Ruffini20101}:
\be{
\mathcal{L}_{eff} = \frac{\vec E^2- \vec B^2}{2} + \frac{\alpha}{90 \pi E_{crit}^2} \br{\br{\vec E^2- \vec B^2}^2 + 7 \br{\vec E \cdot \vec B}^2},
\label{eq1_1:1}
}
with $\alpha$ the fine structure constant and $E_{crit}$ the critical field strength for particle pair production
\be{
E_{crit} = \frac{m_e^2 c^3}{\hbar e}.
}

The effective Lagrangian \eqref{eq1_1:1} describes effects, which are not covered by Maxwell equations. This lies in the fact that the electromagnetic fields appearing in the Lagrangian are of quartic order
and thus Maxwell equations in the vacuum become modified by non-linear terms. This results in the capability of describing phenomena like vacuum polarization or light-light scattering. In particular the amplitude for this scattering in the low energy limit is \cite{hep-th/0406216}
\be{
S = \frac{\alpha}{90 \pi E_{crit}^2} \int d^4x \br{ \br{\vec E^2 - \vec B^2}^2 + 7 \br{\vec E \cdot \vec B}^2}.
}

In the following years, attempts to formulate a consistent theory of the interaction between electromagnetic fields and matter were made. However, they all failed in various aspects. It ultimately took physicists up to the 1950s for a theory that could describe the electromagnetic force relativistically and could cope with quantum corrections. This theory, whose inventors R. Feynman, J. Schwinger and S. Tomanaga received the Physics Nobel Prize in 1965, is referred to as QED or quantum electrodynamics, nowadays. 

One of the Nobel laureates, J. Schwinger, also worked on particle pair creation in the 1950s. At that time he calculated the probability for particle creation in a spatially homogeneous and time-independent electric field using a proper time method \cite{PhysRev.82.664}. This allowed him to derive the probability $\Gamma$ of electron-positron pair creation per unit time and volume
\be{
\Gamma = 2 \imag{\mathcal{L}_{eff}},
}
where the imaginary part of the Lagrangian is given by
\be{
\imag{\mathcal L_{eff}}  \sim \frac{e^2 E^2}{8 \pi^3} \sum_{n=1}^{\infty} \frac{1}{n^2} \mathrm{exp} \br{- \frac{m_e^2 \pi n}{eE}}.
}

The following decades were dominated by the quest for obtaining a standard model of particle physics, whereas only minor progress in the field of the Schwinger effect was made.
Things changed, however, in the 1990s due to the advent of fast computers and high-intensity LASERs. Hence within ten years of development striking breakthroughs have been achieved in numerical simulations, for instance, for particle production \cite{PhysRevD.44.1825,PhysRevLett.67.2427,PhysRevD.58.125015}. In addition new methods, like the quantum kinetic approach \cite{PhysRevD.60.116011,PhysRevD.61.117502,hep-ph/9809227,hep-ph/9809233} have been devised based on previous work on quantum field theory and transport theory.  
Recently also experimental physicists have shown an increased interest in high-intensity LASER experiments \cite{PhysRevLett.79.1626} investigating fundamental issues like the Schwinger effect. Most notably, both the XFEL and ELI facilities are supposed to explore physics at new scales \cite{springerlink:10.1140/epjd/e2009-00113-x,springerlink:10.1140/epjd/e2009-00169-6}.
In this respect, they are going to provide the most intense electric fields ever produced in a laboratory in upcoming experiments. 

\section{Aspects of Pair Production}
A direct observation of the Schwinger effect has been considered as impossible for a long time as electric field strengths of the order of $10^{18} V/m$ seemed out of reach.
However a modern LASER can provide field strengths of approximately $10^{15}V/m$ and even stronger LASERs are planned. Experimental physicists clearly benefit from this development as they are provided with a new powerful tool for experiments. The challenge from a theorist's perspective is now to perform calculations for realistic field parameters. It is obvious, for example, that a LASER field shows a time dependence. In addition, the assumption of a spatially homogeneous field simplifies the equations a lot, but is also not very realistic.  

The first calculations concerning the pair creation process were formulated as a scattering problem. The possibility of an electron tunneling through a potential barrier characterizes this effect also nowadays when we talk about the vacuum tunneling probability for an electron. This method for calculation gave a first estimate, but soon novel and more reliable approaches have been established. Schwinger for example calculated the analytic solution for a time-independent field with the aid of a proper time technique, which was for years the most advanced method. 
Moreover, semi-classical approaches like the WKB formalism proved to be successful in describing the pair production process.
In addition for the case of completely uniform electric fields more advanced algebraic approaches have been introduced \cite{1104.0468} and calculations based on solitons have been made to describe special field configurations \cite{1110.4684,PhysRevD.84.125028}.
In order to describe more realistic field configurations, in the 1990s and in the beginning of the 21st century new formalisms to calculate the Schwinger effect numerically have been introduced, for instance the quantum kinetic approach \cite{hep-ph/9809227}. Other approaches are a closely related scattering-like formalism in terms of the Riccati equation \cite{PhysRevD.79.065027}, the Dirac-Heisenberg-Wigner formalism \cite{PhysRevD.44.1825,Hebenstreit:2011wk,Hebenstreit:2011cr} or the numerical worldline formalism \cite{PhysRevD.72.065001}. 

Once these methods were established, it became feasible to investigate various aspects of pair creation for the first time.
A large and growing body of literature has investigated time-dependent electric fields. One prominent example is the sinusoidal field $E = \varepsilon E_{crit} \mathrm{sin} \br{t/\tau}$ \cite{PhysRevLett.87.193902,PhysRevLett.96.140402,PhysRevLett.89.153901}, which was first discussed in the 1970s \cite{PhysRevD.2.1191}. Another field configuration widely explored in various computations is the Sauter field \\
$E = \varepsilon E_{crit} \mathrm{sech}^2 \br{t/\tau}$ \cite{HebDip,OrtDip}, which was also first investigated in the 1970s \cite{Nikishiv} and further analyzed in an S-matrix formalism \cite{springerlink:10.1134/1.1788038} as well as in the worldline approach \cite{PhysRevD.84.125023,PhysRevD.72.065001}. 
Moreover, the question has been raised whether it is possible to lower the threshold for particle production in order to abet the experimental verification of the Schwinger effect.
Therefore it has been suggested \cite{OrtDip,PhysRevLett.101.130404,PhysRevD.80.111301,PhysRevD.85.025004,Orthaber201180} that the particle production rate could be significantly enhanced by the superposition of electric fields with different time scales. Moreover there have been first investigations of the Schwinger effect in space-time dependent electric fields \cite{PhysRevLett.107.180403,HebDiss,PhysRevLett.102.080402}.
Furthermore there was theoretical progress as well: For instance, it was shown \cite{PhysRevD.78.036008} that there is a difference between particle production rate and the rate associated with the decay of the vacuum.

As already mentioned,the WKB approach has turned out to be a very successful tool for describing pair production since its first application to the Schwinger effect. 
It is still widely used and gives new insight into this process \cite{PhysRevD.79.065027,PhysRevD.82.045007,PhysRevLett.104.250402}. For example, the interference effects occurring in pulsed fields with subcycle structure, first observed in F. Hebenstreit et al. \cite{PhysRevLett.102.150404}, could be explained conclusively within the WKB approach \cite{PhysRevLett.108.030401,PhysRevD.83.065028}.
It is also important to state here, that the different approaches give similar results \cite{PhysRevD.78.061701} and are sometimes even equivalent \cite{PhysRevD.79.065027,PhysRevLett.107.180403,PhysRevD.83.025011}.

The interested reader is referred to the articles \cite{hep-th/0406216,springerlink:10.1140/epjd/e2009-00113-x,springerlink:10.1140/epjd/e2009-00169-6,1202.1557,1111.5192} for further information on the topic.
If one searches for an extensive review on particle production, high-intensity LASERs and its applications we encourage the reader to have a look at \cite{Ruffini20101,1111.3886}.

\section{Objective of this Thesis}

We will entirely focus on the quantum kinetic approach in this thesis, as it provides a flexible and well-established formalism. This thesis is divided into three parts. In the first part, we introduce our notation and state our assumptions. Afterwards we derive a quantum kinetic equation for the particle distribution function  
\be{
\dot{F} \qt = W \qt \IVac W \qts \br{1- F \qts} \tcos{2 \thetaT}, 
}
with $W \qt ,~ \thetaT$ as input functions. Due to the fact, that the formulation in terms of this integro-differential equation is cumbersome in numerical simulations, we will also derive the equivalent set of ordinary differential equations of first order \cite{PhysRevD.60.116011}
\be{
\ma{\dot{F} \\ \dot{G} \\ \dot{H}} = \ma{0 & W & 0 \\ -W & 0 & -2\omega \\ 0 & 2\omega & 0} \ma{F \\ G \\ H} + \ma{0 \\ W \\ 0},
} 
where $G,~ H$ are auxiliary quantities and $W ,~ \omega$ input functions.
Then we will calculate an analytical solution for the electric field 
\be{
E = \varepsilon E_{crit} \mathrm{sech}^2 \br{t/\tau}.
} 
The chapter ends with a presentation of an optimization algorithm. 

Throughout the second and third part we shall work with the field configuration
\be{
A(t) = \sum_i -\va_i \tau_i \tanh \br{\frac{t-t_{0,i}}{\tau_i}} ,\qquad
E(t) = \sum_i \va_i \mathrm{sech}^2 \br{\frac{t-t_{0,i}}{\tau_i}}.
}
The second part assesses different numerical methods, such that one can use the results presented as guideline data. 

In the final part all results are shown in a structured way. Firstly, we try to find new enhancement effects resulting in threshold lowering. To do that we give an account of our attempts to combine different time scales to obtain optimal pulse shapes. Second, we point out that one can observe interferences in the particle distribution depending on the diverse input parameters. Consequently we are able to perform calculations, which show the possibility for an indirect measurement of the Schwinger effect. To conclude, the results obtained from applying optimization procedures to our system are presented.   
\chapter{Quantum Kinetic Equation} 

Various methods have been developed to calculate the particle distribution function for the created electrons until today. 
In this chapter the quantum kinetic formalism is presented, because it has a number of attractive features. First and foremost it is highly flexible and easy to implement in computer programs.
Secondly, it is a fast way to obtain results for different setups in the parameter region we are interested in.

The derivation of the quantum kinetic equation follows closely the PhD-thesis of F. Hebenstreit \cite{HebDiss}. 
It can also be found in the publications \cite{hep-ph/9809227,HebDip,OrtDip}.
However, we do not employ their compact form, but write it down in a more elaborate way.  
Hence we will explicitly examine the derivation of the quantum Vlasov equation and subsequently an analytical solution for the Sauter field configuration is presented.
At the end of this chapter we shall also give an introduction to optimal control theory and its applications to pair production.

\section{Notation}

We start with introductory remarks on the notation for the purpose of a clear structure and better readability of the mathematical intense part of this thesis.

We use the common notation for indices. This means that for Latin letters $i,j = 1,2,3$ hold, while for Greek letters $\mu,\nu = 0,1,2,3$ is assumed. Moreover we use Einstein summation convention 
\be{
\de{\mu} \ga{\mu} = \sum_{\mu=0}^3 \de{\mu} \ga{\mu}
}
to improve readability. 

In the following we will define basis vectors and introduce the gamma matrices $\ga{\mu}$. We specify the unit vectors
\be{
\vec e_1 = \ma{1 \\ 0 \\ 0 \\ 0} ,\qquad \vec e_2 = \ma{0 \\ 1 \\ 0 \\ 0} ,\qquad \vec e_3 = \ma{0 \\ 0 \\ 1 \\ 0} ,\qquad \vec e_4 = \ma{0 \\ 0 \\ 0 \\ 1}
}
and denote the unit matrix as $\unit$. At certain points we write $\unit_d$ to denote the unit matrix in a specific dimension.
Moreover, we choose the Weyl or chiral basis \cite{Ryder1996} for gamma matrices:
\be{
\ga{0} = \ma{0 & \unit_2  \\ \unit_2 & 0} ,\qquad 
\ga{i} = \ma{0 & -\sigma^i \\ \sigma^i & 0} ,\qquad 
\ga{5} = \ii \ga{0} \ga{1} \ga{2} \ga{3} = \ma{\unit_2 & 0 \\ 0 & -\unit_2} }
with the Pauli matrices $\sigma^i$
\be{
\sigma^1 = \ma{0 & 1 \\ 1 & 0} ,\qquad
\sigma^2 = \ma{0 & -\ii \\ \ii & 0} ,\qquad
\sigma^3 = \ma{1 & 0 \\ 0 & -1}.}
The first four gamma matrices form a Clifford algebra, satisfying the important anti-commutator relation 
\be{
\acom{\ga{\mu}, \ga{\nu}} = 2g^{\mu \nu}
}
with the metric tensor 
\be{
g^{\mu \nu} = \textrm{diag} \br{1 ,-1, -1 ,-1}.
}
Accordingly, they are hermitian and anti-hermitian, respectively
\be{
\br{\ga{0}}^{\dag} =  \ga{0} ,\qquad
\br{\ga{i}}^{\dag} =  -\ga{i} ,\qquad
\br{\ga{5}}^{\dag} =  \ga{5}
}
and fulfill
\be{
\br{\ga{0}}^2 = \unit ,\qquad
\br{\ga{i}}^2 = -\unit ,\qquad
\br{\ga{5}}^2 = \unit.
}

We denote 
\be{
x^{\mu} = \br{x^0, x^1, x^2, x^3} ,\qquad
x^0 = t
}
and introduce the shorthand notation
\be{ 
\de{t} = \frac{\partial}{\partial{t}} ,\qquad
\de{\mu} = \frac{\partial}{\partial x^{\mu}} ,\qquad
\br{\de{t} \Psi} = \dot{\Psi}.
}


To distinguish scalars from vectors we write an extra $~\vec {}~$ above a vector.
For adjoint variables we use the notation $\bar{\Psi}$, which is equal to $\bar{\Psi} = \Psi^{\dag} \ga{0} = \br{\Psi^*}^T \ga{0}$, where ${}^T$ means transpose and ${}^*$ complex conjugate.

The preferred system of units in high-energy physics is the Heaviside Lorentz system. Additionally we use natural units and set $\hbar = c = 1$. In chapter $3$ we also set $m_e=1$ and express all variables in terms of the electron mass.
The electron charge $e$ in this system of units is given by $e=\sqrt{4 \pi \alpha} \sim 0.3028$, where $\alpha$ is the fine structure constant $\alpha \approx 1/137$. Last but not least the Maxwell equations take the form
\begin{subequations}
 \begin{align}
	\vnab \cdot \vec E &= \rho, \\
	\vnab \cdot \vec B &= 0, \\
	\vnab \times \vec E + \br{\de{t} \vec B} &= 0, \\
	\vnab \times \vec B - \br{\de{t} \vec E} &= \vec j.
 \end{align}
\end{subequations}


It is not possible to solve the problem of electron-positron pair creation in the framework of QED for any given field configuration exactly. As a consequence, we have to employ several approximations: 
In recent studies theorists concentrated on spatially homogeneous time-dependent fields and so do we. If we intended to abandon the assumption of spatial homogeneity, we would have to use a more
general approach such as the DHW-formalism \cite{PhysRevD.44.1825,HebDiss}. 
As it is our aim to describe processes occurring at field strengths of about $E_{cr}=1.3 \cdot 10^{18} V/m$ \cite{springerlink:10.1007/BF01343663} a mean field description of the electric field should be a good approximation. In further agreement with the literature we treat the electric field classically and quantize only the matter field.

Motivated by standing waves, we will consider only electric fields whereas magnetic fields are totally neglected
\be{
\vec B = \vnab \times \vec A = 0 ,\qquad
\vec E = - \vnab \phi - \frac{\partial \vec A}{\partial t},
}
with $\vec A$ the vector potential and $\phi$ the scalar potential.
Then we exploit the gauge freedom
\be{
A_{\mu} \to A_{\mu} + \de{\mu} \rho
}
to choose Weyl or temporal gauge $\phi = 0$ giving us
\be{
\vec A \rightarrow \vec A + \vnab \rho = \vec A ,\qquad
\phi \rightarrow \phi - \frac{\partial \rho}{\partial t} = 0.
}

We assume the electric field points into the same direction all the time.
Thus potential and electric field take on the form
\begin{subequations}
	\label{eq2_3:1}
	\begin{align}
	A \brt &= \br{0,\vec A \brt} = \br{0,0,0,A_z \br{t}} ,\\
	E \brt &= \br{0,\vec E \brt} = \br{0,0,0,E \br{t}}.
	\end{align}
\end{subequations}

\section{Quantum Vlasov Equation}

As we will see in the following, the description of the whole pair creation process is similar to characterizing a many-body problem by a Boltzmann equation
\be{
\de{t} F \qt = \mathcal{S} \qt + \mathcal{C} \qt. 
}
On the left-hand side the temporal change of the distribution of the particles is written. On the right-hand side we have both a source term $\mathcal{S} \qt$ for particle production and another one that describes the collisions $\mathcal{C} \qt$.

The derivation of the source term $~\mathcal{S} \qt~$ is our main goal in the next sections, while we will disregard the collision term. This neglect
can be justified by applying a simple analysis based on the relaxation time method \cite{PhysRevD.60.116011}.
The idea is to model the collision term as
\be{
\mathcal{C} \qt = \frac{F_{eq} \qt - F \qt}{\tau_r},
}
with $F_{eq} \qt$ the equilibrium distribution function for fermions.
The relaxation time $\tau_r$ is interpreted as the average time between two particle collisions. Due to the fact that the density of the created particles is sufficiently low for subcritical electric fields, one finds
$\tau_r \gg 1$ and, as a consequence, $\mathcal{C} \qt \ll \mathcal{S} \qt$. The dominance of the source term over the collision term legitimizes the neglect of collisions in the parameter region under consideration.



\section{Equation of Motion}

In this section we derive the quantum Vlasov equation. As we do not give an introduction to quantum field theory, we refer the interested reader for details to books on this subject \cite{Ryder1996,Itzykson2006,Maggiore2005,Peskin1995}.

A good starting point for the derivation is the QED Lagrangian
\be{
\mc{L} = \ii \bar {\Psi} \br{\ga{\mu} \de{\mu} - \ii e \ga{\mu} A_{\mu}} \Psi - m \bar{\Psi} \Psi - \frac{1}{4} F_{\mu \nu} F^{\mu \nu},
}
with charged spinor fields $\Psi \br{\vec x,t}$, $~\bar{\Psi} \br{\vec x,t} = \Psi^{\dag} \gamma^0$. The mass term $m$ explicitly denotes $m_e \unit$.
Our first objective is to find an equation of motion for the fields $\Psi \br{\vec x,t}$ and $\bar{\Psi} \br{\vec x,t}$. This is achieved by applying a variational principle to the Lagrangian. This gives the famous Euler-Lagrange equation of motion
\be{
\de{\mu} \br{\frac{\partial \mc{L}}{\partial \br{\de{\mu} \Psi}}} - \frac{\partial \mc{L}}{\partial \Psi} = 0.
\label{eq2_4:1}
}
For the adjoint spinor field this leads to
\be{
\de{\mu} \br{\ii \bar{\Psi} \ga{\mu}} - e \bar{\Psi} \ga{\mu} A_{\mu} + m \bar{\Psi} = 0. 
}
Complex conjugation yields for the spinor field 
\al{
\ii \ga{\mu} \de{\mu} \Psi + e \ga{\mu} A_{\mu} \Psi - m \Psi = 0 ,\\
\br{\ii \ga{0} \de{t} + \ii \vga \cdot \vnab + e \ga{\mu} A_{\mu} - m} \Psi = 0.
}
Now we use the specific form of the vector potential in \eqref{eq2_3:1} and obtain an equation of motion for $\Psi \br{\vec x,t}$
\be{
\br{\ii \ga{0} \de{t}+ \ii \vga \cdot \br{ \vnab - \ii e \vec A\br{t}} - m} \Psi \br{\vec x,t} = 0.
\label{eq2_4:3}
} 
The spinor fields $\Psi \br{\vec x,t}$ and $\bar{\Psi} \br{\vec x,t}$ are treated as dynamically independent fields. 
For the sake of completeness we can write down the canonical momenta
\be{
\Pi = \frac{\partial \mc{L}}{\partial \dot{\Psi}} = \ii \Psi^{\dag} 
\label{eq2_4:2}
}
and the equal time anti-commutation relation(ETAR)
\be{
\acom{\Psi \br{\vec x, t}, \Pi \br{\vec x',t}} = \ii \delta \br{\vec x - \vec x'}.
}
Because of the spatial homogeneity of the vector potential, it is indicated to decompose both the spinor field $\Psi \br{\vec x,t}$ and its conjugate field $\Pi \br{\vec x,t}$ into their Fourier modes 
\be{
\Psi \br{\vec x,t} = \int \frac{d^3q}{\br{2 \pi}^3} \ee^{\ii \vec q \cdot \vec x} \psi \qt ,\qquad
\Pi \br{\vec x,t} = \int \frac{d^3q}{\br{2 \pi}^3} \ee^{-\ii \vec q \cdot \vec x} \pi \qt.
} 
Similar to \eqref{eq2_4:2} we have for the mode functions $\pi \qt = \ii \psi^{\dag} \qt$.
The mode functions also fulfill an ETAR 
\be{
\acom{\psi \br{\vec q, t}, \pi \br{\vec q',t}} = \ii \br{2 \pi}^3 \delta \br{\vec q - \vec q'}.
} 
Inserting the Fourier expansion into \eqref{eq2_4:3} and using the relations
\begin{subequations}
\al{
\ii \ga{0} \br{\de{t} \Psi \br{\vec x,t}} &= \ii \ga{0} \int \frac{d^3q}{\br{2\pi}^3} \ee^{\ii \vec q \cdot \vec x} \br{\de{t} \psi \qt} ,\\
\ii \br{\vga \cdot \vnab_x} \Psi \br{\vec x,t} &= \ii \vga \int \frac{d^3q}{\br{2\pi}^3} \br{\ii \vec q} \ee^{\ii \vec q \cdot \vec x} \psi \qt,
}
\end{subequations}
we obtain
\be{
\int \frac{d^3q}{\br{2\pi}^3} \ee^{\ii \vec q \cdot \vec x} \br{\ii \ga{0} \br{\de{t} \psi \qt} - \vga \cdot \br{\vec q - e \vec A (t)} \psi \qt - m \psi \qt} = 0.
}
This equation is not only true for the whole integration but for every integrand. Hence it is sufficient to work with
\be{
\br{\ii \ga{0} \de{t} - \vga \cdot \br{\vec q - e \vec A \brt} - m } \psi \qt = 0.
}
Introducing the kinetic momentum $\vpi \qt = \vec q - e \vec A \brt$ and especially $\pi_3 \qt = q_3 - e A_z \brt$ the equation of motion for the Fourier modes reads
\be{
\br{\ii \ga{0} \de{t} - \vga \cdot \vpi \qt - m} \psi \qt = 0. 
\label{eq2_4:4}
}
Apparently this is an ordinary differential equation of first order. In order to solve it, we introduce a quasi-particle representation in the following.

\subsection{Quasi-Particle Representation}

The first step in transforming equation \eqref{eq2_4:4} is to introduce an ansatz for the mode function
\be{
\psi \qt = \br{\ii \ga{0} \de{t} - \vga \cdot \vpi \qt + m} \Phi \qt.
}
The equation of motion then takes the form
\be{
\br{\ii \ga{0} \de{t} - \vga \cdot \vpi \qt - m} \br{\ii \ga{0} \de{t} - \vga \cdot \vpi \qt + m} \Phi \qt = 0.
}
Explicit calculation of the left hand side gives
\begin{multline}
\left( -\unit \de{t}^2 - \ii \vga \cdot \vpi \ga{0} \de{t} - \ii m \ga{0} \de{t} - \ii \ga{0} \de{t} \br{\vga \cdot \vpi} \right. \\
\left. + \br{\vga \cdot \vpi}^2 + m \vga \cdot \vpi + \ii m \ga{0} \de{t} - m \vga \cdot \vpi - m^2  \right) \Phi \qt = 0. 
\label{eq2_5:5}
\end{multline}
We see that some of the terms cancel each other upon using the relations 
\al{
\de{t} \br{\vga \cdot \vpi \qt} = \ga{3} \de{t} \br{q_3 - e A_z \brt} = e \ga{3} E \brt ,
} 
\al{
\br{\vga \cdot \vpi \qt}^2 = -\unit \pi_1^2 \brq - \unit \pi_2^2 \brq - \unit \pi_3^2 \qt = - \unit \vpi^2 \qt.
}
Together with\footnote[1]{
$
-\ii \ga{0} \de{t} \br{\vga \cdot \vpi \qt} \Phi \qt 
  = -\ii \ga{0} \br{\vga \cdot \vpi \qt} \br{\de{t} \Phi \qt} -\ii \ga{0} \br{\de{t} \br{\vga \cdot \vpi \qt}} \Phi \qt \\ 
  ~~~~~~ = \ii \br{\vga \cdot \vpi \qt} \ga{0} \br{\de{t} \Phi \qt} + \ii e \ga{0} \ga{3} E \brt \Phi \qt
$
}
\be{
-\ii \ga{0} \de{t} \br{\vga \cdot \vpi \qt} \Phi \qt 
 = \ii \br{\vga \cdot \vpi \qt} \ga{0} \br{\de{t} \Phi \qt} + \ii e \ga{0} E \brt \Phi \qt
}
equation \eqref{eq2_5:5} is simplified to
\be{
\br{\unit \de{t}^2 +\ii e \ga{0} \ga{3} E \brt + \unit \vpi^2 \qt + m^2} \Phi \qt = 0.
}
Introducing the notation $\omega^2 \qt = \vpi^2 \qt + m_e^2$ this equation is written as 
\be{
\br{\unit \de{t}^2 + \ii e \ga{0} \ga{3} E \brt + \unit \omega^2 \qt} \Phi \qt = 0, 
\label{eq2_5:1}
}
which is formal equivalent to an oscillator-type equation with an additional time-dependent term. 
Now we expand the function $\Phi \qt$ in a basis of eigenvectors of $\ga{0} \ga{3}$.
The matrix representation of $\ga{0} \ga{3}$ is 
\be{
\ga{0} \ga{3} = \ma{0 & \unit \\ \unit & 0} \ma{0 & -\sigma^3 \\ \sigma^3 & 0} = \ma{\sigma^3 & 0 \\ 0 & -\sigma^3},
}
which makes it easy to determine the eigenvectors. They are given by
\be{
R_1 = \ma{1 \\ 0 \\ 0 \\ 0} ,\qquad
R_2 = \ma{0 \\ 0 \\ 0 \\ 1} ,\qquad
R_3 = \ma{0 \\ 1 \\ 0 \\ 0} ,\qquad
R_4 = \ma{0 \\ 0 \\ 1 \\ 0} ,\qquad
}
where $R_1$,$~R_2$ have eigenvalue $\lambda = 1$ and $R_3$, $R_4$ have eigenvalue $\lambda = -1$.
This gives us the possibility to expand $\Phi \qt$ in the basis of the eigenvectors 
\be{
\Phi \qt = \sum_{r=1}^4 \chi_r \qt R_r
}
with $\chi_r \qt$ as time-dependent coefficients\footnote[2]{We will refer to $\chi_r \qt$ as mode functions from this point on.}. Inserting this expansion into \eqref{eq2_5:1} leads to
\begin{multline}
\br{\unit \de{t}^2 + \unit \omega^2 \qt + \ii e \ga{0} \ga{3} E\brt} \sum_{r=1}^4 \chi_r \qt R_r \\ 
  = \br{\unit \de{t}^2 + \unit \omega^2 \qt} \sum_{r=1}^4 \chi_r \qt R_r + \ii e E\brt \sum_{r=1}^4 \chi_r \qt \underbrace{\br{\ga{0} \ga{3} R_r}}_{(*)}.
\end{multline}
The term $(*)$ is an eigenvalue equation yielding $R_{1,2}$ and $-R_{3,4}$, respectively.
We obtain one equation for every component, which leads to two different scalar equations
\begin{subequations}
	\begin{align}
	\label{eq2_5:4a}
	\br{\de{t}^2 + \omega^2 \qt + \ii e E\brt} \chi_r \qt = 0 \qquad r=1,2, \\
	\br{\de{t}^2 + \omega^2 \qt - \ii e E\brt} \chi_r \qt = 0 \qquad r=3,4.  
	\label{eq2_5:4}
	\end{align}
\end{subequations}
Each of these equations appears twice. Note that the Dirac equation \eqref{eq2_4:4} is a first order differential equation having four linear independent solutions. On the other hand,
\eqref{eq2_5:1} is a second order differential equation such that we obtain eight linear independent solutions in total. However, this redundancy can be removed by choosing only one complete
set of solutions corresponding to either \eqref{eq2_5:4a} or \eqref{eq2_5:4}.

Hence we decided to skip the second equation. This leaves us with one differential equation of second order with fundamental system $\chi_r^{\pm} \qt$ and 
coefficients $c_r^+ \brq,~ c_r^- \brq$ where $r=1,2$. 
Now we have four different components and can write 
\be{
\Phi \qt = \sum_{r=1}^2 \br{c_r^+ \brq \chi_r^+ \qt R_r + c_r^- \brq \chi_r^- \qt R_r}.
}
Note that the summation goes from one to two and not up to four anymore. Using the equation above one obtains
\begin{multline}
\psi \qt = \br{\ii \ga{0} \de{t} - \vga \cdot \vpi \qt + m} \Phi \qt \\
 = \sum_{r=1}^2 \br{\ii \ga{0} \de{t} - \vga \cdot \vpi \qt + m} \br{c_r^+ \brq \chi_r^+ \qt R_r + c_r^- \brq \chi_r^- \qt R_r} 
\end{multline}
and thus the spinor field takes the form
\be{
\psi \qt  
 = \sum_{r=1}^2 \br{u_r \qt c_r^+ \brq + v_r \mqt c_r^- \brq}
}
with
\begin{subequations}
	\begin{align}
	u_r \qt = \br{\ii \ga{0} \de{t} - \vga \cdot \vpi \qt + m} \br{\chi_r^+ \qt R_r} ,\\
	v_r \mqt = \br{\ii \ga{0} \de{t} - \vga \cdot \vpi \qt + m} \br{\chi_r^- \qt R_r}. 
	\end{align}
\end{subequations}
Moreover, we redefine the coefficients 
\be{
a_r \brq = c_r^+ \brq ,\qquad
b_r^{\dag} \brmq = c_r^- \brq ,\qquad
}
such that we can write
\be{
\psi \qt = \sum_{r=1}^2 \br{u_r \qt a_r \brq + v_r \mqt b_r^{\dag} \brmq}.
}
In the equation above $u_r \qt$ and $v_r \mqt$ are time-dependent spinors, whereas $a_r \brq$ and $b_r^{\dag} \brmq$ can be interpreted as annihilation and creation operators.
These creation/annihilation operators are fermionic operators and thus the following ETAR hold
\begin{subequations}
\al{
\acom{a_r \brq, a_s^{\dag} \brk } &= \acom{b_r \brq, b_s^{\dag} \brk} = \br{2 \pi}^3 \delta_{rs} \delta \br{\vec q - \vec {q'}} , \label{eq2_5:3b}\\
\acom{a_r \brq, a_s \brk} &= \acom{b_r \brq, b_s \brk} = 0 ,\label{eq2_5:3} \\
\acom{a_r^{\dag} \brq, a_s^{\dag} \brk} &= \acom{b_r^{\dag} \brq, b_s^{\dag} \brk} = 0.
} 
\end{subequations}
Investigating the Hamiltonian of the system \cite{OrtDip} we observe non-vanishing off diagonal elements, which account for particle creation and annihilation. In order to calculate the spectrum, we have to diagonalize the Hamiltonian. This is achieved by a basis transformation, where we introduce new time-dependent operators $A_r \qt$ and $B_r \mqt$, which also show fermionic behaviour 
\begin{subequations}
\al{ 
\acom{A_r \qt, A_s^{\dag} \kt} &= \acom{B_r \qt, B_s^{\dag} \kt} = \br{2 \pi}^3  \delta_{rs} \delta \br{\vec q - \vec {q'}} , \\
\acom{A_r \qt, A_s \kt} &= \acom{B_r \qt, B_s \kt} = 0 ,\\
\acom{A_r^{\dag} \qt, A_s^{\dag} \kt} &= \acom{B_r^{\dag} \qt, B_s^{\dag} \kt} = 0. 
}
\end{subequations} 
The relation between the operators $a_r \brq ,~ b_r^{\dag} \brq$ and $A_r \qt ,~ B_r^{\dag} \mqt$ is given by a Bogoliubov transformation
\begin{subequations}
	\begin{align}
	\label{eq2_5:2b}
	A_r\qt &= \alpha \qt a_r \brq - \beta^* \qt b_r^{\dag} \brmq , \\
	\label{eq2_5:2}
	B_r^{\dag} \mqt &= \beta \qt a_r \brq + \alpha^* \qt b_r^{\dag} \brmq.  
	\end{align}
\end{subequations}
Note that the coefficients $\alpha \qt$ and $\beta \qt$ in the equation above satisfy the relation 
\be{
\bet{\alpha \qt}^2 + \bet{\beta \qt}^2 = 1.
} 
We will refer to the new basis as ``quasi-particle representation'' and to the old one as ``particle representation'' from here on. This notion is introduced, because for vanishing electric fields the operators $a \brq$ and $b^{\dag} \brq$ correspond to single particles, in the sense of asymptotic states. On the other hand, the operators $A \qt$ and $B \mqt$ are mixtures of the original operators $a \brq$ and $b^{\dag} \brq$. However, we call them quasi-particle operators due to the fact that the quanta corresponding to these operators behave like particles with dispersion relation $\omega \qt$.

As the Bogoliubov transformation gives us the desired change of basis, we may express $\psi \qt$ in terms of new operators as well
\be{
\psi \qt = \sum_{r=1}^2 \br{U_r \qt A_r \qt + V_r \mqt B_r^{\dag} \mqt}.
}
The spinors in the quasi-particle basis $U_r \qt,~ V_r \mqt$ are denoted by upper case letters to distinguish them from the spinors $u_r \qt$ and $v_r \mqt$. They are given by
\begin{subequations}
	\begin{align}
	U_r \qt &= \br{\ga{0} \omega \qt - \vga \cdot \vpi \qt + m} \br{\kappa_r^+ \qt R_r} ,\\
	V_r \mqt &= \br{-\ga{0} \omega \qt - \vga \cdot \vpi \qt + m} \br{\kappa_r^- \qt R_r}.  
	\end{align}
\end{subequations}
The functions $~\kappa_r^{\pm}~$ are the mode functions in the quasi-particle representation, which are chosen according to the ansatz
\begin{subequations}
	\begin{align}
	\kappa^+ \qt = \frac{\ee^{-\ii \theta \qt}}{\sqrt{2\omega \qt \br{\omega \qt - \pi_3 \qt}}} ,\\
	\kappa^- \qt = \frac{\ee^{\ii \theta \qt}}{\sqrt{2\omega \qt \br{\omega \qt + \pi_3 \qt}}} ,
	\end{align}
\end{subequations}
with the dynamical phase $\theta \qt = \int_{t_0}^t \omega \br{\vec q,\tau} d\tau$. Note that $\omega \br{\vec q,\tau}$ is time-dependent due to its dependence on the vector potential $A \brt$.

The two functions $\kappa_r^{\pm}$ are chosen such that they coincide with the mode functions $\chi_0^{\pm}$ in the case of a vanishing vector potential. The calculation of $\chi_0^{\pm}$, which is done in \ref{append:normMode} yields
\be{
\chi_0^{\pm} \qt = \frac{1}{\sqrt{2 \omega \brq \br{\omega \brq {\mp} q_3}}} \ee^{\mp \ii \omega \brq t}.
}
In the case of vanishing electric field and vector potential the following simplification is obtained
\be{
\omega \qt \rightarrow \omega \brq = m_e^2 + \vec q^2. 
} 
Hence the integration in the dynamical phase $\theta \qt$ yields
\be{ 
\theta \qt = \int_{t_0}^t \omega \br{\vec q,\tau} d\tau \rightarrow \omega \br{\vec q} \br{t - t_0}.
}
We skip an arbitrary phase $\ee^{\mp \ii \omega t_0}$ and obtain
\be{
\kappa_0^{\pm} = \chi_0^{\pm} \qt = \frac{\ee^{\mp \ii \omega \brq t}}{\sqrt{2 \omega \brq \br{\omega \brq \mp q_3}}},
}
proofing that our ansatz for $\kappa^{\pm} \qt$ is valid. 

To summarize, we can express the spinor $\psi \qt$ in both the particle and in the quasi-particle representation
\al{
\psi \qt =& \sum_{r=1}^2 \br{u_r \qt a_r \brq + v_r \mqt b_r^{\dag} \brmq} \\
 =& \sum_{r=1}^2 \br{U_r \qt A_r \qt + V_r \mqt B_r^{\dag} \mqt}.
}
In order to fully specify the Bogoliubov transformation, we still have to derive an equation for $\alpha \qt$ and $\beta \qt$.
However a calculation of these two coefficients is rather challenging but done explicitly in \ref{append:bogoliubov}. Here it is sufficient to know that these coefficients are given by
\begin{subequations}
	\begin{align}
	\alpha \qt &= \ii \sqrt{\omega^2 \qt - \pi_3^2 \qt} \kappa^- \qt \br{\de{t} - \ii \omega \qt} \chi^+ \qt ,\\
	\beta \qt &= -\ii \sqrt{\omega^2 \qt - \pi_3^2 \qt} \kappa^+ \qt \br{\de{t} + \ii \omega \qt} \chi^+ \qt.
	\end{align}
\end{subequations}

An important aspect of the mode functions in the two different representations is their similar behaviour for vanishing potentials and fields at asymptotic times $t \to -\infty$.
In this limit, both types of mode functions $\kappa^{\pm} \qt$ and $\chi^{\pm} \qt$ coincide with the free mode function $\chi_0^{\pm} \qt$ up
to an arbitrary phase. To check this we may calculate the Bogoliubov coefficients in this limit and find 
\begin{multline}
\alpha\br{\vec q, t \to -\infty} = \ii \sqrt{\omega^2 - q_3^2} \chi_0^- \br{\de{t} - \ii \omega} \chip \\
  = \ii \sqrt{\omega^2 - q_3^2} \br{-2\ii \omega} \frac{1}{2\omega \sqrt{\omega^2 - q_3^2}} = 1 
\end{multline}
and 
\be{
\bet{\beta \br{\vec q, t \to -\infty}}^2 = 1 - \bet{\alpha \br{\vec q, t \to -\infty}}^2 = 0.
}
This shows that the two basis indeed coincide at asymptotic times $t \to -\infty$.



\subsection{Deriving the Source Term}

This is now the final part of the derivation, namely the calculation of the equation of motion for the one-particle distribution function $F \qt$. This function gives the quasi-particle number density for any canonical momentum $\vec q$. 
It is a time-dependent quantity and defined as the expectation value of the quasi-particle number operator\footnote{ 
In other words the one-particle distribution function holds as an expectation value of the quasi-particles in our system divided by its volume. This is due to the spatial homogeneity of the system.}:
\be{
F \qt = \lim_{V \to \infty} \frac{1}{V} \sum_{r=1}^2 \obs{ A_r^{\dag} \qt A_r \qt }.
}
We start with the definition of the particle number density in the time-independent basis
\be{
N_r \brq = \lim_{V \to \infty} \frac{1}{V} \obs{a_r^{\dag} \brq a_r \brq} = \lim_{V \to \infty} \frac{1}{V} \obs{b_r^{\dag} \brmq b_r \brmq}. 
}
Using the operators $A_r \qt$ and $A_r^{\dag} \qt$, which have been defined in \eqref{eq2_5:2b}  
\begin{subequations}
	\begin{align}
		A_r \qt = \alpha \qt a_r \brq - \beta^* \qt b_r^{\dag} \brmq ,\\
		A_r^{\dag} \qt = \alpha^* \qt a_r^{\dag} \brq - \beta \qt b_r \brmq.
	\end{align}
\end{subequations}
the computation of their product gives
\be{
A_r^{\dag} A_r = \bet{\alpha}^2 a_r^{\dag} a_r - \beta \alpha b_r a_r - \alpha^* \beta^* a_r^{\dag} b_r^{\dag} + \bet{\beta}^2 b_r b_r^{\dag}.
}
Correspondingly the expectation value is given by
\be{
\obs{A_r^{\dag} A_r} = \bet{\alpha}^2 \obs{a_r^{\dag} a_r} + \bet{\beta}^2 \obs{b_r b_r^{\dag}}.
}
As the order of the operators is important we have to rearrange the terms in the equation above. For that reason we use the anti-commutator relation defined in \eqref{eq2_5:3b} to get
\be{
\obs{b_r b_r^{\dag}} = \obs{b_r b_r^{\dag} + b_r^{\dag} b_r - b_r^{\dag} b_r} = \obs{\acom{b_r, b_r^{\dag}}} - \obs{b_r^{\dag} b_r} = V - \obs{b_r^{\dag} b_r}.
}
We proceed with the calculation
\begin{multline}
\frac{1}{V} \obs{A_r^{\dag} A_r} = \frac{\bet{\alpha}^2}{V} \obs{a_r^{\dag} a_r} + \frac{\bet{\beta}^2}{V} \obs{b_r b_r^{\dag}} 
  = \bet{\alpha}^2 N_r + \frac{\bet{\beta}^2}{V} \br{V - \obs{b_r^{\dag} b_r}} \\
  = \br{1 - \bet{\beta}^2} N_r + \bet{\beta}^2 - \frac{\bet{\beta}^2}{V} \obs{b_r^{\dag} b_r} 
  = N_r + \bet{\beta}^2 \br{1 - 2 N_r},
\end{multline}
leading to the distribution function
\be{
F \qt = \sum_{r=1}^2 \br{ N_r \brq + \bet{\beta \qt}^2 \br{1 - 2 N_r \brq} }.
}

The time-independent quantity $N_r \brq$ specifies the initial conditions of our system. We choose $N_r \brq = 0$ to get pure vacuum for $t \to -\infty$. Thus the one-particle distribution function takes the form
\be{
F \qt = \sum_{r=1}^2 \bet{\beta \qt}^2 = 2 \bet{\beta \qt}^2.
\label{eq2_6:3}
}
Another useful quantity is the quasi-particle correlation function
\be {
C \qt = \lim_{V \to \infty} \frac{1}{V} \sum_{r=1}^2 \obs{ B_r^{\dag} \br{-\vec q,t} A_r^{\dag} \qt}.
}
Again we reformulate this equation to express $C \qt$ in terms of the coefficients $\alpha \qt$ and $\beta \qt$. This time we work with the operators
\begin{subequations}
	\begin{align}
		A_r^{\dag} \qt &= \alpha^* \qt a_r^{\dag} \brq - \beta \qt b_r \brmq ,\\
		B_r^{\dag} \br{-\vec q,t} &= \beta \qt a_r \brq + \alpha^* \qt b_r^{\dag} \brmq
	\end{align}
\end{subequations}
and multiply them to obtain
\be{
B_r^{\dag} A_r^{\dag} = \alpha^* \beta a_r a_r^{\dag} + \alpha^* \alpha^* b_r^{\dag} a_r^{\dag} - \beta \beta a_r b_r - \alpha^* \beta b_r^{\dag} b_r.
}
Then we take the expectation value
\be{
\obs{B_r^{\dag} A_r^{\dag}} = \beta \alpha^* \obs{a_r a_r^{\dag}} - \alpha^* \beta \obs{b_r^{\dag} b_r}.
}
Reordering of the operators and division by the volume gives
\be{
\frac{1}{V} \obs{B_r^{\dag} A_r^{\dag}} = \beta \alpha^* \br{1 - \frac{1}{V} \obs{a_r^{\dag} a_r}} - \frac{\alpha^* \beta}{V} \obs{b_r^{\dag} b_r}.
}
Taking the limit on both sides of the equation results in
\be{
\lim_{V \to \infty} \frac{1}{V} \obs{B_r^{\dag} A_r^{\dag}} = \beta \alpha^* - N_r - \alpha^* \beta N_r.
}
Finally we use the initial condition $N_r \brq = 0$ again, which leads to
\begin{multline}
C \qt = \lim_{V \to \infty} \frac{1}{V} \sum_{r=1}^2 \obs{ B_r^{\dag} \br{-\vec q,t} A_r^{\dag} \qt} \\
 = \sum_{r=1}^2 \beta \qt \alpha^* \qt = 2 \alpha^* \qt \beta \qt.
\end{multline}

A formulation of $F \qt ,~ C \qt$ in terms of $\alpha \qt$ and $\beta \qt$ is not very advantageous as one would have to know the exact mode functions $\chi^+ \qt$. Therefore, we try to obtain an equation for $F \qt$, which is easier to compute. After another tedious calculation, performed in \ref{append:dot_bogoliubov}, we obtain
\begin{subequations}
	\begin{align}
		\dot{\alpha} \qt &= \frac{1}{2} W \qt \beta \qt \ee^{2\ii \theta \qt} ,\\
		\dot{\beta} \qt &= -\frac{1}{2} W \qt \alpha \qt \ee^{-2\ii \theta \qt},
	\end{align}
\end{subequations}
with 
\al{
W \qt = \frac{e E \brt \enorm \brq}{\omega^2 \qt}, \qquad \omega^2 \qt = m_e^2 + \vpi^2 \qt ,\\
\enorm^2 \brq = m_e^2 + q_1^2 + q_2^2.
} 

Taking into account these equations, we find\footnote[1]{
$
\dot{C} = 2\br{\dot{\alpha}^* \beta + \alpha^* \dot{\beta}} = 
	2\br{\frac{1}{2} W \beta^* \ee^{-2\ii \theta} \beta - \frac{1}{2} \alpha^* W \alpha \ee^{-2\ii \theta}} = -W \br{\alpha^* \alpha - \beta^* \beta} \ee^{-2\ii \theta} \\
  ~~~~~~~~~~~~= -W \br{\bet{\alpha}^2 - \bet{\beta}^2} \ee^{-2\ii \theta} 
  = -W \br{1-2\bet{\beta}^2} \ee^{-2 \ii \theta} 
  = -W \br{1-F} \ee^{-2\ii \theta}
$
} 
\be{
\dot{C} \qt = -W \qt \br{1-F \qt} \ee^{-2\ii \theta \qt}
}
and\footnote[2]{
$
\dot{F} = 2 \de{t} \bet{\beta}^2 = 4 \real{\beta \dot{\beta}^*} = 4\real{-\frac{1}{2} \beta W \alpha^* \ee^{2\ii \theta}} 
  = -W \real{2\alpha^* \beta \ee^{2\ii \theta}} 
  = -W \real{C \ee^{2 \ii \theta}}
$
}
\be{
\dot{F} \qt = -W \qt \real{C \qt \ee^{2 \ii \theta \qt}}.
\label{eq2_6:1}
}
Integration of $\dot{C} \qt$ with respect to $t$ gives an expression we can insert into \eqref{eq2_6:1}
\begin{multline}
\dot{F} \qt = -W \qt \real{C \qt \ee^{2\ii \theta \qt}} \\
  = W \qt \real{\int dt' W \qts \br{1-F \qts} \ee^{2\ii \br{\theta \qt - \theta \qts}}}.
\end{multline}
Introducing the shorthand notation
\be{
\theta \qt - \theta \qts = \int_{t'}^t \omega \br{\vec q, t''} dt'' = \thetaT.
}
we finally arrive at
\be{
\dot{F} \qt = W \qt \IVac W \qts \br{1- F \qts} \tcos{2 \thetaT}.
\label{eq2_6:2}
}
Given the one-particle distribution function, the asymptotic particle number per unit volume can be calculated according to
\be{
N = \int_{-\infty}^{\infty} d^3q F \br{\vec q, \infty}.
}

Equation \eqref{eq2_6:2} is an integro-differential equation, where the right hand side acts as a source term for the system. The factor $\br{1- F \qts}$ under the integral gives evidence for non-Markovian behaviour. Furthermore, this term represents Pauli exclusion principle, thus we refer to this term as Pauli-blocking factor. The second interesting part is the $\tcos{2 \thetaT}$, which is an indicator for a highly non-local problem in time.

We have not taken into account the electric fields, which are created by the produced particles, so far. This disregard is based upon previous investigations \cite{PhysRevD.60.116011,HebDip}, 
where it has been shown, that those internal fields only play a role for critical field strengths or high particle densities. However, their contribution
to the final particle density is rather low even in these regimes.
Therefore we decided to entirely neglect this back reaction effect in this thesis. 

\subsection{Differential Equation}

In order to solve the integro-differential equation \eqref{eq2_6:2}
we have to calculate the occurring integrals as exact as possible. In this respect the highly oscillating term 
$\tcos{2 \thetaT}$ is numerically challenging. Moreover, due to the
non-Markovian behaviour, the complete time-history of the one-particle distribution function has to be taken into account, which is potentially error-prone. To avoid these complications we rewrite this equation as a coupled differential equation as proposed by J. C. R. Bloch et al. \cite{PhysRevD.60.116011}. In close analogy to this paper we introduce the auxiliary functions
\begin{subequations}
	\begin{align}
	  G \qt &= \IVac W \qts \br{1 - F \qts} \tcos{2\thetaT} ,\\ 
	  H \qt &= \IVac W \qts \br{1 - F \qts} \tsin{2\thetaT}, 
	\end{align}
\label{eq3_1:3}
\end{subequations}
where
\al{
W \qt = \frac{e E \brt \enorm \brq}{\omega^2 \qt} ,\qquad \enorm^2 \brq = m_e^2 + q_1^2 + q_2^2 ,\\
\thetaT = \int_{t'}^t \omega \br{\vec q, t''} dt'' ,\qquad \omega^2 \qt = m_e^2 + \br{\vec q - e \vec A \brt}^2.
}
The derivatives of the auxiliary functions with respect to $t$ yield
\begin{subequations}
	\begin{align}
	\label{eq3_1:3c}
	\dot{G} \qt &= W \qt \br{1 - F \qt} - 2 \omega \qt H \qt ,\\
	\label{eq3_1:3b}
	\dot{H} \qt &= 2 \omega \qt G \qt ,  
	\end{align}
\end{subequations}
where we have used
\be{
\dot{\theta} \br{\vec q, t, t'} = \de{t} \br{\int_{t'}^t \omega \br{\vec q, t''} dt''} = \omega \qt.
}
Combining the equations \eqref{eq3_1:3b}\eqref{eq3_1:3c} with \eqref{eq2_6:2} gives us now the possibility to work with a coupled differential equation of first order
\al{
\dot{F} \qt &= W \qt \cdot G \qt ,\\
\dot{G} \qt &= W \qt \br{1- F \qt} - 2 \omega \qt \cdot H \qt ,\\
\dot{H} \qt &= 2 \omega \qt G \qt.
\label{eq3_1:2}
}
Its matrix form is
\be{
\ma{\dot{F} \\ \dot{G} \\ \dot{H}} = \ma{0 & W & 0 \\ -W & 0 & -2\omega \\ 0 & 2\omega & 0} \ma{F \\ G \\ H} + \ma{0 \\ W \\ 0},
}
which indicates a special structure of the system. Initial conditions are chosen such that one starts with a pure vacuum at asymptotic time $t \to -\infty$
\be{
F \br{\vec q, t \to -\infty} = G \br{\vec q, t \to -\infty} = H \br{\vec q, t \to -\infty} = 0.
}

\section{Analytical Solutions}

The relationship between time-dependent electric fields and the particle number density has been widely investigated in the last decade.
One of the most-studied field configurations is the pulsed Sauter field
\be{
A \brt = -\va E_{cr} \tau \Tanh ,\qquad
E \brt = \va E_{cr} \textrm{sech}^2 \ttau,
}
as it allows for an analytical solution.
As nearly all of our investigations in chapter 3 and chapter 4 will be based on combinations of such simple pulses, we give the complete calculation of the one-particle distribution function $F \qt$ for a single pulse in this section.

First we set $m_e=1$ and introduce the notation
\al{
\tilde A \brt &= eA \brt = - \va \tau \Tanh ,\\
\tilde E \brt &= eE \brt = \va \textrm{sech}^2 \ttau = \va \br{1 - \textrm{tanh}^2 \ttau},
}
For convenience we redefine $A \brt = \tilde A \brt$ and $E \brt = \tilde E \brt$ to keep the notation of the variables simple. This redefinition holds for the rest of this section, giving
\be{
\pi_3^2 \qt = \br{q_3 - A_z \brt}^2 ,\qquad
\omega^2 \qt = 1 + q_1^2 + q_2^2 + \pi_3^2 \qt.
}
Consequently we obtain for the equation of motion \eqref{eq2_5:4a}
\be{
\br{\de{t}^2 + \omega^2 \qt + \ii \va \textrm{sech}^2 \ttau} \chi \qt = 0,
}
where we have skipped the irrelevant index of the mode function $\chi \qt$.
For convenience, we transform the time variable
\be{
t \mapsto u=\frac{1}{2} \br{1 + \Tanh},  \qquad [-\infty, \infty] \to [0,1]
} 
which leads to 
\be{
A \br{u} = -\va \tau \br{2u-1} ,\qquad
E \br{u} = 4 u \va \br{1-u}. 
}
Additionally, we calculate
\al{
t = \tau \textrm{arctanh} \br{2u-1} ,
}
\be{
u \br{1-u} = \frac{1}{4} \br{1 - \textrm{tanh}^2 \ttau}
  = \frac{1}{4} \textrm{sech}^2 \ttau,
}
\al{
\de{t} = \br{\de{t} u} \de{u} = \frac{1}{2\tau} \textrm{sech}^2 \ttau \de{u} = \frac{2}{\tau} u \br{1-u} \de{u} ,
}
\al{
\de{t}^2 = \frac{4}{\tau^2} u \br{1-u} \de{u} u \br{1-u} \de{u}.
}
Expressing the equation of motion in terms of the new time variable $u$ yields
\be{
\br{\frac{4}{\tau^2} u \br{1-u} \de{u} u \br{1-u} \de{u} + \omega^2 \qu + 4 \ii \va u \br{1-u}} \chi \qu = 0.
}
For convenience we divide by $~\frac{4}{\tau^2}~$ and obtain
\be{
\br{u \br{1-u} \de{u} u \br{1-u} \de{u} + \frac{\tau^2}{4} \omega^2 \qu + \ii \va \tau^2 u \br{1-u}} \chi \qu = 0.
\label{eq2_7:1}
}
In the next step we use an ansatz for the mode function $\chi \qu$:
\be{
\chi \qu = \uu \muu \eta \qu
\label{eq2_7:9}
}
with independent parameters $\alpha$ and $\beta$\footnote[1]{
As $\alpha ,~ \beta$ are constant they can be easily distinguished from the Bogoliubov coefficients.
}
. Inserting this ansatz into \eqref{eq2_7:1} gives after a tedious calculation, which 
is carried out in \ref{append:hyper}, the differential equation 
\begin{alignat}{2}
	  \uua{+2} \muub{+2} &\times \de{u}\br{\de{u}\eta \qu} \notag \\
	  + \left( 2 \alpha \uua{+1} \muub{+2} - 2\beta \uua{+2} \muub{+1} \right. & \notag \\
	  \left. + \uua{+1} \muub{+2} - \uua{+2} \muub{+1} \right) & \times \br{\de{u} \eta \qu} \notag \\
	  + \left( \alpha \br{\alpha-1} \uu \muub{+2} \right. & \notag \\
	  + \beta \br{\beta-1} \uua{+2} \muu - 2\alpha \beta \uua{+1} \muub{+1} & \notag \\
	  + \alpha \uu \muub{+2} - \beta \uua{+1} \muub{+1} & \notag \\
	  \left. - \alpha \uua{+1} \muub{+1}  + \beta \uua{+2} \muu \right) & \times \eta \qu \notag \\
	  + \frac{\tau^2}{4} \left( 1 + q_1^2 + q_2^2 + q_3^2 + 2q_3 \va \tau u - 2q_3 \va \tau \br{1-u} + \va^2 \tau^2 u \right. & \notag \\
	  \left. + \va^2 \tau^2 \br{1-u}^2 - 2\va^2 \tau^2 u \br{1-u}  +4\ii \va u \br{1-u} \right) \uu \muu & \times \eta \qu \label{eq2_7:2a} \\
	  & = 0. \notag 
\end{alignat}

We shall simplify this lengthy equation by rewriting it as a hypergeometric differential equation \cite{Mathematics1964}\footnote[1]{
The hypergeometric differential equation is a special form of Riemann's differential equation. 
}
\be{
\br{u\br{1-u} \de{u}^2 + \br{c -\br{a+b+1}u} \de{u} -ab} \eta \qu = 0.
\label{eq2_7:10}
}
To this end we make an ansatz for $a,~ b$ and $c$
\al{
c &= 1-\ii \tau \omega_0 , \notag \\
b &= 1+\ii \va \tau^2 - \frac{\ii \tau \omega_0}{2} + \frac{\ii \tau \omega_1}{2} ,\label{eq2_7:3} \\
a &= -\ii \va \tau^2 - \frac{\ii \tau \omega_0}{2} + \frac{\ii \tau \omega_1}{2}, \notag
}
where we have used
\begin{subequations}
	\begin{align}
		\omega_0^2 = \omega^2 \br{u=0}=1 + q_1^2 + q_2^2 + \br{q_3 - \va \tau}^2 ,\\
		\omega_1^2 = \omega^2 \br{u=1} = 1 + q_1^2 + q_2^2 + \br{q_3 + \va \tau}^2.
	\end{align}
\end{subequations}
Using these definitions, the prefactors of $\de{\mu} \eta$ and $\eta$ are given by
\begin{alignat}{3}
\br{\de{u} \eta \qu}:& & \notag \\
&	&\br{c - \br{1+b+a} u} \notag \\ 
&	&= \br{1 - \ii \tau \omega_0 - 2u + \ii \tau \omega_0u - \ii \tau \omega_1 u}
\label{eq2_7:4}
\end{alignat}
and
\begin{alignat}{3}
\eta \qu:& & \notag \\
& &-ab = -\va^2 \tau^4 + \ii \va \tau^2 + \frac{\tau^2 \omega_0^2}{4} + \frac{\tau^2 \omega_1^2}{4} - \frac{\tau^2 \omega_0 \omega_1}{2} + \frac{\ii \tau \omega_0}{2} - \frac{\ii \tau \omega_1}{2}.
\label{eq2_7:6}
\end{alignat}
The hard part is now to check the prefactors in \eqref{eq2_7:2a}. We have to divide our whole equation by $\uua{+1} \muub{+1}$. 
Comparing the prefactor yields the coefficients
\be{
\alpha = -\frac{\ii \tau \omega_0}{2} ,\qquad
\beta = \frac{\ii \tau \omega_1}{2}. 
}
Given these expressions, we finally check the equivalence of the terms proportional to $\eta \qu$.
The prefactor of $\eta \qu$ in \eqref{eq2_7:2a} and the right-hand side of \eqref{eq2_7:6} have to be equivalent. Thus we want to check whether this is the case or not. Writing both expression on one side we obtain
\begin{alignat}{2}
	-\omega_0^2 \br{1-u}^2 - \omega_1^2 u^2 + 1 +\vec{q}^2 + 2q_3 \va \tau u - 2q_3 \va \tau \br{1-u} & \notag \\
	  + \va^2 \tau^2 u^2 - 2\va^2 \tau^2 u \br{1-u} + \va^2 \tau^2 \br{1-u}^2 + 4\va^2 \tau^2 u\br{1-u} &  \notag \\
	  - \omega_0^2 u - \omega_1^2 u + \omega_0^2 u^2 + \omega_1^2 u^2 & \stackrel{!}{=}  0.
\label{eq2_7:7}
\end{alignat}
We pick out all terms with $~\omega_0 ,~ \omega_1~$ and find\footnote[2]{
$
-\omega_0^2 + 2\omega_0^2 u - \omega_0^2 u^2 - \omega_0^2 u^2 - \omega_1^2 + \omega_0^2 \\
  ~~~~~~= -1 - \vec{q}^2 + 2q_3 \va \tau - \va^2 \tau^2 +u + \vec{q}^2 u - 2q_3 \va \tau u + \va^2 \tau^2 u - 
  u - \vec{q}^2 u - 2q_3 \va \tau u - \va^2 \tau^2 u \\
  ~~~~~~= -1 - \vec{q}^2 + 2q_3 \va \tau - \va^2 \tau^2 - 4q_3 \va \tau u.
$
}
\be{
-\omega_0^2 + 2\omega_0^2 u - \omega_0^2 u^2 - \omega_0^2 u^2 - \omega_1^2 + \omega_0^2 \\
  = -1 - \vec{q}^2 + 2q_3 \va \tau - \va^2 \tau^2 - 4q_3 \va \tau u.
}
Inserting this expression in \eqref{eq2_7:7} leads to
\be{
\va^2 \tau^2 u^2 - 2\va^2 \tau^2 u + 2\va^2 \tau^2 u^2 + \va^2 \tau^2 - 2\va^2 \tau^2 u + \va^2 \tau^2 u^2 + 4\va^2 \tau^2 u - 4\va^2 \tau^2 u^2 = 0 
}
and as all terms vanish we find
\be{
0 = 0,
}
which is true and completes the check. 

Our next goal is to derive an expression for the one-particle
distribution function $F \qt$ in terms of the solutions of the hypergeometric differential equation \eqref{eq2_7:10}.
This equation has three singular points $u=\{0 ,~ 1 ,~ \infty\}$. A convergent solution for all $u<1$ can be found for all variables $a ,~ b$
and $-c ~ {\not\in} ~ \mathbb{N}_0$. 
As $c \in \mathbb{C}$ in our case, the ansatz
\be{
\eta \qu = \sum_{n=0}^{\infty} a_n u^n ,\qquad 
a_{n+1} = \frac{\br{a+n}\br{b+n}}{\br{1+n}\br{c+n}} a_n
}
is well defined. For $a_0=1$ we get the so called hypergeometric function $\Hyper{a,b,c;u}$\footnote[2]{
	It is a special case of the general hypergeometric function $~{}_pF_q(a_1,\dots,a_p;b_1,\dots,b_q;z)$. Its radius of convergence is $~R=1~$ and its serious expansion is given by
	\be{
	\Hyper{a,b,c;u} = \sum_{n=0}^{\infty} \frac{\br{a}_n \br{b}_n}{\br{c}_n} \frac{z^n}{n!},
	} 
	where $()_n$ denotes the Pochhammer symbol.
}.  
The function $\eta^+$ can be identified with the so-called ``regular solution''.
\be{
\eta^+ \qu = \Hyper{a,b,c;u}.
}

The second linearly independent solution of the hypergeometric differential equation is given by
\be{
u^{1-c} \Hyper{1+a-c,1+b-c,2-c,u} = u^{1-c} \br{1-u}^{c-a-b} \Hyper{1-a,1-b,2-c,u}. 
}
As a matter of fact, this function can be identified with
\be{
\eta^- \qu = u^{-2\alpha} \br{1-u}^{-2\beta} \Hyper{1-a,1-b,2-c;u}.
}
To get the mode functions $\chi^{\pm} \qu$ we have to resubstitute $\eta^{\pm} \qu$ in our ansatz \eqref{eq2_7:9}
\begin{subequations}	
	\begin{align}
		\chi^+ \qu &= N^+ u^{\alpha} \br{1-u}^{\beta} \Hyper{a,b,c;u} ,\\
		\chi^- \qu &= N^- u^{-\alpha} \br{1-u}^{-\beta} \Hyper{1-a,1-b,2-c;u},
	\end{align}
\end{subequations}
with $N^{\pm}$ as normalization constants. These constants are calculated in \ref{append:norm_an}
\begin{subequations}	
	\begin{align}
		N^+ \brq = \frac{\ee^{-\ii \tilde{\theta} \brq}}{\sqrt{2 \omega \br{\vec q,0} \br{\omega \br{\vec q,0} - \pi_3 \br{\vec q,0}}}}, \\
		N^- \brq = \frac{\ee^{\ii \tilde{\theta} \brq}}{\sqrt{2 \omega \br{\vec q,0} \br{\omega \br{\vec q,0} + \pi_3 \br{\vec q,0}}}}.
	\end{align}
\end{subequations}

We are now able to calculate the one-particle distribution function according to \eqref{eq2_6:3}
\be{
F \qt = 2\bet{\beta \qt}^2,
}
as we know each term appearing in the definition of $\beta \qt$
\be{
\beta \qt = -\ii \enorm \kappa^+ \qt \br{\de{t} + \ii \omega \qt} \chi^+ \qt.
}
To proceed we transform all functions in $\beta \qt$ to the new time variable $u$. This gives 
\al{
\de{t} = \de{u} \br{\de{t} u} = \frac{2}{\tau} u \br{1-u} \de{u}, \\
\enorm^2 \brq = \omega^2 \qu - \pi_3^2 \qu = \br{\omega \qu -\pi_3 \qu} \br{\omega \qu + \pi_3 \qu}
}
and thus\footnote[3]{
$F \qu = 2 \bet{\beta \qu}^2 = \frac{\enorm^2}{\omega \br{\omega - \pi_3}} 
  \bet{\br{ \frac{2}{\tau} u \br{1-u} \de{u} + \ii \omega} N^+ \qu \uu \muu \Hyper{a,b,c;u}}^2 \\
  ~~~~~~= \br{1+\frac{\pi_3}{\omega}} \bet{N^+}^2 \bet{\br{\frac{2}{\tau} u \br{1-u} \de{u} + \ii \omega} \uu \muu \Hyper{a,b,c;u}}^2.
$
}
\begin{multline}
F \qu = \br{1+\frac{\pi_3 \qu}{\omega \qu}} \bet{N^+}^2 \\
  \times \underbrace{\bet{\br{\frac{2}{\tau} u \br{1-u} \de{u} + \ii \omega \qu} \uu \muu \Hyper{a,b,c;u}}^2}_{(++)}.
\end{multline}
We focus on the term $(++)$ first. The derivative term gives
\begin{multline}
\de{u} \br{\uu \muu \Hyper{a,b,c;u}} = \alpha \uua{-1} \muu \Hyper{a,b,c;u} \\
  - \beta \uu \muub{-1} \Hyper{a,b,c;u} + \uu \muu \frac{ab}{c} \Hyper{1+a,1+b,1+c;u}, 
\end{multline}
where we have used the relation
\be{
\de{u} \Hyper{a,b,c;u} = \frac{ab}{c} \Hyper{1+a,1+b,1+c;u}.
}
The term $(++)$ can thus be written as
\begin{multline}
\left \lvert \frac{2}{\tau} u \br{1-u} \left( \alpha \uua{-1} \muu \Hyper{a,b,c;u} - \beta \uu \muub{-1} \Hyper{a,b,c;u} \right. \right. \\
   \left. \left. + \uu \muu \frac{ab}{c} \Hyper{1+a,1+b,1+c;u} \right) + \ii \omega \qu \uu \muu \Hyper{a,b,c;u} \right \rvert^2.
\end{multline}
As $\alpha ,~ \beta$ are purely imaginary we get 
\be{ 
\bet{\uu \muu} = \bet{\ee^{\alpha \tln{u}} \ee^{\beta \tln{1-u}}} = 1.
}
The remaining terms will be denoted as 
\al{
F_1 \qu &= \frac{2}{\tau} u \br{1-u} \frac{ab}{c} \Hyper{1+a,1+b,1+c;u} ,\\
F_2 \qu &= \br{\omega \qu - \br{1-u} \omega_0 - u \omega_1} \Hyper{a,b,c;u} ,
}
such that the one-particle distribution function as a function of the transformed time variable $u$ is given by
\be{
F \qu = \bet{N^+}^2 \br{1+ \frac{\pi_3 \qu}{\omega \qu}} \bet{F_1 \qu + \ii F_2 \qu}^2.
}

According to the previous discussion of the quasi-particle picture, we are mainly interested in the asymptotic behaviour of the one-particle distribution function. 
Hence we investigate $F \qu$
in the limit $u \to 1^-$ corresponding to $t \to \infty$. The one-particle distribution function for asymptotic times is given by
\begin{multline}
F \br{\vec q, u \to 1^-} = \\
  \bet{N^+ \brq}^2 \br{1 + \frac{\pi_3 \br{\vec q, 1}}{\omega_1}} \bet{F_1 \br{\vec q, u \to 1^-} + \ii F_2 \br{\vec q, u \to 1^-}}^2.
\end{multline}
In \ref{append:an_final} we explicitly show that this can be simplified to
\be{
F \br{\vec q, u \to 1^-} = \frac{2 \tsinh{\pi \tau \br{2 \va \tau + \omega_0 - \omega_1}/2} \tsinh{\pi \tau \br{2 \va \tau - \omega_0 + \omega_1}/2}}
  {\tsinh{\pi \tau \omega_0} \tsinh{\pi \tau \omega_1}}.
\label{eq2_7:11}
}
\section{Optimal Control Theory}
 
In this section we shall give an introduction to optimization followed by a presentation of important equations. These relations will be derived step by step, beginning with
a cost functional and ending with a discussion about the gradient and the optimization direction. Finally, the implementation on a computer is discussed.   

Optimal control theory is a mathematical optimization method operating under certain control laws. These laws are necessary
 to have control over the calculation and force the system into a state for with an optimality criterion is achieved. To perform these calculations one introduces
a cost functional $J$, which depends on input parameters, constraints and control variables. Usually the constraints are inequalities, chosen such that they force 
the system away from unwanted configurations. The control variables on the other hand are governed by differential equations and evaluated such that they do not violate the boundary conditions.
The goal is then to minimize the cost functional under the given constraints. 

In our case, optimal control theory forces a given potential $A \brt$ into a stationary point concerning the asymptotic particle number density.
In order to derive the corresponding equations, we start with the differential equation \eqref{eq3_1:3}. Due to the fact, that $F \qt ,~ G \qt$ and  $H \qt$ form a complete system, we have to take all of them into account. 
Furthermore we introduce the Lagrangian parameters $\mu_F \qt ,~ \mu_G \qt ,~ \mu_H \qt$ needed for the adjoint differential equation. In addition we define the inner products
\al{
\langle f \brt ,g \brt \rangle_t &= \int_{\mathbb{R}} f \qt g \qt dq, \\
\langle f,g \rangle &= \int_{\mathbb{R}} \langle f \brt ,g \brt \rangle_t dt.
}
The most important part regarding optimal control theory is to find an appropriate cost functional. In our case, we want to maximize the asymptotic particle number, so we start with the expression
\be{
J \br{F \qt, G \qt, H \qt, A} = -\gamma \int_{\mathbb{R}} F \br{\vec q, T} dq + p[A],
}
with $A(t)$ as control variable and $p[A]$ being an additional functional. 
As we try to keep the derivation as general as possible we specify the form of $p[A]$ later.

The next step is to introduce constraints given by 
\al{
\lambda_F \qt &= \dot{F} \qt - W \qt G \qt = 0 ,\\
\lambda_G \qt &= \dot{G} \qt + W \qt F \qt + 2 \omega \qt H \qt - W \qt = 0 ,\\
\lambda_H \qt &= \dot{H} \qt - 2 \omega \qt G \qt = 0,
}
where as usual
\al{
W \qt = \frac{E \brt \enorm \brq}{\omega^2 \qt} ,\qquad \enorm^2 \brq = 1 + q_1^2 + q_2^2 ,\\
\omega^2 \qt = \enorm^2 + \br{q_3 - A_z \brt}^2.
}
Then we can write down the Lagrange function
\be{
L = J \br{F, G , H , A} + \langle \lambda_F , \mu_F \rangle + \langle \lambda_G , \mu_G \rangle + \langle \lambda_H , \mu_H \rangle.
}
We search for a stationary point of the Lagrange function
\begin{equation}
\begin{split}
L = J &+ \int dq \int dt \left( \br{\mu_F \de{t} + W \mu_G} F \right. \\
  &\left. + \br{-W \mu_F + \mu_G \de{t} - 2\omega \mu_H} G + \br{2\omega \mu_G + \mu_H \de{t}} H - W\mu_G \right) \\
  = J &+ \int dq \left( \mu_F F \mid_{\mathbb{R}} - \int dt \br{\dot{\mu}_F F} + \int dt W \mu_G F \right. \\
  &+ \mu_G G \mid_{\mathbb{R}} - \int dt \br{\dot{\mu}_G} G + \int dt \br{-W \mu_F - 2\omega \mu_H} G \\
  &\left. + \mu_H H \mid_{\mathbb{R}} - \int dt \br{\dot{\mu}_H H} + \int dt 2\omega \mu_G H - \int dt W \mu_G \right) ,
\end{split}
\end{equation}
thus 
\be{
\frac{\delta L}{\delta F} = \frac{\delta L}{\delta G} = \frac{\delta L}{\delta H} \stackrel{!}{=} 0,
}
which gives the adjoint differential equation
\al{
\dot{\mu}_F \qt &=  W \qt \mu_G \qt ,\\
\dot{\mu}_G \qt &= - W \qt \mu_F \qt - 2\omega \qt \mu_H \qt ,\\
\dot{\mu}_H \qt &=  2\omega \qt \mu_G \qt.
\label{eq2_8:1}
}
One has to solve this equation backwards in time with the initial conditions
\al{
\frac{\delta L}{\delta F \br{\vec q, T}} &= \mu_F \br{\vec q, T} - \gamma \stackrel{!}{=} 0,\\
\frac{\delta L}{\delta G \br{\vec q, T}} &= \mu_G \br{\vec q, T}  \stackrel{!}{=} 0,\\
\frac{\delta L}{\delta H \br{\vec q, T}} &= \mu_H \br{\vec q, T}  \stackrel{!}{=} 0.
}
Finally, we proceed with the derivation of the gradient for a given potential. First we split the Lagrange function into two parts
\begin{multline}
L \br{F,G,H,\mu_F,\mu_G,\mu_H,A} = p[A] + \int dq \int dt W \br{\mu_G F - \mu_F G - \mu_G} \\
  + 2 \int dq \int dt \omega \br{-\mu_H G + \mu_G H} + \tilde{L} \br{F,G,H,\dot{\mu}_F,\dot{\mu}_G,\dot{\mu}_H} ,
\end{multline}
where all irrelevant terms are absorbed into $\tilde{L}$. 

Now we define the reduced cost functional from the relevant part
\begin{multline}
\tilde J \br{A} = p[A] + \int dq \int dt W \br{\mu_G F - \mu_F G - \mu_G} \\
  + 2 \int dq \int dt \omega \br{-\mu_H G + \mu_G H}
\end{multline}
which gives us the gradient
\begin{equation}
\begin{split}
\nabla \tilde J = \frac{\delta L}{\delta A}
  = \frac{\delta p}{\delta A} &- 2 \int dq \br{-\mu_H G + \mu_G H} \frac{q-A}{\omega} \\
  &+ \int dq \frac{\enorm}{\omega^2} \de{t} \br{\mu_G F - \mu_F G - \mu_G}.
\end{split}
\label{eq2_8:2}
\end{equation}

So far we have not specified the functional $p[A]$. Our idea is to use this term as a punishment term to prevent field configurations with higher field strengths from dominating the calculations. Therefore, we introduce an upper limit $E_{max}$ and a lower limit for the field strength $E_{min}$ and choose  
\be{
p[A] = -\mu_1 \int_{\mathbb{R}} dt \mathrm{ln} \br{E_{max}\brt + \dot{A}} - \mu_2 \int_{\mathbb{R}} dt \mathrm{ln} \br{-E_{min}\brt - \dot{A}},
\label{e2_8:3}
}
where $\mu_1 ,~ \mu_2$ assesses the importance of the constraints.
This fixes the additional term for the gradient
\be{
\frac{\delta p}{\delta A} = -\mu_1 \frac{\dot E_{max} + \ddot{A}}{\br{E_{max} + \dot{A}}^2} -\mu_2 \frac{\ddot{A} + \dot E_{min}}{\br{E_{min} + \dot{A}}^2}.
}
Note that such a term for the constraints still allows us to implement negative field strengths. 

During calculation we have to perform a line search to minimize the cost functional. We proceed by stepwise calculation of the problem
\be{
J \br{A_i - \alpha_i \nabla \tilde J \br{A_i}},
}
where the gradient $-\nabla \tilde J \br{A_i}$ gives the search direction and $\alpha_i$ the step length. A crucial point here is to find $\alpha_i's$ such that convergence can be achieved. In our case we simple use bisection method and apply the so-called Wolfe conditions to the algorithm
\begin{subequations}
	\begin{align}
		J \br{ A_k - \alpha_k \nabla \tilde J \br{A_k}} \le J \br{A_k} - c_1 \alpha_k \langle\nabla \tilde J \br{A_k}, \nabla \tilde J \br{A_k} \rangle ,\\
		\langle \nabla \tilde J \br{ A_k - \alpha_k \nabla \tilde J \br{A_k}}, \nabla \tilde J \br{A_k} \rangle \le c_2 \langle \nabla \tilde J \br{A_k}, \nabla \tilde J \br{A_k} \rangle,
	\end{align}
\end{subequations}
where $0 < c_1 < c_2 < 1$. The first inequality implies that the cost functional becomes smaller for each step whereas the second one states that the gradient vanishes at a stationary point. 

 
\chapter{Numerics}

Analytical solutions, like the one presented in chapter 2, are only available for a very restricted number of field configurations. Hence, one has to rely on numerical approaches in general. As we have employed a numerical approach in our calculations 
we have included this chapter for several reasons: It explains the effectiveness of the differential equation formalism and introduces several other transformations. Moreover, it provides information about the performance of different algorithms, 
shows the results of various numerical benchmark tests and demonstrates the pros and cons of discretized potentials.

\section{Differential Equation}

As already shown in chapter 2, we benefit from transforming the integro-differential equation
\be{
\dot{F} \qt = W \qt \IVac W \qts \br{1- F \qts} \tcos{2 \thetaT}
\label{eq3_1:1b}
}
into a coupled differential equation of first order
\begin{subequations}
	\label{eq3_1:2b}
	\begin{align}
		\dot{F} \qt &= W \qt \cdot G \qt ,\\
		\dot{G} \qt &= W \qt \br{1- F \qt} - 2 \omega \qt \cdot H \qt ,\\
		\dot{H} \qt &= 2 \omega \qt G \qt,
	\end{align}
\end{subequations}
with $W \qt = \frac{e E \brt \enorm \brq}{\omega^2 \qt} ,~ \enorm^2 \brq = m_e^2 + q_1^2 + q_2^2 ,~ \omega^2 \qt = m_e^2 + \br{\vec q - e \vec A \brt}^2$ for computational reasons.
In the following subsections we introduce additional transformations and discuss their applicability.

\subsection{Transformation 1}

In this subsection we show a transformation, which, we hope, could prevent the calculation 
from producing underflow errors. This should result in a more stable code.

We start again from the integro-differential equation \eqref{eq3_1:1b}. However, we write $F \brt = a I \brt$ this time with $a$ being a simple constant. 
This gives 
\be{
a \dot{I} \qt = W \qt \IVac W \qts \br{1- a I \qts} \tcos{2 \thetaT}.
}
Then we introduce the two auxiliary functions
\al{
G &= \IVac W \qts \br{1 - aI \qts} \tcos{2\thetaT} ,\\
H &= \IVac W \qts \br{1 - aI \qts} \tsin{2\thetaT}. 
}
As the system is basically the same as before, the results for the derivatives are similar to the previous calculation
\al{
\dot{G} \qt &= W \qt \br{1 - aI \qt} - 2 \omega \qt H \qt ,\\
\dot{H} \qt &= 2 \omega \qt G \qt.
}
We have now an equation for $I \qt$ instead of $F \qt$
\be{
\ma{\dot{I} \\ \dot{G} \\ \dot{H}} = \ma{0 & W/a & 0 \\ -aW & 0 & -2\omega \\ 0 & 2\omega & 0} \ma{I \\ G \\ H} + \ma{0 \\ W \\ 0}.
\label{eq3_3:1}
}
Using the initial conditions
\be{
I \br{\vec q, t \to -\infty} = G \br{\vec q, t \to -\infty} = H \br{\vec q, t \to -\infty} = 0
}
we calculate $I \qt$ and recover $F \qt$ in the end by 
\be{
F \qt = a I \qt.
}
The main advantage of this transformation is that $I \qt$ is larger than the one-particle distribution function for $a < 1$.

\subsection{Transformation 2} 

The idea of this transformation is to rewrite the differential equation \eqref{eq3_1:2b} in such a way that it is solved on a compact interval. This should help saving computer time and, moreover, the calculations should become more stable as one needs fewer evaluation steps.
Thus, we start from the coupled differential equation \eqref{eq3_1:2b}
and employ a transformation, which will be adopted for field configurations of the type
\be{
A(t) = \sum_i -\va_i \tau_i \tanh \br{\frac{t-t_{0,i}}{\tau_i}} ,\qquad
E(t) = \sum_i \va_i \mathrm{sech}^2 \br{\frac{t-t_{0,i}}{\tau_i}}.
}
We will distinguish the dominant electric field (index $1$) from all other ones (index $i\neq1$) until the end of this subsection.
The introduced transformation is similar to the one used to derive the analytic solution for the single pulse potential in chapter 2
\begin{align}
t \mapsto u=\frac{1}{2} \br{1 + \textrm{tanh}\br{\frac{t}{\tau_1}}}  ,\\ [-\infty, \infty] \to [0,1].
\end{align}
At that point one gets an impression of the disadvantage of this transformation. 
The most important parameter in the transformation is the pulse length $\tau_1$, which restricts our calculations to setups with a dominant long pulse.  
The adjective 'dominant' means in our notion $\tau_1 \gg \tau_i$ and $\varepsilon_1 \gg \varepsilon_i$.
Using the new variable $u$ we obtain the following transformed vector potential
\begin{multline}
A(u) = A_1 \br{u} + \sum_i A_i \br{u} \\
= -\tau_1 \varepsilon_1 \left(2u-1 \right) - \sum_i \tau_i \varepsilon_i \mathrm{tanh} \left( \frac{\tau_1 \mathrm{arctanh} \left(2u-1 \right) - t_{0,i}}{\tau_i} \right)
\label{eq3_4:2}
\end{multline}
and transformed electric field 
\begin{multline}
E (u) = E_1 \br{u} + \sum_i E_i \br{u} \\
= 4 \varepsilon_1 u(1-u) + \sum_i \varepsilon_i \left(1- \mathrm{tanh}^2 \left( \frac{\tau_1 \mathrm{arctanh} \left(2u-1 \right) - t_{0,i}}{\tau_i} \right) \right),
\label{eq3_4:3}
\end{multline}
where $t_{0,i}$ is the time lag between the dominant pulse and pulse $i$.
It may look like a big disadvantage that the terms in the sum become harder to compute. Analysis of the computation time, however, shows that the compression of the time interval from $\left[-\infty, \infty \right]$ to $\left[0, 1 \right]$ outweighs the drawbacks.

The resulting differential equation looks similar to \eqref{eq3_1:2b}
\begin{subequations}
	\label{eq3_4:1}
	\begin{align}
		\de{u} F \qu &= W \qu \cdot G \qu, \\
		\de{u} G \qu &= W \qu \cdot (1-F \qu) - 2 \omega \qu \cdot H \qu, \\
		\de{u} H \qu &= 2 \omega \qu \cdot G \qu,
	\end{align}
\end{subequations}
however, the input functions are different
\al{
\omega^2 \qu = \frac{\tau_1 r^2 \qu}{2u \br{1-u}} ,\qquad r \qu = \sqrt{\enorm^2 \brq + \br{q_3 - A \bru}^2} ,\\
\enorm^2 \brq = m_e^2 + q_1^2 + q_2^2 ,\qquad W \qu = \frac{E \bru \enorm \brq}{r^2 \qu}.
}

In the end we want to give some hints concerning numerical efficiency. One can observe in \eqref{eq3_4:2} and \eqref{eq3_4:3} that the term $\tau_1 \mathrm{arctanh} \left(2u-1 \right) - t_{0,i}$ remains unchanged for a given value of $u$. Therefore it is much faster to calculate this term once and read it from the memory instead of calculating it several times. The same reasoning applies to $r \qu$. 

\subsection{Low Density Approximation}

Low density indicates that the particle number is low such that 
$F \qts \ll 1$ for all $t \in \mathbb{R}$. Accordingly, we approximate the Pauli-blocking term
\be{
\br{1- F \qts} \sim 1
}
such that the integro-differential equation takes the form
\be{
\dot{F} \qt = W \qt \IVac W \qts \tcos{2 \thetaT}.
}
The benefit of this approximation is that the history of the process itself is ignored, which leads to a simple integral. 
N.B.: This advantage vanishes, however, if one deals with the equivalent differential equation
\al{
\dot{F} \qt &= W \qt \cdot G \qt ,\\
\dot{G} \qt &= W \qt - 2 \omega \qt \cdot H \qt ,\\
\dot{H} \qt &= 2 \omega \qt G \qt,
\label{eq3_5:1}
}
as the neglect of the term $\br{1- F \qts}$ gives only worse results. 
\section{Comparison of Transformations}
\label{sec:numerics}

After deriving three distinct differential equations \eqref{eq3_1:2}, \eqref{eq3_3:1}, \eqref{eq3_4:1} to describe pair creation theoretically, the question of the ``best'' method arises. Thus we first have to identify the favorable transformation and correspondingly a well-adapted solver. In our computation we use the configuration
\be{
A(t) = \sum_i -\va_i \tau_i \tanh \br{\frac{t-t_{0,i}}{\tau_i}} ,\qquad
E(t) = \sum_i \va_i \mathrm{sech}^2 \br{\frac{t-t_{0,i}}{\tau_i}},
}
where we choose specific configurations to investigate three aspects of the different transformations:
The first and most crucial one is runtime\footnote{The values given for the runtime should only hold as an indicator whether a transformation works well or not, because for a detailed analysis one would have to perform hundreds of simulation runs and analyze all data.}, the others are accuracy and stability. Usually accuracy is the most important specification, but according to the thesis of M. Orthaber \cite{OrtDip} we know that a sufficiently good accuracy can be achieved. Nevertheless we will also test it on our own, especially with configurations that are comparable with the analytical solution. 

In the following we perform runtime and accuracy tests for the different transformations and refer to them as 
\begin{description}
  \item[Standard] differential equation \eqref{eq3_1:2}, 
  \item[Standard(I)] differential equation \eqref{eq3_1:2} with an additional implicit solver, 
  \item[Transformation 1] differential equation \eqref{eq3_3:1} with additional parameter $~a=10^{-4}$,
  \item[Transformation 2] differential equation \eqref{eq3_4:1},
  \item[Transformation 2(I)] differential equation \eqref{eq3_4:1} with an additional implicit solver.
\end{description}

We have used a Runge-Kutta 8 solver from Numerical Recipes 3 \cite{Press:2007:NRE:1403886} for non-stiff problems and the RADAU5 solver from E. Hairer \cite{Hairer1993} for stiff differential equations. 
In numerical calculations, we cannot start at asymptotic times $t \to -\infty$ nor can we end at $t \to \infty$. In our simulations, the starting point was chosen such that
the vector potential was numerically constant up to a relative precision of $10^{-7}$, corresponding to a vanishing electric field. The end point was reached once the potential becomes constant again.

For the system ``Standard(I)'' computations using the implicit solver have been performed from initial time up to a maximum of $10$ per cent of the pulse length of the first pulse. In contrast, for the setup ``Transformation2(I)'' the implicit calculation has been done for $u$ starting at $10^{-8}$ and ending at $0.1$. Then the solver has changed. 

Choosing the right end point for the implicit solver is a very tricky issue. If one works too early with an explicit solver, the computation time will increase. If one waits too long with a change of the solvers one runs into troubles regarding accuracy. Even worse, it could happen that the solver produces completely wrong results, although the calculation looks stable 
\footnote{A possible explanation for this effect could be the need of a higher precision for an implicit solver.}.

For the purpose of a systematic evaluation of the various transformations, we have picked out a sample of the later used field configurations and compared the obtained results.
We analyze the runtime performance in Tab.\ref{Tab1} and the accuracy performance in Tab.\ref{Tab2}. Comparing these results one gets the impression that the Standard solver is slow,
but works for every field configuration under consideration as one can see in Tab.\ref{Tab2}.  
The more elaborated solvers ``Standard(I)'' and ``Transformation2(I)'', on the other hand, are the better choice for certain configurations. One has to be aware, however, that these solvers cannot be used for every field configuration as presented in Tab.\ref{Tab2}.

\ctable[
notespar,
caption = {Runtime comparison for different field configurations and solvers. The results are normalized to the results of the Standard solver to get a better impression of the pros and cons of the various transformations. The solver for St(I), Tr2 and Tr2(I) are faster in every calculation except those with ten small pulses.\footnotemark[3]},
cap = {Runtime comparison for different field configurations and systems},
label = Tab1,
pos = h,
]{lccccc}{
  \tnote[1]{Single pulse}
  \tnote[2]{Double pulse}
  \tnote[3]{Ten identical pulses with time lag $t_0$}
  \tnote[4]{One long pulse and ten short pulses}
  } {
    \toprule
      Field Parameters & St & St(I) & Tr2(I) & Tr2 & Tr1   \\
      ($\va$[1], $\tau$[1/m], $t0$[1/m]) & & & & & \\
    \midrule
      $\va=0.1, \tau=100$\footnotemark[1] 	& 1.000 & 0.675 & 0.689 & 0.910  & 2.212 \\
    \midrule
      $\va=0.005, \tau=3$\footnotemark[1] 	& 1.000	 & 0.746   & 0.682      & 0.312  & 1.511 \\
    \midrule
      $\va_1=0.1, \tau_1=100,$ 	& 1.000 & 0.743 & 0.611 & 0.893 & 2.060 \\
      $\va_2=0.01, \tau_2=2$\footnotemark[2] & & & & & \\
    \midrule
      $\va_1=0.1, \tau_1=100,$ 	& 1.000 & 0.674 & 0.630 &   \wh{-}     & 2.138 \\
      $\va_2=0.01, \tau_2=2, t_0=30$\footnotemark[2] & & & & & \\
    \midrule
      $\va= \pm{0.005}, \tau=3, t_0=5$\footnotemark[3] 	& 1.000 & 0.839 & 1.706 & 1.458 & 1.802 \\ 
    \midrule
      $\va= 0.005, \tau=3, t_0=5$\footnotemark[3] 	& 1.000 & 0.827  & 1.843  & 1.769  & 2.203 \\
    \midrule
      $\va_1=0.1, \tau_1=100,$ 	& 1.000 & 0.654  & 0.422 & 0.558 & 2.213 \\ 
      $\va_i= \pm{0.005}, \tau_i=3, t_0=9$\footnotemark[4] & & & & & \\
    \midrule
      $\va_1=0.1, \tau_1=100,$ 	& 1.000 & 0.650 & 0.430 & 0.543 & \wh{-} \\
      $\va_i= 0.005, \tau_i=3, t_0=9$\footnotemark[4] & & & & & \\
    \midrule
      $\va_1=0.1, \tau_1=100, t_{0,1} = 200,$ & 1.000 & 0.926 & 0.306 & \wh{-} & \wh{-} \\
      $\va_i= 0.005, \tau_i=3, t_0=24.95$\footnotemark[4] & & & & & \\
    \bottomrule
}



\ctable[
notespar,
caption = {Comparison of the particle number for different solvers and field configurations. The solver St works for any field configuration, whereas the implicit/explicit solver St(I) surprisingly fails for a single pulse configuration. The solver Tr2(I), on the other hand, performs pretty well for all field configurations except those it was not designed for\footnotemark[3], i.e. configurations without a dominant pulse.},
cap = {Accuracy comparison for different field configurations and systems},
label = Tab2,
pos = h,
]{lccccc}{
  \tnote[1]{Single pulse}
  \tnote[2]{Double pulse}
  \tnote[3]{Ten identical pulses with time lag $t_0$}
  \tnote[4]{One long pulse and ten short pulses}
  } {
    \toprule
      Field Parameters & St & St(I) & Tr2(I) & Tr2 & Tr1\\
      ($\va$[1], $\tau$[1/m], $t0$[1/m]) & & & & & \\
    \midrule
      $\va=0.1, \tau=100$\footnotemark[1] 	& 1.529 & 1.668 & 1.529 & 1.540 & 1.529  \\
    \midrule
      $\va=0.005, \tau=3$\footnotemark[1] 	& 5.856	& 5.856 & 5.856 & 5.933 & 5.856  \\
    \midrule
      $\va_1=0.1, \tau_1=100,$ 	& 4.240 & 4.240 & 4.240 & 4.240 & 4.240  \\
      $\va_2=0.01, \tau_2=2$\footnotemark[2] & & & & & \\
    \midrule
      $\va_1=0.1, \tau_1=100,$ 	& 4.021 & 4.021 & 4.021 &   \wh{-}   & 4.021   \\
      $\va_2=0.01, \tau_2=2, t_0=30$\footnotemark[2] & & & & & \\
    \midrule
      $\va= \pm{0.005}, \tau=3, t_0=5$\footnotemark[3] 	& 8.700 & 8.700 & 30.99 & 31.03 & 8.700  \\ 
    \midrule
      $\va= 0.005, \tau=3, t_0=5$\footnotemark[3] 	& 3.782 & 3.782 & 5.241 & 5.267 & 3.782 \\
    \midrule
      $\va_1=0.1, \tau_1=100,$ 	& 1.421 & 1.421 & 1.421 & 1.420 & 1.421 \\ 
      $\va_i= \pm{0.005}, \tau_i=3, t_0=9$\footnotemark[4] & & & & & \\
    \midrule
      $\va_1=0.1, \tau_1=100,$ 	& 1.353 & 1.353 & 1.353 & 1.353 &   \wh{-}  \\
      $\va_i= 0.005, \tau_i=3, t_0=9$\footnotemark[4] & & & & & \\
    \midrule
      $\va_1=0.1, \tau_1=100, t_{0,1} = 200,$ & 2.259 & 2.259 & 2.259 & \wh{-} & \wh{-}  \\
      $\va_i= 0.005, \tau_i=3, t_0=24.95$\footnotemark[4] & & & & & \\
    \bottomrule
}

\newpage
The tables Tab.\ref{Tab1} and Tab.\ref{Tab2} provide much information with regard to coding fast and reliable computer program. We can conclude that the use of an implicit solver at the beginning of a calculation is a competitive alternative compared to the standard solver. The surprising failures are possibly driven by numerical instabilities, which could probably be overcome by using a different implicit solver. A good option would be a Gauss-Runge-Kutta solver of 4th or higher order.

These tables further illustrate that the solver ``Standard'' produces correct results for any given configuration above a certain numerical limit, which is set by a combination of the field strength parameter and the pulse length. Additionally, we observe that the solver ``Transformation2'' gives the same results as the solver ``Standard'' for all field configurations possessing a dominant pulse, however, performs substantially faster in these cases.

\newpage

In order to conduct stability tests, we have performed calculations upon changing the parameters only slightly. In this way we can identify parameter regions in which our numerics work properly. 
These calculations indicate that the particle distribution function is more difficult to calculate for
negative values of $q_3$ than for positive ones. Probably this happens, because
particle pair creation occurs for negative values of $q_3$ at earlier times such that one would have to start the calculations at earlier initial times, too. Another aspect worth mentioning is that a change $t \to -t$ in the potential gives a mirrored distribution function $F \br{q_3,t \to \infty} \to F \br{-q_3,t \to \infty}$. This is a further monitoring option for our solvers and in perfect agreement with ideas presented in C. K. Dumlu et al. \cite{PhysRevD.83.065028}.

\section{Full Momentum Space}

In the previous section we only investigated the performance of the various differential equations.
In this section we shall give ideas for further optimization, especially for the calculation of the particle spectrum in full momentum space. An option, for example, is to use an interpolation algorithm avoiding calculations of less important data points. Another idea would be to use an improved integration routine such as Romberg integration.

First we investigate the issue of interpolation routine. To test them we have used single pulse configurations involving configurations for $q_1 = q_2 = 0$. For the purpose of testing we use the differential equation \eqref{eq3_4:1}, but the results remain valid for the other transformations as well. 

We tried a $1D$ cubic spline interpolation \cite{Press:2007:NRE:1403886} in direction of fixed $q_3$. On the other hand, we use a linear spline interpolation \cite{Press:2007:NRE:1403886} in the direction of fixed $q_{\perp}$ as well as in case of errors in direction of fixed $q_3$. As a matter of fact, we have also tested other routines, but they all suffer from severe disadvantages. A $2D$ cubic spline for example does not work properly, because of the typically irregular data grid. Working with such a method would cause a lot of additional computer time without the certainty that the result would be worth it. A $2D$ linear spline interpolation as well as Laplace interpolation are not able to produce accurate results as higher order interpolation routines are a necessity at least in direction of fixed $q_3~$.

We investigate the performance for calculating the particle number for various parameters. 
It can be seen from the data in Tab.\ref{Tab3}B that the results for $\Delta q_{\perp} \in \com{0.1,0.5}m$ and $\Delta q_{\perp} = \Delta q_3$ are in good agreement with the analytical calculation. As expected, the results get better the more data points are involved. Tab.\ref{Tab3}A, on the other hand, shows that the runtime increases significantly by decreasing $\Delta q_3 = 0.05m$ to $\Delta q_3 = 0.01m$ whereas the accuracy of the results is not improved.

A similar analysis is performed for a different parameter set in Tab.\ref{Tab5}. These results indicate that we have to use a step size of at least $\Delta q_{\perp}=0.1m$ for these parameters.
A decrease in the momentum spacing is, however, accompanied by an increase in runtime. As a matter of fact, if we choose the spacing too small, the runtime increases substantially
whereas the accuracy is not affected anymore. Accordingly, finding the appropriate value for $\Delta q$ is an important but nontrivial task.

\newpage
\ctable[
notespar,
caption = {Runtime comparison for different momentum spacings for a single pulse in particle distribution with $~\varepsilon = 0.005 ,~ \tau =3m^{-1}$ in A. The results are normalized to the fastest calculation and all momenta are given in units of the electron mass $q[m]$. In B the results for the particle number are normalized to a calculation based on the analytical result.},
cap = {Runtime and accuracy comparison for different momentum spacings in particle distribution for a short pulse},
label = Tab3,
pos = h,
]{lccc}{
  \tnote[1]{$\Delta q_3 = 0.05,~ 10\%~$ Tolerance, whether a $\Delta q_3$ is calculated is chosen dynamically }
  } {
	\toprule
	  A & $~\Delta q_{\perp} = 0.5~$ & $~\Delta q_{\perp} \in \left[0.1,0.5 \right]~$ & $~\Delta q_{\perp} = \Delta q_3~$ \\
	\midrule
	  Important $~q_3$\tmark[1] & 1.000 & 1.631 & 3.063 \\
	\midrule
	  $~\Delta q_3 = 0.05~$ & 1.210	& 1.621 & 3.263  \\
	\midrule
	  $~\Delta q_3 = 0.01~$	& 3.810 & 6.968 & 62.705 \\
	\bottomrule
}

\ctable[
notespar,
pos = h,
]{lccc}{
  \tnote[1]{$\Delta q_3 = 0.05,~ 10\%~$ Tolerance, whether a $\Delta q_3$ is calculated is chosen dynamically  }
  } {
	\toprule
	  B & $~\Delta q_{\perp} = 0.5~$ & $~\Delta q_{\perp} \in \left[0.1,0.5 \right]~$ & $~\Delta q_{\perp} = \Delta q_3~$ \\
	\midrule
	  	Important $~q_3$\tmark[1] & 1.0255 & 1.0021 & 0.9979 \\
	\midrule
	   	$~\Delta q_3 = 0.05~$ & 1.0253 & 1.0039 & 0.9997  \\
	\midrule
	   	$~\Delta q_3 = 0.01~$ & 1.0245 & 1.0047 & 1.0000 \\
	\bottomrule
}


\ctable[
notespar,
caption = {Runtime comparison for different momentum spacings for a single pulse with $~\varepsilon = 0.1 ,~ \tau =100m^{-1}$ in A. The results for the particle number are normalized to the fastest calculation and all momenta are given in units of the electron mass $~q[m]$. In B the results for the particle number are normalized to a calculation based on the analytical result.},
cap = {Runtime and accuracy comparison for different momentum spacings in particle distribution for a long pulse},
label = Tab5,
pos = h,
]{lccc}{
  \tnote[1]{$\Delta q_3 = 0.05,~ 10\%~$ Tolerance, whether a $\Delta q_3$ is calculated is chosen dynamically  }
  } {
	\toprule
	  A & $~\Delta q_{\perp} = 0.5~$ & $~\Delta q_{\perp} \in \left[0.1,0.5 \right]~$ & $~\Delta q_{\perp} = \Delta q_3~$ \\
	\midrule
	  Important $~q_3$\tmark[1] & 1.000 & 2.129 & 3.461 \\
	\midrule
	  $~\Delta q_3 = 0.05~$ & 1.345 & 2.943 & 4.843  \\
	\midrule
	  $~\Delta q_3 = 0.01~$ & 6.235 & 14.058 & 98.640 \\
	\bottomrule
}

\ctable[
notespar,
pos = h,
]{lccc}{
  \tnote[1]{$\Delta q_3 = 0.05,~ 10\%~$ Tolerance, whether a $\Delta q_3$ is calculated is chosen dynamically  }
  } {
	\toprule
	  B & $~\Delta q_{\perp} = 0.5~$ & $~\Delta q_{\perp} \in \left[0.1,0.5 \right]~$ & $~\Delta q_{\perp} = \Delta q_3~$ \\
	\midrule
	    Important $~q_3$\tmark[1] & 0.771168 & 1.02592 & 1.02585 \\
	\midrule
	    $~\Delta q_3 = 0.05~$ & 0.770376 & 1.0005 & 1.0005  \\
	\midrule
	    $~\Delta q_3 = 0.01~$ & 0.74475 & 0.9998 & 1.00086 \\
	\bottomrule
}

\newpage
We now turn our attention towards integration algorithms. For configurations with $q_{\perp} = 0$ it was sufficient to use the trapezoidal rule for integration, but for calculations in two dimensions this is not true any more. 
We have used Romberg integration instead to obtain better results. The problem with this integration routine is, however, that we need data points outside of the calculated grid. In order to avoid calculation of extra data points, which would increase the runtime tremendously, we have used a $2D$ polynomial interpolation \cite{Press:2007:NRE:1403886} instead. As one can see in Tab.\ref{Tab7}A, the trapezoidal rule is still competitive for short pulses. For the long pulse, however, we need a much smaller step size of $\Delta q_3 = 0.01m$ in the trapezoidal rule integration compared to the Romberg integration.

\ctable[
notespar,
caption = {Comparison of the accuracy when adopting different integration routines for a single pulse with A) $~\varepsilon = 0.005 ,~ \tau =3m^{-1}$ and B) $~\varepsilon = 0.1 ,~ \tau =100m^{-1}$. In addition the influence of different momentum spacings has been investigated. The results for the particle number are normalized to a calculation based on the analytical result. All momenta are given in units of the electron mass $~q[m]$.},
cap = {Comparison of the accuracy when adopting different integration routines for different pulses},
label = Tab7,
pos = h,
]{lcccccc}{
  \tnote[1]{$\Delta q_3 = 0.05,~ 10\%~$ Tolerance, whether a $\Delta q_3$ is calculated is chosen dynamically  }
  \tnote[2]{Trapezoidal rule}
  \tnote[3]{Romberg integration}
  } {
    \toprule
      A & \multicolumn{2}{c}{$~\Delta q_{\perp} = 0.5~$} &  \multicolumn{2}{c}{$~\Delta q_{\perp} \in \left[0.1,0.5 \right]~$} & \multicolumn{2}{c}{$~\Delta q_{\perp} = \Delta q_3~$} \\
    \cmidrule(r){2-3}
    \cmidrule(lr){4-5}
    \cmidrule(l){6-7}
      & TR\tmark[2] &  RI\tmark[3] & TR & RI & TR & RI \\
    \midrule
      Important $~q_3$\tmark[1] & 1.0255 & 1.0210 & 1.0021 & 1.0070 & 0.9979 & 1.0027 \\
    \midrule
      $~\Delta q_3 = 0.05~$ 	& 1.0260 & 1.0253 & 1.0110 & 1.0039 & 1.0068 & 0.9997  \\
    \midrule
      $~\Delta q_3 = 0.01~$ 	& 1.0226 & 1.0245 & 1.0038 & 1.0047 & 0.9997 & 1.0000 \\
    \bottomrule
}

\ctable[
notespar,
pos = h,
]{lcccccc}{
  \tnote[1]{$\Delta q_3 = 0.05,~ 10\%~$ Tolerance, whether a $\Delta q_3$ is calculated is chosen dynamically  }
  \tnote[2]{Trapezoidal rule}
  \tnote[3]{Romberg integration}
  } {
    \toprule
      B & \multicolumn{2}{c}{$~\Delta q_{\perp} = 0.5~$} &  \multicolumn{2}{c}{$~\Delta q_{\perp} \in \left[0.1,0.5 \right]~$} & \multicolumn{2}{c}{$~\Delta q_{\perp} = \Delta q_3~$} \\
    \cmidrule(r){2-3}
    \cmidrule(lr){4-5}
    \cmidrule(l){6-7}
      & TR\tmark[2] &  RI\tmark[3] & TR & RI & TR & RI \\
    \midrule
      Important $~q_3$\tmark[1] & 1.2967 & 0.7711 & 1.0259 & 1.0259 & 1.0258 & 1.0258 \\
    \midrule
      $~\Delta q_3 = 0.05~$ 	& 0.7703 & 0.7703 & 1.0263 & 1.0005 & 1.0262 & 1.0005  \\
    \midrule
      $~\Delta q_3 = 0.01~$ 	& 0.7458 & 0.7447 & 1.0010 & 0.9998 & 1.0008 & 1.0008 \\
    \bottomrule
}
\clearpage
\section{Comparison with Lookup-Table}
\label{sec:LUT}

%

We finally want to check whether simulations with precalculated values of the vector potential and the electric field yield the same results as calculations where these values are calculated on-the-fly.
This investigation is motivated by the request to calculate the pair production process for arbitrary field configurations as well.

Therefore, we recompute all field configurations, which have previously been discussed in Tab.\ref{Tab1} and Tab.\ref{Tab2}. In comparison, we have now precalculated the vector potential
and the electric field for a different number of sampling points. The values for a required time step have been calculated by interpolation routines.

\ctable[
notespar,
caption = {Runtime comparison for different sets of sampling points. Potential and electric field have been discretized first. In order to calculate intermediate values linear interpolation is used. The results for the runtime are normalized to the results of the ``Standard'' solver.},
cap = {Runtime comparison for different sets of sampling points and linear interpolation},
label = Tab9,
pos = h,
]{lccccc}{
  \tnote[1]{Single pulse}
  \tnote[2]{Double pulse}
  \tnote[3]{Ten identical pulses with time lag $t_0$}
  \tnote[4]{One long pulse and ten short pulses}
  } {
    \toprule
      Field Parameters & St & Trans2(I) & 512 & 4096 & 32768   \\
      ($\va$[1], $\tau$[1/m], $t0$[1/m]) & & & Points & Points & Points \\
    \midrule
      $\va=0.1, \tau=100$\footnotemark[1] 	& 1.000 & 0.689 & \wh{0.253} & 0.784 & 2.532 \\
    \midrule
      $\va=0.005, \tau=3$\footnotemark[1] 	& 1.000 & 0.682 & 0.331 & 4.470 & 22.82 \\
    \midrule
      $\va_1=0.1, \tau_1=100,$  & 1.000 & 0.611 & 0.202 & 0.527 & 1.900 \\
      $\va_2=0.01, \tau_2=2$\footnotemark[2] & & & & & \\
    \midrule
      $\va_1=0.1, \tau_1=100,$ 	& 1.000 & 0.630 & 0.220 & 0.526 & 1.872 \\
      $\va_2=0.01, \tau_2=2, t_0=30$\footnotemark[2] & & & & & \\
    \midrule
      $\va= \pm{0.005}, \tau=3, t_0=5$\footnotemark[3] 	& 1.000 &  & 0.099 & 1.406 & 4.441 \\ 
    \midrule
      $\va= 0.005, \tau=3, t_0=5$\footnotemark[3] 	 & 1.000 &  & 0.051 & 1.535 & 5.016 \\
    \midrule
      $\va_1=0.1, \tau_1=100,$ 	& 1.000 & 0.422 & \wh{0.090} & 0.167 & 0.699 \\ 
      $\va_i= \pm{0.005}, \tau_i=3, t_0=9$\footnotemark[4] & & & & & \\
    \midrule
      $\va_1=0.1, \tau_1=100,$ 	& 1.000 & 0.430 & 0.090 & 0.143 & 0.692 \\
      $\va_i= 0.005, \tau_i=3, t_0=9$\footnotemark[4] & & & & & \\
    \bottomrule
}

\newpage
\ctable[
notespar,
caption = {Comparison of the particle number for different sets of sampling points. In order to calculate intermediate values linear interpolation is used.},
cap = {Comparison of the particle number for different sets of sampling points via linear interpolation},
label = Tab10,
pos = h,
]{lcccccc}{
  \tnote[1]{Single pulse}
  \tnote[2]{Double pulse}
  \tnote[3]{Ten identical pulses with time lag $t_0$}
  \tnote[4]{One long pulse and ten short pulses}
  } {
    \toprule
      Field Parameters & St & Tr2(I) & 512 & 4096 & 32768 \\
      ($\va$[1], $\tau$[1/m], $t0$[1/m]) & & & Points & Points & Points \\
    \midrule
      $\va=0.1, \tau=100$\footnotemark[1] 	& 1.529 & 1.529 & \wh{2.1573E+05} & 1.586 & 1.529 \\
    \midrule
      $\va=0.005, \tau=3$\footnotemark[1] 	& 5.856	& 5.856 & 5.733 & 5.859 & 5.857  \\  
    \midrule
      $\va_1=0.1, \tau_1=100,$ 	& 4.240 & 4.240 & 2.218 & 2.748 & 4.212 \\
      $\va_2=0.01, \tau_2=2$\footnotemark[2] & & & & & \\
    \midrule
      $\va_1=0.1, \tau_1=100,$ 	& 4.021 & 4.021 & 1.672 & 2.572 & 3.993 \\
      $\va_2=0.01, \tau_2=2, t_0=30$\footnotemark[2] & & & & & \\
    \midrule
      $\va= \pm{0.005}, \tau=3, t_0=5$\footnotemark[3] 	& 8.700 &  & 8.013 & 8.697 & 8.701 \\
    \midrule
      $\va= 0.005, \tau=3, t_0=5$\footnotemark[3] 	& 3.782 &  & 3.435 & 3.780 & 3.783 \\
    \midrule
      $\va_1=0.1, \tau_1=100,$ 	 & 1.421 & 1.421 & \wh{1.03958E+02} & 1.006 & 1.413 \\ 
      $\va_i= \pm{0.005}, \tau_i=3, t_0=9$\footnotemark[4] & & & & & \\
    \midrule
      $\va_1=0.1, \tau_1=100,$ 	& 1.353 & 1.353 & 7.925 & 9.193 & 1.345 \\
      $\va_i= 0.005, \tau_i=3, t_0=9$\footnotemark[4] & & & & & \\
    \bottomrule
}

The number of sampling points is very important as one can see in Tab.\ref{Tab9} and Tab.\ref{Tab10}. Interestingly, we can observe a lack of accuracy in case of $512$ sampling points.
For most configurations, at least $4096$ sampling points are essential. For very difficult and complex pulse structures even $32768$ points produce an error of $1 \%$.
Additionally, these tables demonstrate that there are field configurations with a comparatively long runtime. 
This is quite surprising as lookup tables are usually used to accelerate calculations.
It seems that error correction during a calculation slows down the whole computation. 
This also indicates that there are problems concerning stability when using discretized potentials in conjunction with linear interpolation.

In order to remedy the difficulties with linear interpolation we switch to a more advanced interpolation routine, namely cubic interpolation. 
This interpolation routine is more advantageous when dealing with a high number of sampling points as one can see in Tab.\ref{Tab11} and Tab.\ref{Tab12}. 
For $512$ sampling points there are empty cells due to a very long runtime, however, the larger the set of sampling point get the better the calculations become. 
For $32768$ sampling points, for example, cubic interpolation outperforms linear interpolation as presented in Tab.\ref{Tab11}. 

 
\newpage
\ctable[
notespar,
caption = {Runtime comparison for different sets of sampling points. In order to calculate intermediate values cubic interpolation has been used. The results for the runtime are normalized to the results of the ``Standard'' solver.},
cap = {Runtime comparison for a different set of sampling points with cubic interpolation},
label = Tab11,
pos = h,
]{lccccc}{
  \tnote[1]{Single pulse}
  \tnote[2]{Double pulse}
  \tnote[3]{Ten identical pulses with time lag $t_0$}
  \tnote[4]{One long pulse and ten short pulses}
  } {
    \toprule
      Field Parameters & St & Trans2(I) & 512 & 4096 & 32768   \\
      ($\va$[1], $\tau$[1/m], $t0$[1/m]) & & & Points & Points & Points \\
    \midrule
      $\va=0.1, \tau=100$\footnotemark[1] 	& 1.000 & 0.689 & 0.9648 & 0.968 & 1.111 \\
    \midrule
      $\va=0.005, \tau=3$\footnotemark[1] 	& 1.000 & 0.682 & 1.183 & 1.585 & 5.734 \\ 
    \midrule
      $\va_1=0.1, \tau_1=100,$ & 1.000 & 0.611 & \wh{0.7038} & 0.7291 & 0.9519 \\
      $\va_2=0.01, \tau_2=2$\footnotemark[2] & & & & & \\
    \midrule
      $\va_1=0.1, \tau_1=100,$ 	& 1.000 & 0.630 & \wh{0.863} & 0.8945 & 1.167 \\
      $\va_2=0.01, \tau_2=2, t_0=30$\footnotemark[2] & & & & & \\
    \midrule
      $\va= \pm{0.005}, \tau=3, t_0=5$\footnotemark[3] & 1.000 &  & 0.2009 & 0.2319 & 0.4572 \\ 
    \midrule
      $\va= 0.005, \tau=3, t_0=5$\footnotemark[3] 	& 1.000 &  & 0.1824 & 0.2396 & 0.5574 \\
    \midrule
      $\va_1=0.1, \tau_1=100,$ 	& 1.000 & 0.422 & \wh{0.2267} & 0.3086 & 0.3273 \\ 
      $\va_i= \pm{0.005}, \tau_i=3, t_0=9$\footnotemark[4] & & & & & \\
    \midrule
      $\va_1=0.1, \tau_1=100,$ 	 & 1.000 & 0.430 & \wh{0.2053} & 0.3076 & 0.3311 \\
      $\va_i= 0.005, \tau_i=3, t_0=9$\footnotemark[4] & & & & & \\
    \bottomrule
}

\ctable[
notespar,
caption = {Comparison of the particle number density for different sets of sampling points. In order to calculate intermediate values cubic interpolation has been used.},
cap = {Comparison of the particle number density for different sets of sampling points with cubic interpolation},
label = Tab12,
pos = h,
]{lcccccc}{
  \tnote[1]{Single pulse}
  \tnote[2]{Double pulse}
  \tnote[3]{Ten identical pulses with time lag $t_0$}
  \tnote[4]{One long pulse and ten short pulses}
  } {
    \toprule
      Field Parameters & St & Tr2(I) & 512 & 4096 & 32768 \\
      ($\va$[1], $\tau$[1/m], $t0$[1/m]) & & & Points & Points & Points \\
    \midrule
      $\va=0.1, \tau=100$\footnotemark[1] 	& 1.529 & 1.529 & 1.45 & 1.519 & 1.528 \\
    \midrule
      $\va=0.005, \tau=3$\footnotemark[1] 	 & 5.856	& 5.856 & 5.617 & 5.838 & 5.85  \\
    \midrule
      $\va_1=0.1, \tau_1=100$, 	& 4.240 & 4.240 & \wh{9.081E+03} & 3.162 & 4.221 \\
      $\va_2=0.01, \tau_2=2$\footnotemark[2] & & & & & & \\
    \midrule
      $\va_1=0.1, \tau_1=100$, 	& 4.021 & 4.021 & \wh{5.150E+03} & 2.970 & 4.00 \\
      $\va_2=0.01, \tau_2=2, t_0=30$\footnotemark[2] & & & & & & \\
    \midrule
      $\va= \pm{0.005}, \tau=3, t_0=5$\footnotemark[3] 	& 8.700 &  & 7.990 & 8.634 & 8.692 \\ 
    \midrule
      $\va= 0.005, \tau=3, t_0=5$\footnotemark[3] 	& 3.782 &  & 3.637 & 3.791 & 3.784 \\
    \midrule
      $\va_1=0.1, \tau_1=100$, 	& 1.421 & 1.421 & \wh{1.54E+01} & 1.127 & 1.415 \\ 
      $\va_i= \pm{0.005}, \tau_i=3, t_0=9$\footnotemark[4] & & & & & & \\
    \midrule
      $\va_1=0.1, \tau_1=100$, 	& 1.353 & 1.353 & \wh{4.206E+02} & 1.04 & 1.347 \\
      $\va_i= 0.005, \tau_i=3, t_0=9$\footnotemark[4] & & & & & & \\
    \bottomrule
}

\chapter{Results}
\label{sec:results}

In this chapter we present our new findings regarding particle pair production, with a focus on the following key aspects:
First an analysis of combinations of two pulses with two different time scales is done. These combination of fields have been discussed recently, but we provide an accurate perspective on setups with new combinations and also on the full momentum spectrum in comparison to these previous investigations \cite{OrtDip,PhysRevLett.101.130404,PhysRevD.85.025004}. 
Secondly an investigation of many-pulse configurations is done. In addition, the differences between fields consisting of equally and alternating signed pulses are examined.
These investigations should produce novel insights for an either direct or indirect measurement of the Schwinger effect. 
Adopting optimal control theory, we finally present results on pulse shaping in particle pair production for the first time.

\section{Dynamically Assisted Schwinger Effect}

As mentioned previously, we focus on combinations of pulses with different time scales. To this end, we employ the electric field:
\be{
E(t) = \sum_i \va_i \mathrm{sech}^2 \br{\frac{t-t_{0,i}}{\tau_i}}.
}
Before ultimately coming to the results, we shall introduce the Keldysh parameter $\gamma$. 
It is defined for a single pulse as
\be{
\gamma = \frac{\tau_T}{\tau} = \frac{m_e}{e E_{crit} \tau} = \frac{1}{m_e \tau \varepsilon},
}
where $\tau$ is the pulse length and $\tau_T$ is the characteristic time for a tunneling process from the Dirac sea to the continuum. 
The Keldysh parameter separates two different regimes. 
For a parameter value of $\gamma \ll 1$ the pair production process shows characteristics of a tunneling process and is therefore associated with the Schwinger effect. On the other hand, a parameter value of $\gamma \gg 1$ indicates that a high number of photons is absorbed. 
This parameter region is referred to as multiphoton regime as the multiphoton absorption process dominates the pair production in this case. 
It is the idea of the dynamically assisted Schwinger effect to combine pulses in the different parameter regimes.
In order to describe such a process properly, we introduce the combined Keldysh parameter $\gamma_C$, which is defined based on properties of both pulses \cite{PhysRevLett.101.130404,PhysRevD.85.025004}.
\be{
\gamma_C = \frac{1}{m_e \tau_2 \varepsilon_1}.
}
Here $\tau_2$ is the pulse length of the short pulse and $\varepsilon_1$ the maximal field strength of the long pulse. 
This enables us to characterize the field pulses in terms of one parameter as illustrated in the following figures.

In Fig.\ref{Fig1} we investigate the enhancement of particle production when combining two pulses, the first one with $\gamma_1 \ll 1$ and the second one with $\gamma_2 \gg 1$.
We further define enhancement as the ratio
\be{
\frac{N_C}{N_1 + N_2},
} 
where $N_1 ,~ N_2~$ denote the particle number in a single pulse of given parameters, whereas $N_C$ denotes the particle number for combined electric fields. 
As can be seen in Fig.\ref{Fig1} the enhancement becomes largest at a combined Keldysh parameter of $\gamma_C \approx 2$. At this point the pair production rate of
the two single pulses is of the same order. Remarkably, there seems to be no qualitative difference between calculations based on the full momentum space Fig.\ref{Fig1}a-d and calculations based only on the parallel momentum Fig.\ref{Fig1}e. 
\begin{figure}[h]
\subfigure[\label{Fig1_1}]{\includegraphics[width=0.49\textwidth]{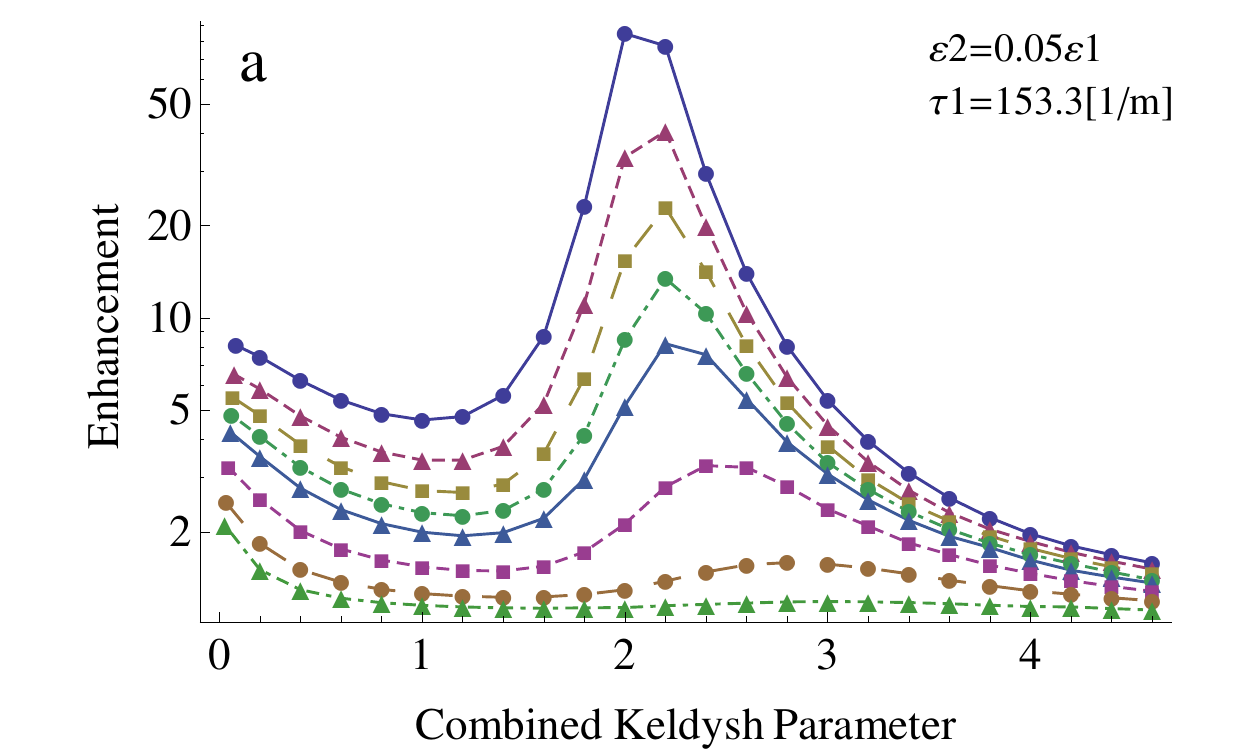}}
\hfill
\subfigure[\label{Fig1_2}]{\includegraphics[width=0.49\textwidth]{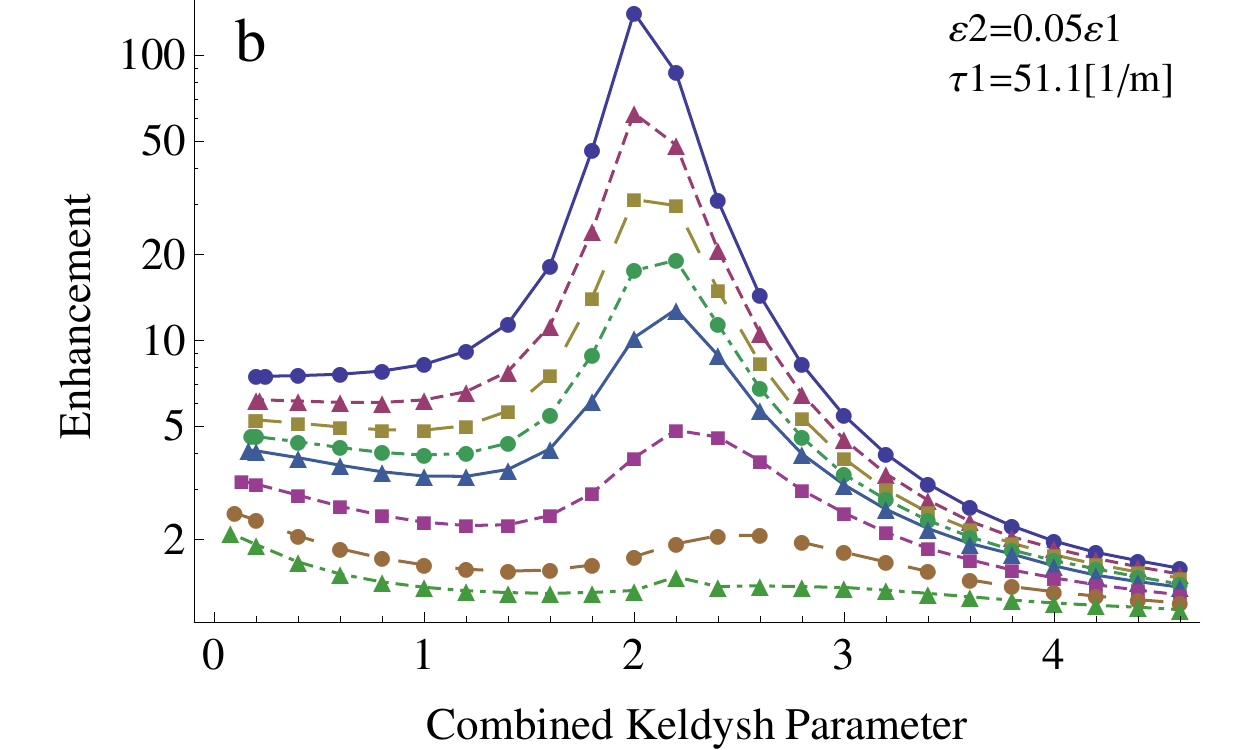}} \\[-1cm]
\subfigure[\label{Fig1_3}]{\includegraphics[width=0.49\textwidth]{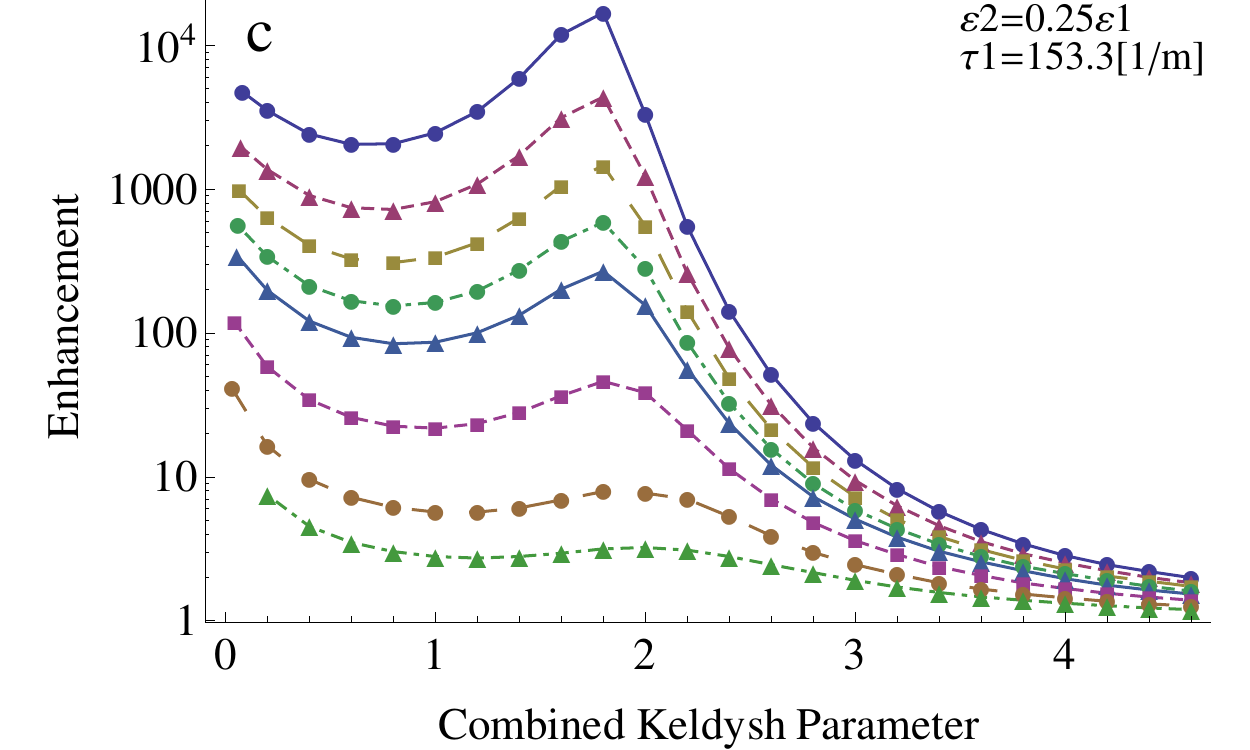}}
\hfill
\subfigure[\label{Fig1_4}]{\includegraphics[width=0.49\textwidth]{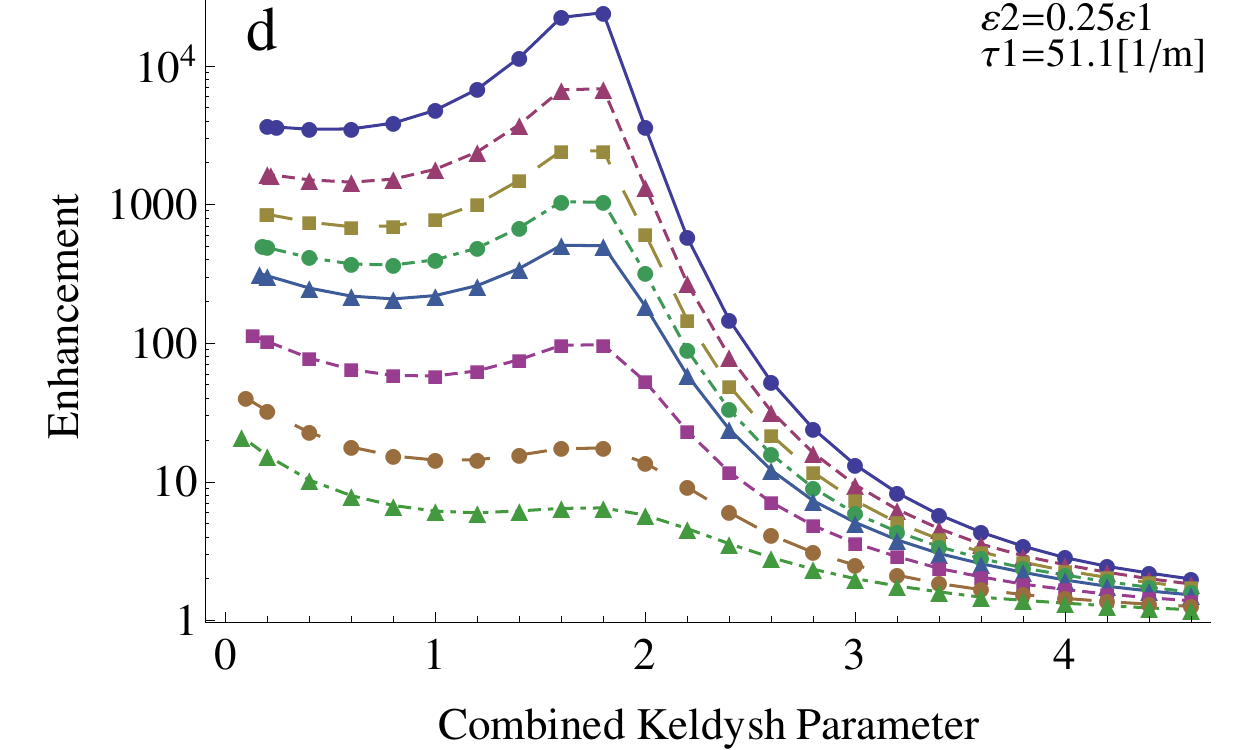}} \\[-1.5cm]
\hfill
\subfigure[\label{Fig1_5}]{\includegraphics[width=0.59\textwidth]{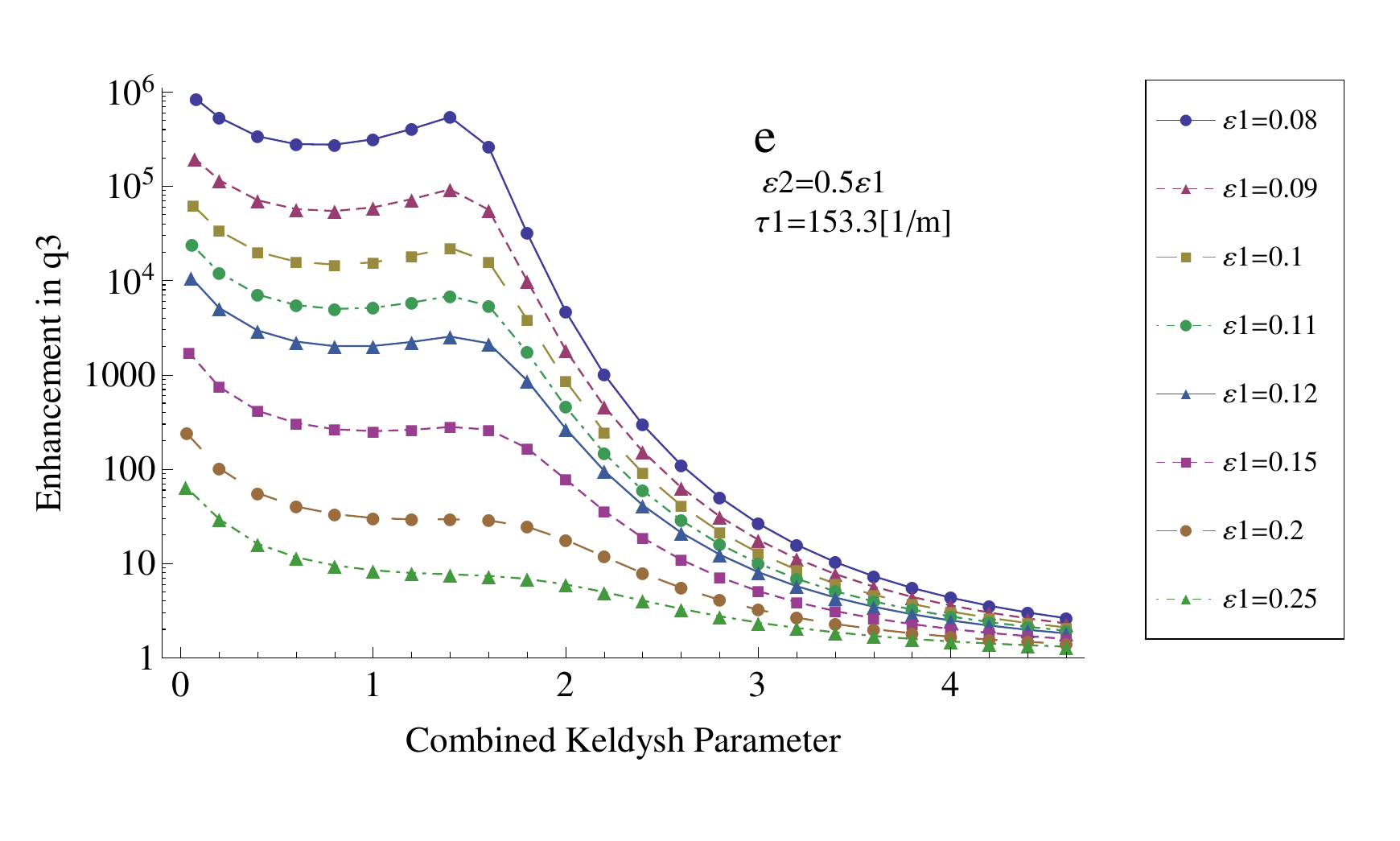}} \\[-1cm]
\caption[Enhancement of the particle number density due to dynamical assistance]{Enhancement of the particle number density due to dynamical assistance. 
The calculations in $~a-d~$ are based on full momentum space calculations whereas the calculation in $~e~$ is done in the parallel direction only. The offset between the two pulses is $~t_0=0$.
The absolute field strength is varied in each plot line. 
The same pulse length $~\tau_1~$ is used in $~a ,~ c~$ and $~b ,~ d$, respectively, whereas the field strength ratio $~\varepsilon_2 / \varepsilon_1~$ is the same in $~a ,~ b~$ and $~c ,~ d$, respectively.}
\label{Fig1}
\end{figure}
 
\begin{figure}[h]
\centering	
  \includegraphics[width=0.85\textwidth]{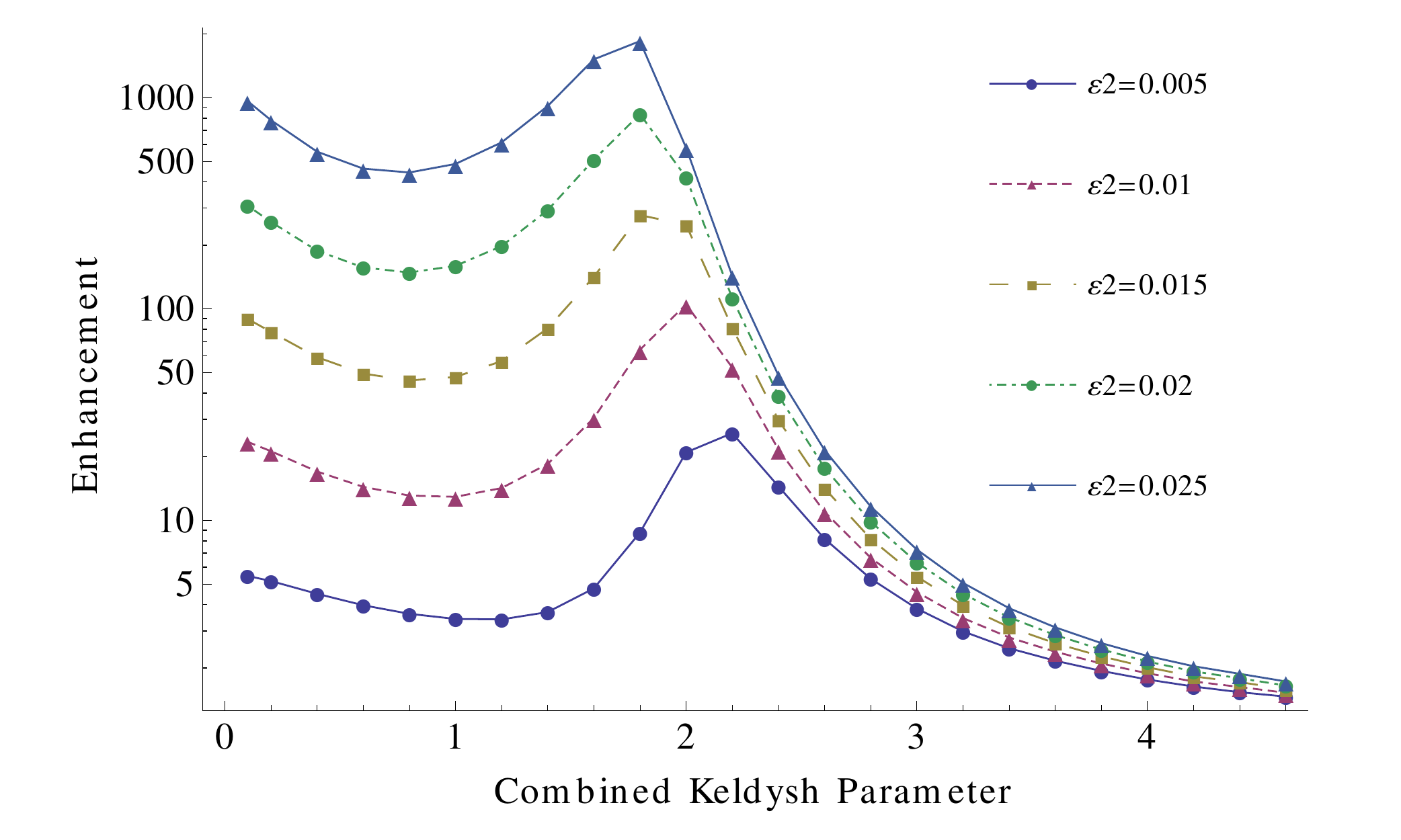}
  \caption[Enhancement of the particle number density for two pulses based on a full momentum space calculation]{Enhancement of the particle number density based on a full momentum space calculation. A short pulse is assisting a long pulse with an offset of $t_0=0$. The field parameters of the short pulse are varied while the parameters for the long pulse are fixed at $~\varepsilon_1 = 0.1,~ \tau_1 = 102.2m^{-1}$.}
  \label{Fig2}
\end{figure}

In the next calculation, illustrated in Fig.\ref{Fig2}, we varied the field strength of the short pulse, while keeping the parameters for the long pulse fixed at $\varepsilon_1 = 0.1,~ \tau_1 = 102.2m^{-1}$.
Again, the effect from the dynamical assistance of photon absorption is most pronounced at $\gamma_C \approx 2$. The highest relative enhancement effect can be seen for the configuration $\va_2=0.005$. 
On the other hand, the highest absolute values are obtained for $\varepsilon=0.025$. The increase in the absolute value of the enhancement for increasing values of
$~\varepsilon_2~$ is simply explained with the higher overall field strength in combining the two pulses.

\newpage

\begin{figure}[h]
\centering	
  \includegraphics[width=0.85\textwidth]{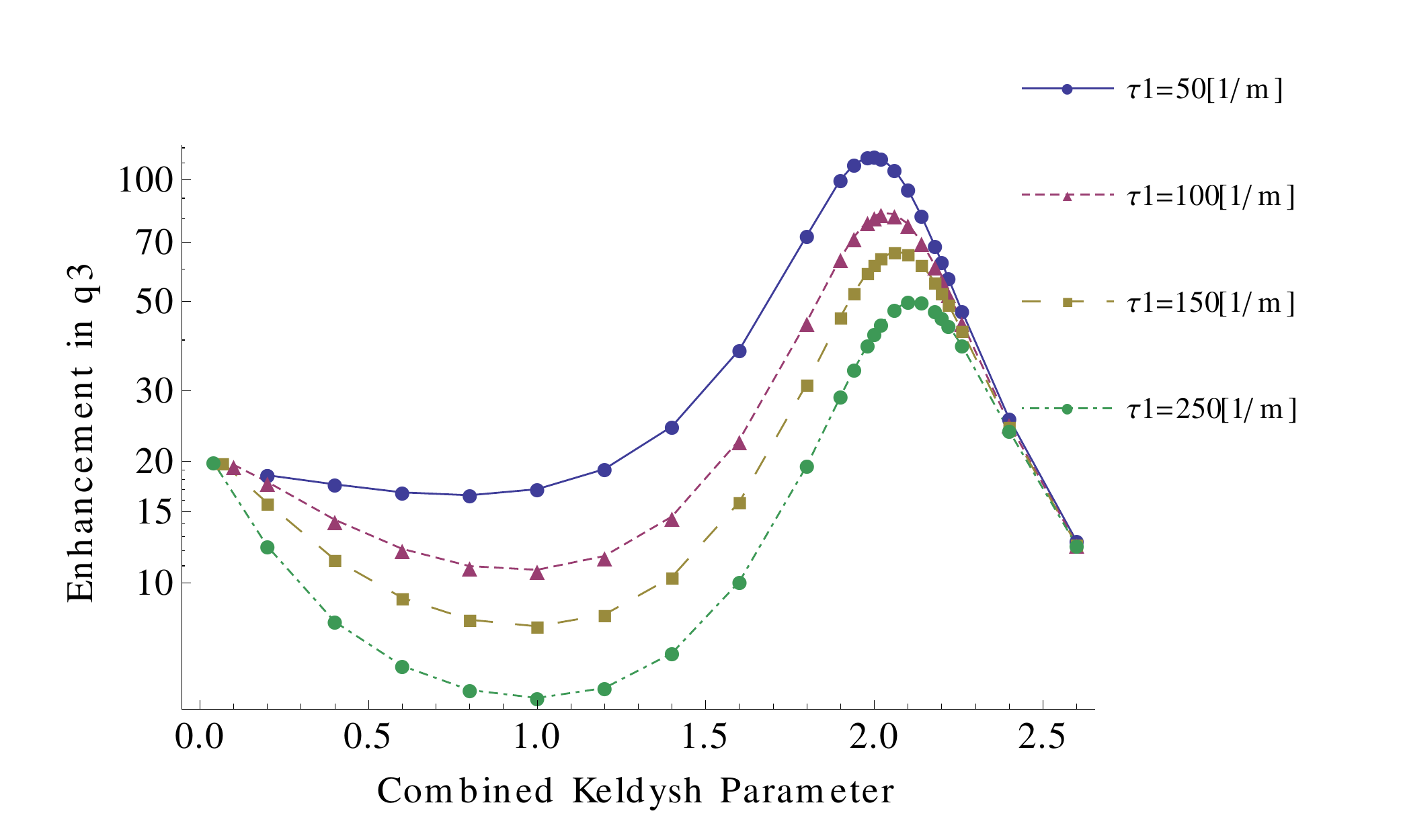}
  \caption[Enhancement of the particle number density by a short pulse assisting a long pulse]{Enhancement of the particle number density based on a calculation in the parallel direction only. The offset between the two pulses is $t_0=0$. The field strengths are  $~\varepsilon_1=0.1~$ for the long pulse and $~\varepsilon_2=0.01~$ for the short pulse.}
  \label{Fig2b}
\end{figure}

Subsequently, we kept the field strength constant at $\varepsilon_1=0.1 ,~ \varepsilon_2=0.01$ and varied just the pulse lengths. The results obtained from these calculations are shown in Fig.\ref{Fig2b}. Interestingly, all lines meet at $\gamma_C > 2.2$ whereas the influence of the different pulse lengths can be observed in the region $0 < \gamma_C < 2$. 
The limit $\gamma_C \to 0$ corresponds to a situation where the first pulse is in the regime $\gamma_C \ll 1$ whereas the second pulse becomes adiabatically constant. Thus the outcome corresponds more or less to 
the outcome obtained from a single pulse with pulse strength $\varepsilon_1 + \varepsilon_2$. In the limit $\gamma_C > 2.2$ the particle production due to the second pulse $\gamma_2 \gg 1$
starts to dominate. In this limit, the effect of the first pulse $\gamma_1 \ll 1$ becomes effectively negligible.

\newpage

\begin{figure}[h]
\subfigure[]{\includegraphics[width=0.49\textwidth]{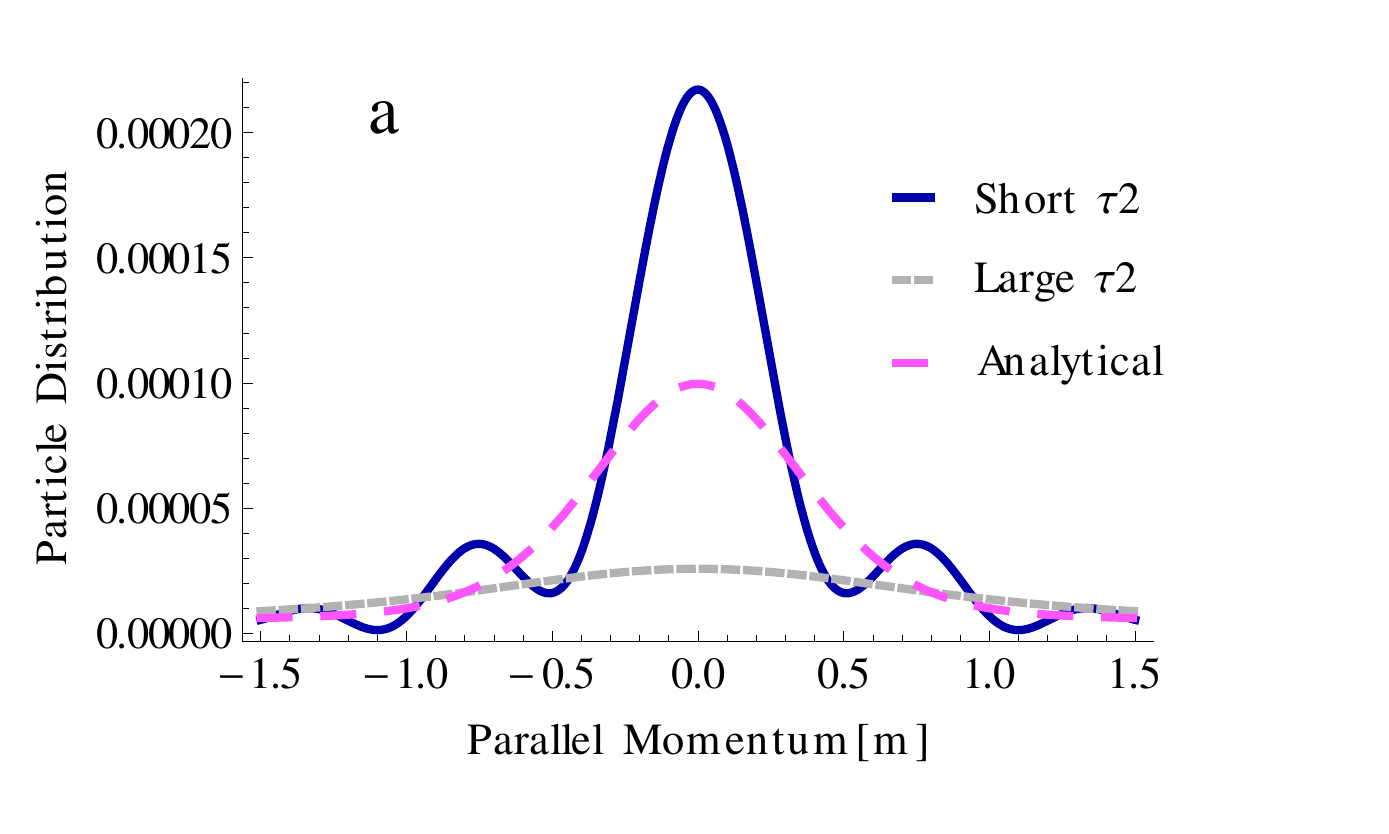}}
\hfill
\subfigure[]{\includegraphics[width=0.49\textwidth]{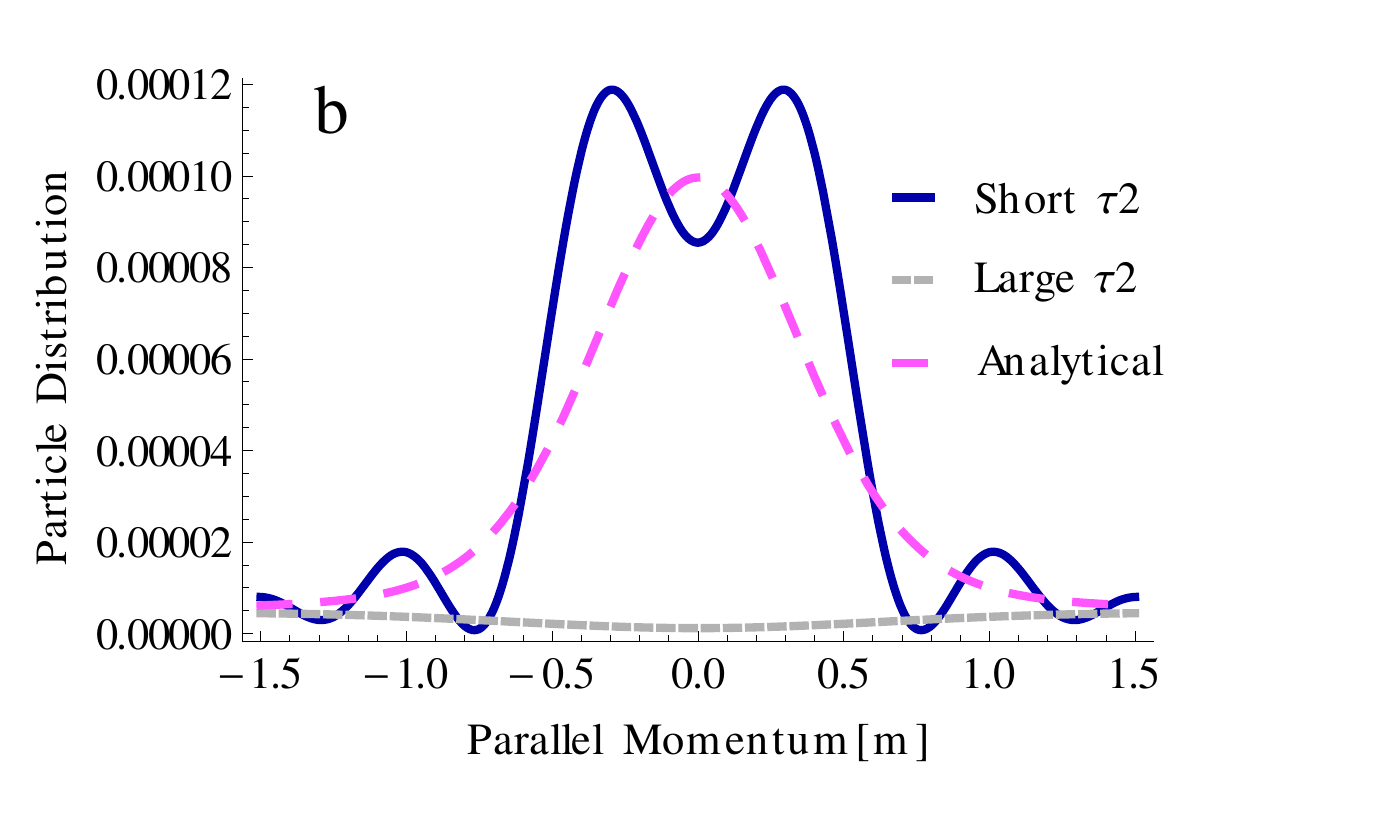}} \\[-1.5cm]
\caption[Oscillations in the particle distribution function caused by combining two pulses with different pulse scales]{Oscillations in the particle distribution function caused by combining two pulses with different pulse scales. We assume $~\varepsilon_1 = 0.25,~ \tau_1 = 50m^{-1}$ for the long pulse. The field strength of the short pulse is $~\varepsilon_2 = 0.025~$ whereas different time scales $~\tau_2 = 1m^{-1}$ or $~\tau_2=5m^{-1}$ have been used. The offset between the two pulses is $~t_0=0$. The dashed line is the simple sum of the long and the shorter second pulse. In $~a~$ the two fields strengths have the same sign whereas they have opposite sign in $~b$.}
\label{Fig3}
\end{figure}

Finally, we also investigate the particle distribution in Fig.\ref{Fig3}. 
We have used a long pulse with parameters $\varepsilon_1 = 0.25,~ \tau_1 = 50m^{-1}$ combined with a short pulse $\varepsilon_2 = 0.025,~ \tau_2 = 1m^{-1}$ and $\varepsilon_2 = 0.025,~ \tau_2 = 5m^{-1}$, respectively. 
For a single pulsed field $\gamma_1 \ll 1$ no oscillations in the distribution function can be observed. This is even true for configurations of two pulses when $\tau_2$ is not too short.
For the even shorter multiphoton pulse $\tau_2=1m^{-1}$, however, the distribution is deformed leading to more than one local maximum. 
Moreover an interesting behaviour of the particle distribution is observed when changing the sign of the second pulse $\varepsilon_2 \to -\varepsilon_2$. This effect will be discussed in more detail in the following sections.

So far we have seen that one additional short pulse can significantly change the distribution and the number density of the created particles. We proceed by investigating configurations with more than one short pulse. 
To be specific, we will investigate dynamical assistance of a long pulse by a train of short pulses as well as the resulting interference pattern in the momentum distribution.

\begin{figure}[h]
\subfigure[]{\includegraphics[width=0.49\textwidth]{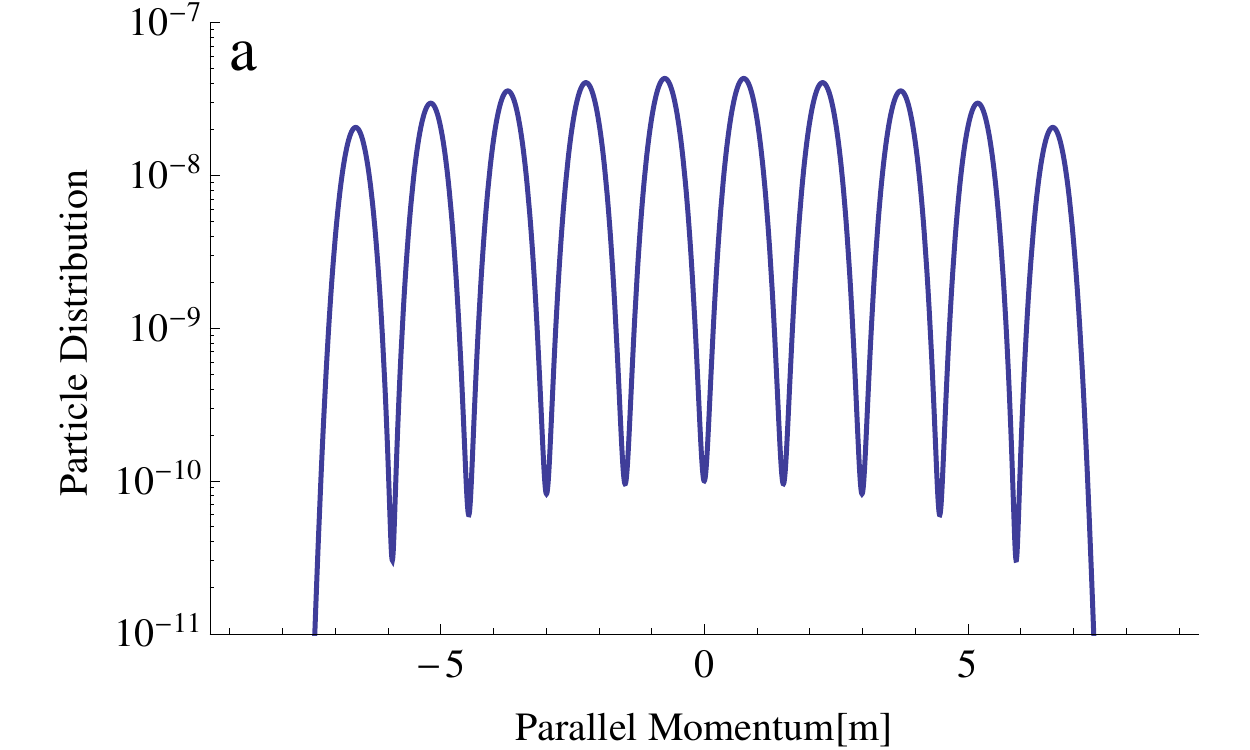}} 
\hfill
\subfigure[]{\includegraphics[width=0.49\textwidth]{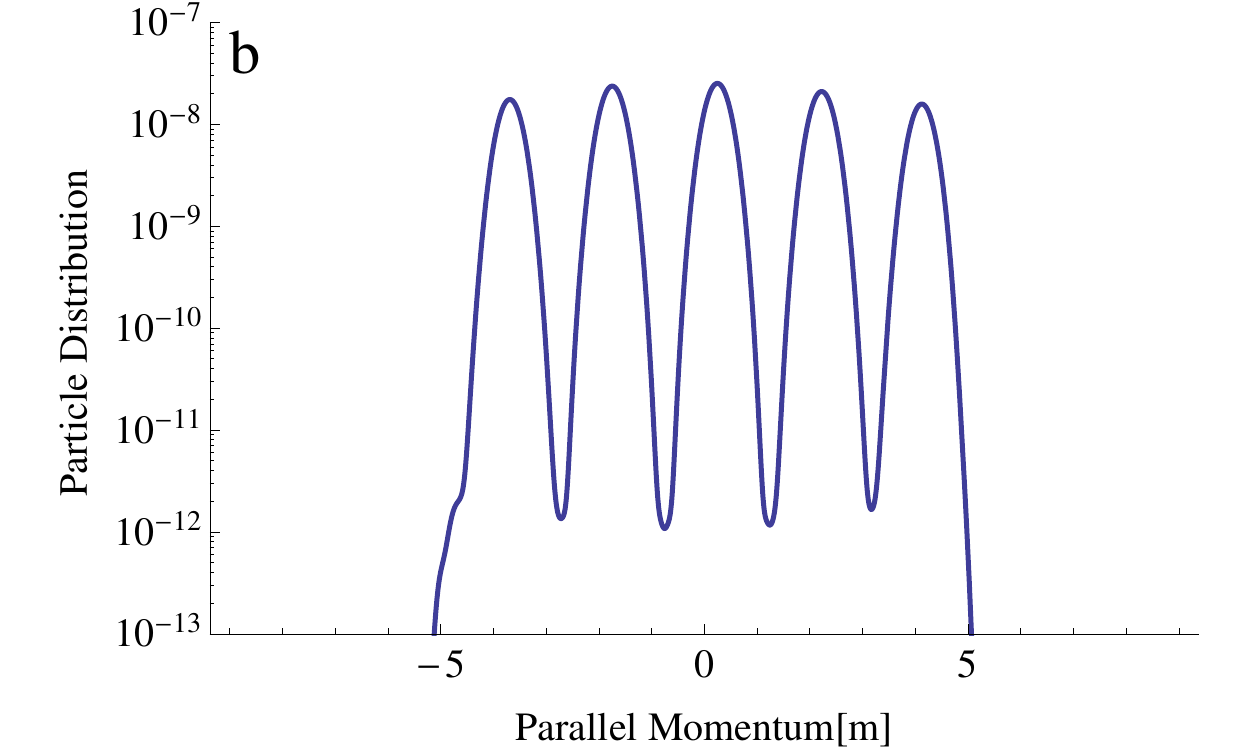}} \\[-1cm]
\caption[Signals for multiphoton pulses in the particle distribution function]{Particle distribution function for a configuration with a long pulse being assisted by ten evenly spaced short pulses. Their field parameters are $~\varepsilon_1 = 0.1,~ \tau_1 = 150m^{-1}$, $~\bet{\varepsilon_2} = 0.05,~ \tau_2 = 5m^{-1}$ and the time lag is chosen to be $~t_0 = 10m^{-1}$. We present the parallel configuration in $~a~$ and the alternating configuration in $~b$.}
\label{Fig4}
\end{figure}

We present results on the momentum distribution for a configuration with one long pulse $\varepsilon_1 = 0.1,~ \tau_1 = 150m^{-1}$ and ten identical small pulses with $\varepsilon_2 = 0.05,~ \tau_2 = 5m^{-1}$ separated by a time lag of $10m^{-1}$ in Fig.\ref{Fig4}.
Note that this is only a shorthand notation. Explicitly, it means that the first small pulse has a time lag of $t_{0,1}=-4.5t_0$ and the last small pulse has a time lag of $t_{0,10}=4.5t_0$.
Moreover, we investigate configurations where the small pulses have the same sign as the long pulse (denoted as parallel) and alternating signs (denoted as alternating), respectively.
In fact one can observe the effect due to each small pulse in the parallel configuration in Fig.\ref{Fig4}a. For the alternating configuration, however, only the pulses
parallel to the long pulse yield an enhancement as seen in Fig.\ref{Fig4}b. Again, this effect is due to the dynamical assistance, as each photon absorption boosts the particle production only in a small momentum range.
Furthermore, we observe that the two different configurations differ by the range occupied in momentum space.
This might be due to the fact that all small pulses act accelerating in the parallel configuration whereas half of the pulses also act decelerating in the alternating configuration.
Finally, the particle number at its local minima shows different characteristics as well: They increase in the center for the parallel configuration whereas they decrease in
the center for the alternating configuration. Unfortunately, we could not find a simple explanation for this effect.

%

\begin{figure}[h]
\subfigure[]{\includegraphics[width=0.49\textwidth]{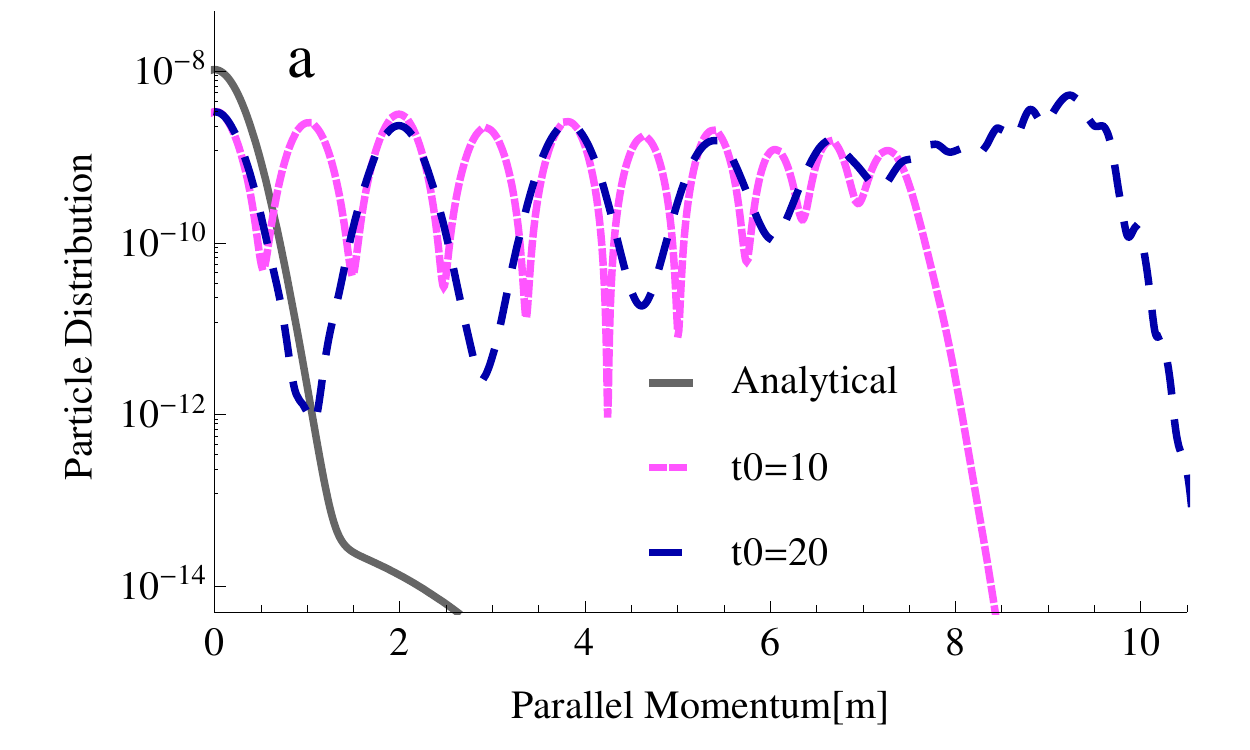}} 
\hfill
\subfigure[]{\includegraphics[width=0.49\textwidth]{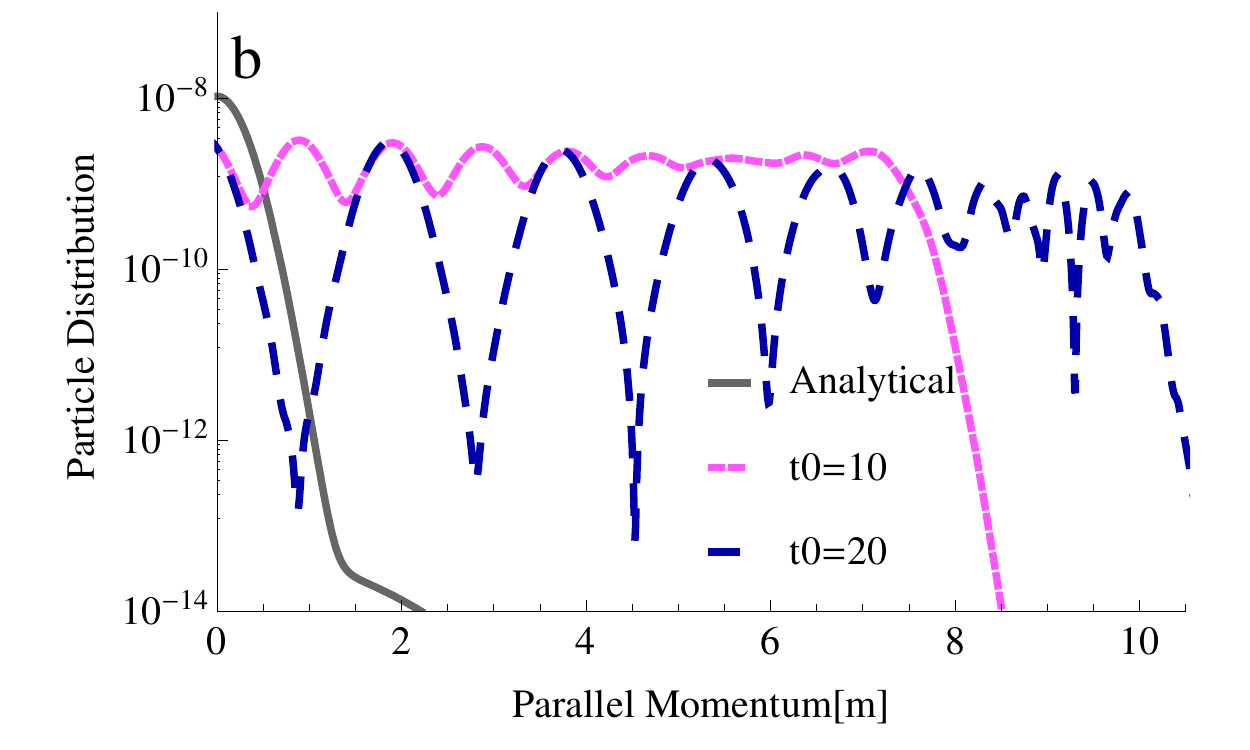}} \\[-1cm]
\caption[Particle distribution function for a configuration with a long pulse being assisted by ten evenly space short pulses.]{Particle distribution function for a configuration with a long pulse being assisted by ten evenly space short pulses. 
The characteristic field parameters for the long pulse are $~\varepsilon_1 = 0.1,~ \tau_1 = 100m^{-1}$ and for the $~10~$ evenly spaced short pulses they are $~\varepsilon_2 = 0.005,~ \tau_2 = 3m^{-1}$.
The first small pulse has a time lag of $t_{0,1}=0$. Again we present the parallel configuration in $~a~$ and the alternating configuration in $~b$. The results are compared with a field consisting of all pulses with positive field strengths placed on top of each other as shown by the gray line.} 
\label{Fig5}
\end{figure}

\newpage
In Fig.\ref{Fig5} we present the results for a setup with one long pulse $\varepsilon_1 = 0.1,~ \tau_1 = 100m^{-1}$ and ten short pulses $\varepsilon_2 = 0.005,~ \tau_2 = 3m^{-1}$. Again, the ten short pulses are evenly spaced, however, the pulse train is only switched on when the long pulse reaches its peak value. Thus, $t_{0,1} = 0$ and $t_{0,i} > 0$ is true. 
Consequently interferences can be seen in the particle distribution function also for higher parallel momenta in Fig.\ref{Fig5}. 

\section{Interferences in Many Pulse Configurations}

In this section we want to analyze the ideas presented in \cite{PhysRevLett.108.030401} within the quantum kinetic approach. In this paper, it has been suggested that setups with alternating fields show characteristics of a Ramsey interferometer. We have performed the corresponding computations and additionally examined configurations consisting of equally signed pulses. 

\begin{figure}[h]
\subfigure[]{\includegraphics[width=0.49\textwidth]{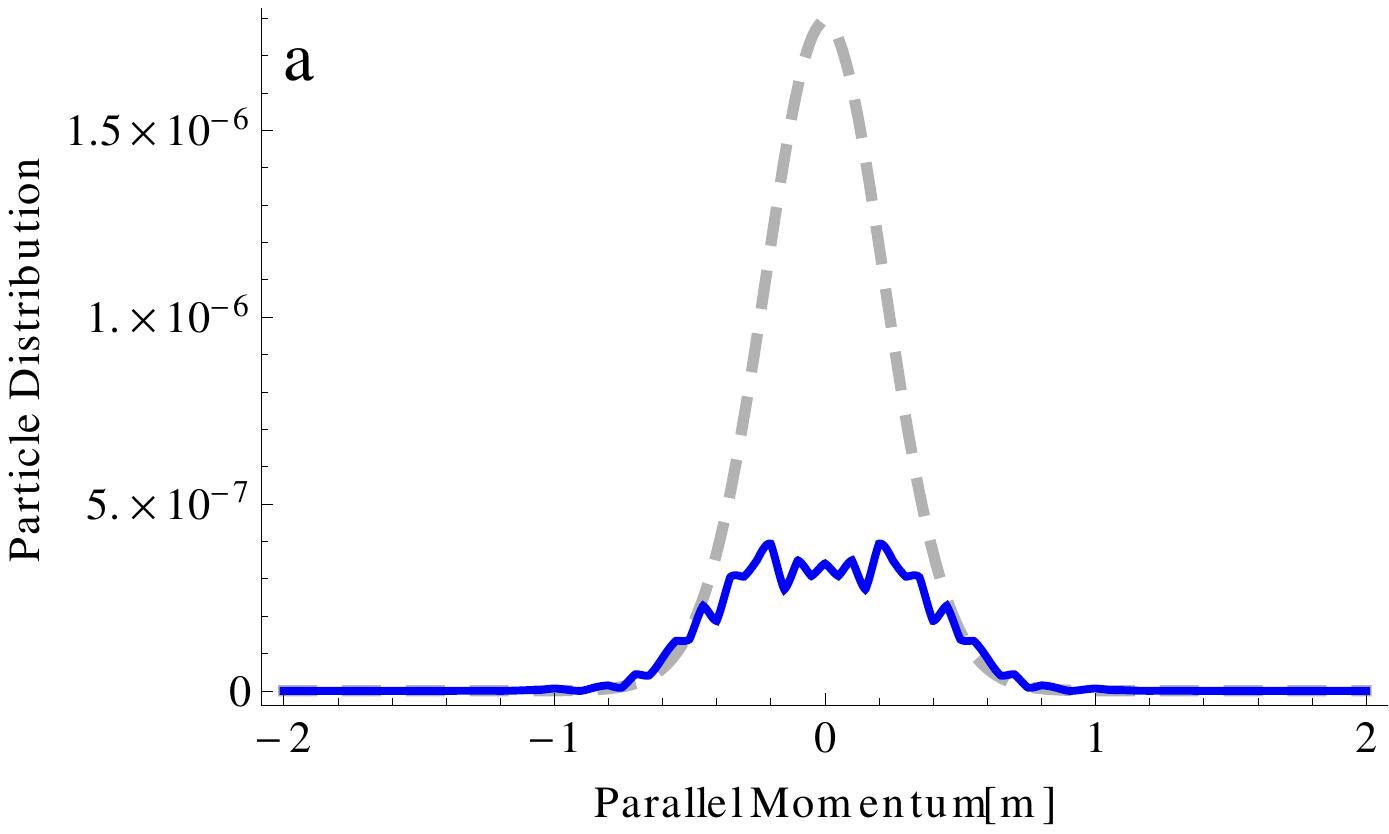}} 
\hfill
\subfigure[]{\includegraphics[width=0.49\textwidth]{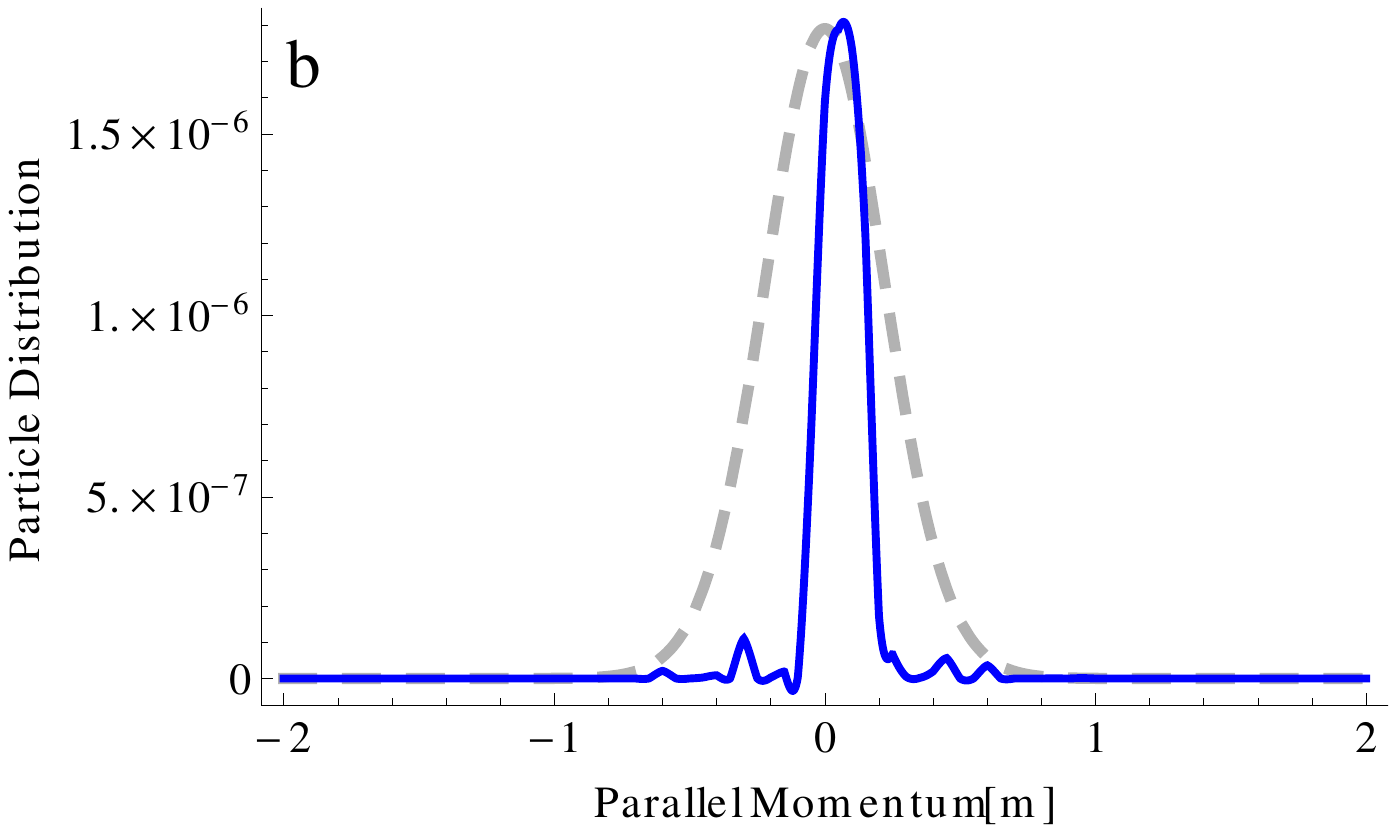}} \\[-1cm]
\caption[Positive interference effects in particle distribution for ten short pulses]{Positive interference effects in particle distribution for $~N=10~$ short pulses with field parameters are $~\varepsilon = 0.02,~ \tau = 3m^{-1}$. 
For comparison, we plot $N^2$ times the particle distribution function of a single pulse with the same parameters. We present the parallel configuration in $~a~$ 
and the alternating configuration in $~b$.}
\label{Fig6}
\end{figure}

The particle distribution function for both cases is shown in Fig.\ref{Fig6}. We consider $N=10$ equally spaced pulses with parameters $\varepsilon = 0.02,~ \tau = 3m^{-1}$ and a specific time lag. Most notably, the two configurations show a very different behaviour:
In the parallel configuration we obtain a width in momentum space corresponding to $N^2$ times the particle distribution function of a single pulse with the same parameters as shown in Fig.\ref{Fig6}a. For the alternating configuration, however, one obtains the corresponding maximum magnitude as shown in Fig.\ref{Fig6}b. Note, that the exact form of the distribution function strongly depends on the choice of the time lag $t_0$.
These findings further support the idea of choosing pulses with an adjusted time lag to enhance the pair production probability substantially.

\begin{figure}[h]
\centering	
  \includegraphics[width=0.55\textwidth]{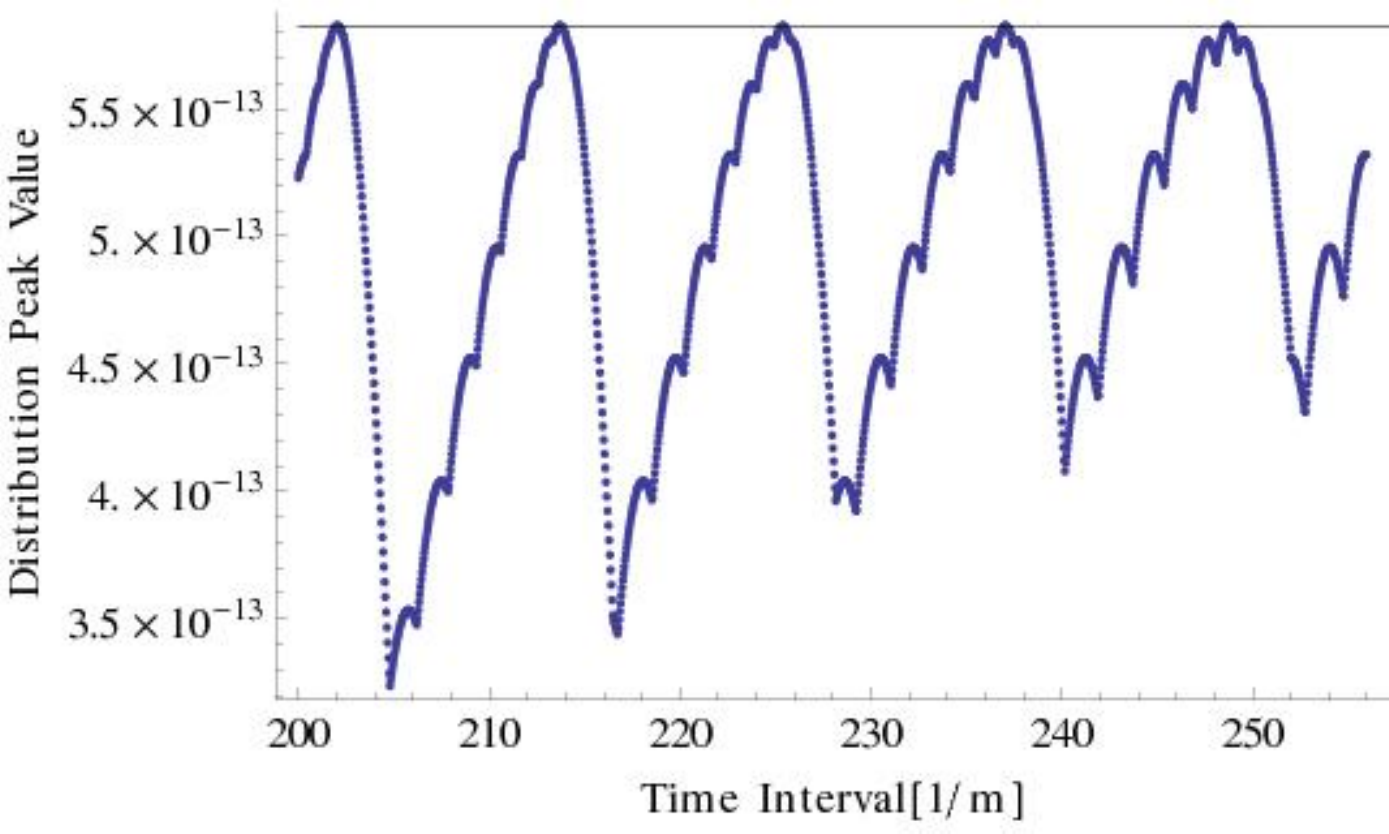}
  \caption[Behaviour of the peak value of the particle distribution function]{Behaviour of the peak value of the particle distribution function for $~N=2~$ short pulses with parameters $~\varepsilon_1 = 0.1,~ \tau_1 = 25m^{-1}$ and $~\varepsilon_2 = -0.1,~ \tau_2 = 25m^{-1}$. The maximal value of the distribution function is shown as a function of the time lag and compared with $~N^2~$ times the peak value of the distribution function of a single pulse with the same parameters.}
  \label{Fig7}
\end{figure}

Similarly to \cite{PhysRevLett.108.030401} we have also calculated the peak value of the particle distribution function for an alternating field configuration consisting of $N=2$ short pulses with parameters $~\varepsilon = 0.1,~ \tau = 25m^{-1}$. The results are illustrated in Fig.\ref{Fig7} and are consistent with the previous discussion.

\section{Oscillations in Particle Density}

Based on the discussion of the previous section, it has been our main goal to combine interference effects with the dynamical assistance effect. 
Obviously, it is unrealistic to do calculations for the whole configuration space, so we have computed only the most intuitive setups. 
Accordingly, we will examine the dynamical assistance of a long pulse with a train of short pulses in both the parallel and the alternating configuration.
Finally, an additional antisymmetric configuration has been used for comparison. This setup has been tweaked such that one has only parallel pulses in the beginning and anti-parallel pulses in the end.

\begin{figure}[h]
\subfigure[]{\includegraphics[width=0.49\textwidth]{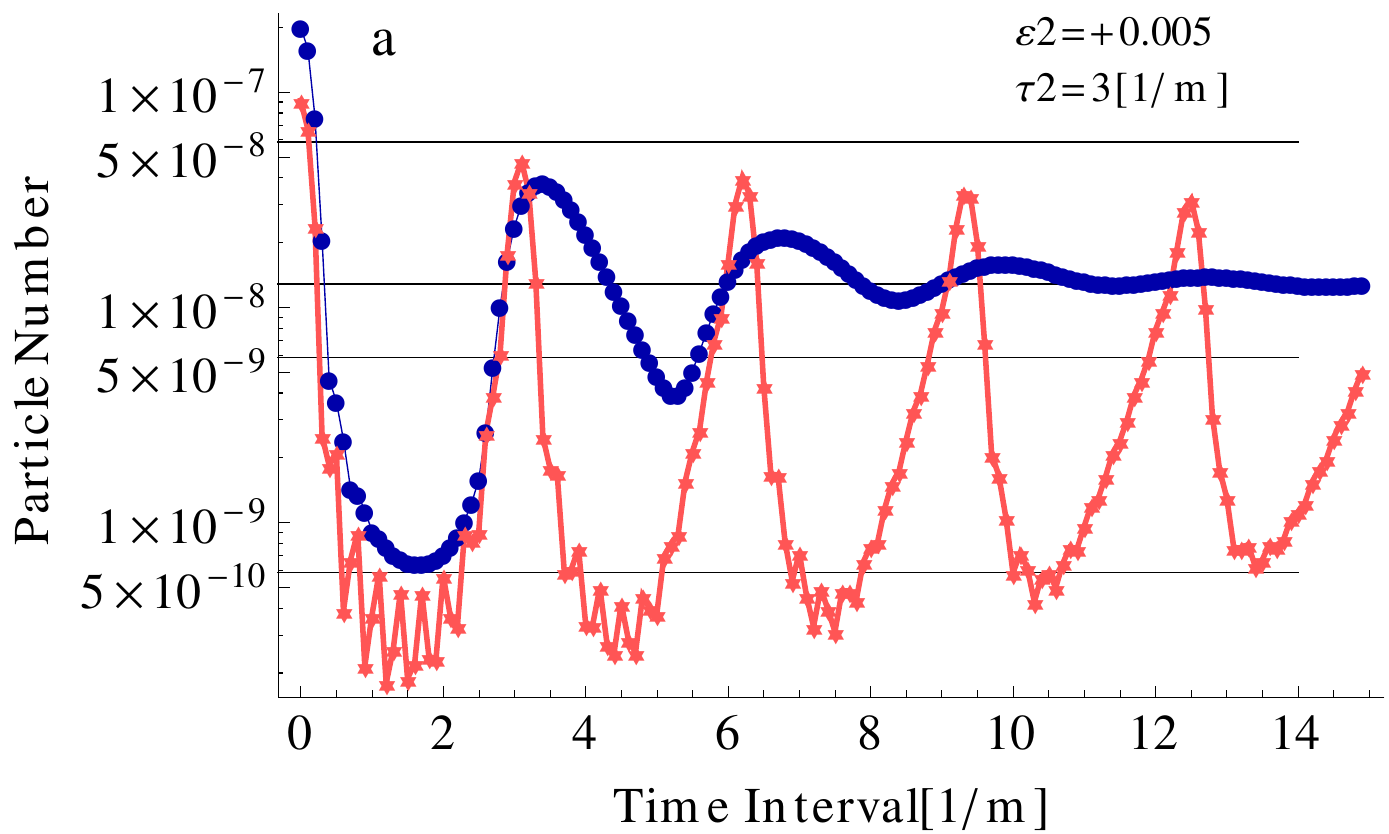}} 
\hfill
\subfigure[]{\includegraphics[width=0.49\textwidth]{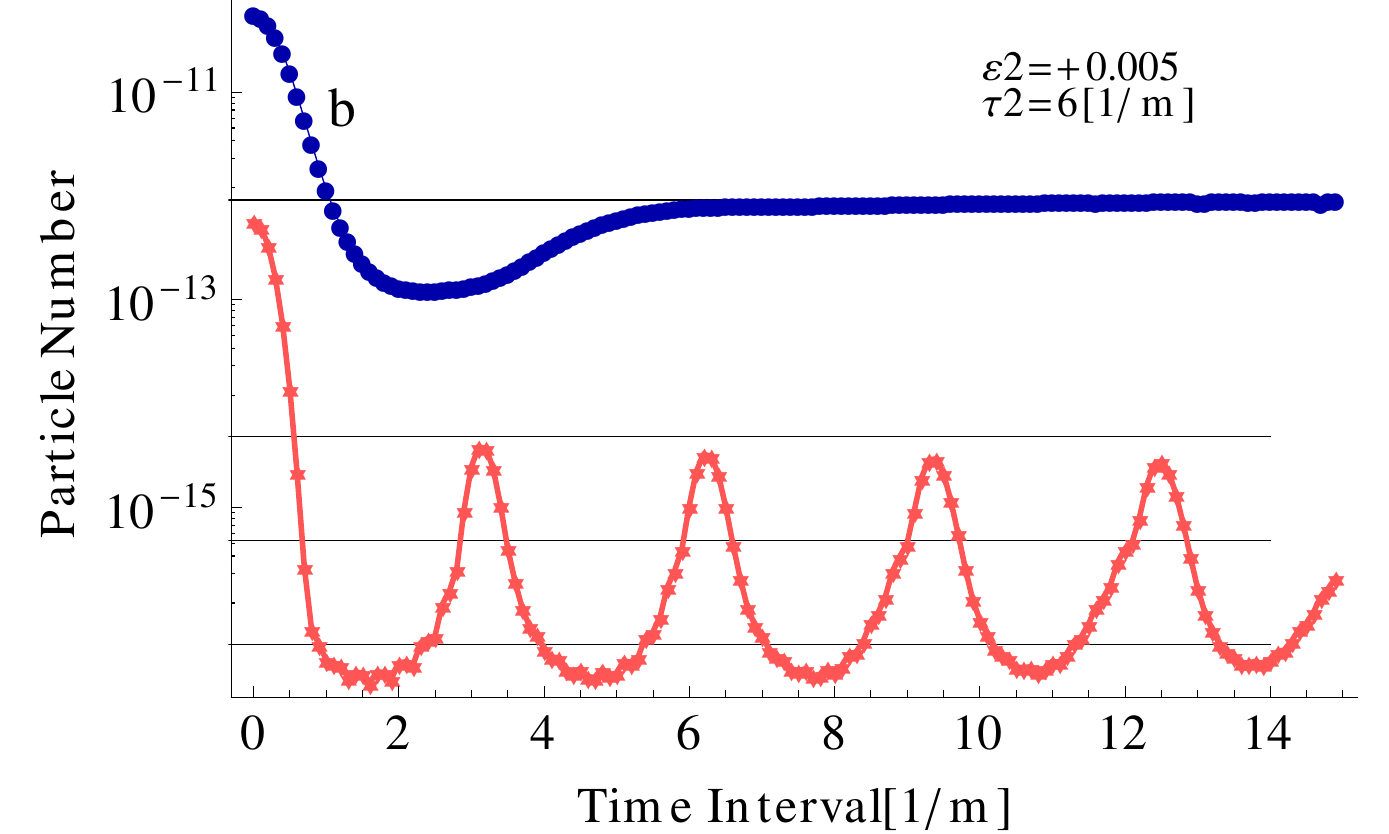}} \\[-1cm]
\subfigure[]{\includegraphics[width=0.49\textwidth]{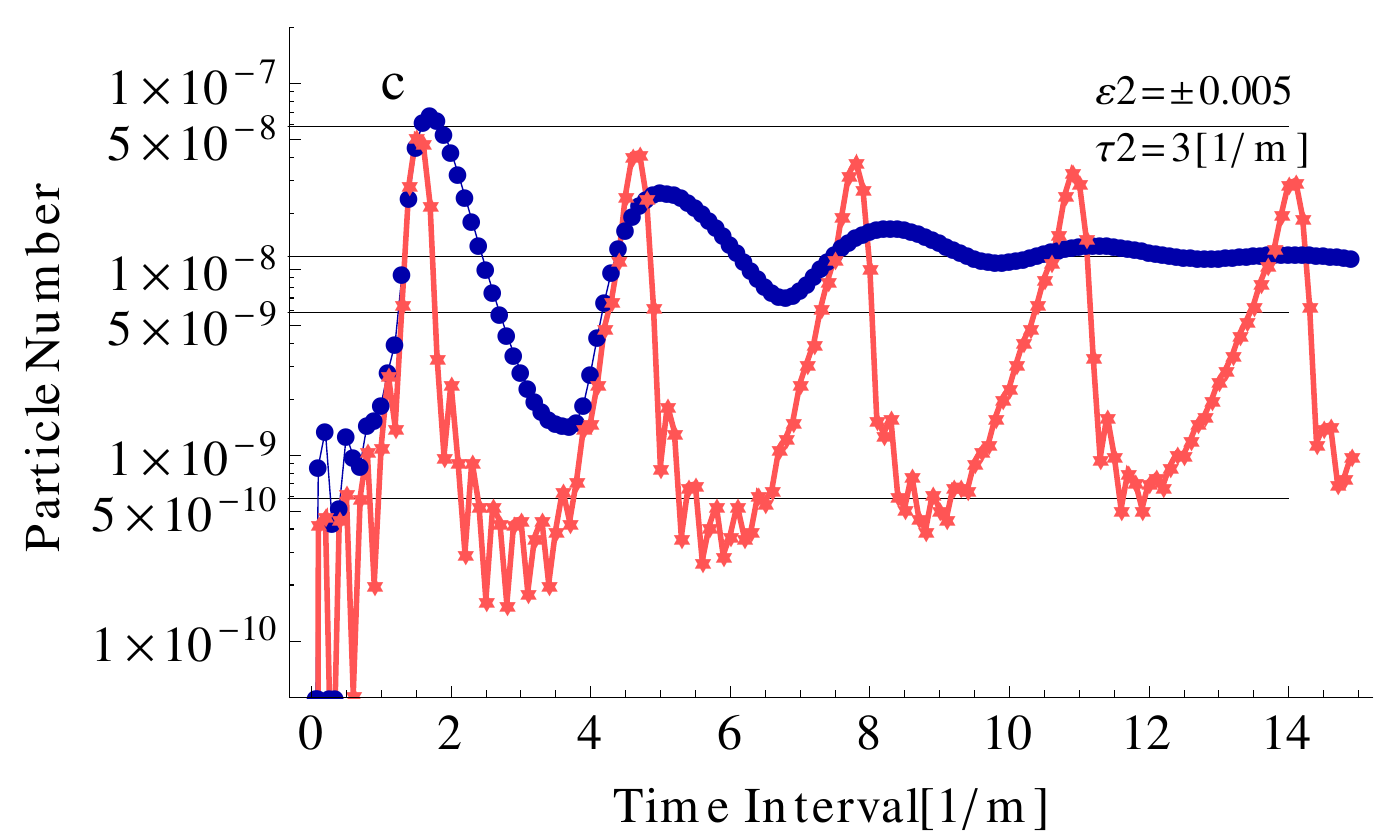}}
\hfill
\subfigure[]{\includegraphics[width=0.49\textwidth]{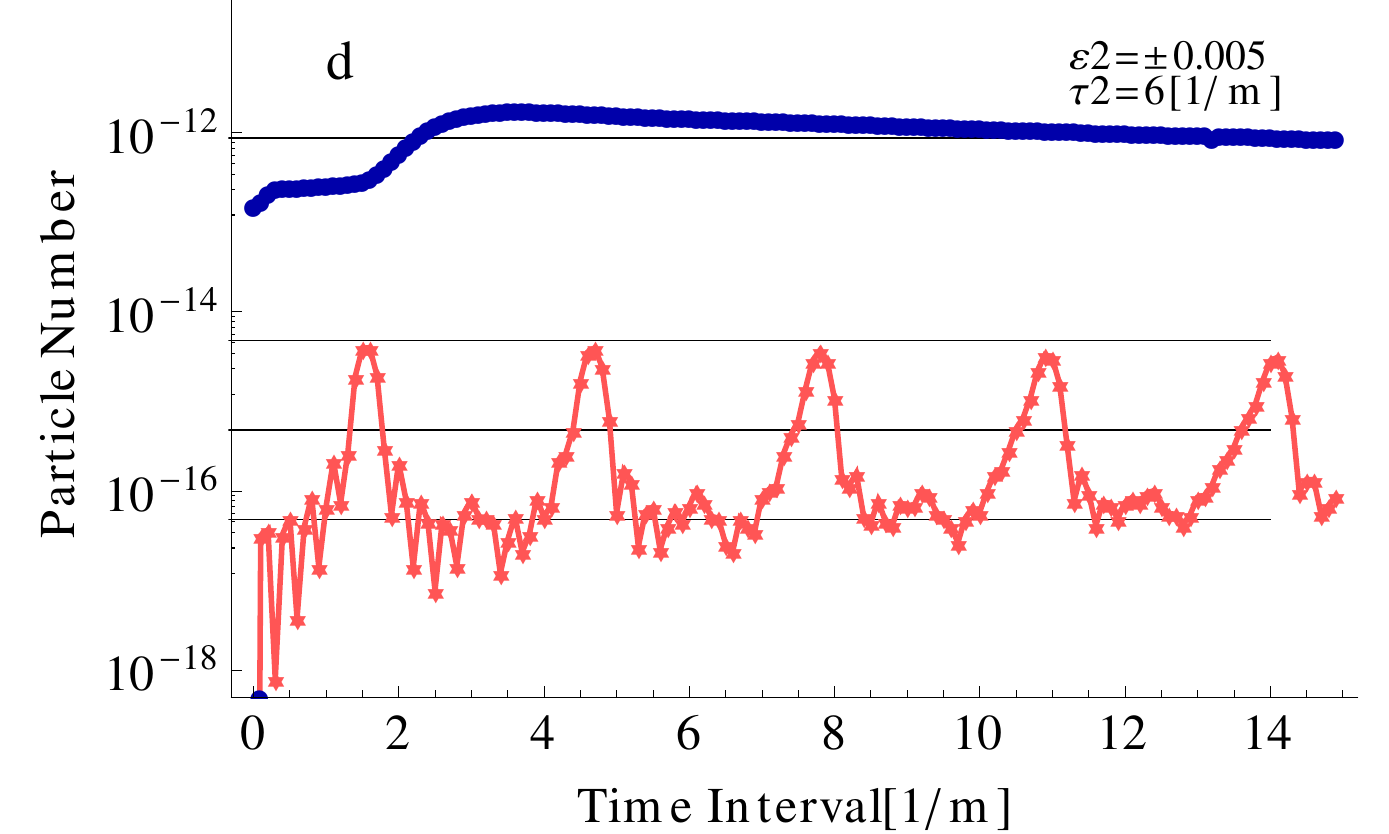}} \\[-1cm]
\subfigure[]{\includegraphics[width=0.49\textwidth]{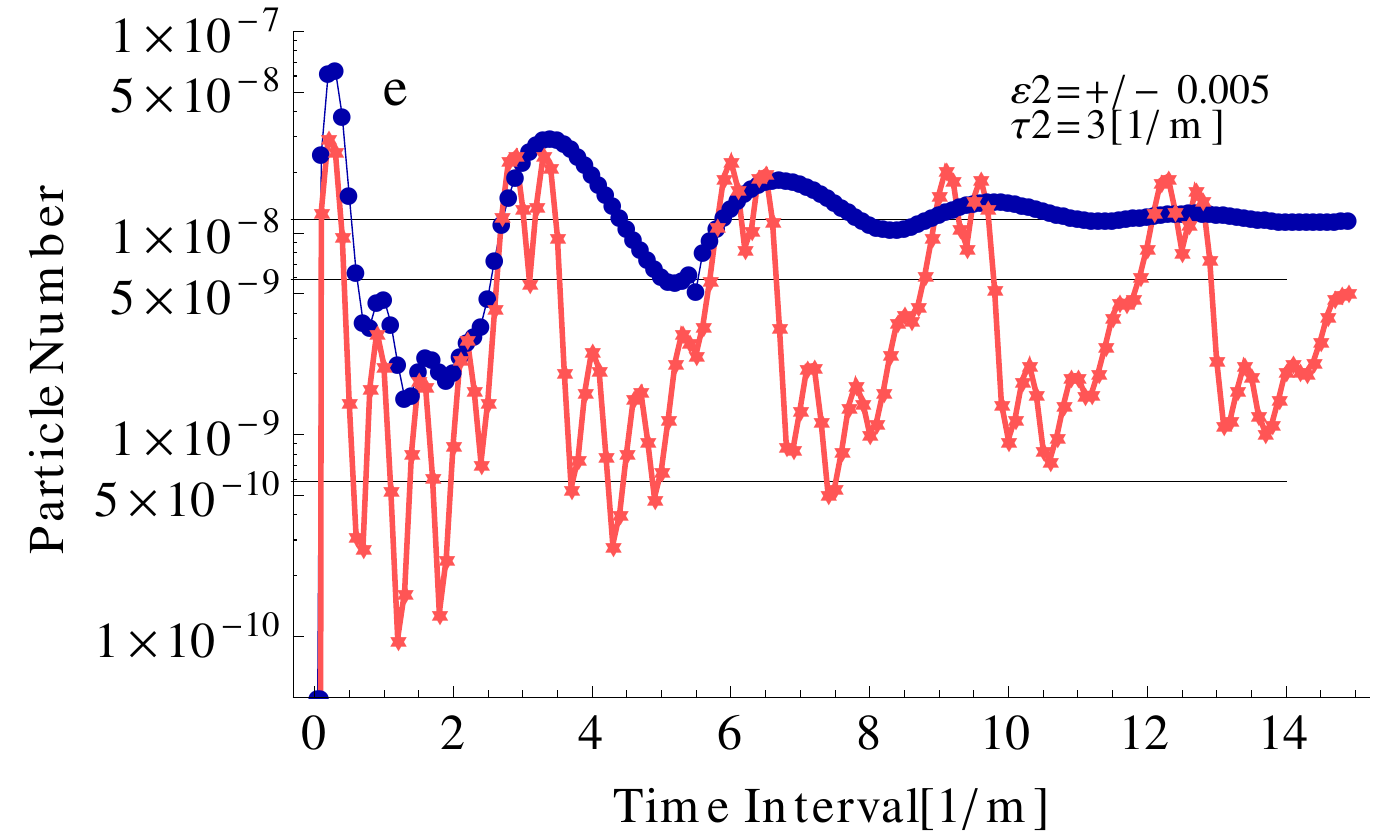}} 
\hfill
\subfigure[]{\includegraphics[width=0.49\textwidth]{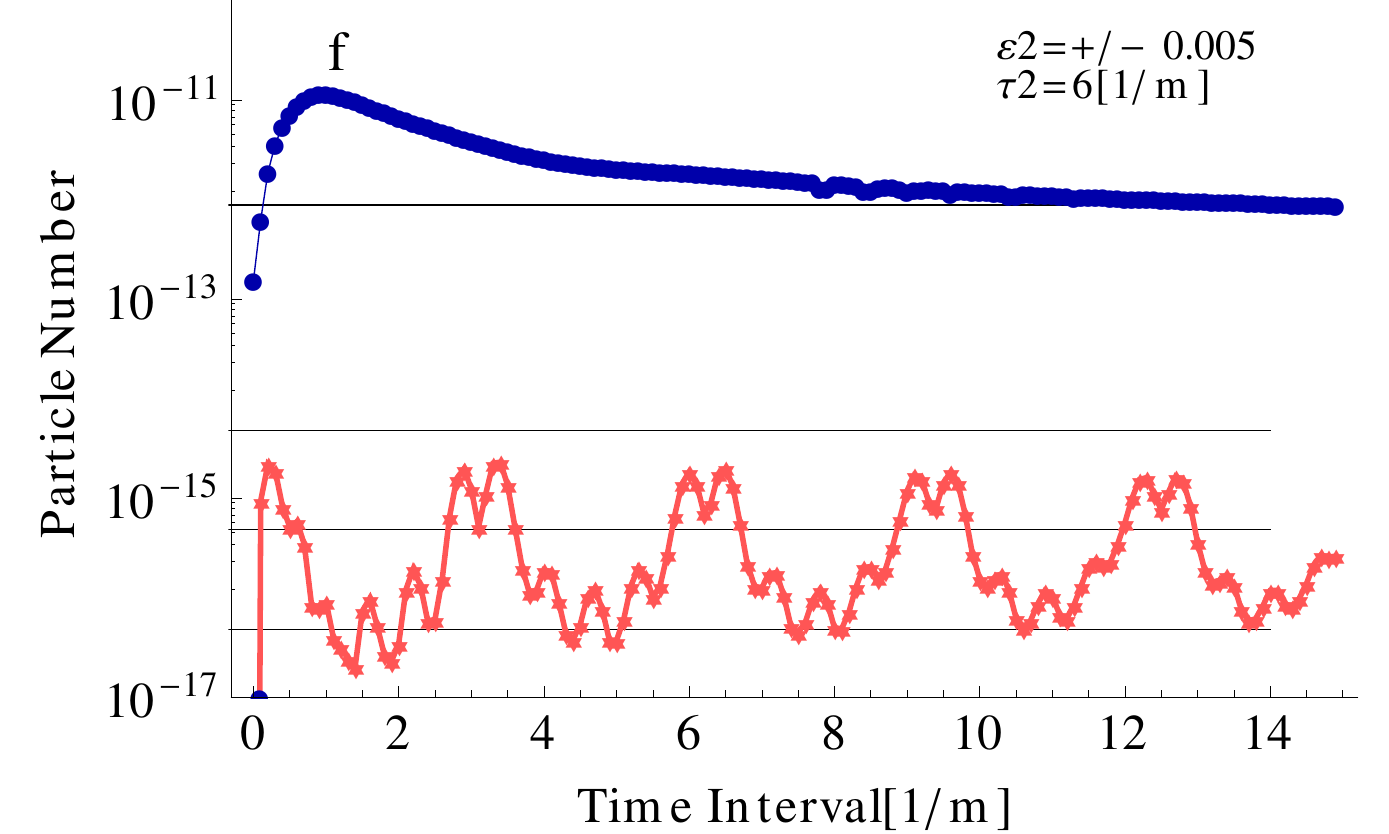}} \\[-1cm]
\caption[Importance of the time lag between short pulses]{Particle number as function of the time lag $t_0$ between $~N=10~$ short pulses with pulse length $\tau_2$ assisting a single long pulse. The thick blue line shows the particle number for the superposition of a single long pulse with the pulse train whereas the red line shows the particle number for the pulse train only. The maximal field strength is limited to $~\varepsilon_{max}=0.1~$ and the maximal energy to $~1.33m$. We present results for the parallel configuration in $~a,b~$ and for alternating fields in $~c,d$. In $~e,f~$ we started with positive fields and ended with negative fields. For a better orientation the values of $~1 ,~ 10 ,~ 100~$ times the particle number of a single small pulse as well as the results for asymptotic time lags are shown.}
\label{Fig8}
\end{figure}

The results presented in Fig.\ref{Fig8} exhibit several surprising details of electron positron pair production. 
In this investigation, we defined a maximal field strength $\varepsilon_{max} = 0.1$ and a maximal energy of the configuration $E_n = 1.33m$ to avoid that field strength effects spoil the result.
Moreover, we consider pulse trains consisting of $10$ equally spaced short pulses. The thick blue line shows the particle number for the superposition of a single long pulse with the pulse train whereas the red line shows the particle number for the pulse train only.

We first focus on the results of the pulse train without a long pulse. We observe that the particle number oscillates as function of the time lag. Most notably, big local 
maxima and comparatively less pronounced local extrema arise, at least for the parallel and the alternating configuration. In the anti-symmetric configuration, however, a double peak structure appears and the local extrema become much more pronounced.
It is quite remarkable, however, that the oscillatory behaviour as a function of the time lag is very regular and stable in all possible configurations, at least for large values of $t_0$.

Adding a single long pulse to the pulse train, we find in Fig.\ref{Fig8} that the pattern of local extrema due to the pulse train is inherited by the full solution. 
We observe, however, that a long pulse can keep the small pulses from causing oscillations in the distribution function for rather large values of $\tau_2$. For decreasing values of $\tau_2$,
however, the oscillations are damped for higher time lags but stay observable. It is also remarkable that the superposition of a long pulse results in a delay of the oscillation frequency.
Thus, the surprising effect arises that the additional long pulse can both increase and decrease the particle number depending on the time lag $t_0$.

\begin{figure}[h]
\subfigure[]{\includegraphics[width=0.49\textwidth]{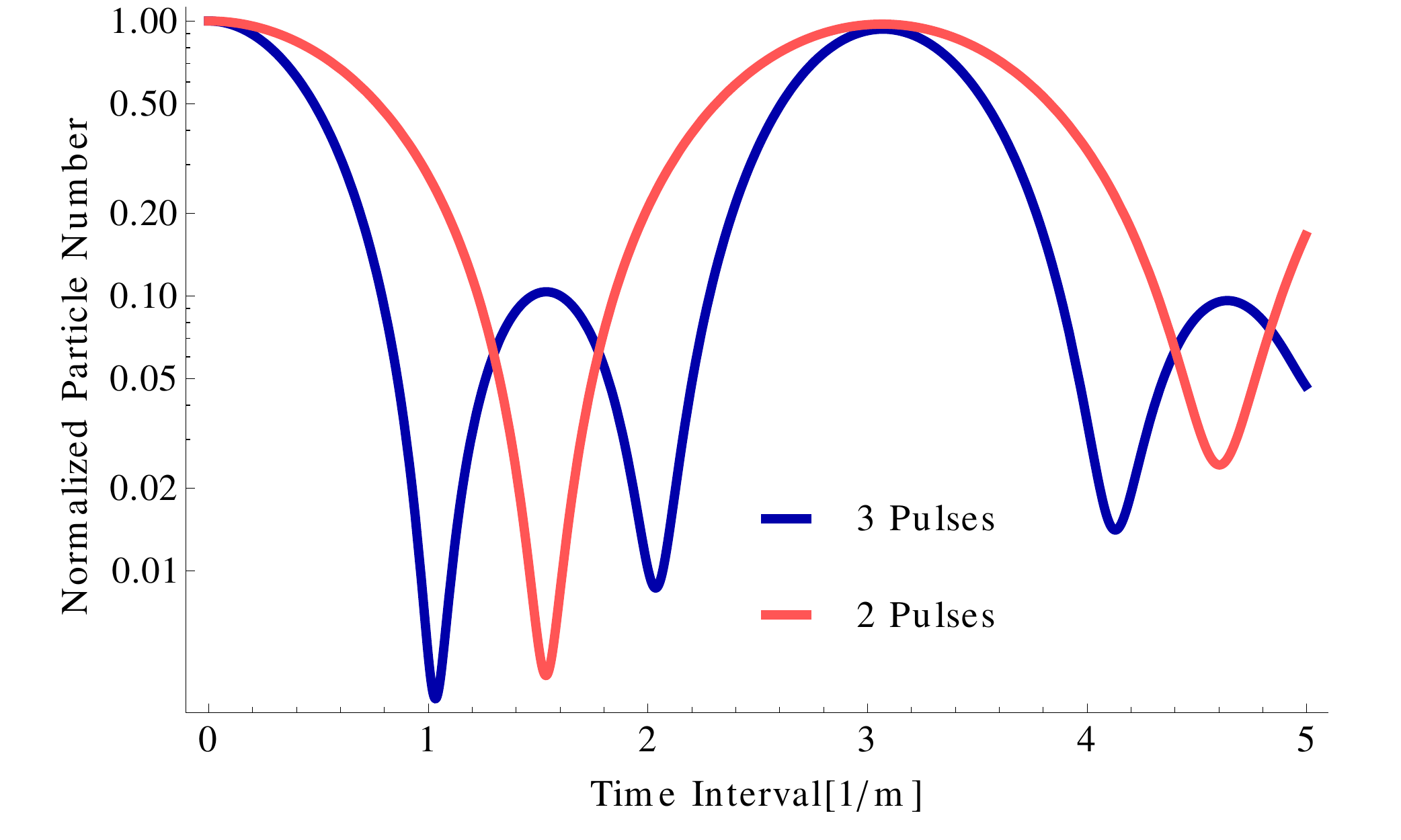}} 
\hfill
\subfigure[]{\includegraphics[width=0.49\textwidth]{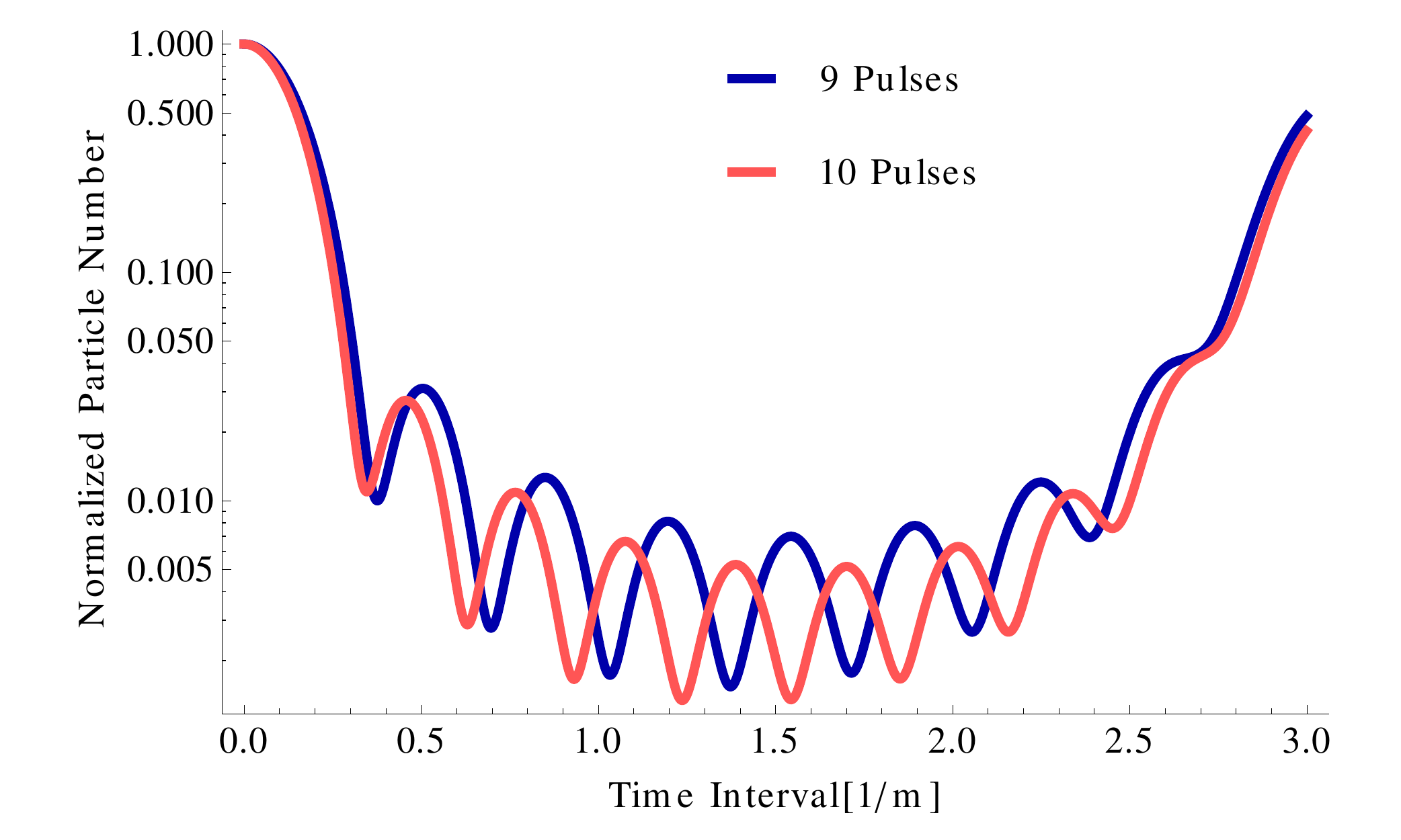}} \\[-1cm]
\caption[Pattern of the local extrema in the particle number for a different number of pulses in the pulse train]{Pattern of the local extrema in the particle number for a different number of pulses in the pulse train. A different number of pulses produces
 a different number of local maxima between the peak values. We show the results for pulse trains with parameters $~\varepsilon=0.005~$ and $~\tau=3m^{-1}$ in the parallel configuration.}
 \label{Fig9}
\end{figure}

\begin{figure}[h]
\centering	
  \includegraphics[width=0.85\textwidth]{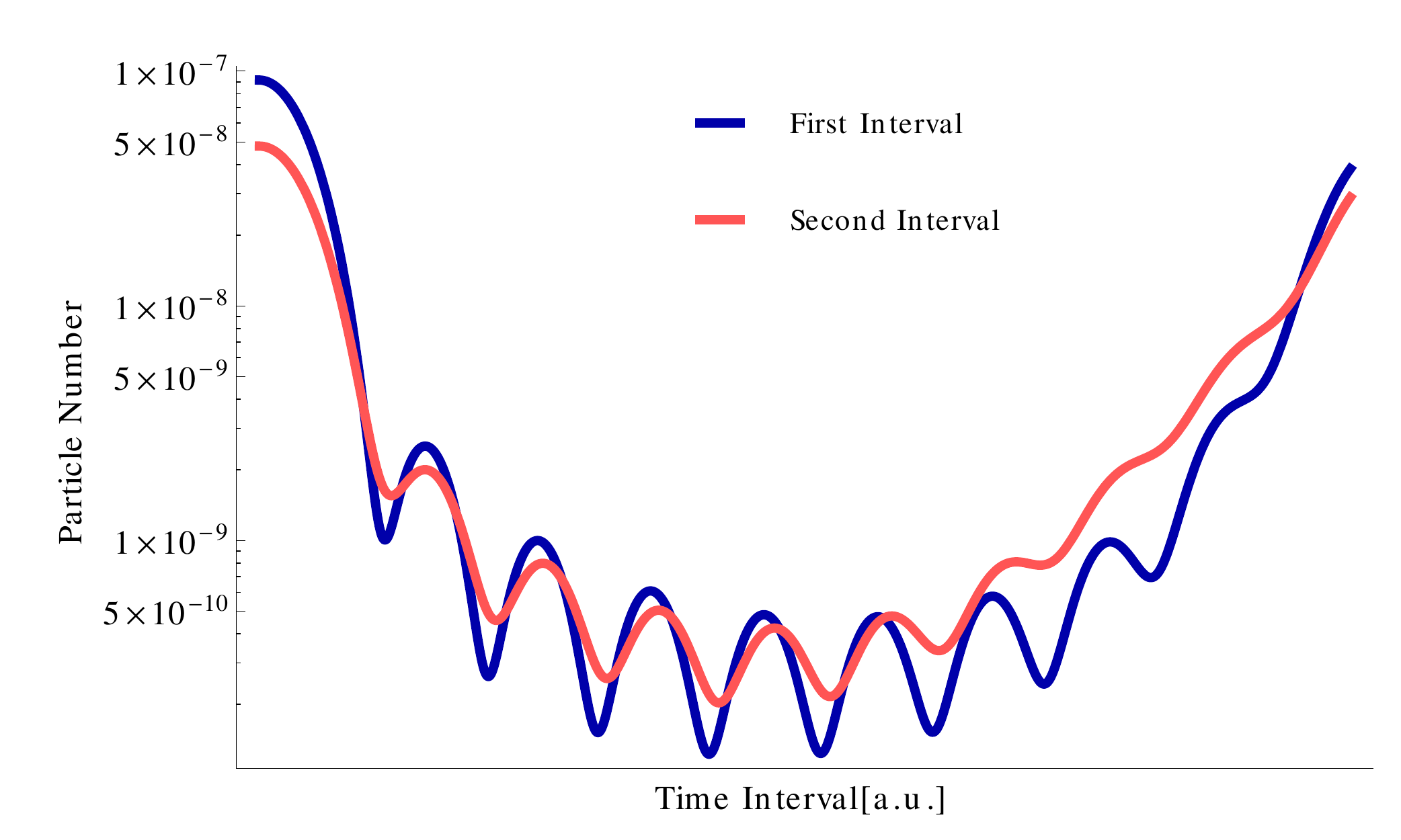}
  \caption[Damping of the local extrema pattern in the particle number upon changing the time interval]{Damping of the local extrema pattern in the particle number upon changing the time interval. 'First interval' refers to a time lag of $t_0 \in \com{0, 3}$ whereas 'second interval' refers to a time lag of $t_0 \in \com{3, 6}$. We show the results for a pulse train of $N=10$ pulses with parameters $~\varepsilon=0.005~$ and $~\tau=3m^{-1}$. The height of the oscillations decreases for higher time intervals.}
   \label{Fig10}
\end{figure}

\begin{figure}[h]
\centering	
  \includegraphics[width=0.85\textwidth]{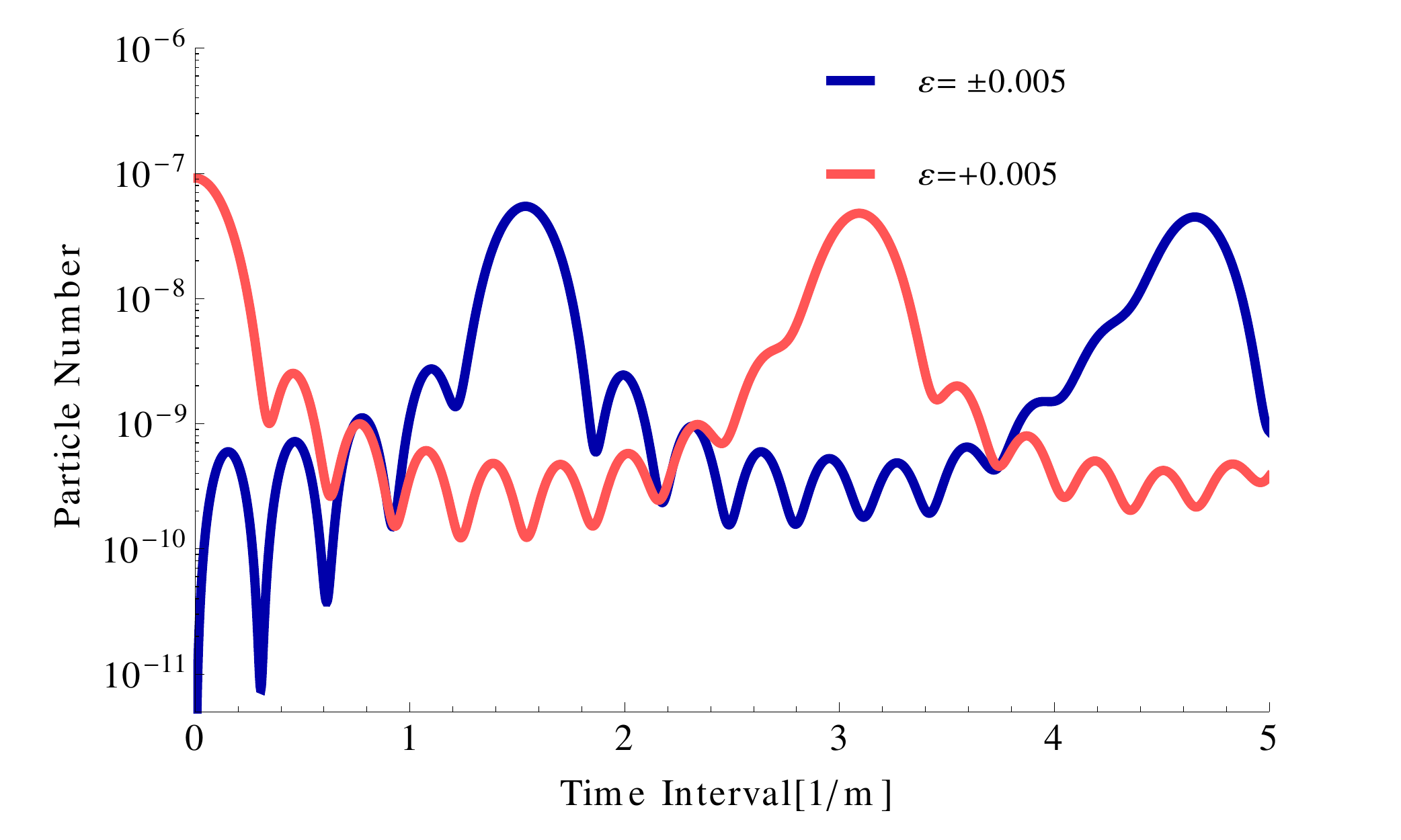}
  \caption[Dependence of the particle number on the used field configuration]{Dependence of the particle number on the used field configuration. Again, we show the results for a pulse train of $N=10$ pulses with parameters $~\bet{\varepsilon}=0.005~$ and $~\tau=3m^{-1}$. The red line corresponds to a parallel field configuration, while the blue line corresponds to the alternating configuration.}
\label{Fig11}
\end{figure}

\begin{figure}[h]
\centering	
\includegraphics[width=0.85\textwidth]{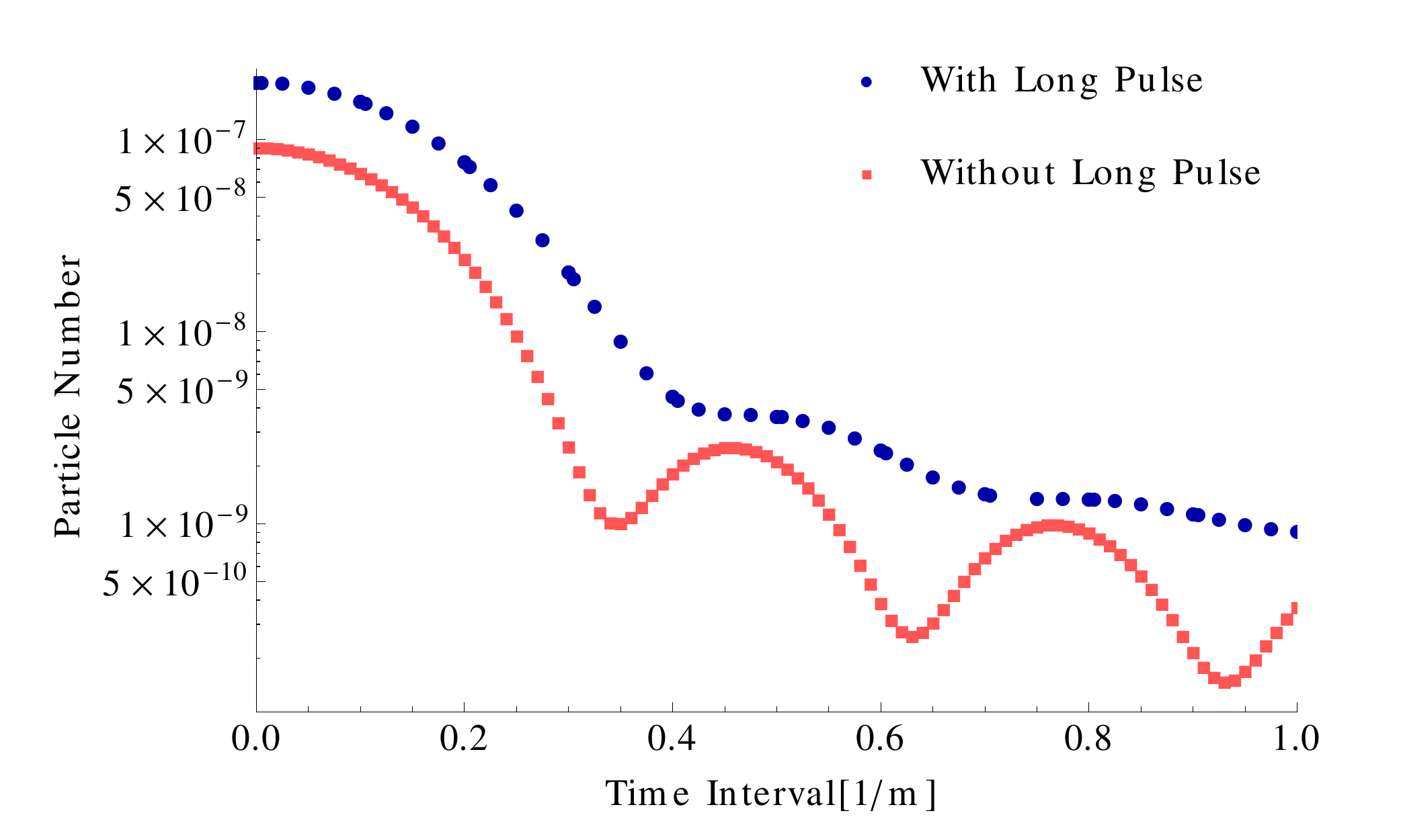}
\caption[Inheritance of local oscillations in the particle number]{Inheritance of local oscillations in the particle number from a pulse train of $N=10$ pulses with parameters $~\varepsilon=0.005~$ and $~\tau=3m^{-1}$ compared with a single long pulse in addition to the same pulse train.}
\label{Fig12}
\end{figure}

Based on Fig.\ref{Fig8}, we may observe interesting aspects.
We start our analysis with a detailed description of the results for the time interval $t_0 \in \com{0,3m^{-1}}$ and have a look at the pattern of local minima in Fig.\ref{Fig9}.
These results indicate that there is a strong correspondence between the number of the local minima and the number of pulses used. Besides comparing the behaviour of the particle distribution function
for the first two time intervals in Fig.\ref{Fig10}, a more detailed comparison between the parallel and the alternating configuration is shown in Fig.\ref{Fig11}.
This surprising result raises the question whether there is a deeper connection between these two configurations. In Fig.\ref{Fig12}, we finally resolve that the local maxima pattern in a pulse train can cause oscillations in setups with an additional long pulse.

\subsection{Time-scale Comparison}

Similar to the last section we focus on the influence of the chosen configuration on the particle number and provide further data for
pulse trains in parallel and alternating configurations. Again, we use pulse trains consisting of $N=10$ equally space pulses and vary their pulse length $\tau_2$.  

\begin{figure}[h]
\subfigure[]{\includegraphics[width=0.49\textwidth]{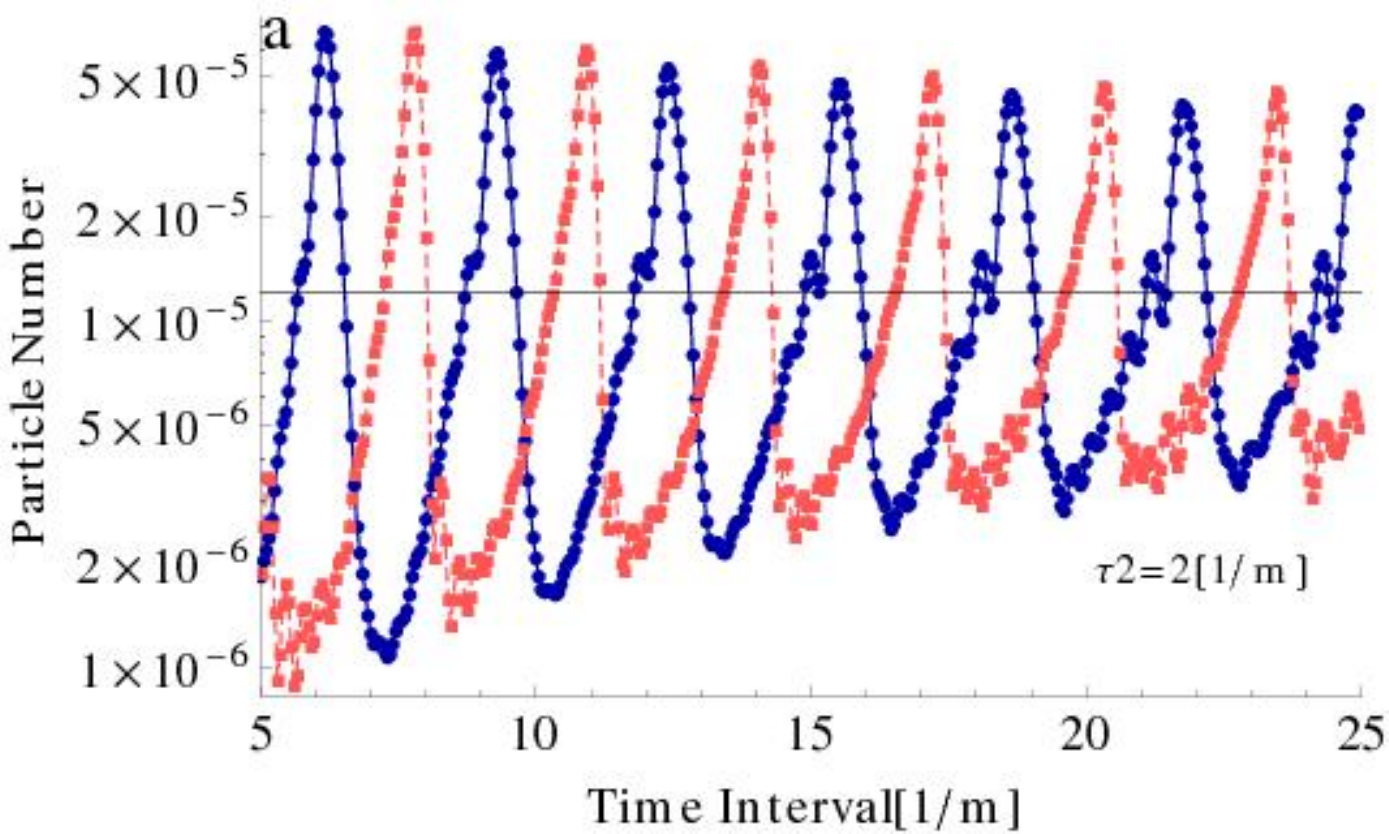}} 
\hfill
\subfigure[]{\includegraphics[width=0.49\textwidth]{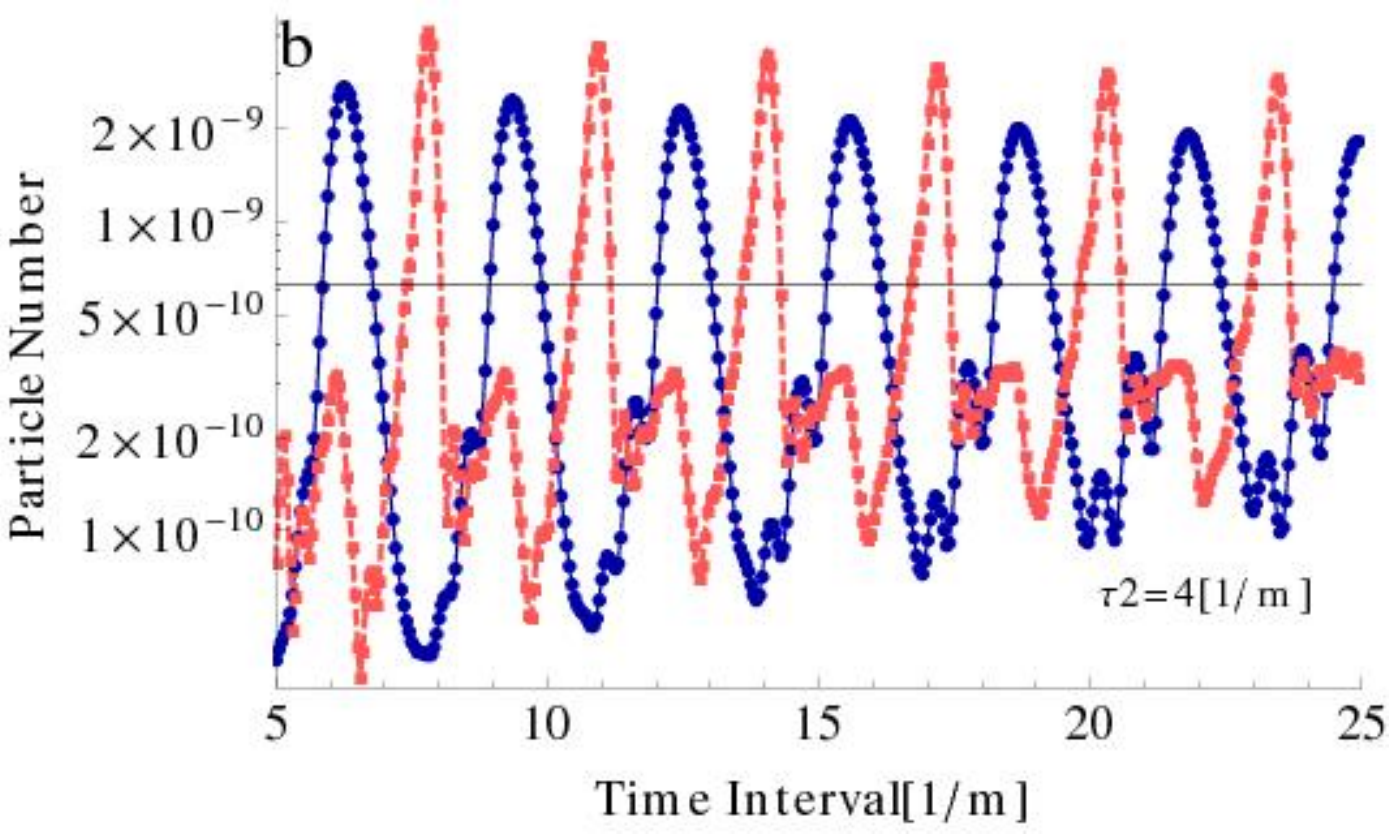}} \\[-1cm]
\subfigure[]{\includegraphics[width=0.49\textwidth]{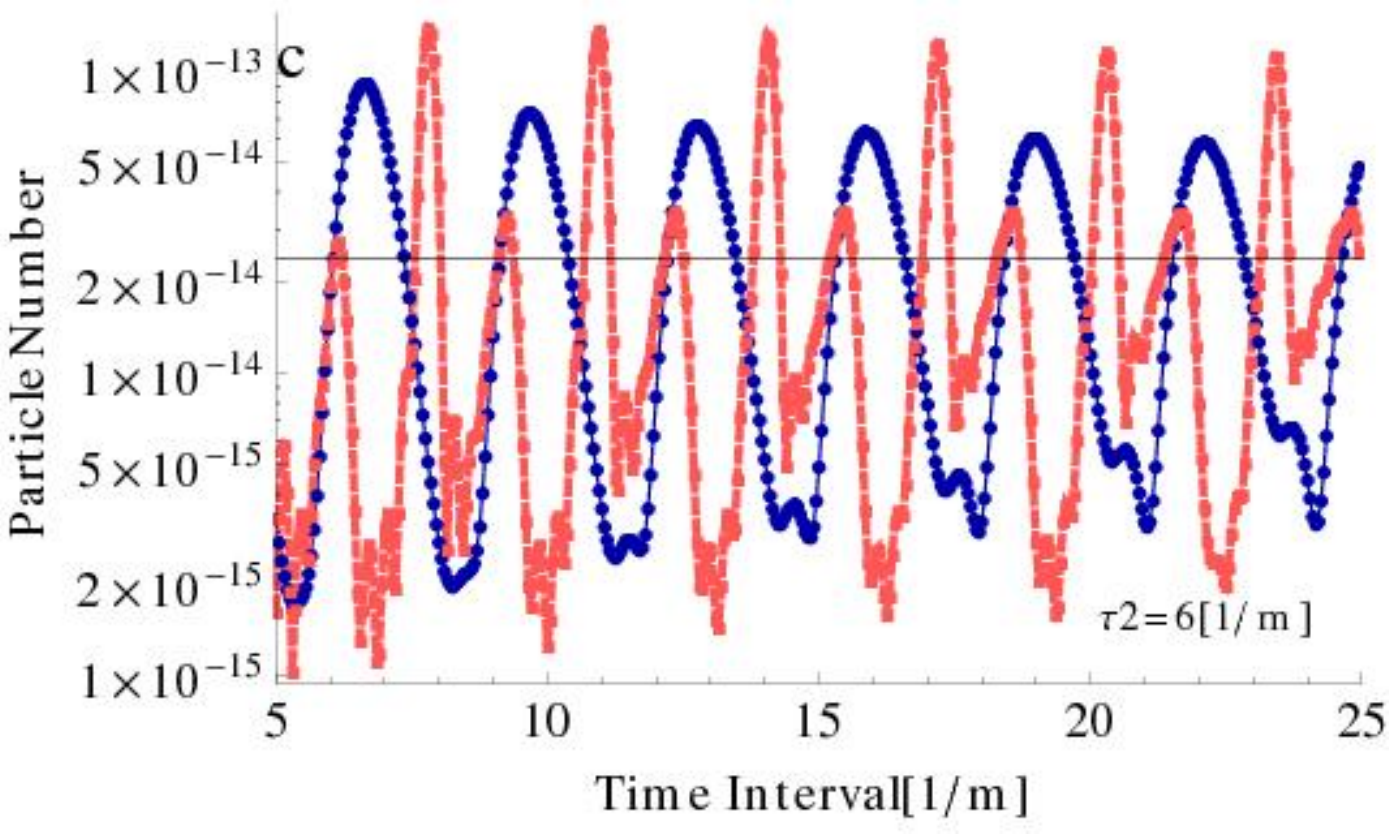}} 
\hfill
\subfigure[]{\includegraphics[width=0.49\textwidth]{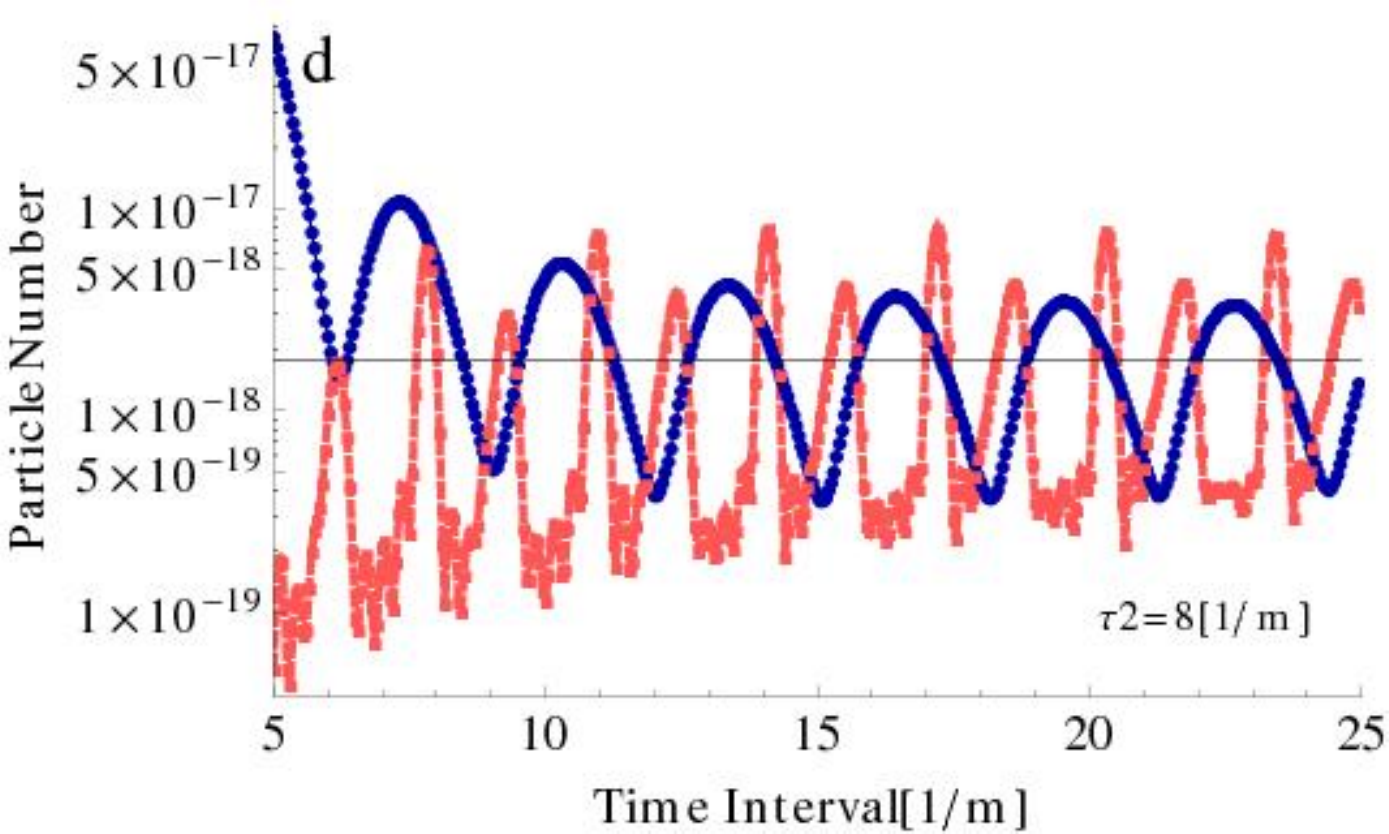}} \\[-1cm]
\caption[Particle number for a pulse train of $~N=10~$ pulses and different pulse lengths]{Particle number for $N=10$ pulses with field strength $~\varepsilon = 0.02~$ and increasing pulse length. The blue line corresponds to parallel configurations and the red line to anti-parallel configurations, both are shown as a function of the time lag $t_0>5$.}
\label{Fig13}
\end{figure}

It can be observed in Fig.\ref{Fig13}, that the oscillations in the parallel case become smaller in magnitude the longer the pulse gets. For the alternating configuration, however, additional peaks arise which become even more pronounced for longer pulses. For long pulses it even looks as if the frequency of the oscillations has doubled in the alternating case. On the contrary, the weak local extrema in the symmetric configuration seem to vanish.
 
\newpage


\begin{figure}[h]
\subfigure[]{\includegraphics[width=0.49\textwidth]{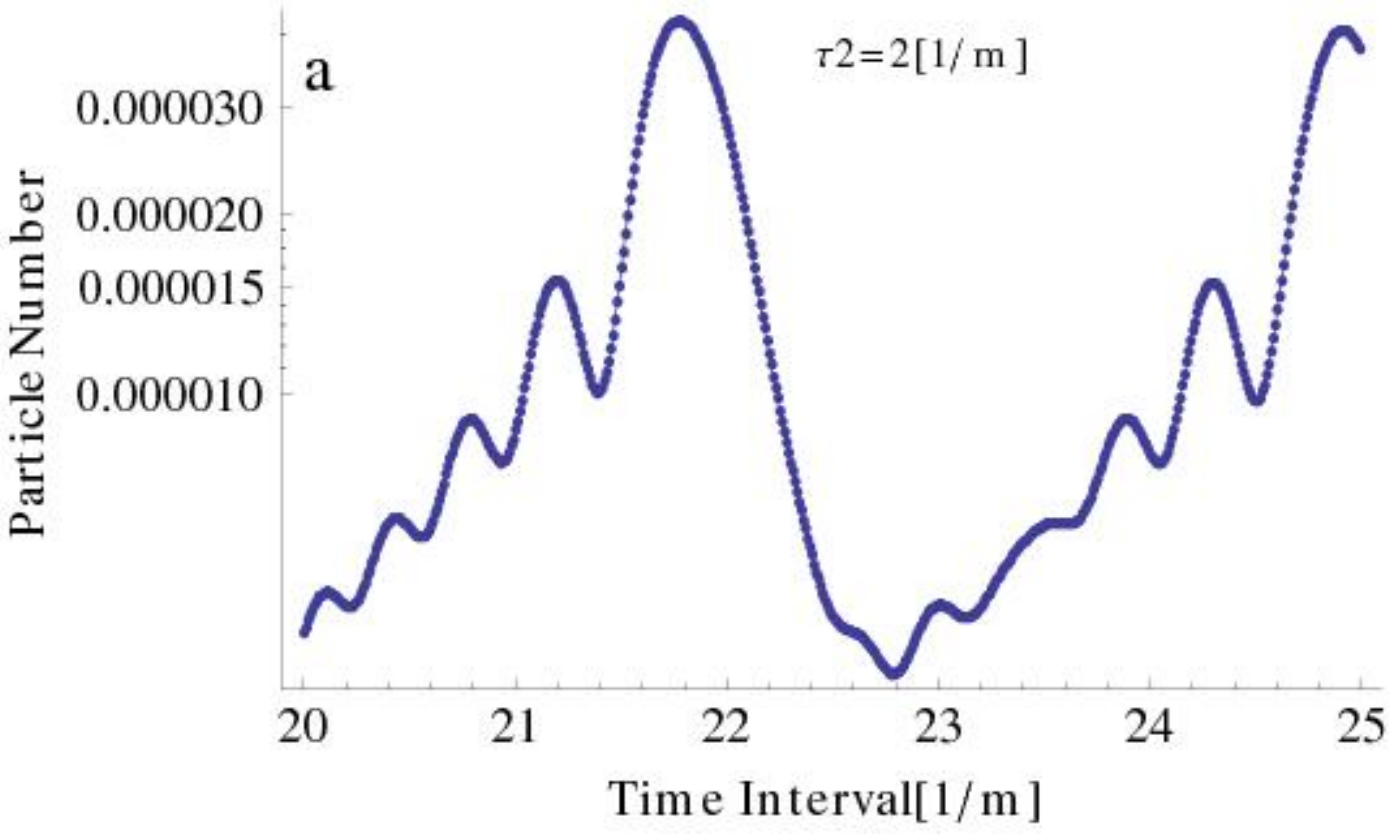}} 
\hfill
\subfigure[]{\includegraphics[width=0.49\textwidth]{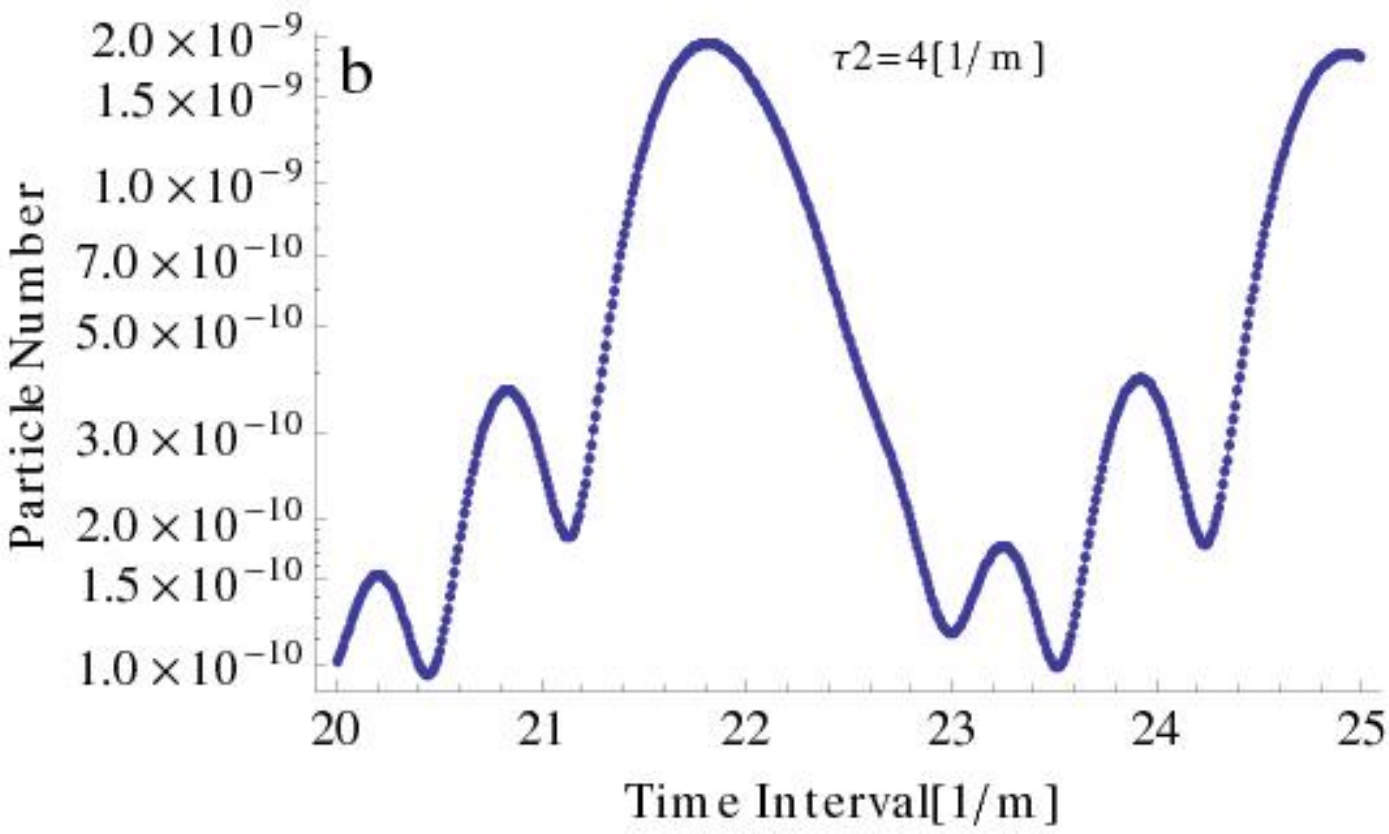}} \\[-1cm]
\subfigure[]{\includegraphics[width=0.49\textwidth]{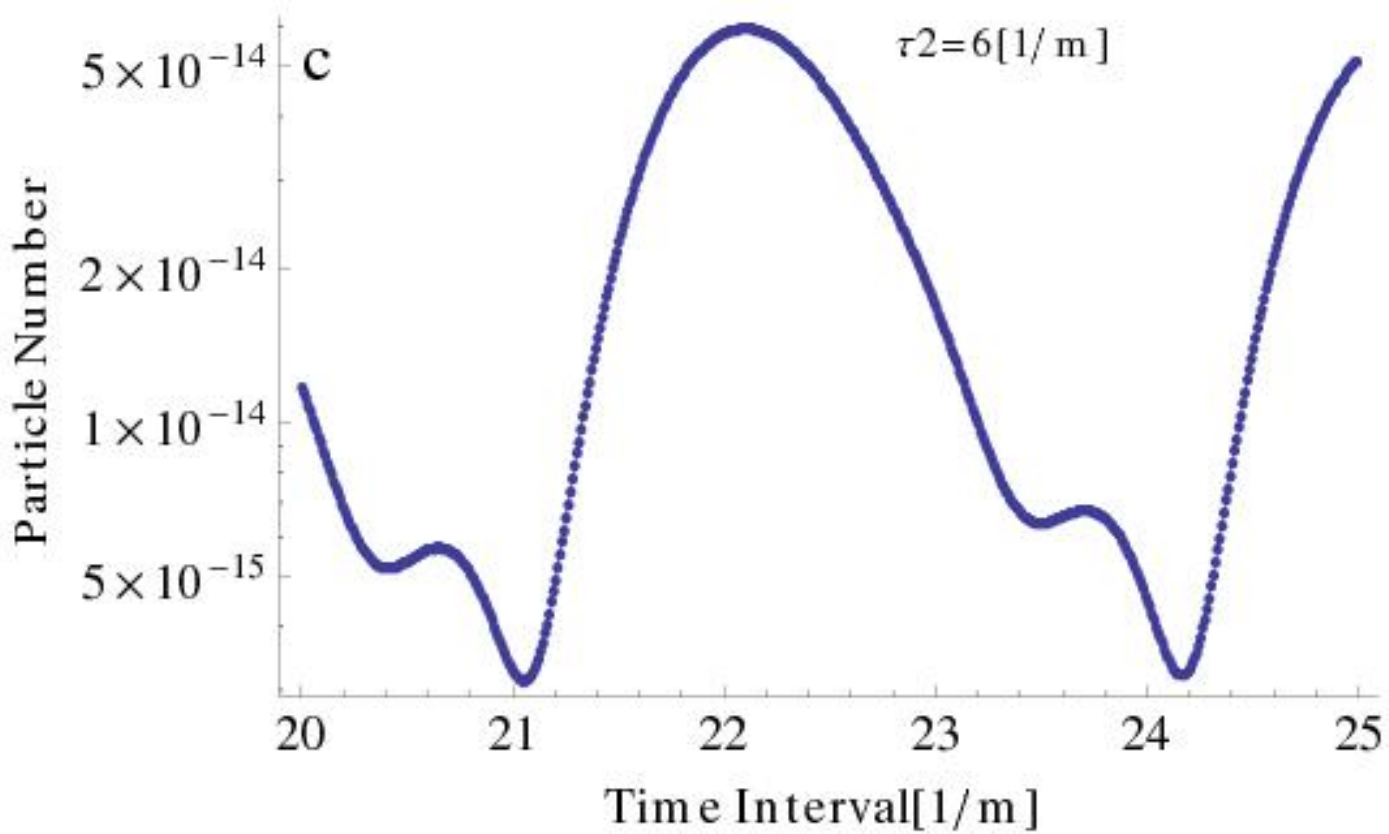}} 
\hfill
\subfigure[]{\includegraphics[width=0.49\textwidth]{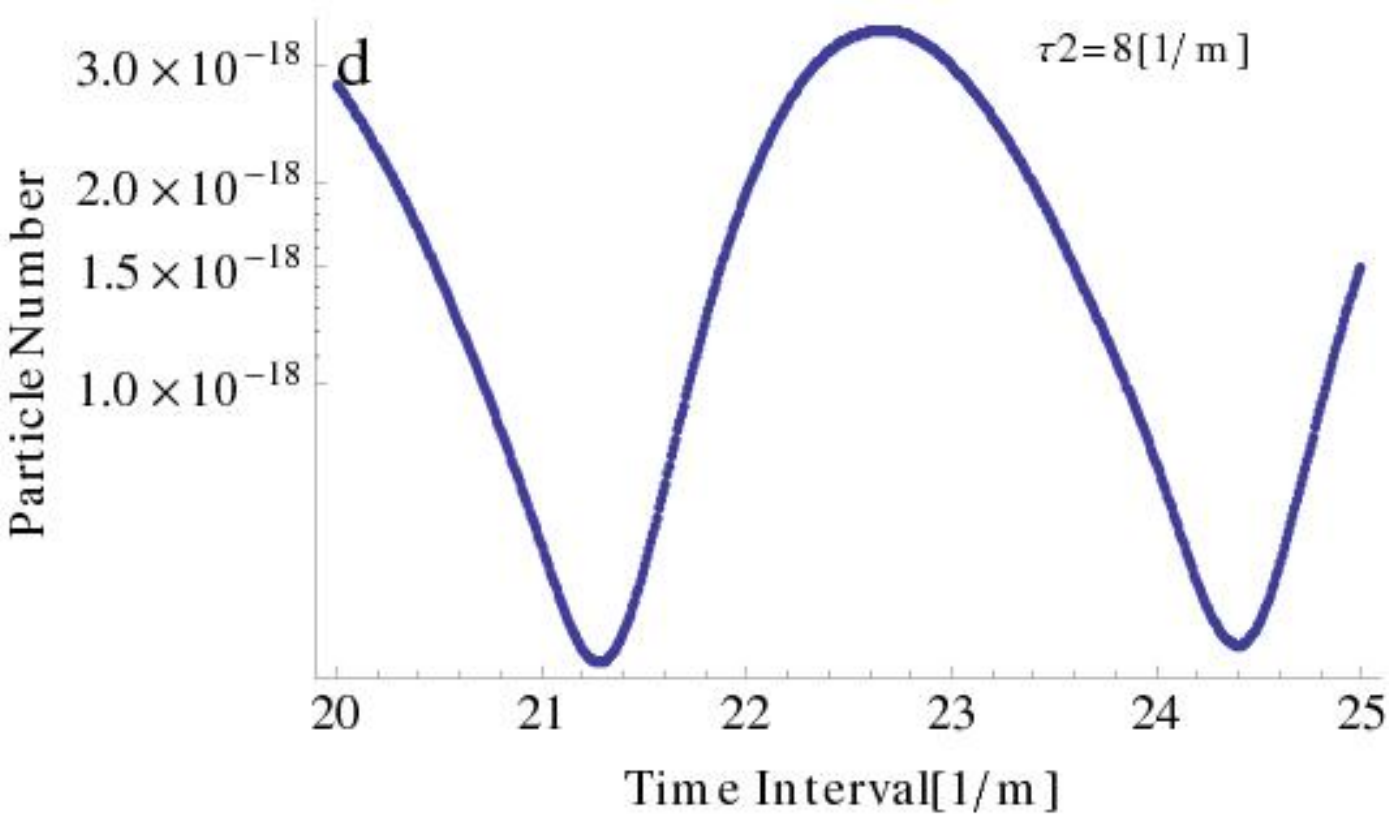}} \\[-1cm]
\caption[Behaviour of local fluctuations in the particle number with increasing pulse length for parallel configuration]{Behaviour of local fluctuations in the particle number with increasing pulse length for parallel configuration. The used field strength has been $~\varepsilon=0.02$. The pattern of the local extrema vanishes for longer pulse lengths.}
\label{Fig14}
\end{figure}

We further investigate the parallel configuration and focus on a time lag of about $t_0 \sim 20$. 
The first interesting observation, as illustrated in Fig.\ref{Fig14}, is that the small local extrema vanish entirely for longer pulses. The main maximum, however, becomes broader. 

\newpage
\begin{figure}[h]
\subfigure[]{\includegraphics[width=0.49\textwidth]{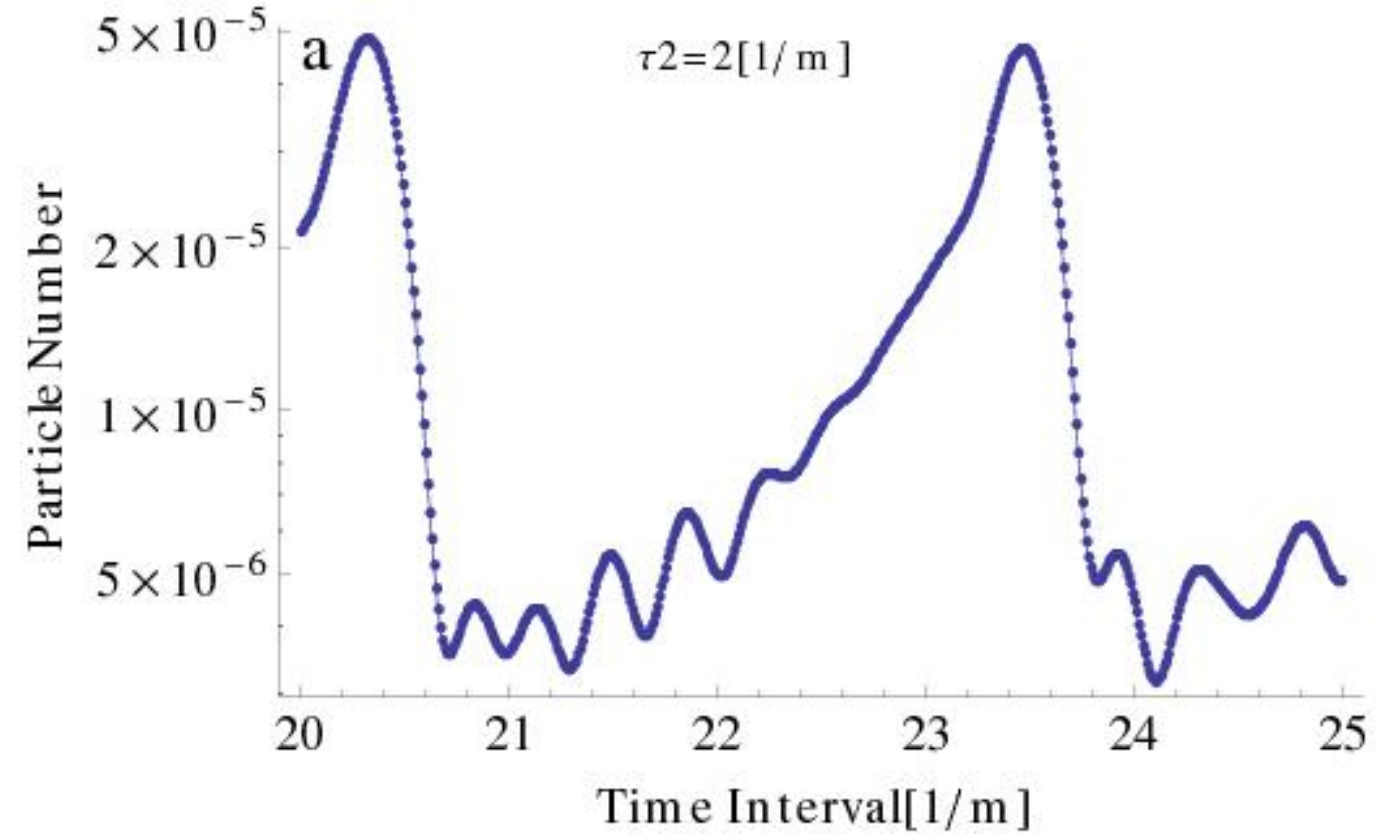}} 
\hfill
\subfigure[]{\includegraphics[width=0.49\textwidth]{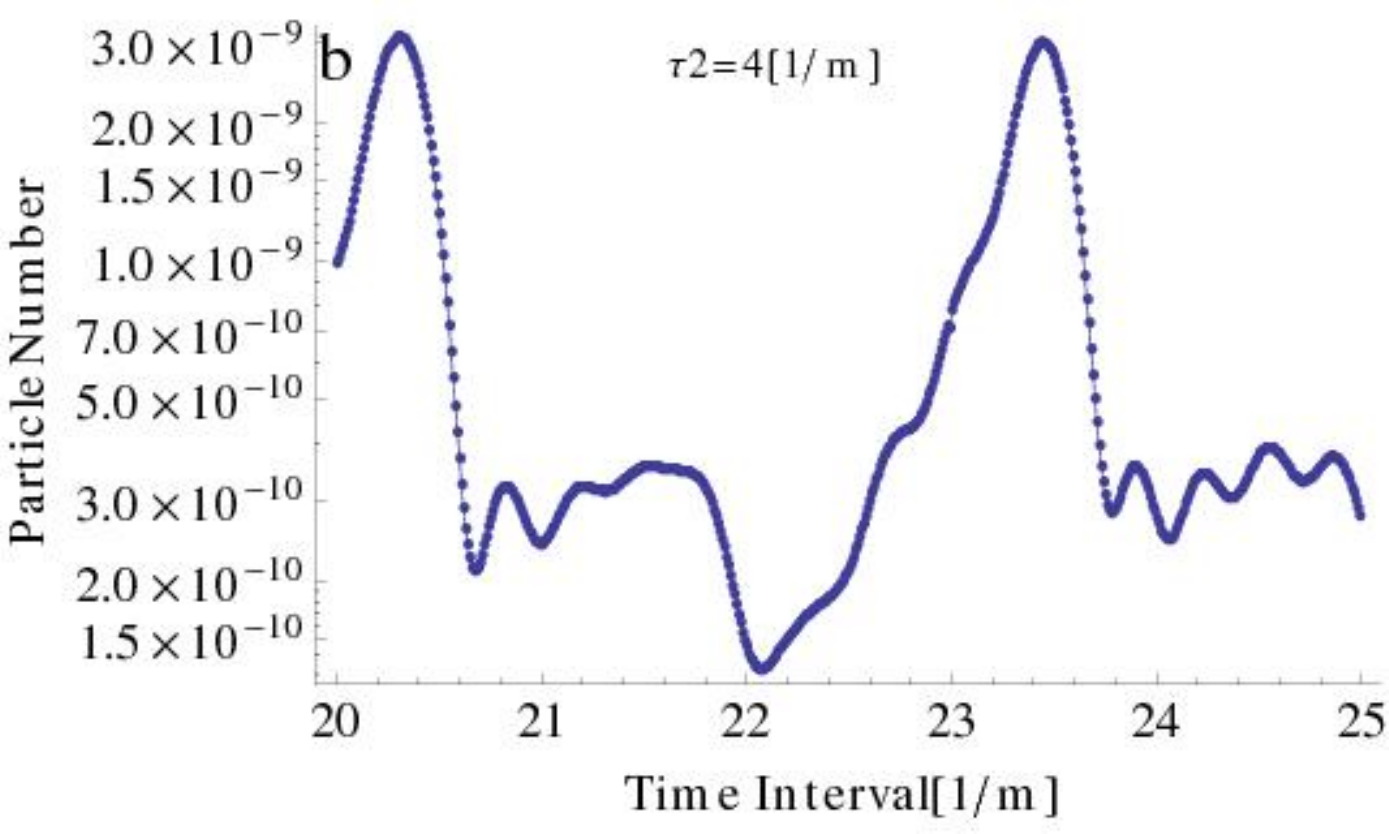}} \\[-1cm]
\subfigure[]{\includegraphics[width=0.49\textwidth]{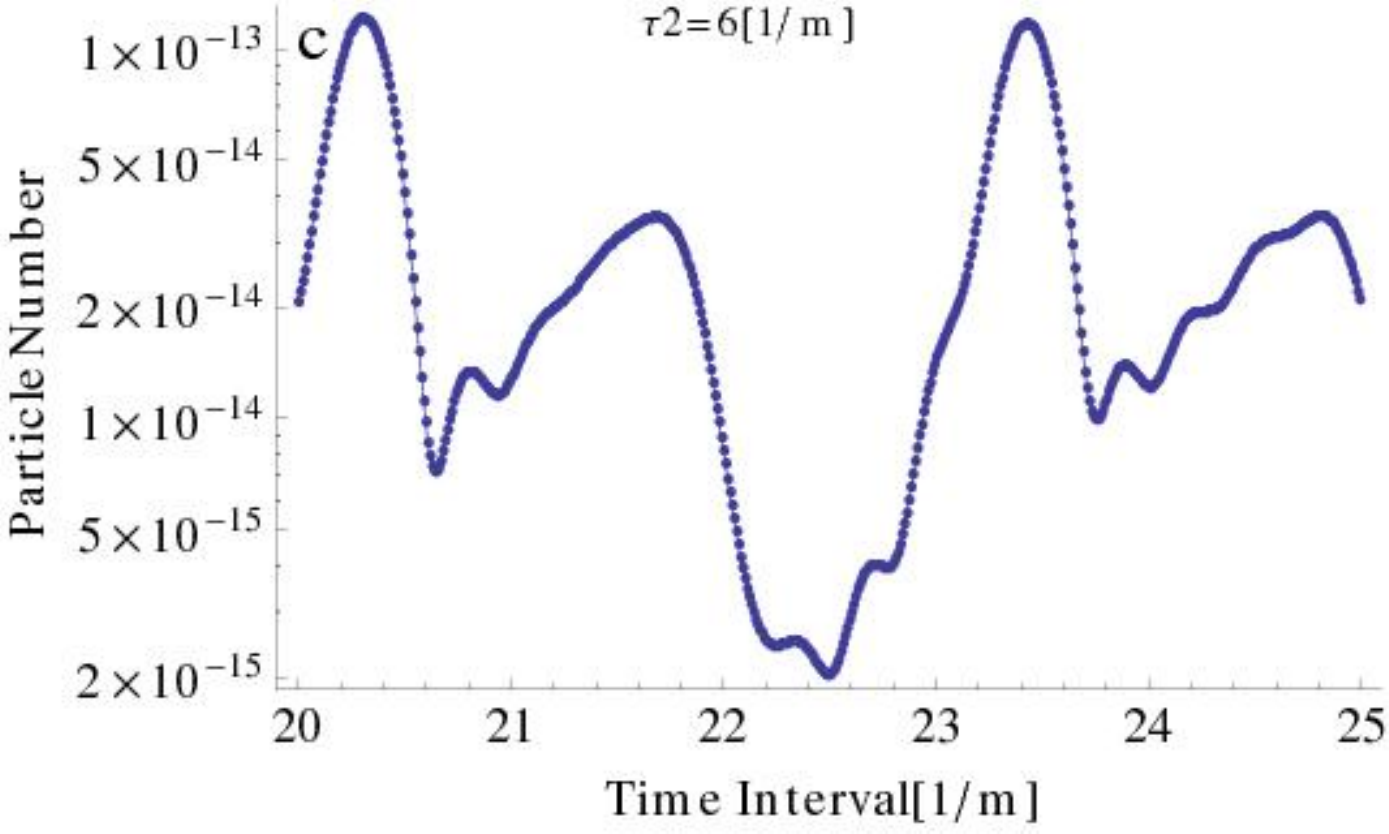}} 
\hfill
\subfigure[]{\includegraphics[width=0.49\textwidth]{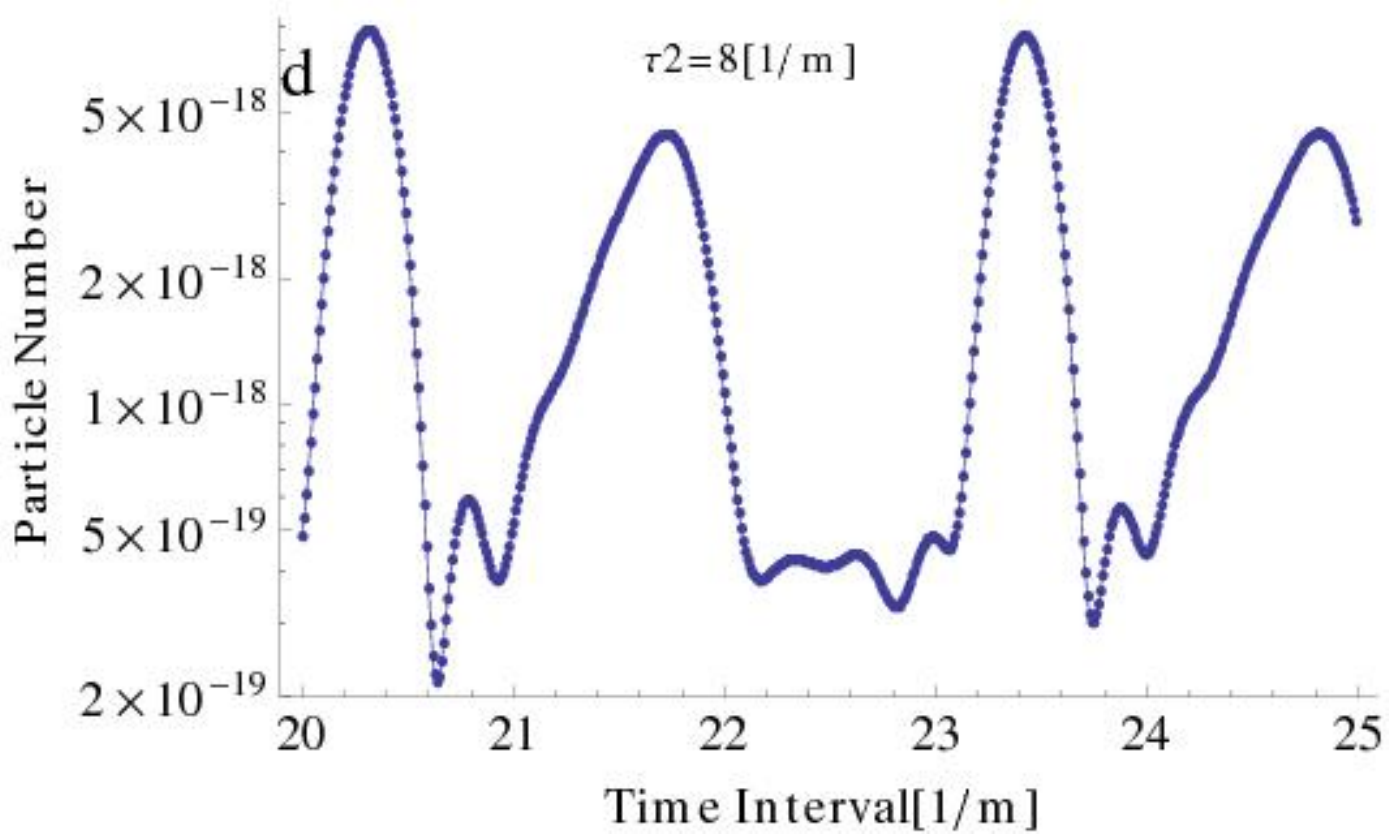}} \\[-1cm]
\caption[Behaviour of local fluctuations in the particle number with increasing pulse length for alternating configuration]{Behaviour of local fluctuations in the particle number with increasing pulse length for alternating configuration. The used field strength has been $~\varepsilon=0.02$. An additional local extremum arises for longer pulse lengths.}
\label{Fig15}
\end{figure}

In Fig.\ref{Fig15} we have done the same detailed analysis of local extrema, but this time for the alternating configuration. The main difference compared to the parallel configuration is that the local extrema do not vanish for longer pulse lengths. Instead another local extremum raises for longer pulses. Therefore, one gets the impression that the rate of oscillations has doubled. This gives rise to a distinctive difference between the parallel and the alternating configuration.

\subsection{Field Strength Comparison}

\begin{figure}[h]
\subfigure[]{\includegraphics[width=0.49\textwidth]{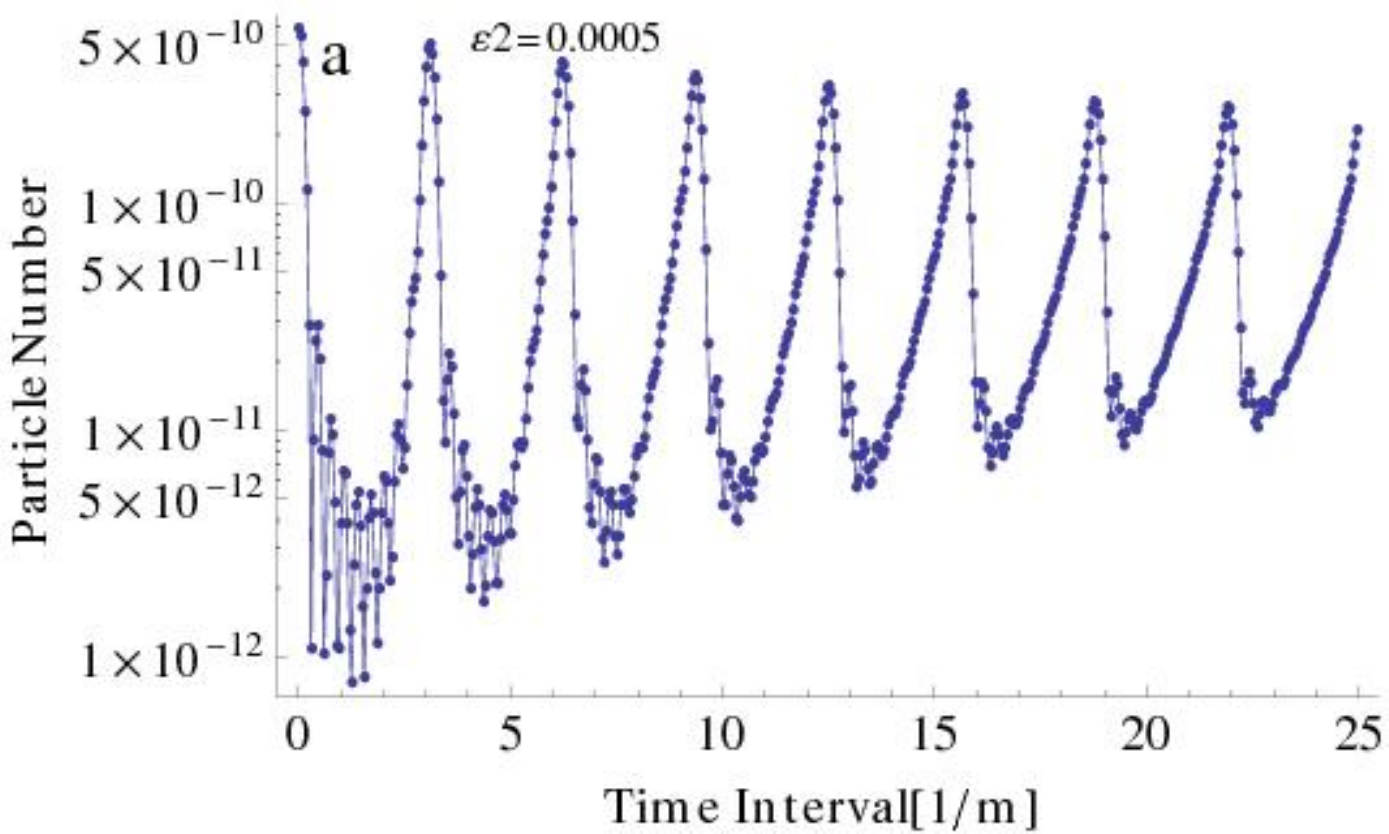}} 
\hfill
\subfigure[]{\includegraphics[width=0.49\textwidth]{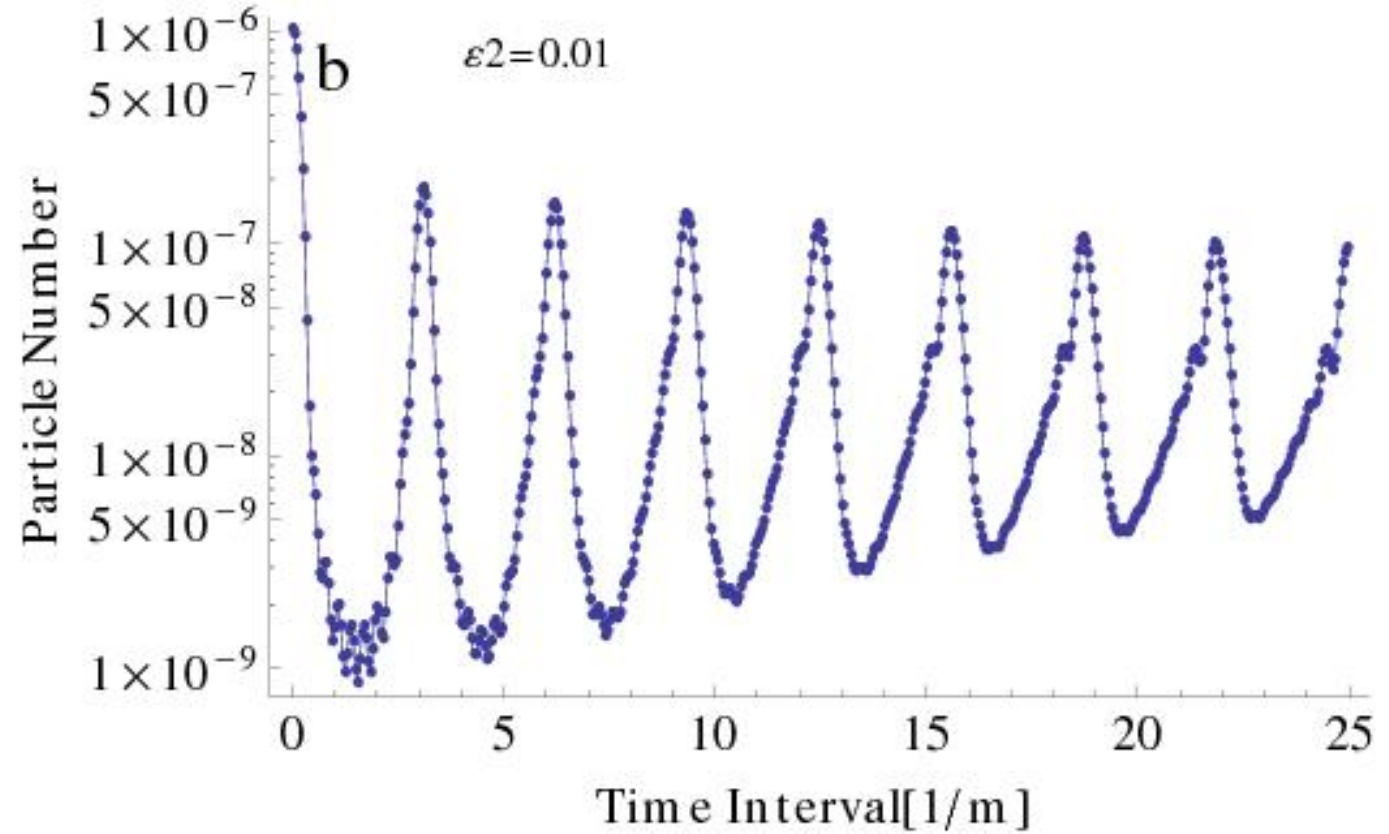}} \\[-1cm]
\subfigure[]{\includegraphics[width=0.49\textwidth]{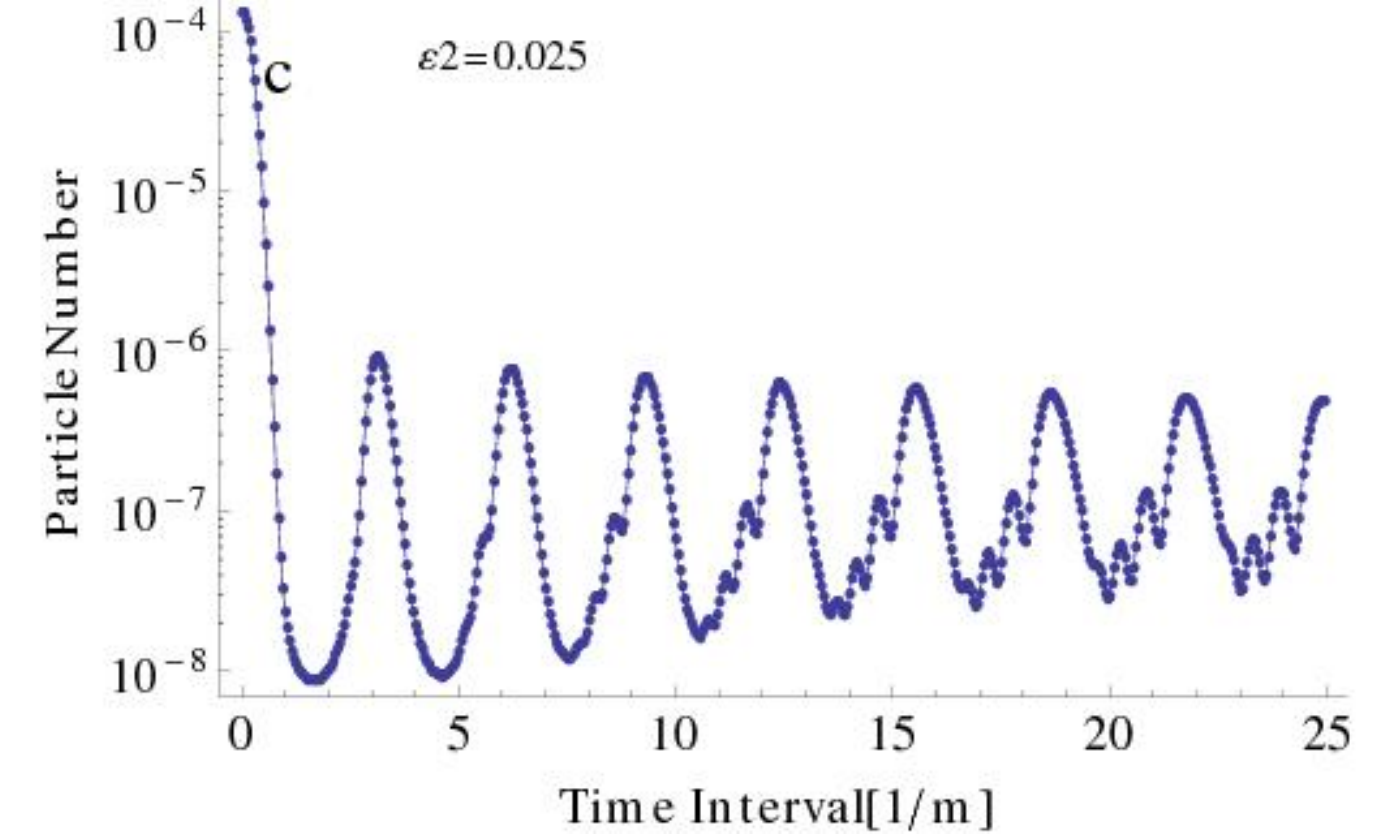}}
\hfill
\subfigure[]{\includegraphics[width=0.49\textwidth]{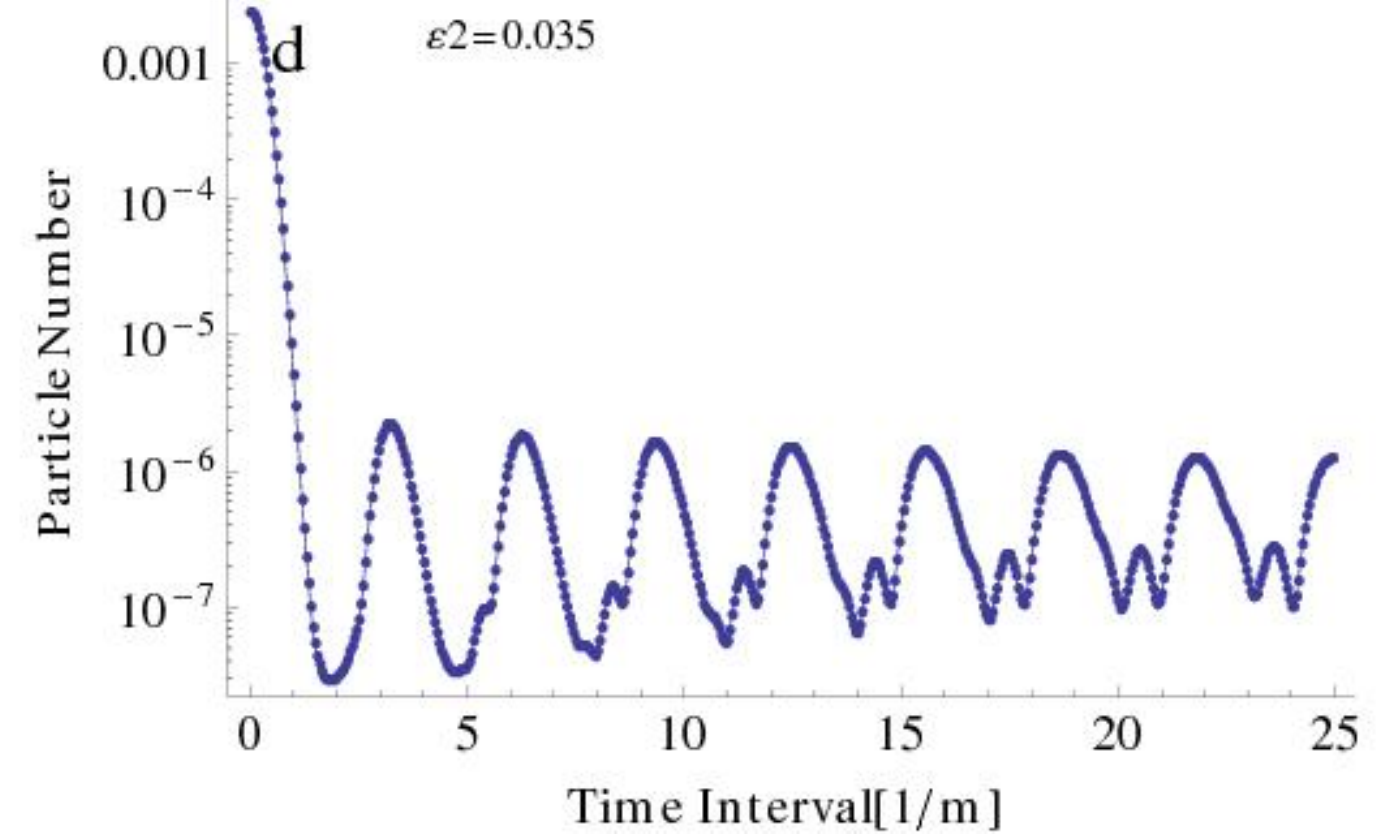}} \\[-1cm]
\caption[Particle number for a pulse train of $N=10$ pulses with increasing pulse strength]{Particle number for a pulse train of $N=10$ pulses with pulse length $~\tau=3m^{-1}$ and increasing pulse strength.}
\label{Fig16}
\end{figure}

In the last part of this section, we investigate the impact of the field strength on the particle number for the parallel configuration. In this case, we use a pulse train with $N=10$ single pulses with pulse length $\tau=3m^{-1}$ and different field strengths $\varepsilon_2$. It can be seen in Fig.\ref{Fig16} that there is a qualitative change in the particle number for increasing field strengths. 

\begin{figure}[h]
\subfigure[]{\includegraphics[width=0.49\textwidth]{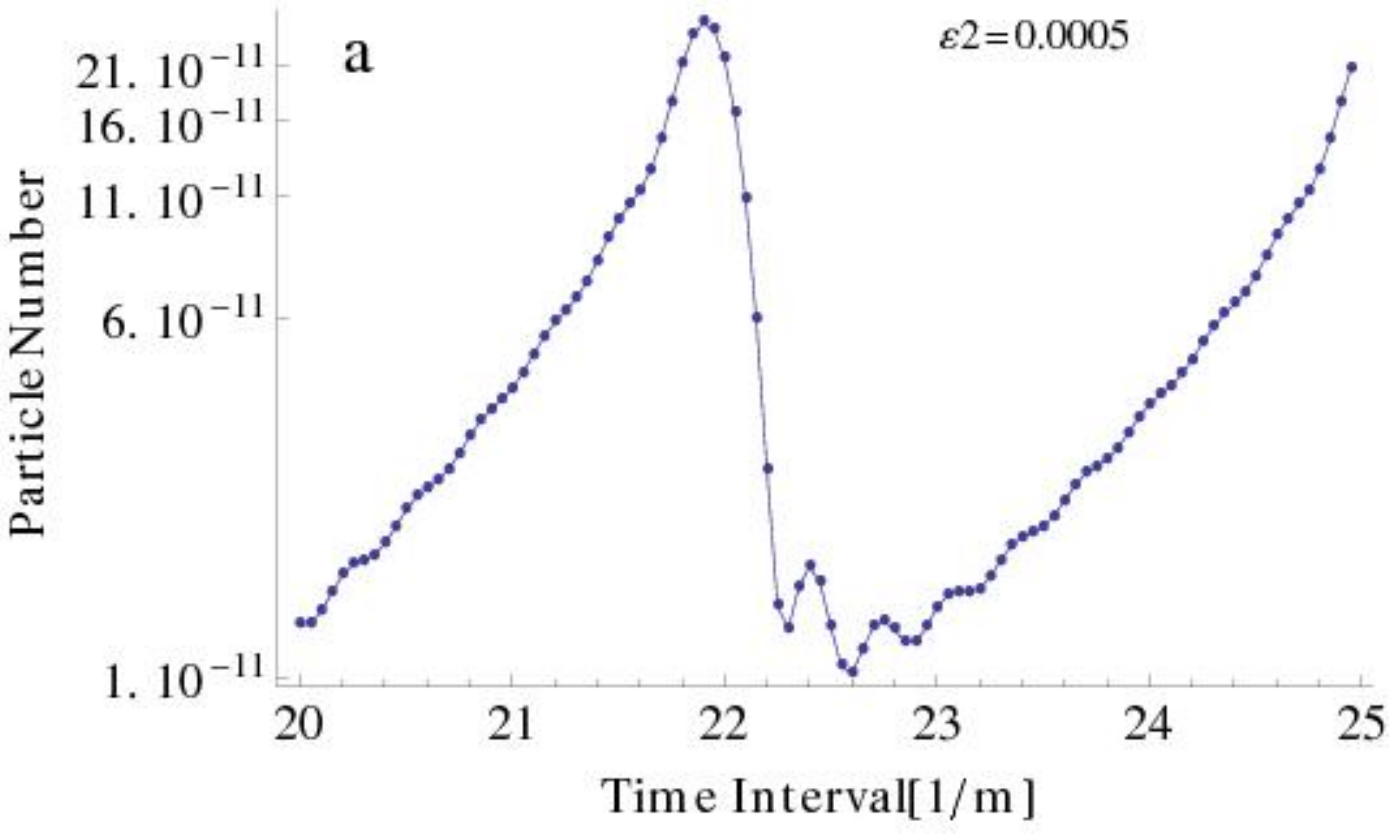}} 
\hfill
\subfigure[]{\includegraphics[width=0.49\textwidth]{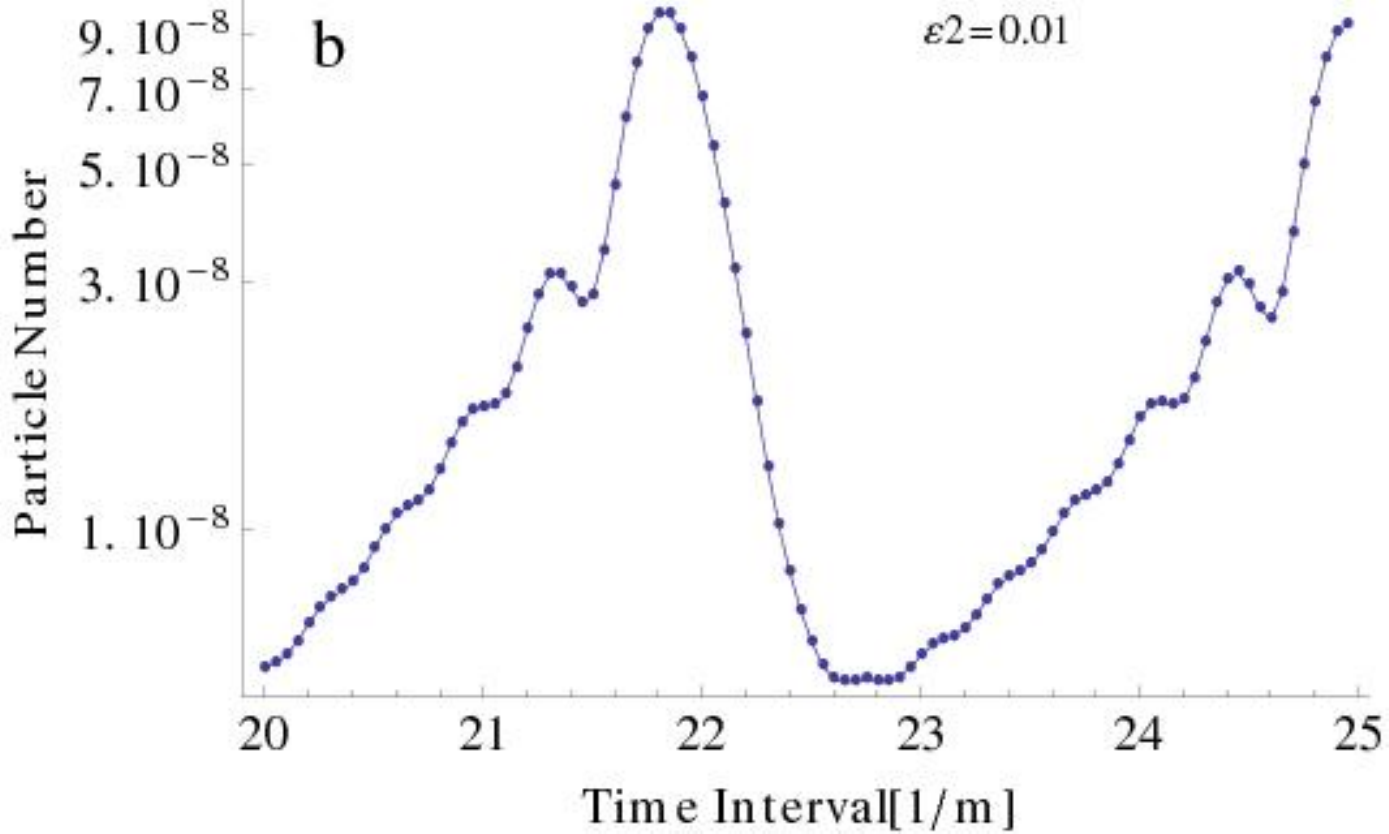}} \\[-1cm]
\subfigure[]{\includegraphics[width=0.49\textwidth]{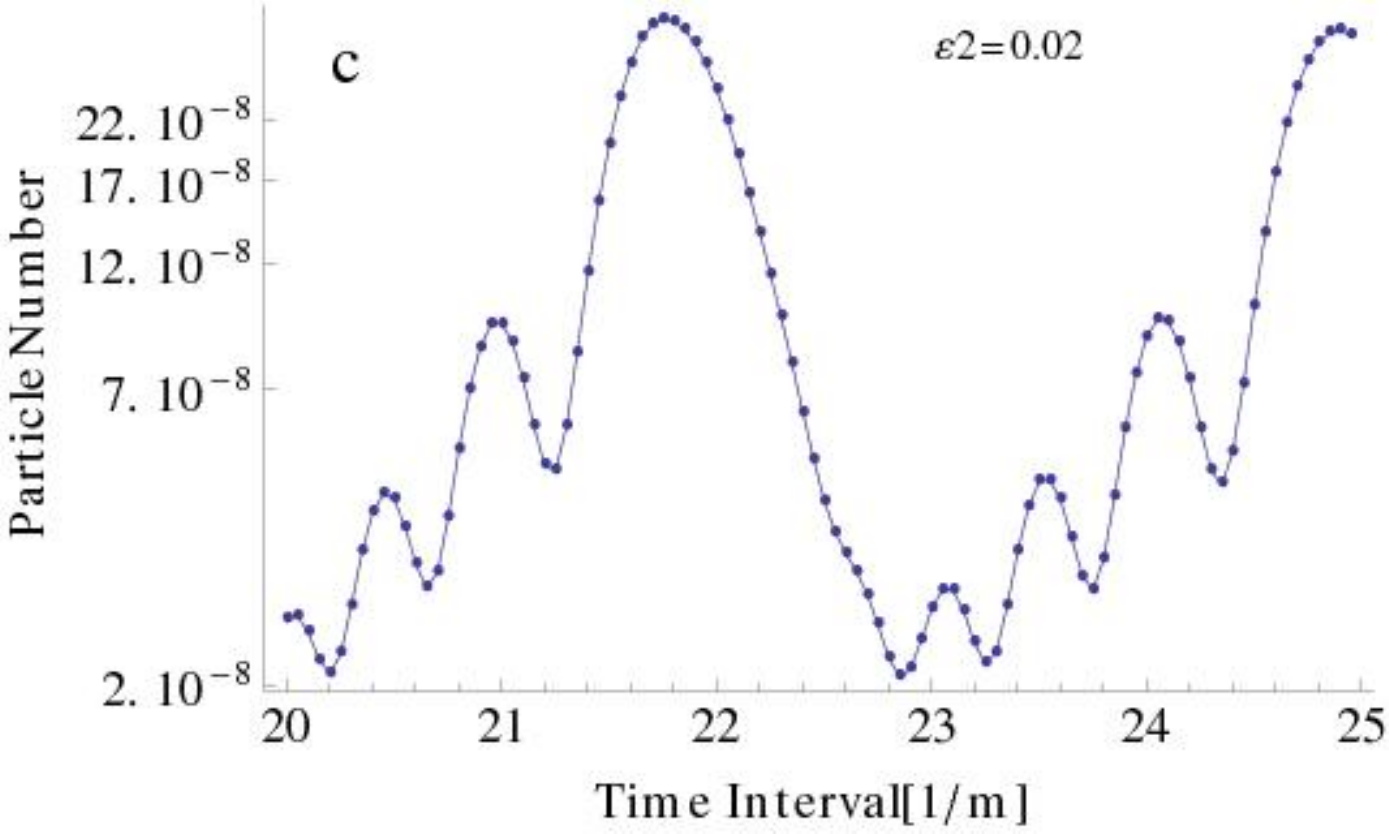}} 
\hfill
\subfigure[]{\includegraphics[width=0.49\textwidth]{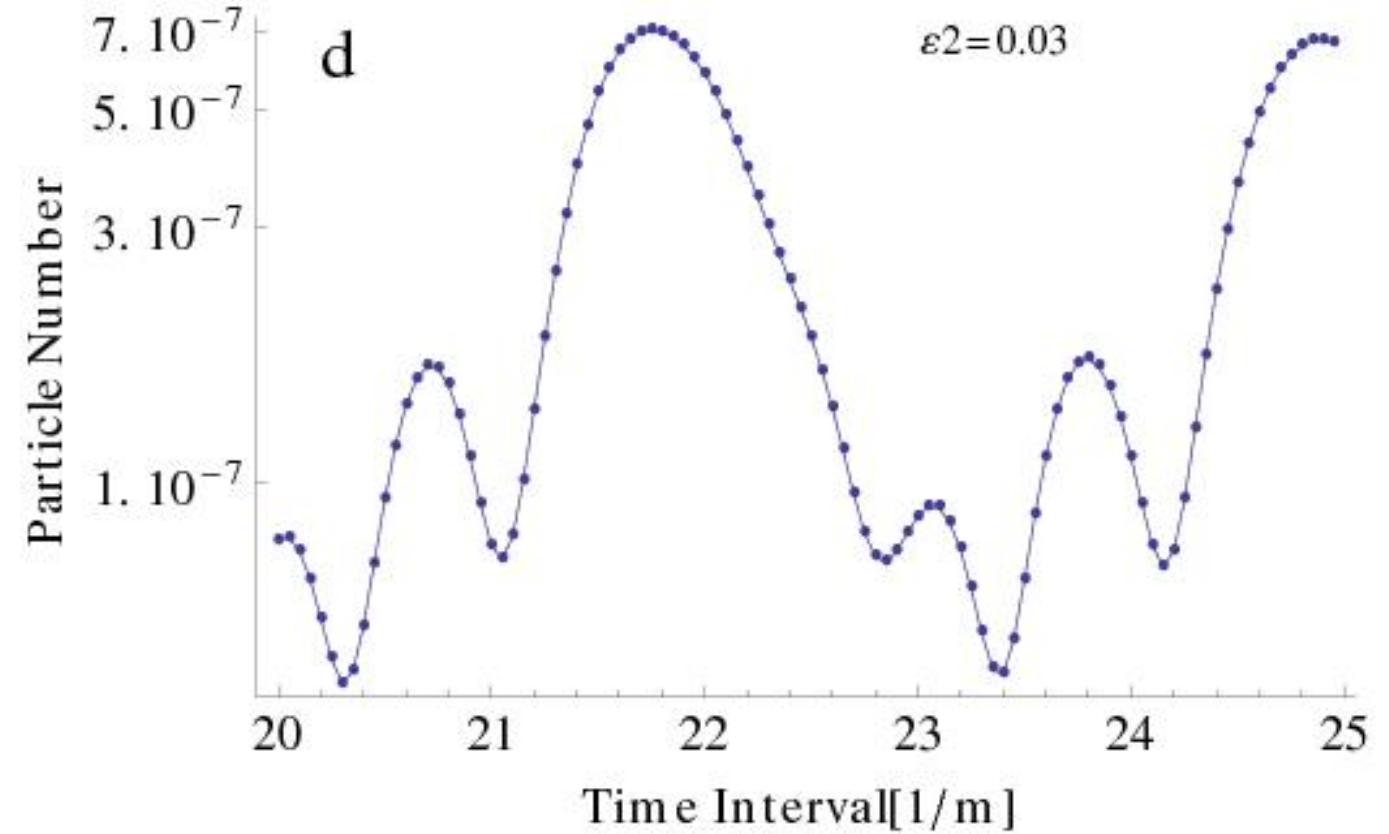}} \\[-1cm]
\subfigure[]{\includegraphics[width=0.49\textwidth]{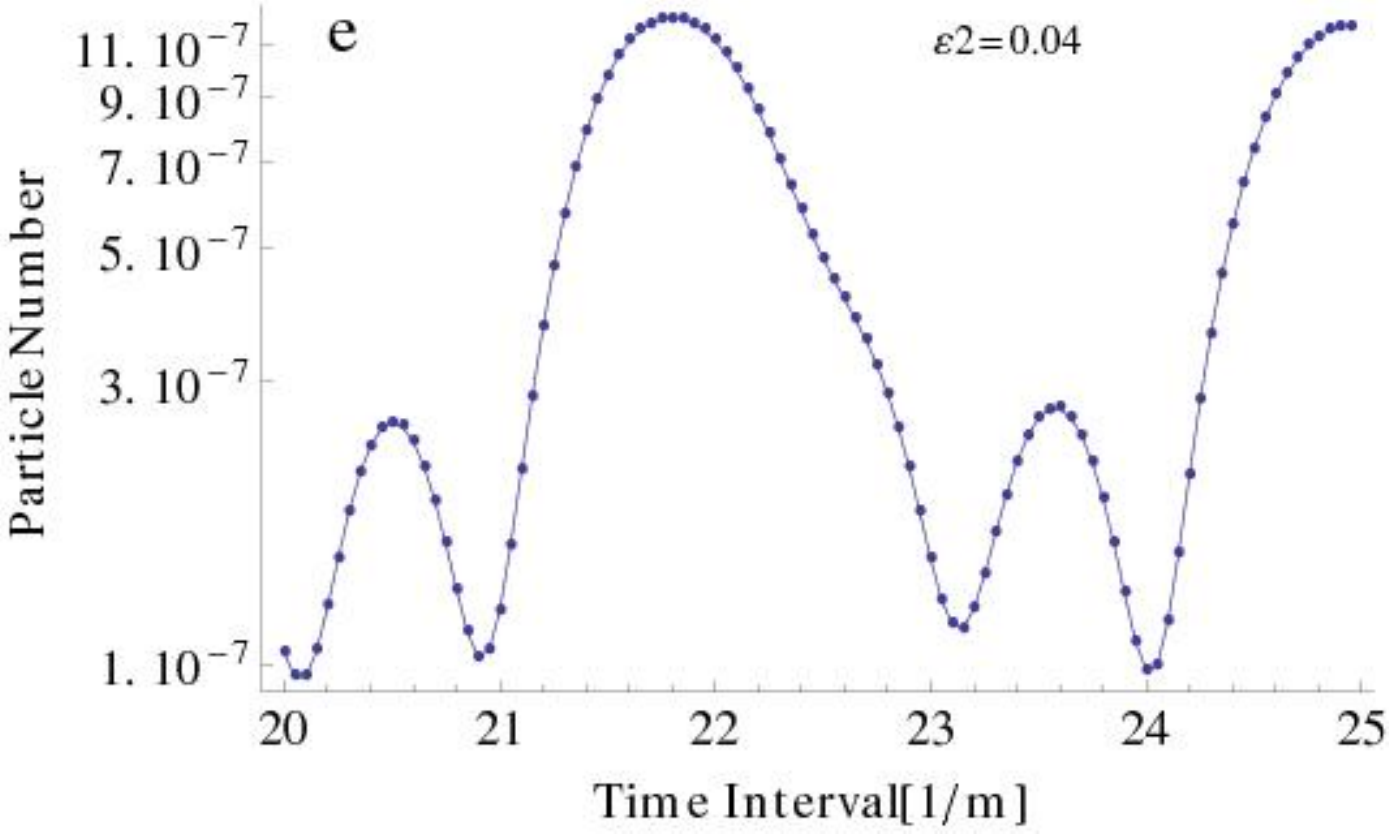}} 
\hfill
\subfigure[]{\includegraphics[width=0.49\textwidth]{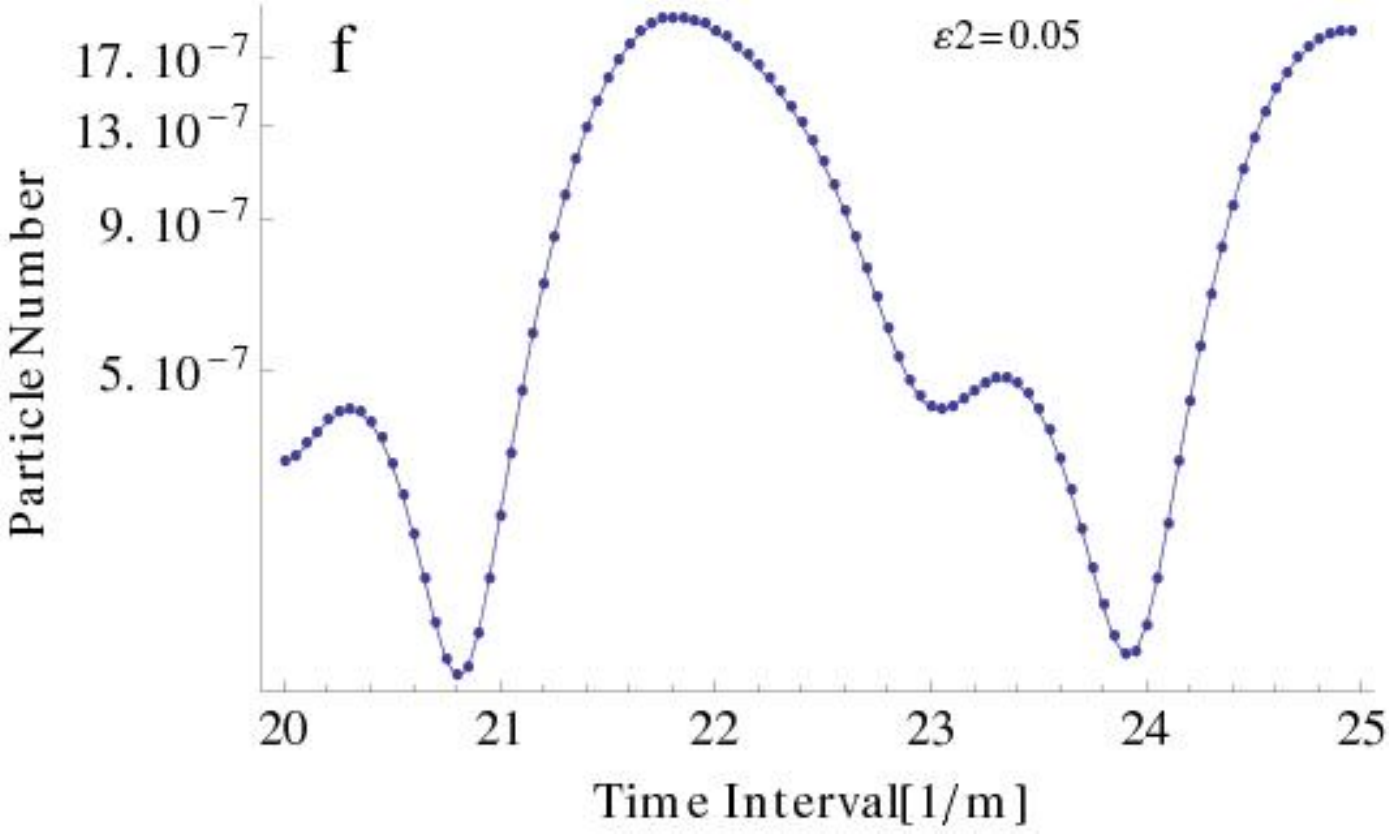}} \\[-1cm]
\caption[Particle number for a pulse train of $N=10$ pulses with increasing pulse strength for $t_0>20$]{Particle number for a pulse train of $N=10$ pulses with increasing pulse strength for $t_0>20$ with pulse length $~\tau=3m^{-1}$ and different field strength. }
\label{Fig17}
\end{figure}

Again, we further investigate the details in Fig.\ref{Fig17}. We can see that the shape of the particle distribution function shows significant distinctions for different field strengths. 


\section{Pulse Shaped Fields}

In this section, we explain the basic aspects of pulse shaping. 
The goal of these calculations is to maximize the particle number by deforming the electric field according to the optimal control algorithm. 
At the beginning, we introduce the procedure and then go ahead and present the data.
First, we define an initial electric field, which has the form
\be{
E(t) = \va \ \textrm{sech}^2 \br{t/ \tau}.
}
Then we discretize the vector potential and the electric field with $4096$ non-equidistant points. 
These sampling points are obtained from a set of equidistant sampling points by means of an $atanh$ mapping.
This transformation needs to be performed in order to obtain the necessary accuracy later on. Additionally,
we will compute the gradient for the search direction at these sampling points as well.
Subsequently, we solve both the standard differential equations of first order \eqref{eq3_1:2}
and the adjoint differential equation \eqref{eq2_8:1}.
We do not only save the results at asymptotic times, but keep the values at all sampling points.
The computation of the gradient is straightforward \eqref{eq2_8:2} and gives us the search direction via steepest descent method
\be{
A_{i+1}(t) = A_i(t) - \alpha_i \nabla J \br{A_i(t)} \cdot \textrm{sech}^2 \br{t/\tau}.
\label{eq4_8:1}
}
The $sech^2 \br{t/\tau}$-term is introduced to ensure sufficient damping of the gradient so that the electric field vanishes at asymptotic times for every iteration step. 
Another crucial task is to fix the constraints for the optimal control algorithm. Generally spoken, the electric field strength increases ad infinitum if there is no upper bound. 
To ensure that this will not happen in our simulations we introduced bounds for the field strength.

\begin{figure}[h]
\centering
\includegraphics[width=0.8\textwidth]{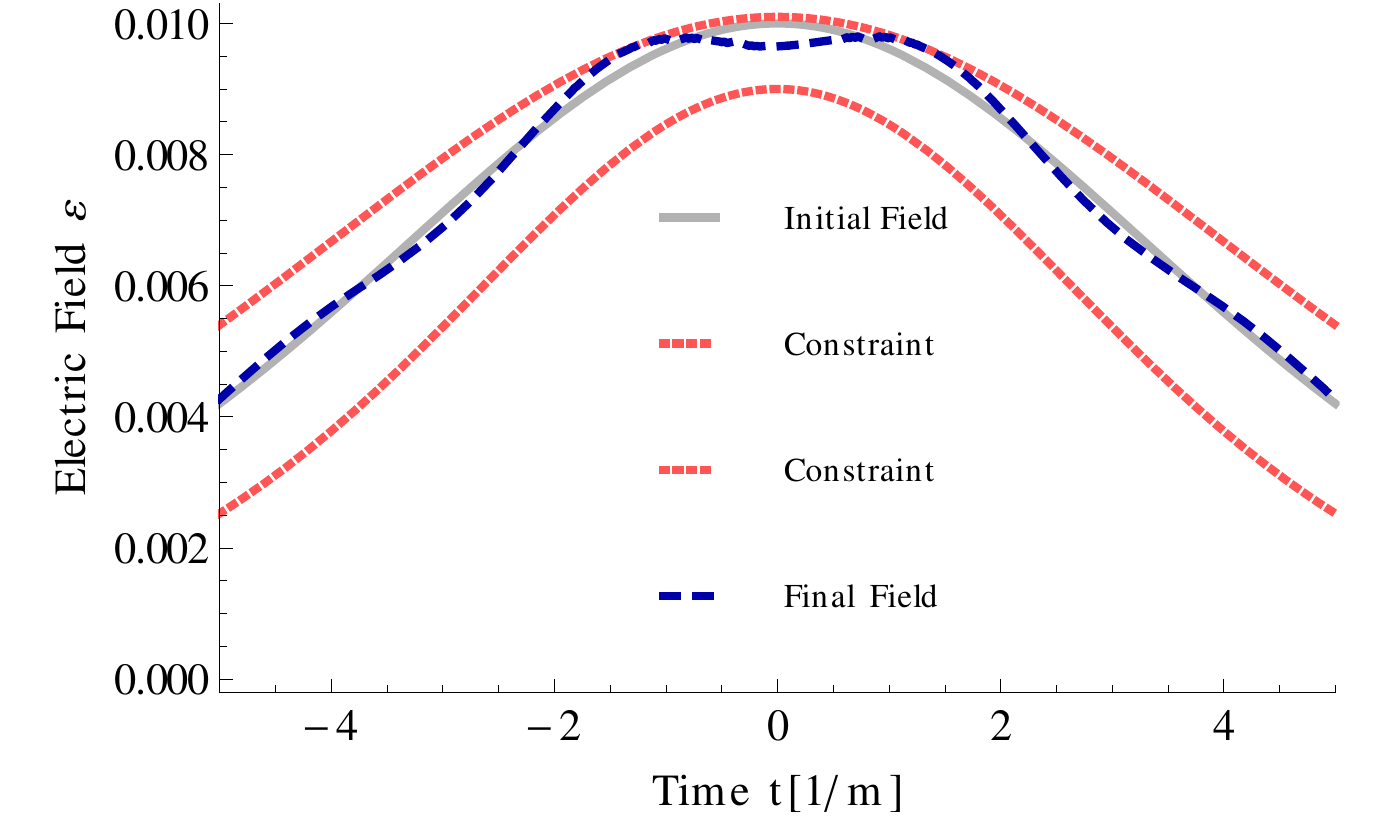}
\caption[Deformation of an optimized electric field for time-dependent constraints]{Deformation of an optimized electric field for time-dependent constraints. The upper limit in the field strength is obtained at two points, whereas the lower limit does not affect the result. The initial field parameters are $~\varepsilon = 0.01,~ \tau = 5m^{-1}$ whereas the bounds are $~\tau = 4m^{-1}$ and $~\tau = 6m^{-1}$ and the offset $~\Delta \va = \pm 0.0001$.}
\label{Fig22}
\end{figure}

The easiest way to do this is to define a maximum field strength for all times. However, we also tried to use time-dependent constraints. In Fig.\ref{Fig22} we show the deformation of the initial pulse $\va = 0.01 ,~ \tau = 5m^{-1}$. In this calculation the upper limit is set at $\va = 0.01 ,~ \tau = 6m^{-1}$ whereas the lower limit is set at $\va = 0.009 ,~ \tau = 4m^{-1}$. Both limits hold with a small offset to ensure that there is free space to explore. 

\begin{figure}[h]
\centering
\includegraphics[width=0.65\textwidth]{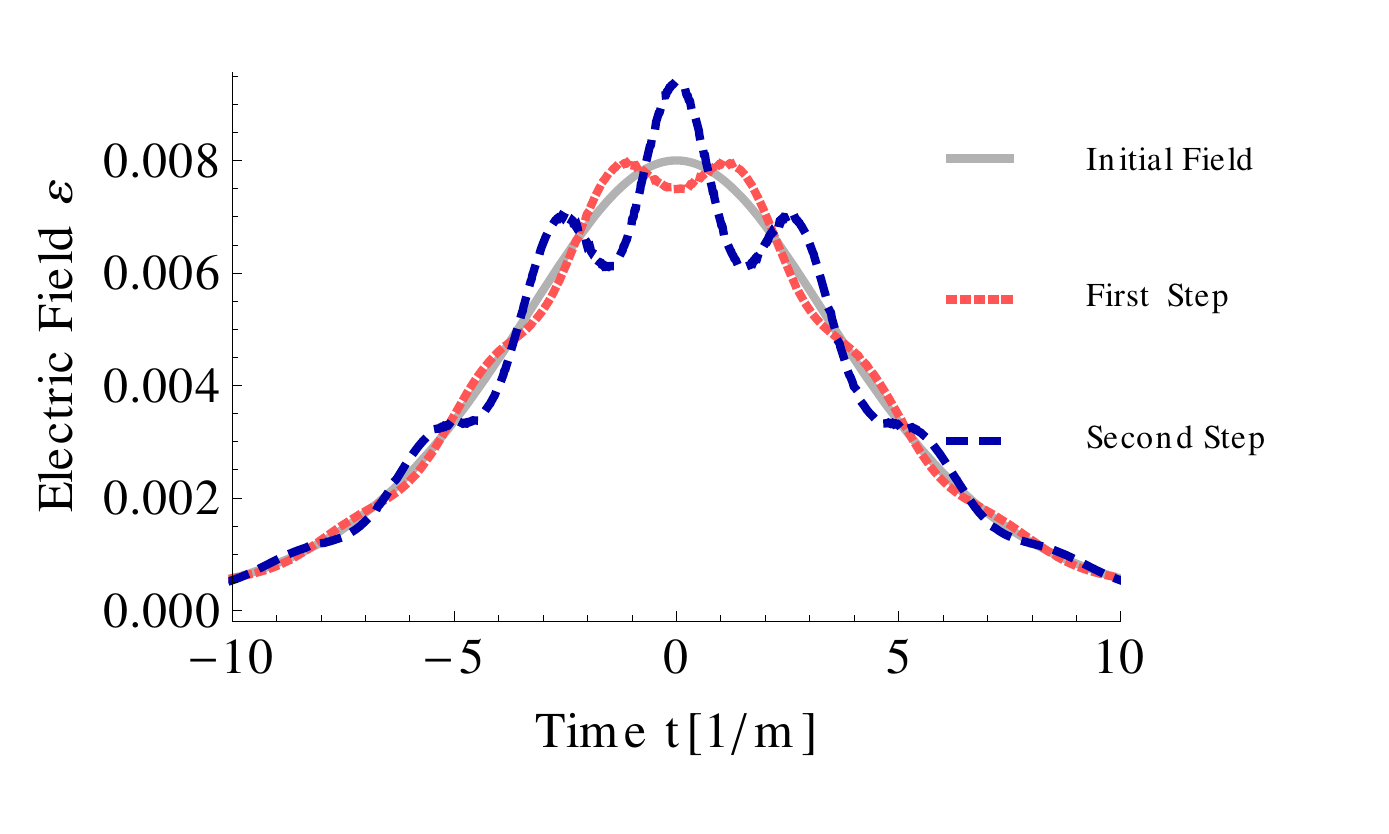}
\caption[Deformation of an unbounded electric field]{Deformation of an initial electric field without constraints. The initial field parameters are $~\varepsilon = 0.008,~ \tau = 5m^{-1}$.}
\label{Fig21}
\end{figure}

We investigated the impact of weaker constraints, too. The corresponding results are shown in Fig.\ref{Fig21}. It can be seen that the preferred substructure appearing after optimization has periodic form. 

\begin{figure}[h]
\subfigure[]{\includegraphics[width=0.49\textwidth]{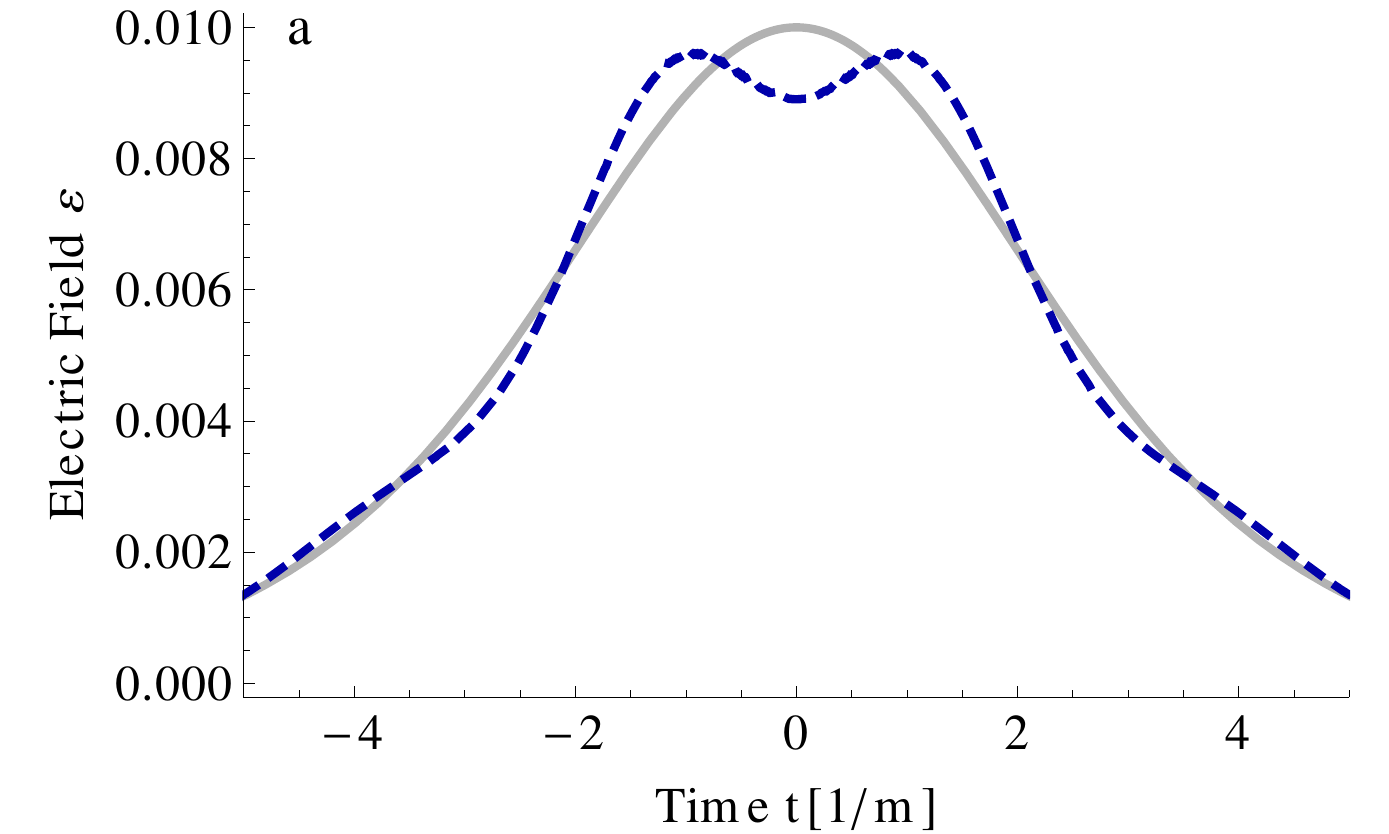}} 
\hfill
\subfigure[]{\includegraphics[width=0.49\textwidth]{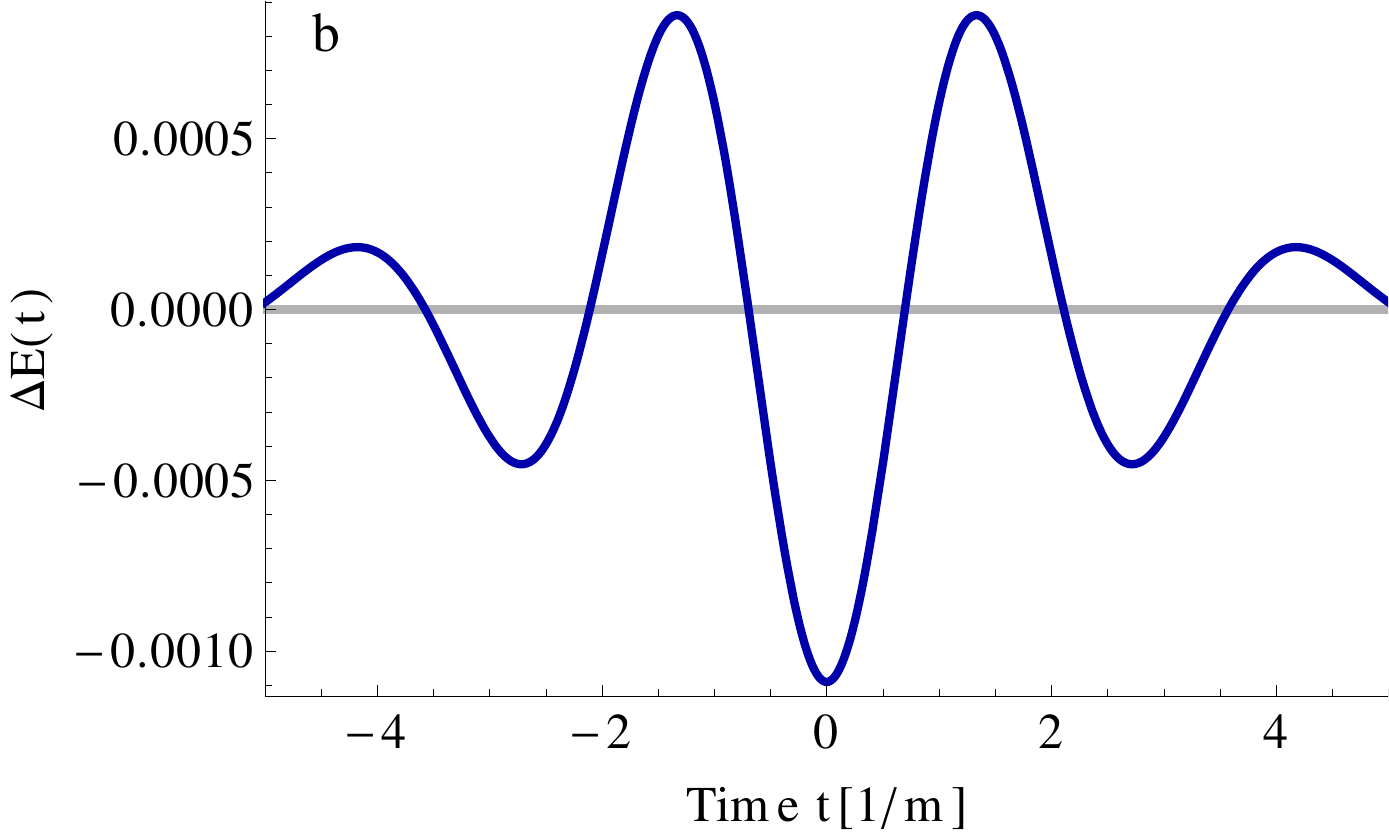}}  \\[-1cm]
\subfigure[]{\includegraphics[width=0.49\textwidth]{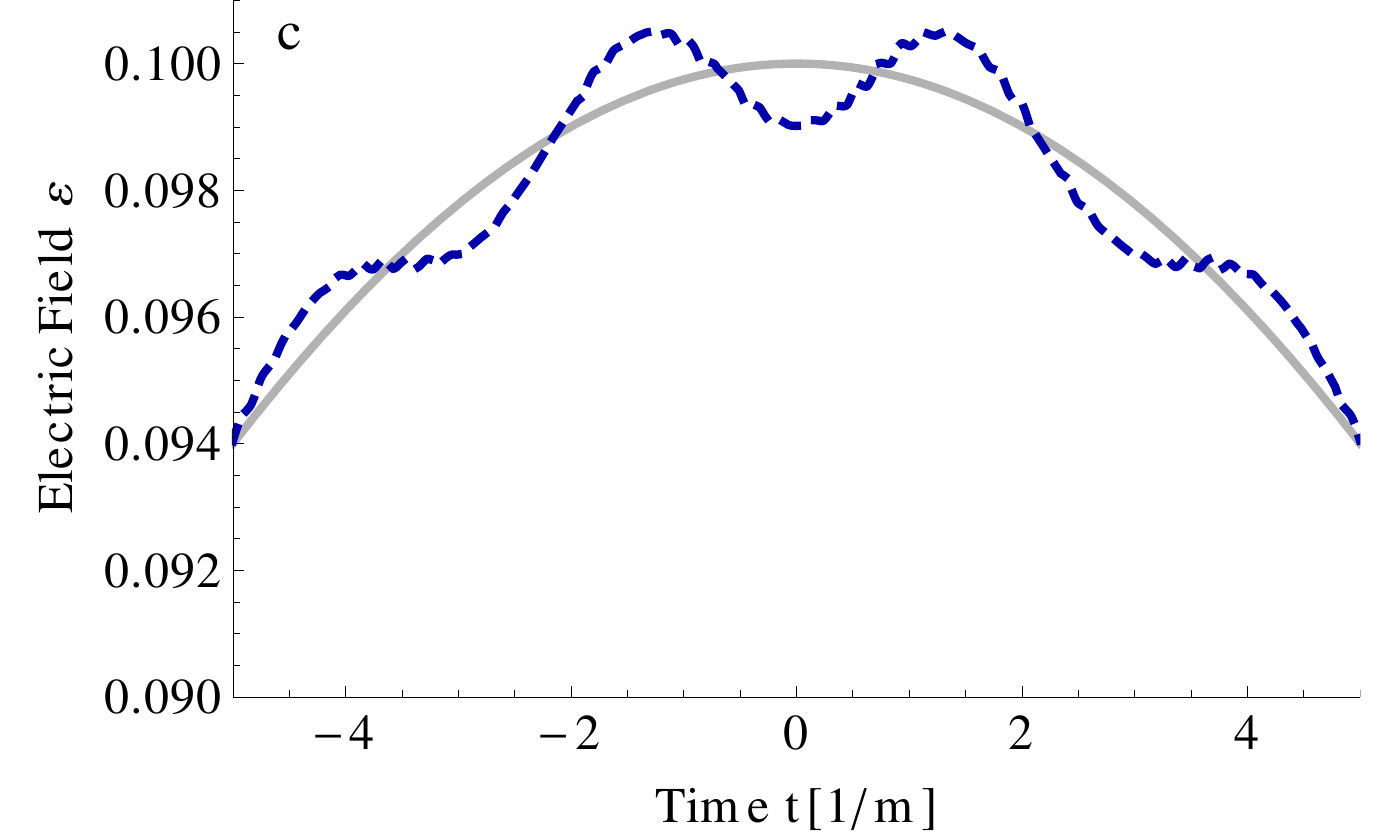}} 
\hfill
\subfigure[]{\includegraphics[width=0.49\textwidth]{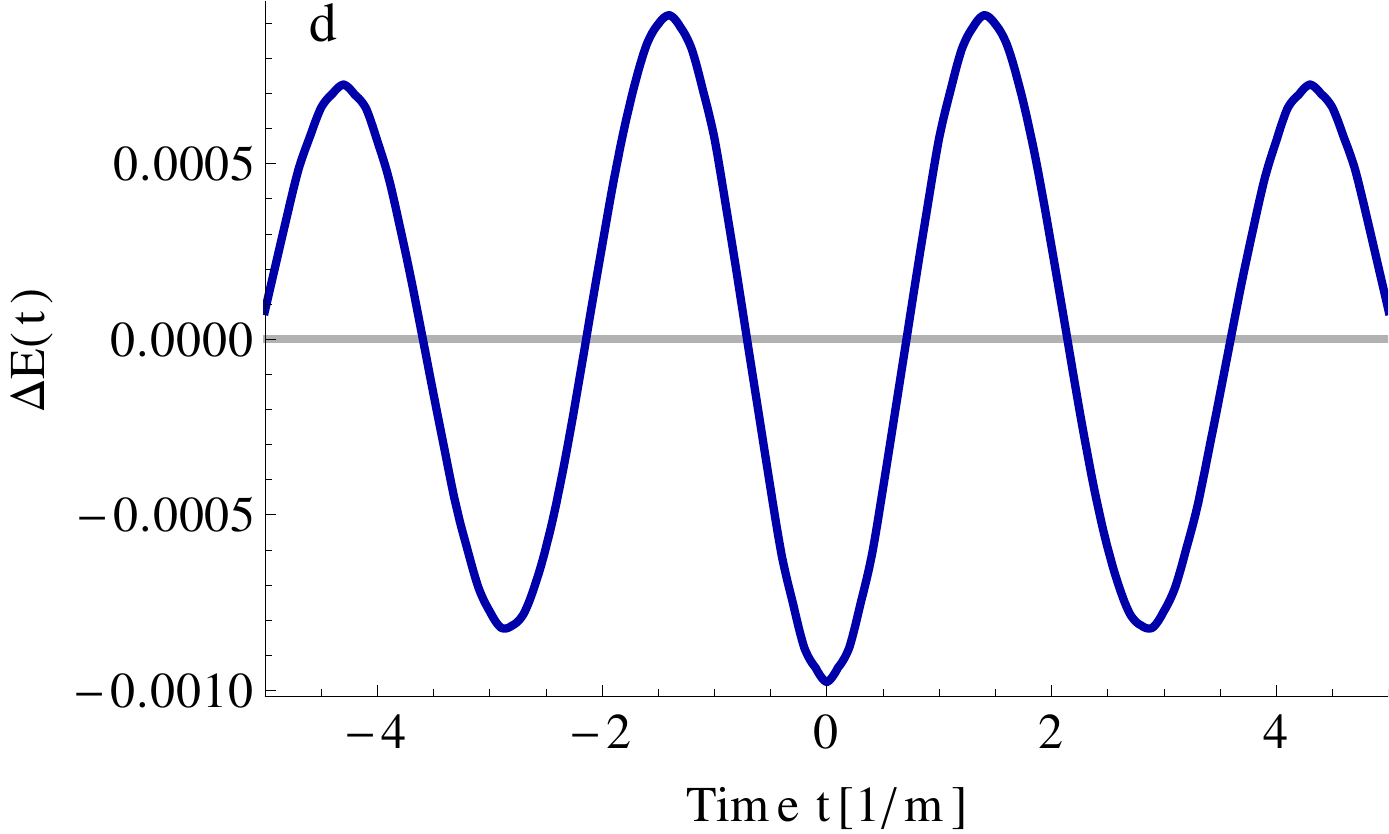}}  \\[-1cm]
\caption[Deformation of an optimized field for different initial configurations]{Deformation of an optimized field for different initial configurations and a constant constraint. One obtains similarities in the structure of optimized fields even for different initial fields. The initial configurations are $~\varepsilon = 0.01,~ \tau = 3m^{-1}$ in $~a~$ and $~\varepsilon = 0.1,~ \tau = 20m^{-1}$ in $~c~$. In $~b~$ and $~d~$ the corresponding substructures are shown. Most notably, the oscillation rate and the peak height of the substructure resemble each other.}
\label{Fig23}
\end{figure}

In order to understand this pulse form we have further investigated setups with two different initial conditions in Fig.\ref{Fig23}. These initial parameters are $\va = 0.01 ,~ \tau=3m^{-1}$ in Fig.\ref{Fig23}a and $\va = 0.1 ,~ \tau=20m^{-1}$ in Fig.\ref{Fig23}c. In both cases the upper limit for the field strength is specified at $101$ percent of the initial field strength. 
As one can see from Fig.\ref{Fig23}b and Fig.\ref{Fig23}d, the substructures favored by the optimal control algorithm are very similar. 
Most notably, both the oscillation rate and the peak height of the substructure resemble each other. Accordingly, one may conclude the existence of an optimal field configuration. 

\begin{figure}[h]
\centering
\includegraphics[width=0.85\textwidth]{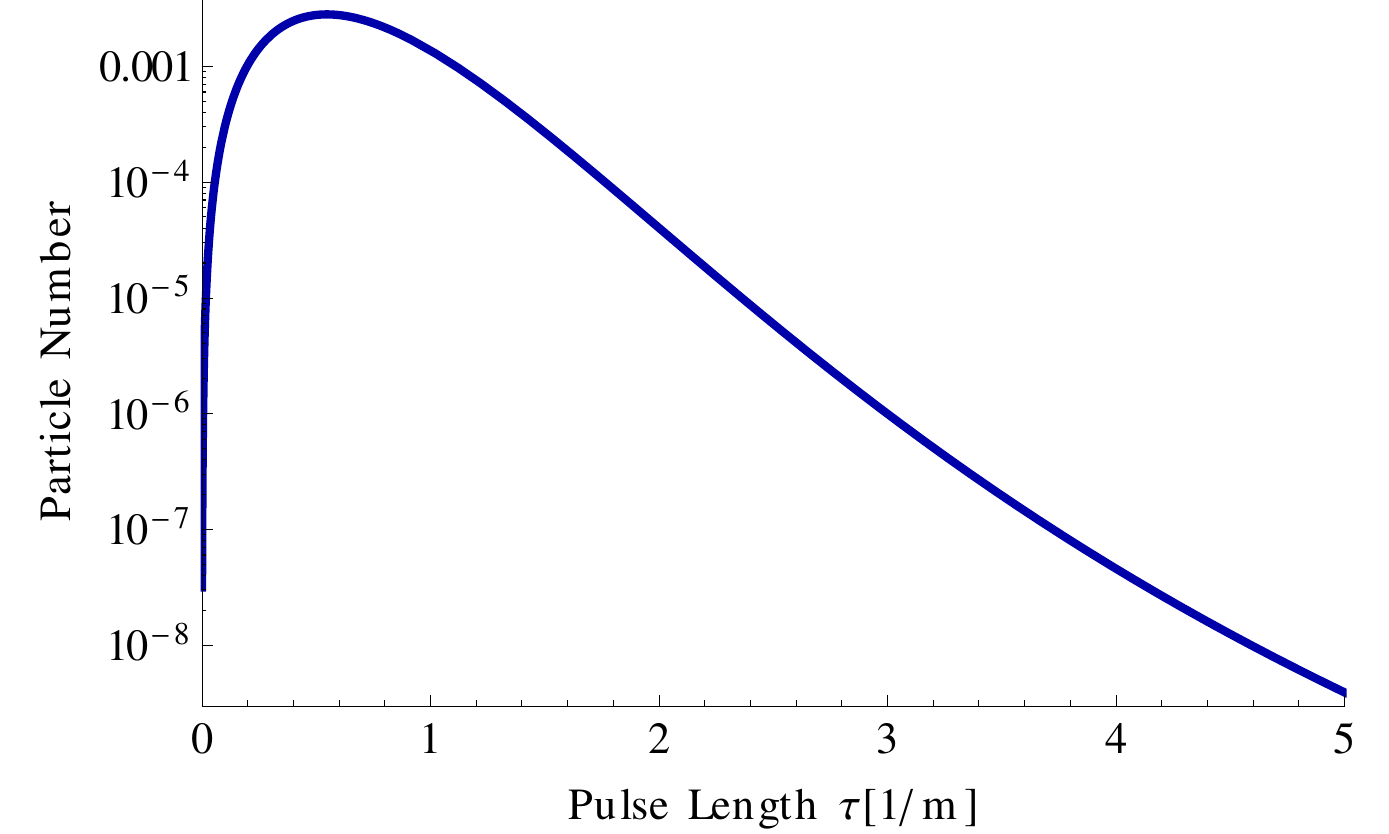}
\caption[Dependence of the pulse length on the particle number for a single pulse in the multiphoton regime]{Dependence of the particle number on the pulse length for a single pulse in the multiphoton regime. For any given field strength the particle number increases with decreasing $~\tau~$ until it reaches $~\tau \approx 0.5m^{-1}$, corresponding to the threshold for creating the rest mass of the an electron and a positron.}
\label{Fig20}
\end{figure}

According to Fig.\ref{Fig23} we conclude that the optimal setup shows substructures of alternating fields with pulse lengths of $\tau \approx 0.5-0.6m^{-1}$.
As a matter of fact, this is what one would expect from an investigation of the particle number of a single pulse as shown in Fig.\ref{Fig20}. 
The particle number shows a maximum at $\tau \approx 0.5m^{-1}$\cite{OrtDip} corresponding to the threshold for creating the rest mass of the electron-positron system. 
Due to the fact that we have not chosen any constraint with respect to the oscillation frequency, it seems that the optimization procedure drives the field configuration towards the multiphoton optimum.





\begin{figure}[h]
\centering
\includegraphics[width=0.85\textwidth]{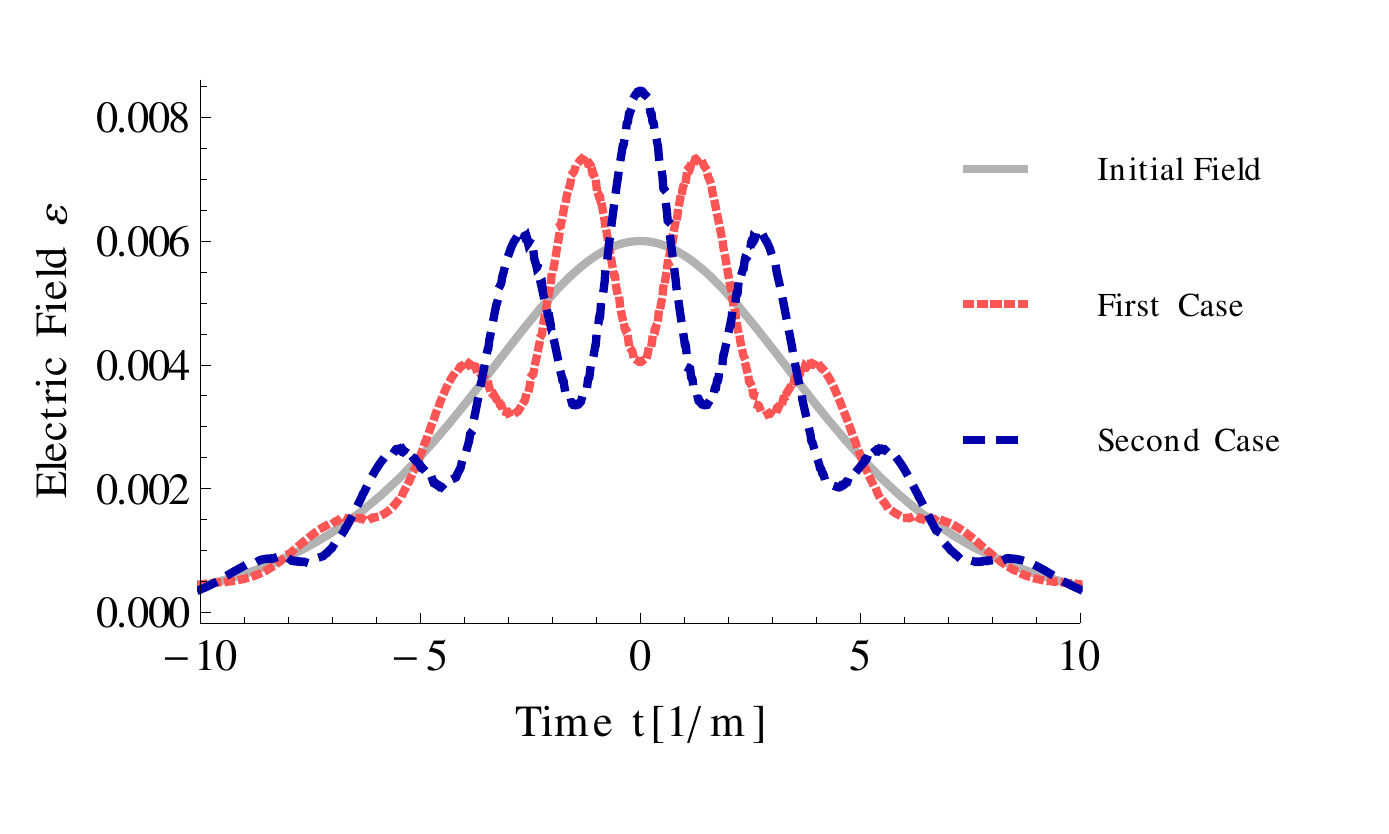}
\caption[Possible cases for a local maximum in the particle number within our approach]{The two possible cases for a local maximum in the particle number within our approach. The oscillations are on top of the initial field with either a positive or a negative sign. The global maximum in the particle number then depends on the chosen constraints.}
\label{Fig24}
\end{figure}

In order to find a maximum of the particle number for any initial field configuration, it seems to be best to add short pulses in alternating configuration with $\tau \approx 0.5m^{-1}$.
Hence, the search for an optimal field configuration can be reduced to the two cases presented in Fig.\ref{Fig24}. It can be seen that either the peak or the valley of the optimized short pulse configuration is at $t=0$. Nevertheless, one has to be aware of the fact that the short-pulsed substructure will dominate the production rate and thus, mainly the multiphoton effect will contribute to the particle number.
Accordingly, the optimal field configuration with regard to particle production does not correspond to the dynamically assisted Schwinger effect, but to the multiphoton effect.
This is in agreement with previous studies, where multiphoton pair production becomes perturbative for $\tau \approx 0.5m^{-1}$\cite{PhysRevD.2.1191} whereas the dynamically assisted Schwinger effect is still exponentially suppressed\cite{PhysRevLett.101.130404}.

\chapter{Conclusions}

In this thesis we have presented an investigation of particle pair production in strong electric fields within the framework of quantum kinetic theory. We focused on field configurations consisting of superpositions of Sauter pulses. Our main findings are the following: \\

We further investigated the dynamically assisted Schwinger effect. In comparison to
previous work, we focused on the effect of superimposing a strong and long pulse with a different number of short and weak pulses.
In this respect, we have also examined the idea of a similarity between Ramsey-interferometry and the Schwinger effect.
Our simulations revealed that the particle distribution function is sensitive to the specific form of the electric field. A further result of this analysis suggests that an indirect measurement of the Schwinger effect
in well-adjusted experimental setups might be possible. \\

In the course of our simulations it turned out that the combination of different solvers might enhance the numerical performance significantly.
In this respect, we want to emphasize that a thorough selection of both solution and interpolation algorithms can save a large amount of computer time. \\

Most notably we have derived equations to do pulse-shaping within the framework of optimal control theory systematically. 
The results of this investigation have revealed that the optimization procedure drives the system into a well-defined direction. 
Starting from a given initial field the system evolves into a state, in which the initial field is superimposed by an oscillating field. 
The oscillation rate of the superimposed field seems to correspond to the perturbative threshold for particle production. \\

This research has raised questions, which need to be further investigated. Unfortunately, an explanation for the deformations in the particle distribution as observed in our simulations cannot be given at the moment.
Further studies on this subject, probably in complementary approaches, could reveal this puzzle.
Moreover, further research might explore configurations with more realistic parameters in the optical or X-ray regime. Another focus should be laid on the inclusion of magnetic field effects and spatial inhomogeneities. 
Finally, we think that the idea of optimization to enhance pair production should be pursued in the future. Further investigations could offer new opportunities in this as well as in related research areas.  \\

All in all one can conclude there are still unsolved mysteries hidden in the area of strong field QED and, particularly, pair production. 
The rapid advances in LASER technology give us hope that an experimental observation based on explicit theoretical predictions comes into reach within the next decade.


\appendix
\chapter{Additional Derivations}
\section{Normalization of Mode Functions}
\label{append:normMode}
The mode functions $\chi^+ \qt$ and $\chi^- \qt$ form the fundamental system of the differential equation
\be{
\br{\de{t}^2 + \omega^2 \qt + \ii e E \brt} \chi \qt = 0.
\label{eqA_1:x1a}
}
For a vanishing electric field $E \brt \to 0$ the term $\omega \qt$ becomes $\omega(q)$ and we obtain the simpler oscillator equations
\be{
\br{\de{t}^2 + \omega^2 \brq} \chi_0 \qt = 0.
}
The linear independent solutions of this equation are clearly given by plane waves and thus the mode functions are given by
\be{
\chi_0^+ \qt = \frac{1}{N^+} \ee^{-\ii \omega \brq t} ,\qquad
\chi_0^- \qt = \frac{1}{N^-} \ee^{\ii \omega \brq t},
\label{eqA_1:x2a}
}
where $N^{\pm}$ are normalization factors. They can be found by adopting the orthogonality relations for the spinors $u_r \qt$ and $v_r \mqt$ 
\al{
u_r^{\dag} \qt \cdot u_s \qt &= v_r^{\dag} \mqt \cdot v_s \mqt = \delta_{rs},  \\
\bar{u}_r \qt \cdot u_s \qt &= - \bar{v}_r \mqt \cdot v_s \mqt = \frac{m}{\omega} \delta_{rs}, \\
u_r^{\dag} \qt \cdot v_s \mqt &= 0,  
} 
where $r=1,2 \quad s=1,2$.
For the sake of a simple presentation we just show the calculation of $u_1^{\dag} \cdot u_1$ in explicit form here. The spinor $u_r$ was defined by
\be{
u_r \qt = \br{\ii \ga{0} \de{t} - \vga \cdot \vpi \qt + m} \chi^+ \qt R_r \qquad r=1,2
}
so we can write for vanishing electric fields
\be{
u_r \qt = \br{\ii \ga{0} \de{t} - \vga \cdot \vpi \brq + m} \chip \qt R_r.
}
Using the definitions for the gamma matrices we obtain
\al{
\vga \cdot \vpi &= \ma{0 & 0 & -\pi_3 & -\pi_1+\ii \pi_2 \\ 0 & 0 & -\pi_1-\ii \pi_2 & \pi_3 \\ \pi_3 & \pi_1-\ii \pi_2 & 0 & 0 \\ \pi_1 + \ii \pi_2 & -\pi_3 & 0 & 0} ,\\
\ii \ga{0} &= \ma{0 & 0 & \ii & 0 \\ 0 & 0 & 0 & \ii \\ \ii & 0 & 0 & 0 \\ 0 & \ii & 0 & 0}.
}
Accordingly, the spinor takes the form
\al{
u_1 \qt = \ma{m \\ 0 \\ \ii \de{t} - q_3 \\ -q_1 - \ii q_2} \chip \qt ,\\
u_1^{\dag} \qt = \ma{m & 0 & -\ii \de{t} - q_3 & -q_1 + \ii q_2} \chis \qt.
}
Using the orthogonality relation between the spinors $u^{\dag} u = 1$, their scalar product gives
\begin{multline}
u_1^{\dag} \cdot u_1 = m^2 \bet{\chip}^2 - \br{\ii \de{t} + q_3} \chis \br{\ii \de{t} - q_3} \chip \\
  - \br{-q_1 + \ii q_2} \chis \br{q_1 + \ii q_2} \chip
\end{multline}
and using the relations
\begin{multline}
\br{q_3 + \ii \de{t}} \chis \br{q_3 - \ii \de{t}} \chip \\
  = q_3^2 \bet{\chip}^2 - \ii q_3 \br{\de{t} \chip} \chis + 
  \ii q_3 \chip \br{\de{t} \chis} + \bet{\de{t} \chip}^2 
\end{multline}
\be{
-\br{q_1 + \ii q_2} \chip \br{-q_1 + \ii q_2} \chis = q_1^2 \bet{\chip}^2 + q_2^2 \bet{\chip}^2
}
one ends up with
\begin{multline}
u_1^{\dag} \cdot u_1 = m^2 \bet{\chip}^2 + \vec{q}^2 \bet{\chip}^2 + \ii q_3 \br{\chip \br{\de{t} \chis}} \\
  - \br{\br{\de{t} \chip} \chis} + \bet{\de{t} \chip}^2.
\label{eqA_1:1}
\end{multline}
In combination with equation \eqref{eqA_1:x2a} this leads to an equation for the normalization constant $~N^+~$ 
\be{
\omega^2 \bet{\frac{\ee^{-\ii \omega t}}{N^+}}^2 + \ii q_3 \br{\frac{\ee^{-\ii \omega t}}{N^+} \br{\frac{\ii \omega \ee^{\ii \omega t}}{\br{N^+}^*}}- \br{\frac{-\ii \omega \ee^{-\ii \omega t}}{N^+}} \frac{\ee^{\ii \omega t}}{\br{N^+}^*}} + 
  \bet{\frac{-\ii \omega}{N^+} \ee^{-\ii \omega t}}^2 \stackrel{!}{=} 1.
}
This is equal to 
\be{
\omega^2 - q_3 \omega - q_3 \omega + \omega^2 = \bet{N^+}^2
}
and yields to the result
\be{
N^+ \brq = \sqrt{2 \omega \brq \br{\omega \brq - q_3}}. 
}
Using the orthogonality relation between the spinors $v^{\dag} v = 1$, one obtains the normalization constant $N^- = \sqrt{2 \omega \br{\omega + q_3}}$.
Thus we can express both mode functions for a vanishing electric field as
\be{
\chi_0^{\pm} \qt = \frac{1}{\sqrt{2 \omega \brq \br{\omega \brq {\mp} q_3}}} \ee^{\mp \ii \omega \brq t}.
}
It can be checked explicitly that all other orthogonality relations hold when we use this normalization.

\section{Bogoliubov Coefficients}
\label{append:bogoliubov}
We start with the spinor $~\psi \qt~$ in the particle and in the quasi-particle representation
\al{
\psi \qt =& \sum_{r=1}^2 \br{u_r \qt a_r \brq + v_r \mqt b_r^{\dag} \brmq} \\
 =& \sum_{r=1}^2 \br{U_r \qt A_r \qt + V_r \mqt B_r^{\dag} \mqt}.
}
By inserting the expressions \eqref{eq2_5:2b} and \eqref{eq2_5:2}
\begin{subequations}
	\begin{align}
		A_r\qt &= \alpha \qt a_r \brq - \beta^* \qt b_r^{\dag} \brmq ,\\
		B_r^{\dag} \mqt &= \beta \qt a_r \brq + \alpha^* \qt b_r^{\dag} \brmq
	\end{align}
\end{subequations}
into the equation above one gets the expression
\begin{equation}
\begin{split}
\sum_{r=1}^2& \left( \br{\ga{0} \omega - \vga \cdot \vpi + m} \kappa^+ R_r \br{\alpha a_r - \beta^* b_r^{\dag}} \right.  \\
  &\left. + \br{-\ga{0} \omega - \vga \cdot \vpi + m} \kappa^- R_r \br{\beta a_r + \alpha^* b_r^{\dag}} \right) \label{eqA_2:9}\\
  = \sum_{r=1}^2& \left( \br{\ii \ga{0} \de{t} - \vga \cdot \vpi + m} \chi^+ R_r a_r \right. \\
  &\left. + \br{\ii \ga{0} \de{t} - \vga \cdot \vpi + m} \chi^- R_r b_r^{\dag} \right). 
\end{split}
\end{equation}

We split this equation now in different parts for better readability. The left hand side gives for $r=1$ 
\be{
\ma{m_e \\ 0 \\ \omega-\pi_3 \\ -\pi_1 - \ii \pi_2} \kappa^+ \br{\alpha a_1 - \beta^* b_1^{\dag}} + 
  \ma{m_e \\ 0 \\ -\omega-\pi_3 \\ -\pi_1 - \ii \pi_2} \kappa^- \br{\beta a_1 + \alpha^* b_1^{\dag}}.
}
Consequently we find for $r=2$
\be{
\ma{\pi_1 - \ii \pi_2 \\ \omega - \pi_3 \\ 0 \\ m_e} \kappa^+ \br{\alpha a_2 - \beta^* b_2^{\dag}} + 
  \ma{\pi_1 - \ii \pi_2 \\ -\omega - \pi_3 \\ 0 \\ m_e} \kappa^- \br{\beta a_2 + \alpha^* b_2^{\dag}}.
}
The right hand side gives for $r=1$
\be{
\ma{m_e \\ 0 \\ \ii \de{t} -\pi_3 \\ -\pi_1 - \ii \pi_2} \chi^+ a_1 + 
  \ma{m_e \\ 0 \\ \ii \de{t}-\pi_3 \\ -\pi_1 - \ii \pi_2} \chi^- b_1^{\dag}
}
and for $r=2$
\be{
\ma{\pi_1 - \ii \pi_2 \\ \ii \de{t} - \pi_3 \\ 0 \\ m_e} \chi^+ a_2 + 
  \ma{\pi_1 - \ii \pi_2 \\ \ii \de{t} - \pi_3 \\ 0 \\ m_e} \chi^- b_2^{\dag}.
}
The obtained expression is a vector equation with four components. To keep the treatment simple, we do not deal with all four components at the same time. Rather we start with the first component
\begin{equation}
\begin{split}
m_e \kappa^+ &\br{\alpha a_1 - \beta^* b_1^{\dag}} + m_e \kappa^- \br{\beta a_1 + \alpha^* b_1^{\dag}} \\
  &+ \br{\pi_1 - \ii \pi_2} \kappa^+ \br{\alpha a_2 - \beta^* b_2^{\dag}} + \br{\pi_1 - \ii \pi_2} \kappa^- \br{\beta a_2 + \alpha^* b_2^{\dag}} \\
  = m_e &\chi^+ a_1 + m_e \chi^- b_1^{\dag} 
  + \br{\pi_1 - \ii \pi_2} \chi^+ a_2 + \br{\pi_1 - \ii \pi_2} \chi^- b_2^{\dag}
\end{split}
\end{equation}
and find via comparison of coefficients
\begin{alignat}{3}
a_1:& &\quad \kappa^+ \alpha + \kappa^- \beta =& \chi^+ 	,\qquad m_e &\ne 0, \label{eqA_2:1} \\
a_2:& &\quad  \kappa^+ \alpha + \kappa^- \beta =& \chi^+ 	,\qquad \br{\pi_1- \ii \pi_2} &\ne 0, \label{eqA_2:2} \\
b_1^{\dag}:& &\quad  -\kappa^+ \beta^* + \kappa^- \alpha^* =& \chi^- ,\qquad m_e &\ne 0, \label{eqA_2:3} \\
b_2^{\dag}:& &\quad  -\kappa^+ \beta^* + \kappa^- \alpha^* =& \chi^- ,\qquad \br{\pi_1 - \ii \pi_2} &\ne 0. \label{eqA_2:4}
\end{alignat}
We are dealing with massive particles so we can use equations \eqref{eqA_2:1} \eqref{eqA_2:3} and neglect the restrictions of the equations \eqref{eqA_2:2} \eqref{eqA_2:4}. We proceed with the second component
\begin{multline}
\br{\omega - \pi_3} \kappa^+ \br{\alpha a_2 - \beta^* b_2^{\dag}} + \br{-\omega - \pi_3} \kappa^- \br{\beta a_2 + \alpha^* b_2^{\dag}} \\
  = \br{\ii \de{t} - \pi_3} \chi^+ a_2 + \br{\ii \de{t} - \pi_3} \chi^- b_2^{\dag}
\end{multline}
and again we find via comparison of coefficients
\begin{alignat}{3}
a_2:& &\quad \br{\omega - \pi_3} \kappa^+ \alpha - \br{\omega + \pi_3} \kappa^- \beta =& \br{\ii \de{t} - \pi_3} \chi^+ \label{eqA_2:5} \\
b_2^{\dag}:& &\quad -\br{\omega - \pi_3} \kappa^+ \beta^* - \br{\omega + \pi_3} \kappa^- \alpha^* =& \br{\ii \de{t} - \pi_3} \chi^-. \label{eqA_2:6}
\end{alignat}
We get four different equations \eqref{eqA_2:1}, \eqref{eqA_2:3}, \eqref{eqA_2:5} and \eqref{eqA_2:6}, as the lower two components in \eqref{eqA_2:9} do not contain additional information.
Now we restrict ourselves to the coefficients of the operators $a_i$ and pick out the two equations
\begin{subequations}
	\begin{align}
		\label{eqA_2:x1a}
		\quad \kappa^+ \alpha + \kappa^- \beta =& \chi^+, \\
		\label{eqA_2:x1b}
		\quad \br{\omega - \pi_3} \kappa^+ \alpha - \br{\omega + \pi_3} \kappa^- \beta =& \br{\ii \de{t} - \pi_3} \chi^+.
	\end{align}
\end{subequations}
Multiplying equation \eqref{eqA_2:x1a} with $\br{\omega + \pi_3}$ and taking the sum of both equations gives
\be{
\br{\omega + \pi_3 + \omega - \pi_3} \kappa^+ \alpha = \br{\omega + \pi_3} \chi^+ + \br{\ii \de{t} - \pi_3} \chi^+,
}  
which yields
\be{
2 \omega \kappa^+ \alpha = \omega \chi^+ + \ii \br{\de{t} \chi^+}. \label{eqA_2:7}
}
Then we multiply this equation by $\br{2 \omega \kappa^+}^{-1}$.
As $\kappa^{\pm}$ was defined as
\begin{subequations}
	\begin{align}
	\kappa^+ \qt = \frac{\ee^{-\ii \theta \qt}}{\sqrt{2\omega \qt \br{\omega \qt - \pi_3 \qt}}} ,\\
	\kappa^- \qt = \frac{\ee^{\ii \theta \qt}}{\sqrt{2\omega \qt \br{\omega \qt + \pi_3 \qt}}} ,
	\end{align}
\end{subequations}
we can write
\al{
\br{2 \omega \kappa^+}^{-1} &= \br{\frac{\sqrt{2\omega} \ee^{-\ii \theta}}{\sqrt{\omega - \pi_3}}}^{-1} \\
  &= \frac{\omega^2 - \pi_3^2}{\sqrt{2\omega \br{\omega + \pi_3}}} \ee^{\ii \theta} = \sqrt{\omega^2 - \pi_3^2} \kappa^-.
}
Thus, we obtain for $\alpha \qt$
\be{
\alpha \qt = \ii \sqrt{\omega^2 \qt - \pi_3^2 \qt} \kappa^- \qt \br{\de{t} - \ii \omega \qt} \chi^+ \qt.
\label{eqA_2:x2a}
}


A similar calculation holds for $\beta \qt$, too. Again we start with the equations \eqref{eqA_2:x1a} and \eqref{eqA_2:x1b},
but this time we multiply equation \eqref{eqA_2:x1a} with $-\br{\omega - \pi_3}$. The resulting equation is then
\be{
2 \omega \kappa^- \beta = \br{\omega - \ii \de{t}} \chi^+.
\label{eqA_2:8}
}
After a calculation similar to the one above we obtain
\be{
\beta \qt = -\ii \sqrt{\omega^2 \qt - \pi_3^2 \qt} \kappa^+ \qt \br{\de{t} + \ii \omega \qt} \chi^+ \qt.
}
 

\section{Derivatives of the Bogoliubov Coefficients}
\label{append:dot_bogoliubov}
We start with the definition of $\alpha \qt$ as given in \eqref{eqA_2:x2a} but explicitly insert the expression for $\kappa^-$
\be{
\alpha \qt = \ii \br{\br{\de{t} \chi^+ \qt} - \ii \omega \qt \chi^+ \qt} \sqrt{\frac{\omega \qt - \pi_3 \qt}{2\omega \qt}} \ee^{\ii \theta \qt}.
} 
Then we differentiate $\alpha \qt$ with respect to $t$
\begin{multline}
\dot{\alpha} = \ii \br{\de{t} \br{\de{t} \chi^+} - \ii \de{t} \br{\omega \chi^+}} \sqrt{\frac{\omega - \pi_3}{2\omega}} \ee^{\ii \theta} \\
  + \ii \br{ \br{\de{t} \chi^+}  -\ii \omega \chi^+} \de{t} \underbrace{\br{\sqrt{\frac{\omega - \pi_3}{2\omega}} \ee^{\ii \theta}}}_{(*)}.
\label{eqA_3:2}
\end{multline}
The time derivative of the term $(*)$ gives
\begin{equation}
\begin{split}
  \de{t}& \br{\sqrt{\frac{\omega - \pi_3}{2\omega}} \ee^{\ii \theta}} \\
  =& \frac{1}{2} \sqrt{\frac{2\omega}{\omega - \pi_3}} \frac{\br{\dot{\omega} - \dot{\pi_3}} 2 \omega - \br{\omega - \pi_3}2 \dot{\omega}}{4\omega^2} \ee^{\ii \theta} + \sqrt{\frac{\omega - \pi_3}{2 \omega}} \ee^{\ii \theta} \ii \omega. \\
\end{split}
\end{equation}	  
Taking into account the expressions
\al{
\dot{\omega} &= \frac{\pi_3 e E}{\omega} ,\\
\dot{\pi_3} &= eE,
}
we can further rewrite the term $(*)$ as
\begin{equation}
\begin{split}
& \frac{1}{2} \sqrt{\frac{2\omega}{\omega- \pi_3}} \frac{eE\pi_3 - eE\omega - eE\pi_3 + eE\pi_3^2/\omega}{2\omega^2} \ee^{\ii \theta} + \sqrt{\frac{\omega - \pi_3}{2\omega}} \ee^{\ii \theta} \ii \omega \\
	  = -\sqrt{\frac{\omega - \pi_3}{2\omega}}& \frac{eE}{2\omega^2} \br{\omega + \pi_3} \ee^{\ii \theta} + \sqrt{\frac{\omega - \pi_3}{2\omega}} \ee^{\ii \theta} \ii \omega.
\label{eqA_3:1}
\end{split}
\end{equation}	
Inserting \eqref{eqA_3:1} into \eqref{eqA_3:2} we obtain upon taking into account the differential equation \eqref{eqA_1:x1a}
\begin{equation}
\begin{split}
	\dot{\alpha} &= \ii \sqrt{\frac{\omega - \pi_3}{2\omega}} \ee^{\ii \theta} \left( -\omega^2 \chi^+ - \ii e E \chi^+ - \ii \frac{eE\pi_3}{\omega} \chi^+ - \ii \omega \br{\de{t} \chi^+} \right. \\
	  &\left. - \br{\de{t} \chi^+ - \ii \omega \chi^+} \br{\frac{eE}{2\omega} + \frac{eE \pi_3}{2\omega^2} + \ii \omega} \right). \\ 
\end{split}
\end{equation}
This can be simplified to
\begin{equation}
\begin{split}
	  \dot{\alpha} &= \ii \sqrt{\frac{\omega - \pi_3}{2\omega}} \ee^{\ii \theta} \br{\br{-\ii e E - \ii \frac{eE \pi_3}{\omega} + \frac{\ii e E}{2} + \frac{\ii eE \pi_3}{2\omega}} \chi^+ 
	  -\frac{eE}{2\omega^2} \br{\omega + \pi_3} \br{\de{t} \chi^+}} \\
	  &= \ii \sqrt{\frac{\omega - \pi_3}{2\omega}} \ee^{\ii \theta} \br{-\frac{eE}{2\omega^2} \br{\omega + \pi_3} \br{\de{t} + \ii \omega} \chi^+} \\
	  &= -\ii \sqrt{\frac{\omega - \pi_3}{2 \omega}} \ee^{-\ii \theta} \frac{eE}{2\omega^2} \sqrt{\omega^2 - \pi_3^2} \br{\de{t} + \ii \omega} \chi^+ \ee^{2\ii \theta} \\
	  &= \frac{1}{2} W \beta \ee^{2\ii \theta},
\end{split}
\end{equation}
with $W \qt = \frac{e E \brt \enorm \brq}{\omega^2 \qt}$ and $\enorm^2 \brq = m_e^2 + q_1^2 + q_2^2$.

Analogously, we perform the same lengthy but straightforward calculation for $\beta \qt$
\be{
\beta \qt = -\ii \br{\br{\de{t} \chi^+ \qt} + \ii \omega \qt \chi^+ \qt} \sqrt{\frac{\omega \qt + \pi_3 \qt}{2\omega \qt}} \ee^{-\ii \theta \qt}.
} 
Differentiation with respect to $t$ gives
\begin{multline}
\dot{\beta} = -\ii \br{\de{t} \br{\de{t} \chi^+} + \ii \de{t} \br{\omega \chi^+}} \sqrt{\frac{\omega + \pi_3}{2\omega}} \ee^{-\ii \theta} \\
  - \ii \br{ \br{\de{t} \chi^+}  +\ii \omega \chi^+} \de{t} \underbrace{\br{\sqrt{\frac{\omega + \pi_3}{2\omega}} \ee^{-\ii \theta}}}_{(**)}.
\end{multline}
The time derivative of the term $(**)$ gives
\begin{equation}
\begin{split}
  \de{t}& \br{\sqrt{\frac{\omega + \pi_3}{2\omega}} \ee^{-\ii \theta}} \\
  =& \frac{1}{2} \sqrt{\frac{2\omega}{\omega + \pi_3}} \frac{\br{\dot{\omega} + \dot{\pi_3}} 2 \omega - \br{\omega + \pi_3}2 \dot{\omega}}{4\omega^2} \ee^{-\ii \theta} + \sqrt{\frac{\omega + \pi_3}{2 \omega}} \ee^{-\ii \theta} \br{-\ii \omega} \\
\end{split}
\end{equation}	  
Taking into account the expressions $\dot{\omega} = \frac{\pi_3 e E}{\omega} ,~ \dot{\pi_3} = eE$ again, 
we can further rewrite the term $(**)$ as
\begin{equation}
\begin{split}
& \frac{1}{2} \sqrt{\frac{2\omega}{\omega + \pi_3}} \frac{eE\pi_3 + eE\omega - eE\pi_3 - eE\pi_3^2/\omega}{2\omega^2} \ee^{-\ii \theta} + \sqrt{\frac{\omega + \pi_3}{2\omega}} \ee^{-\ii \theta} \br{-\ii \omega} \\
	  = -\sqrt{\frac{\omega + \pi_3}{2\omega}}& \frac{eE}{2\omega^2} \br{\omega - \pi_3} \ee^{-\ii \theta} + \sqrt{\frac{\omega + \pi_3}{2\omega}} \ee^{-\ii \theta} \br{-\ii \omega}.
\end{split}
\end{equation}
We put everything together and get
\begin{equation}
\begin{split}
	\dot{\beta} &= - \ii \sqrt{\frac{\omega + \pi_3}{2\omega}} \ee^{-\ii \theta} \left( -\omega^2 \chi^+ - \ii e E \chi^+ + \ii \frac{eE \pi_3}{\omega} \chi^+ + \ii \omega \br{\de{t} \chi^+} \right. \\ 
	  &\left. +\br{\br{\de{t} \chi^+} + \ii \omega \chi^+} \br{\frac{eE}{2\omega^2} \br{\omega - \pi_3} - \ii \omega} \right) \\
	  &= -\ii \sqrt{\frac{\omega + \pi_3}{2\omega}} \ee^{-\ii \theta} \left( \br{-\ii e E + \ii e E \frac{\omega}{\pi_3} + \ii \frac{eE}{2} - \ii \pi_3 \frac{eE}{2\omega}} \chi^+ \right. \\
	  &\left. +\br{\br{\de{t} \chi^+} \frac{eE}{2\omega^2} \br{\omega - \pi_3}} \right) \\
	  &= \ii \sqrt{\frac{\omega - \pi_3}{2\omega}} \ee^{\ii \theta} \frac{eE}{2\omega^2} \sqrt{\omega^2 - \pi_3} \br{\br{\de{t} \chi^+} - \ii \omega \chi^+} \ee^{-2\ii \theta} \\
	  &= -\frac{1}{2} W \alpha \ee^{-2\ii \theta}.
\end{split}
\end{equation}
This completes the check on the Bogoliubov coefficients.

\section{Hypergeometric Differential Equation}
\label{append:hyper}
We start with the differential equation
\be{
\br{u \br{1-u} \de{u} u \br{1-u} \de{u} + \frac{\tau^2}{4} \omega^2 \qu + \ii \va \tau^2 u \br{1-u}} \chi \qu = 0
\label{eqA_4:1}
}
and the ansatz
\be{
\chi \qu = \uu \muu \eta \qu.
\label{eqA_4:4}
}
In order to solve the differential equation \eqref{eqA_4:1}, we rewrite it by using the ansatz \eqref{eqA_4:4} starting with the
derivatives. The first derivative with respect to $u$ is given by
\al{  
\br{\de{u}\chi} = \br{\alpha \uua{-1} \muu - \beta \uu \muub{-1}} &\times \eta \notag \\
  + \br{\uu \muu} &\times \br{\de{u} \eta}
}
and the second one is given by
\al{ 
\de{u}\br{\de{u}\chi}
  = \left( \alpha \br{\alpha-1} \uua{-2} \muu \right. \notag \\
  - 2\br{\alpha \uua{-1} \beta \muub{-1}} \notag \\
  \left. + \beta \uu \br{\beta-1} \muub{-2} \right) &\times \eta \notag \\
  + 2 \left( \alpha \uua{-1} \muu - \beta \uu \muub{-1} \right) &\times \br{\de{u}\eta} \notag \\
  + \left( \uu \muu \right) &\times  \de{u}\br{\de{u}\eta}.
\label{eqA_4:2}
}
Note that we have to apply the product rule in the first term in \eqref{eqA_4:1}
\begin{multline}
u\br{1-u} \de{u} u\br{1-u} \br{\de{u}\chi} \\
 = u \br{1-u}^2 \br{\de{u}\chi} - u^2 \br{1-u} \br{\de{u}\chi} + u^2 \br{1-u}^2 \br{\de{u}\br{\de{u}\chi}}.
\label{eqA_4:3}
\end{multline}
We then insert these derivatives into the differential equation \eqref{eqA_4:1} resulting in a lengthy expression. The prefactor of the second derivative $\de{u}\br{\de{u}\eta}$ is given by
\be{
\uua{+2} \muub{+2},
}
whereas the prefactor of the first derivative $\br{\de{u} \eta \qu}$ reads
\begin{multline}
2 \alpha \uua{+1} \muub{+2} - 2\beta \uua{+2} \muub{+1} \\
  + \uua{+1} \muub{+2} - \uua{+2} \muub{+1}.
\end{multline}
Finally, the prefactor of $\eta \qu$ is the most complicated one. There are three contributions in total, one arising from \eqref{eqA_4:2}
\be{
\alpha \br{\alpha-1} \uu \muub{+2} + \beta \br{\beta-1} \uua{+2} \muu - 2 \alpha \beta \uua{+1} \muub{+1},
}
another one stemming from \eqref{eqA_4:3}
\be{
\alpha \uu \muub{+2} - \beta \uua{+1} \muub{+1} - \alpha \uua{+1} \muub{+1}  + \beta \uua{+2} \muu
}
and a last one originating from the remaining terms in \eqref{eqA_4:1}
\begin{equation}
\begin{split}
\frac{\tau^2}{4} &\br{\omega^2 + 4 \ii \va u \br{1-u}} \uu \muu \\
  = \frac{\tau^2}{4} &\left( 1 + q_1^2 + q_2^2 + q_3^2 + 2q_3 \va \tau u - 2q_3 \va \tau \br{1-u} + \va^2 \tau^2 u \right. \\
  &\left. + \va^2 \tau^2 \br{1-u}^2 - 2\va^2 \tau^2 u \br{1-u}  +4\ii \va u \br{1-u} \right)  \uu \muu , 
\end{split}
\end{equation}
where we have used 
\be{
\br{2u-1}^2 = 4u^2 - 4u +1= u^2 -2u \br{1-u}  +\br{1-u}^2.
}
Accordingly, the differential equation eventually is given by
\begin{alignat}{2}
	\uua{+2} \muub{+2} &\times \de{u}\br{\de{u}\eta \qu} \\
	  + \left( 2 \alpha \uua{+1} \muub{+2} - 2\beta \uua{+2} \muub{+1} \right. & \notag \\
	  \left. + \uua{+1} \muub{+2} - \uua{+2} \muub{+1} \right) & \times \br{\de{u} \eta \qu} \notag \\
	  + \left( \alpha \br{\alpha-1} \uu \muub{+2} \right. & \notag \\
	  + \beta \br{\beta-1} \uua{+2} \muu - 2\alpha \beta \uua{+1} \muub{+1} & \notag \\
	  + \alpha \uu \muub{+2} - \beta \uua{+1} \muub{+1} & \notag \\
	  \left. - \alpha \uua{+1} \muub{+1}  + \beta \uua{+2} \muu \right) & \times \eta \qu \notag \\
	  + \frac{\tau^2}{4} \left( 1 + q_1^2 + q_2^2 + q_3^2 + 2q_3 \va \tau u - 2q_3 \va \tau \br{1-u} + \va^2 \tau^2 u \right. & \notag \\
	  \left. + \va^2 \tau^2 \br{1-u}^2 - 2\va^2 \tau^2 u \br{1-u}  +4\ii \va u \br{1-u} \right) \uu \muu & \times \eta \qu \label{eq2_7:2} \\
	  & = 0. \notag 
\end{alignat}

\section{Normalization of Analytic Mode Functions}
\label{append:norm_an}
The normalization constants $N^{\pm}$ are chosen such that
\be{
\chi^{\pm} \bru = \kappa^{\pm} \bru ,\qquad u \to 0^+.
}
Accordingly we have to investigate the behaviour of the mode functions $\chi \qu$ in this limit
\begin{equation}
\begin{split}
\chi^+ \qu &= 
  N^+ u^{\alpha} \br{1-u}^{\beta} \Hyper{a,b,c;u} \\
  &= N^+ \ee^{\alpha \tln{u}} \ee^{\beta \tln{1-u}} \Hyper{a,b,c;u} \\
  &\stackrel{u \to 0^+}{=} N^+ \ee^{\alpha \tln{u}} = 
  N^+ u^{\alpha}, 
\label{eqA5_x1a}
\end{split}
\end{equation}
as $\Hyper{a,b,c;u \to 0^+} = 1$. A similar calculation can be done for $\chi^- \qu$, which results in
\be{
\chi^- \qu = N^- u^{-\alpha} ,\qquad u \to 0^+.
\label{eqA5_x1b}
}
It is possible to calculate $\kappa \qu$ in the same limit. From the definition of the functions, we have
\be{
\kappa^{\pm} \qu = \frac{\ee^{\mp \ii \theta \qu}}{\sqrt{2 \omega \qu \br{\omega \qu \mp \pi_3 \qu}}},
}
with the transformed dynamical phase $\theta \qu$. Using $\de{t} = \frac{2}{\tau} u \br{1-u} \de{u}$ we find
\be{
\theta \qt = \int_{t_0}^t \omega \br{\vec q,\tau} d\tau,
} 
which takes the form
\be{
\theta \qu = \frac{2}{\tau} \int_{u_0}^u du' \frac{\omega \br{\vec q, u'}}{u \br{1-u}}.
}
The integration can be performed analytically
\begin{equation}
\begin{split}
\int &\frac{\omega \qu}{u \br{1-u}} du = - 2\va \tau \tln{q_3 + \va \tau u - \va \tau \br{1-u} + \omega \qu} \\
  &+ \omega_0 \br{\tln{u} - \tln{1+q_1^2+q_2^2+ \br{q_3 - \va \tau} \br{q_3 + \br{2u-1} \va \tau} + \omega_0 \omega \qu}} \\
  &- \omega_1 \br{\tln{1-u} - \tln{1+q_1^2+q_2^2 + \br{q_3 + \va \tau} \br{q_3 + \br{2u-1} \va \tau} + \omega_1 \omega \qu}}, 
\end{split}
\end{equation}
where $\omega_0 = \omega \br{\vec q,0} ,~ \omega_1 = \omega \br{\vec q,1}$. 
In the limit $u \to 0^+$, the dynamical phase splits into a regular part $\tilde{\theta} \br{\vec q, 0}$ as well as into a divergent part
\be{
\theta \qu \stackrel{u \to 0^+}{\to} \frac{\tau}{2} \omega_0 \tln{u} + \tilde{\theta} \br{\vec q, 0}.
} 
Denoting $\pi_3 \br{\vec q,0} = q_3 - \va \tau$ the mode functions $\kappa$ read
\be{
\kappa^{\pm} \br{\vec q,0} = \frac{\ee^{\mp \ii \tilde{\theta} \br{\vec q, 0}}}{\sqrt{2 \omega \br{\vec q, 0} \br{\omega \br{\vec q, 0} \mp \pi_3 \br{\vec q, 0}}}} \ee^{\mp \frac{\ii \tau}{2} \omega_0 \tln{u}}.
\label{eqA5_x1c}
}
Comparison of \eqref{eqA5_x1a}, \eqref{eqA5_x1b} and \eqref{eqA5_x1c} fixes the normalization constants
\al{
N^+ \brq = \frac{\ee^{-\ii \tilde{\theta} \brq}}{\sqrt{2 \omega \br{\vec q, 0} \br{\omega \br{\vec q, 0} - \pi_3 \br{\vec q, 0}}}} \\
N^- \brq = \frac{\ee^{\ii \tilde{\theta} \brq}}{\sqrt{2 \omega \br{\vec q, 0} \br{\omega \br{\vec q, 0} + \pi_3 \br{\vec q, 0}}}}.
}

\section{Asymptotic Distribution Function}
\label{append:an_final}
We found an analytical expression for the distribution function 
\be{
F \qu = \bet{N^+ \brq}^2 \br{1+ \frac{\pi_3 \qu}{\omega \qu}} \underbrace{\bet{F_1 \qu + \ii F_2 \qu}^2}_{(*)},
\label{eqA_6:1}
}
where
\al{
F_1 \qu &= \frac{2}{\tau} u \br{1-u} \frac{ab}{c} \Hyper{1+a,1+b,1+c;u} ,\\
F_2 \qu &= \br{\omega \qu - \br{1-u} \omega_0 - u \omega_1} \Hyper{a,b,c;u} .
}
We now want to derive an asymptotic expression for the distribution function in the limit $u \to 1^-$. As we will encounter the gamma function $\Gamma(x)$ in the following, we introduce it here. 
This gamma function is defined for all complex numbers except the non-positive integers. For $n \in \mathbb{N}$ the gamma function coincides with the factorial function
\be{
\Gamma \br{n} = \br{n-1}!.
}
The gamma function obeys, among others, the following relations
\be{
\Gamma \br{1+z} = z \Gamma \br{z} ,\qquad
\bet{\Gamma \br{1+ \ii z}}^2 = \frac{\pi z}{\tsinh{\pi z}}.
}
In order to determine the limit $u \to 1^-$ of the hypergeometric function $\Hyper{a,b,c;z}$, we may transform it according to linear transformation rules\cite{Mathematics1964}
\begin{multline}
\Hyper{a,b,c;z} = \frac{\Gamma \br{c} \Gamma \br{c-a-b}}{\Gamma \br{c-a} \Gamma \br{c-b}} \Hyper{a,b,a+b-c+1;1-z} \\
  + \br{1-z}^{c-a-b} \frac{\Gamma \br{c} \Gamma \br{a+b-c}}{\Gamma \br{a} \Gamma \br{b}} \Hyper{c-a, c-b, c-a-b+1; 1-z} .\\
\bet{\textrm{arg(1-z)}} < \pi
\end{multline}

We will focus on the term $(*)$ in \eqref{eqA_6:1} for the moment as it is the most complicated one. In the limit $u \to 1^-$ we obtain
\begin{multline}
F_1 \qu = \frac{2}{\tau} u \br{1-u} \frac{ab}{c} \Hyper{1+a,1+b,1+c;u} \\
\stackrel{u \to 1^-}{=} \frac{2}{\tau} u \br{1-u} \frac{ab}{c} \br{1-u}^{c-a-b-1} \frac{\Gamma \br{1+c} \Gamma \br{1+a+b-c}}{\Gamma \br{1+a} \Gamma \br{1+b}} 
\end{multline}
and 
\begin{multline}
F_2 \qu = \br{\omega \qu - \br{1-u} \omega_0 - u \omega_1} \Hyper{a,b,c;u} \\
\stackrel{u \to 1^-}{=} \br{\br{1-u} \omega_1 - \br{1-u} \omega_0} \br{1-u}^{c-a-b} \frac{\Gamma \br{c} \Gamma \br{a+b-c}}{\Gamma \br{a} \Gamma \br{b}}.
\end{multline}
In the equation $\bet{F_1 + \ii F_2}^2$ we can single out the term $\bet{\br{1-u}^{c-a-b}}^2$. 
This term gives always $1$, even in the limit $~u \to 1^-$. 
The second term in the equation above vanishes completely in this limit (because of the remaining $\br{1-u}$ term) leading to
\be{
\bet{F_1 + \ii F_2}^2 \stackrel{u \to 1^-}{=} \frac{4}{\tau^2} \bet{\frac{ab}{c} \frac{\Gamma \br{1+c} \Gamma \br{1+a+b-c}}{\Gamma \br{1+a} \Gamma \br{1+b}}}^2.
}
This then yields 
\be{
F \stackrel{u \to 1^-}{=} \frac{1}{2 \omega_0 \br{\omega_0 - \pi_3 \br{\vec q, 0}}} \frac{\omega_1 + 
  \pi_3 \br{\vec q, 1}}{\omega_1} \frac{4}{\tau^2} \bet{\frac{ab}{c} \frac{\Gamma \br{1+c} \Gamma \br{1+a+b-c}}{\Gamma \br{1+a} \Gamma \br{1+b}} }^2.
}
We define the following shorthand notations for the arguments of the gamma functions
\be{
1+a = 1- \ii \br{\va \tau^2 + \frac{\tau \omega_0}{2} - \frac{\tau \omega_1}{2}} 
  = 1+ \ii \br{-\va \tau^2 - \frac{\tau \omega_0}{2} + \frac{\tau \omega_1}{2}} = 1+ \ii \alpha ,
}
\al{
b = 1 + \ii \br{\va \tau^2 - \frac{\tau \omega_0}{2} + \frac{\tau \omega_1}{2}} = 1+ \ii \beta ,
}
\al{  
c = 1- \ii \tau \omega_0 = 1+\ii \br{-\tau \omega_0} = 1+ \ii \gamma 
}
\al{
1+a+b-c = 1+ \ii \tau \omega_1 = 1+ \ii \delta.
}
In a next step we rewrite the asymptotic distribution function according to
\al{
F \br{\vec q, u \to 1^{-}} &=
\frac{2}{\tau^2 \omega_0^2 \omega_1^2} \frac{\omega_1 + \pi_1}{\omega_0 - \pi_0} \bet{a}^2 
  \bet{\frac{\Gamma \br{c} \Gamma \br{1+a+b-c}}{\Gamma \br{1+a} \Gamma \br{b}} }^2 \notag \\
  &= \frac{2}{\tau^2 \omega_0^2 \omega_1^2} \frac{\omega_1 + \pi_1}{\omega_0 - \pi_0} \bet{a}^2 
  \frac{\pi \gamma \delta \tsinh{\pi \alpha} \tsinh{\pi \beta}}{\pi^2 \alpha \beta \tsinh{\pi \gamma} \tsinh{\pi \delta}} \notag \\
}
\al{
  F \br{\vec q, u \to 1^{-}} &= \frac{-2 \bet{\alpha}^2}{\alpha \beta} \frac{\omega_1 + \pi_1}{\omega_0 - \pi_0} \rho \notag \\
  &= 2 \frac{\omega_1 + \pi_1}{\omega_0 - \pi_0} \frac{2 \va \tau + \omega_0 - \omega_1}{2 \va \tau - \omega_0 + \omega_1} \rho \notag \\
  &= 2 \rho \frac{2 \va \tau \br{\omega_1 + \pi_1} + \omega_0 \omega_1 - \omega_1^2 + \omega_0 \pi_1 - \omega_1 \pi_1} 
  {2 \va \tau \br{\omega_0 - \pi_0} + \omega_0 \omega_1 - \omega_0^2 + \omega_0 \pi_0 - \omega_1 \pi_0} \notag \\
  &= 2 \rho \frac{\omega_0 \omega_1 - \pi_0 \omega_1 + \pi_1 \omega_0 + 2 \va \tau \pi_1 - \omega_1^2} 
  {\omega_0 \omega_1 - \pi_0 \omega_1 + \pi_1 \omega_0 - 2 \va \tau \pi_0 - \omega_0^2}.
}
where we have used
\begin{alignat}{3}
\pi_0 &=& q_3 - \va \tau ,\qquad
\pi_1 &=& q_3 + \va \tau ,\\
\omega_0^2 &=& \enorm^2 + \pi_0^2 ,\qquad
\omega_1^2 &=& \enorm^2 + \pi_1^2 ,\\
\end{alignat}
and
\be{
\rho = \frac{\tsinh{\pi \tau \br{2 \va \tau + \omega_0 - \omega_1}/2} \tsinh{\pi \tau \br{2 \va \tau - \omega_0 + \omega_1}/2}} 
   {\tsinh{\pi \tau \omega_0} \tsinh{\pi \tau \omega_1}}.
}
We further simplify the following expressions
\begin{multline}
2 \va \tau \pi_1 -\omega_1^2 = -\enorm^2 - q_3^2 - 2 \va \tau q_3 - \br{ \va \tau}^2 + 2 \va \tau \br{q_3 + \va \tau} = \\
  -\enorm^2 - q_3^2 + \br{ \va \tau}^2, 
\end{multline}
\begin{multline}
- 2 \va \tau \pi_0 -\omega_0^2 = -\enorm^2 - q_3^2 + 2 \va \tau q_3 - \br{\va \tau}^2 - 2 \va \tau \br{q_3 - \va \tau} = \\
  -\enorm^2 - q_3^2 + \br{\va \tau}^2 ,
\end{multline}
such that
\be{
2 \va \tau \pi_1 - \omega_1^2 = -2 \va \tau \pi_0 - \omega_0^2.
}
Altogether, we obtain the final result
\be{
F \stackrel{u \to 1^-}{=} 2 \rho = 2 \frac{\tsinh{\pi \tau \br{2 \va \tau + \omega_0 - \omega_1}/2} \tsinh{\pi \tau \br{2 \va \tau - \omega_0 + \omega_1}/2}}
  {\tsinh{\pi \tau \omega_0} \tsinh{\pi \tau \omega_1}}.
} 
\chapter{Runtime Tables}

To provide additional information concerning the numerical performance, we give here the unnormalized data from section 3 on Numerics.
We have done all test calculations on an ``Intel(R) Core(TM)2 Duo CPU     E7400  @ 2.80GHz'' with the GNU-C++-Compiler g++-4.5. For testing we tried to use always two threads at the same time.
The templates for the C++ code were obtained from the books \cite{Press:2007:NRE:1403886,Hairer1993}

\scriptsize

\ctable[
notespar,
caption = {Runtime comparison for different field configurations and solvers (c.f. Tab.\ref{Tab1}).},
cap = {Runtime comparison for different field configurations and systems without normalization},
pos = h,
]{lccccc}{
  \tnote[1]{Single pulse}
  \tnote[2]{Double pulse}
  \tnote[3]{Ten identical pulses with time lag $t_0$}
  \tnote[4]{One long pulse and ten short pulses}
  } {
    \toprule
      Field Parameters & Standard & Standard(I) & Trans2(I) & Trans2 & Trans1   \\
      ($\va$[1], $\tau$[1/m], $t0$[1/m])  & (sec) & (sec) & (sec) & (sec) & (sec) \\
    \midrule
      $\va=0.1, \tau=100$\footnotemark[1] 	& 292.54 & 197.72 & 201.62 & 266.4  & 647.11 \\	
    \midrule
      $\va=0.005, \tau=3$\footnotemark[1] 	& 8.79	 & 6.56   & 6      & 2.75   & 13.29 \\
    \midrule
      $\va_1=0.1, \tau_1=100,$ 	& 72.03	 & 65.63  & 68.38  & 112.67 & 326.75\\
      $\va_2=0.01, \tau_2=2$\footnotemark[2] & & & & & \\
    \midrule
      $\va_1=0.1, \tau_1=100,$ 	& 421.34 & 313.09 & 257.44 & 376.56 & 868.15 \\
      $\va_2=0.01, \tau_2=2, t_0=30$\footnotemark[2] & & & & & \\
    \midrule
      $\va= \pm{0.005}, \tau=3, t_0=5$\footnotemark[3] 	& 343.44 & 231.67 & 216.52 &    -    & 734.49 \\
    \midrule
      $\va= 0.005, \tau=3, t_0=5$\footnotemark[3] 	& 128.67 & 108.08 & 219.61 & 187.66 & 231.99 \\ 
    \midrule
      $\va_1=0.1, \tau_1=100,$ 	& 101.21 & 83.73  & 186.8  & 179.1  & 223.02 \\
      $\va_i= \pm{0.005}, \tau_i=3, t_0=9$\footnotemark[4] & & & & & \\
    \midrule
      $\va_1=0.1, \tau_1=100,$ 	& 830.95 & 543.5  & 350.96 & 463.78 & 1839.54 \\ 
      $\va_i= 0.005, \tau_i=3, t_0=9$\footnotemark[4] & & & & & \\
    \midrule
      $\va_1=0.1, \tau_1=100, T_0 = 200,$ & 829.96 & 540.29 & 357.01 & 450.81 & - \\
      $\va_i= 0.005, \tau_i=3, t_0=24.95$\footnotemark[4] & & & & & \\
    \bottomrule
}

\ctable[
notespar,
caption = {Comparison of the particle number for different solvers and field configurations (c.f. Tab.\ref{Tab2}).},
cap = {Accuracy comparison for different field configurations and systems},
pos = h,
]{llcccccc}{
  \tnote[1]{Single pulse}
  \tnote[2]{Double pulse}
  \tnote[3]{Ten identical pulses with time lag $t_0$}
  \tnote[4]{One long pulse and ten short pulses}
  \tnote[5]{Discrepancy coming from underflow errors}
  } {
    \toprule
      Field Parameters & Standard & Standard(I) & Tr2(I) & Trans2 & Trans1 & An \\
      ($\va$[1], $\tau$[1/m], $t0$[1/m]) & & & & & & \\
    \midrule
      $\va=0.1, \tau=100$\footnotemark[1] 	& 1.529 & 1.668 & 1.529 & 1.540 & 1.529 & 1.529 & $\times E-13$ \\
    \midrule
      $\va=0.005, \tau=3$\footnotemark[1] 	& 5.856	& 5.856 & 5.856 & 5.933 & 5.856 & 5.856 & $\times E-10$  \\
    \midrule
      $\va=0.05, \tau=100$\footnotemark[1] 	& 1.392	& 9856. & 1.193 & 0.599 & 1.392 & 2.4E-09\footnotemark[5] & $\times E-18$ \\  
    \midrule
      $\va_1=0.1, \tau_1=100,$ 	& 4.240 & 4.240 & 4.240 & 4.240 & 4.240 & \wh{-} & $\times E-07$ \\
      $\va_2=0.01, \tau_2=2$\footnotemark[2] & & & & & & & \\
    \midrule
      $\va_1=0.1, \tau_1=100,$ 	& 4.021 & 4.021 & 4.021 &   \wh{-}   & 4.021 & \wh{-} & $\times E-07$ \\
      $\va_2=0.01, \tau_2=2, t_0=30$\footnotemark[2] & & & & & & & \\
    \midrule
      $\va= \pm{0.005}, \tau=3, t_0=5$\footnotemark[3] 	& 8.700 & 8.700 & 30.99 & 31.03 & 8.700 & \wh{-} & $\times E-10$ \\ 
    \midrule
      $\va= 0.005, \tau=3, t_0=5$\footnotemark[3] 	& 3.782 & 3.782 & 5.241 & 5.267 & 3.782 & \wh{-} & $\times E-10$ \\
    \midrule
      $\va_1=0.1, \tau_1=100,$ 	& 1.421 & 1.421 & 1.421 & 1.420 & 1.421 & \wh{-} & $\times E-08$ \\ 
      $\va_i= \pm{0.005}, \tau_i=3, t_0=9$\footnotemark[4] & & & & & \\
    \midrule
      $\va_1=0.1, \tau_1=100,$ 	& 1.353 & 1.353 & 1.353 & 1.353 &   \wh{-}    & \wh{-} & $\times E-08$ \\
      $\va_i= 0.005, \tau_i=3, t_0=9$\footnotemark[4] & & & & & \\
    \midrule
      $\va_1=0.1, \tau_1=100, T_0 = 200,$ & 2.259 & 2.259 & 2.259 & \wh{-} & \wh{-} & \wh{-} & $\times E-08$ \\
      $\va_i= 0.005, \tau_i=3, t_0=24.95$\footnotemark[4] & & & & & \\
    \bottomrule
}

\ctable[
notespar,
caption = {Runtime comparison for different momentum spacings for a single pulse with $\varepsilon = 0.005 ,~ \tau =3m^{-1}$. All momenta are given in units of the electron mass $~q[m]$ (c.f. Tab.\ref{Tab3}).},
cap = {Runtime comparison for different momentum spacings for a single short pulse without normalization},
pos = h,
]{lccc}{
  \tnote[1]{$\Delta q_3 = 0.05,~ 0.1~$ Tolerance }
  } {
	\toprule
	  & $~\Delta q_{\perp} = 0.5~$ & $~\Delta q_{\perp} \in \left[0.1,0.5 \right]~$ & $~\Delta q_{\perp} = \Delta q_3~$ \\
	  & (sec) & (sec) & (sec) \\
	\midrule
	  Important $~q_3$\tmark[1] & 0.95 & 1.55 & 2.91 \\
	\midrule
	  $~\Delta q_3 = 0.05~$ & 1.15	& 1.54 & 3.10  \\
	\midrule
	  $~\Delta q_3 = 0.01~$ & 3.62 & 6.62 & 59.57 \\
	\bottomrule
}

\ctable[
notespar,
caption = {Runtime comparison for different momentum spacings for a single pulse with $~\varepsilon = 0.1 ,~ \tau =100m^{-1}$. All momenta are given in units of the electron mass $~q[m]$ (c.f. Tab.\ref{Tab5}).},
cap = {Runtime comparison for different momentum spacings for a single long pulse without normalization},
pos = h,
]{lccc}{
  \tnote[1]{$\Delta q_3 = 0.05,~ 0.1~$ Tolerance }
  } {
	\toprule
	  & $~\Delta q_{\perp} = 0.5~$ & $~\Delta q_{\perp} \in \left[0.1,0.5 \right]~$ & $~\Delta q_{\perp} = \Delta q_3~$ \\
	  & (sec) & (sec) & (sec) \\
	\midrule
	  Important $~q_3$\tmark[1] & 189.43 & 403.47 & 655.78 \\
	\midrule
	  $~\Delta q_3 = 0.05~$ & 254.82 & 557.61 & 917.52  \\
	\midrule
	  $~\Delta q_3 = 0.01~$ & 1181.22 & 2663.16 & 18685.5 \\
	\bottomrule
}

\ctable[
notespar,
caption = {Runtime comparison for different sets of sampling points. In order to calculate intermediate values linear interpolation is used (c.f. Tab.\ref{Tab9}).},
cap = {Runtime comparison for different sets of sampling points and linear interpolation without normalization},
pos = h,
]{lccccc}{
  \tnote[1]{Single pulse}
  \tnote[2]{Double pulse}
  \tnote[3]{Ten identical pulses with time lag $t_0$}
  \tnote[4]{One long pulse and ten short pulses}
  } {
    \toprule
      Field Parameters & Standard & Trans2(I) & 512 Points & 4096 Points & 32768 Points   \\
      ($\va$[1], $\tau$[1/m], $t0$[1/m]) & (sec) & (sec) & (sec) & (sec) & (sec) \\
    \midrule
      $\va=0.1, \tau=100$\footnotemark[1]  & 292.54 & 201.62 & 74.21 & 229.4 & 740.98 \\	
    \midrule
      $\va=0.005, \tau=3$\footnotemark[1] & 8.79 & 6 & 2.91 & 39.3 & 201.14 \\
    \midrule
      $\va_1=0.1, \tau_1=100,$  & 421.34 & 257.44 & 85.41 & 222.17 & 800.55 \\
      $\va_2=0.01, \tau_2=2$\footnotemark[2] & & & & & \\
    \midrule
      $\va_1=0.1, \tau_1=100,$ 	& 343.44 & 216.52 & 75.65 & 180.82 & 643.2 \\
      $\va_2=0.01, \tau_2=2, t_0=30$\footnotemark[2] & & & & & \\
    \midrule
      $\va= \pm{0.005}, \tau=3, t_0=5$\footnotemark[3] 	& 128.67 &  & 12.8 & 180.94 & 571.54 \\ 
    \midrule
      $\va= 0.005, \tau=3, t_0=5$\footnotemark[3] & 101.21 &  & 5.19 & 155.36 & 507.67\\	 
    \midrule
      $\va_1=0.1, \tau_1=100,$ 	& 830.95 & 350.96 & 75.2 & 138.99 & 581.54 \\ 
      $\va_i= \pm{0.005}, \tau_i=3, t_0=9$\footnotemark[4] & & & & & \\
    \midrule
      $\va_1=0.1, \tau_1=100,$ 	& 829.96 & 357.01 & 74.93 & 119 & 575 \\
      $\va_i= 0.005, \tau_i=3, t_0=9$\footnotemark[4] & & & & & \\
    \bottomrule
}

\ctable[
notespar,
caption = {Comparison of the particle number for different sets of sampling points. In order to calculate intermediate values linear interpolation is used (c.f. Tab.\ref{Tab10}).},
cap = {Comparison of the particle number for different sets of sampling points via linear interpolation},
pos = h,
]{llcccccc}{
  \tnote[1]{Single pulse}
  \tnote[2]{Double pulse}
  \tnote[3]{Ten identical pulses with time lag $t_0$}
  \tnote[4]{One long pulse and ten short pulses}
  } {
    \toprule
      Field Parameters & Standard & Trans2(I) & 512 Points & 4096 Points & 32768 Points \\
      ($\va$[1], $\tau$[1/m], $t0$[1/m]) & & & & & \\
    \midrule
      $\va=0.1, \tau=100$\footnotemark[1] 	& 1.529 & 1.529 & \wh{2.1573E+05} & 1.58659 & 1.52987 & $\times E-13$ \\
    \midrule
      $\va=0.005, \tau=3$\footnotemark[1] 	& 5.856	& 5.856 & 5.73369 & 5.85943 & 5.85716 & $\times E-10$  \\  
    \midrule
      $\va_1=0.1, \tau_1=100,$ 	& 4.240 & 4.240 & 2.21893 & 2.74849 & 4.21281 & $\times E-07$ \\
      $\va_2=0.01, \tau_2=2$\footnotemark[2] & & & & & & \\
    \midrule
      $\va_1=0.1, \tau_1=100,$ 	& 4.021 & 4.021 & 1.67201 & 2.57259 & 3.99372 & $\times E-07$ \\
      $\va_2=0.01, \tau_2=2, t_0=30$\footnotemark[2] & & & & & & \\
    \midrule
      $\va= \pm{0.005}, \tau=3, t_0=5$\footnotemark[3] 	& 8.700 &  & 8.01398 & 8.69707 & 8.70156 & $\times E-10$ \\
    \midrule
      $\va= 0.005, \tau=3, t_0=5$\footnotemark[3] 	& 3.782 &  & 3.43572 & 3.78036 & 3.78314 & $\times E-10$ \\
    \midrule
      $\va_1=0.1, \tau_1=100,$ 	 & 1.421 & 1.421 & \wh{1.03958E+02} & 1.00616 & 1.41359 & $\times E-08$ \\ 
      $\va_i= \pm{0.005}, \tau_i=3, t_0=9$\footnotemark[4] & & & & & & \\
    \midrule
      $\va_1=0.1, \tau_1=100,$ 	& 1.353 & 1.353 & 7.92583 & 9.19399 & 1.34551 & $\times E-08$ \\
      $\va_i= 0.005, \tau_i=3, t_0=9$\footnotemark[4] & & & & & & \\
    \bottomrule
}

\ctable[
notespar,
caption = {Runtime comparison for different sets of sampling points. In order to calculate intermediate values cubic interpolation has been used (c.f. Tab.\ref{Tab11}).},
cap = {Runtime comparison for a different set of sampling points with cubic interpolation},
pos = h,
]{lccccc}{
  \tnote[1]{Single pulse}
  \tnote[2]{Double pulse}
  \tnote[3]{Ten identical pulses with time lag $t_0$}
  \tnote[4]{One long pulse and ten short pulses}
  } {
    \toprule
      Field Parameters & Standard & Trans2(I) & 512 Points & 4096 Points & 32768 Points   \\
      ($\va$[1], $\tau$[1/m], $t0$[1/m]) & (sec) & (sec) & (sec) & (sec) & (sec) \\
    \midrule
      $\va=0.1, \tau=100$\footnotemark[1] & 292.54 & 201.62 & 282.27 & 283.29 & 325.25 \\	
    \midrule
      $\va=0.005, \tau=3$\footnotemark[1] & 8.79 & 6 & 10.4 & 13.94 & 50.41 \\	
    \midrule
      $\va_1=0.1, \tau_1=100,$ & 421.34 & 257.44 & 296.55 & 307.23 & 401.08 \\
      $\va_2=0.01, \tau_2=2$\footnotemark[2] & & & & & \\
    \midrule
      $\va_1=0.1, \tau_1=100,$ 	& 343.44 & 216.52 & 236.8 & 249.4 & 312.73 \\
      $\va_2=0.01, \tau_2=2, t_0=30$\footnotemark[2] & & & & & \\
    \midrule
      $\va= \pm{0.005}, \tau=3, t_0=5$\footnotemark[3]  & 128.67 &  & 25.85 & 29.84 & 58.84 \\ 
    \midrule
      $\va= 0.005, \tau=3, t_0=5$\footnotemark[3] 	& 101.21 &  & 18.47 & 24.25 & 56.42 \\
    \midrule
      $\va_1=0.1, \tau_1=100,$ 	& 830.95 & 350.96 & 188.4 & 256.49 & 271.99 \\ 
      $\va_i= \pm{0.005}, \tau_i=3, t_0=9$\footnotemark[4] & & & & & \\
    \midrule
      $\va_1=0.1, \tau_1=100,$ 	 & 829.96 & 357.01 & 170.44 & 255.34 & 274.85 \\
      $\va_i= 0.005, \tau_i=3, t_0=9$\footnotemark[4] & & & & & \\
    \bottomrule
}

\ctable[
notespar,
caption = {Comparison of the particle number density for different sets of sampling points. In order to calculate intermediate values cubic interpolation has been used (c.f. Tab.\ref{Tab12}).},
cap = {Comparison of the particle number density for different sets of sampling points with cubic interpolation},
pos = h,
]{llcccccc}{
  \tnote[1]{Single pulse}
  \tnote[2]{Double pulse}
  \tnote[3]{Ten identical pulses with time lag $t_0$}
  \tnote[4]{One long pulse and ten short pulses}
  } {
    \toprule
      Field Parameters & Standard & Trans2(I) & 512 Points & 4096 Points & 32768 Points \\
      ($\va$[1], $\tau$[1/m], $t0$[1/m]) & & & & & \\
    \midrule
      $\va=0.1, \tau=100$\footnotemark[1] 	& 1.529 & 1.529 & 1.4597 & 1.51973 & 1.52846 & $\times E-13$ \\
    \midrule
      $\va=0.005, \tau=3$\footnotemark[1] 	 & 5.856	& 5.856 & 5.61761 & 5.83844 & 5.8543 & $\times E-10$  \\
    \midrule
      $\va_1=0.1, \tau_1=100,$ 	& 4.240 & 4.240 & \wh{9.08134E+03} & 3.16232 & 4.22142 & $\times E-07$ \\
      $\va_2=0.01, \tau_2=2$\footnotemark[2] & & & & & & \\
    \midrule
      $\va_1=0.1, \tau_1=100,$ 	& 4.021 & 4.021 & \wh{5.15012E+03} & 2.97061 & 4.0021 & $\times E-07$ \\
      $\va_2=0.01, \tau_2=2, t_0=30$\footnotemark[2] & & & & & & \\
    \midrule
      $\va= \pm{0.005}, \tau=3, t_0=5$\footnotemark[3] 	& 8.700 &  & 7.99011 & 8.63437 & 8.69277 & $\times E-10$ \\ 
    \midrule
      $\va= 0.005, \tau=3, t_0=5$\footnotemark[3] 	& 3.782 &  & 3.63728 & 3.79134 & 3.78425 & $\times E-10$ \\
    \midrule
      $\va_1=0.1, \tau_1=100,$ 	& 1.421 & 1.421 & \wh{1.5481E+01} & 1.12703 & 1.41548 & $\times E-08$ \\ 
      $\va_i= \pm{0.005}, \tau_i=3, t_0=9$\footnotemark[4] & & & & & & \\
    \midrule
      $\va_1=0.1, \tau_1=100,$ 	& 1.353 & 1.353 & \wh{4.20644E+02} & 1.0406 & 1.34787 & $\times E-08$ \\
      $\va_i= 0.005, \tau_i=3, t_0=9$\footnotemark[4] & & & & & & \\
    \bottomrule
}

\normalsize 


\bibliographystyle{unsrt}
\cleardoublepage
\normalbaselines 
\bibliography{References} 

\begin{thebibliography}{10}

\bibitem{Sauter:1931zz}
F.~Sauter.
\newblock {\"Uber das Verhalten eines Elektrons im homogenen elektrischen Feld
  nach der relativistischen Theorie Diracs}.
\newblock {\em Z.Phys.}, 69:742--764, Apr 1931.

\bibitem{springerlink:10.1007/BF01343663}
W.~Heisenberg and H.~Euler.
\newblock {Folgerungen aus der Diracschen Theorie des Positrons}.
\newblock {\em Zeitschrift f\"ur Physik}, 98:714--732, Dec 1936.
\newblock DOI:10.1007/BF01343663.

\bibitem{Ruffini20101}
R.~Ruffini, G.~Vereshchagin, and S.~Xue.
\newblock Electron-positron pairs in physics and astrophysics: From heavy
  nuclei to black holes.
\newblock {\em Physics Reports}, 487(1 4):1 -- 140, Nov 2010.
\newblock DOI:10.1016/j.physrep.2009.10.004.

\bibitem{hep-th/0406216}
Gerald~V. Dunne.
\newblock {Heisenberg-Euler effective Lagrangians: Basics and extensions}.
\newblock Jun 2004.
\newblock hep-th/0406216.

\bibitem{PhysRev.82.664}
J.~Schwinger.
\newblock {On Gauge Invariance and Vacuum Polarization}.
\newblock {\em Phys. Rev.}, 82:664--679, Jun 1951.
\newblock DOI:10.1103/PhysRev.82.664.

\bibitem{PhysRevD.44.1825}
{Bialynicki-Birula, I. and G\'ornicki, P. and Rafelski, J.}
\newblock Phase-space structure of the dirac vacuum.
\newblock {\em Phys. Rev. D}, 44:1825--1835, Sep 1991.
\newblock DOI:10.1103/PhysRevD.44.1825.

\bibitem{PhysRevLett.67.2427}
Y.~Kluger, J.~M. Eisenberg, B.~Svetitsky, F.~Cooper, and E.~Mottola.
\newblock {Pair production in a strong electric field}.
\newblock {\em Phys. Rev. Lett.}, 67:2427--2430, Oct 1991.
\newblock DOI:10.1103/PhysRevLett.67.2427.

\bibitem{PhysRevD.58.125015}
Y.~Kluger, E.~Mottola, and J.~M. Eisenberg.
\newblock {Quantum Vlasov equation and its Markov limit}.
\newblock {\em Phys. Rev. D}, 58:125015, Nov 1998.
\newblock DOI:10.1103/PhysRevD.58.125015.

\bibitem{PhysRevD.60.116011}
J.~C.~R. Bloch, V.~A. Mizerny, A.~V. Prozorkevich, C.~D. Roberts, S.~M.
  Schmidt, S.~A. Smolyansky, and D.~V. Vinnik.
\newblock {Pair creation: Back reactions and damping}.
\newblock {\em Phys. Rev. D}, 60:116011, Nov 1999.
\newblock DOI:10.1103/PhysRevD.60.116011.

\bibitem{PhysRevD.61.117502}
J.~C.~R. Bloch, C.~D. Roberts, and S.~M. Schmidt.
\newblock {Memory effects and thermodynamics in strong field plasmas}.
\newblock {\em Phys. Rev. D}, 61:117502, May 2000.
\newblock DOI:10.1103/PhysRevD.61.117502.

\bibitem{hep-ph/9809227}
S.~M. Schmidt, D.~B. Blaschke, G.~R\"opke, S.~A. Smolyansky, A.~V.
  Prozorkevich, and V.~D. Toneev.
\newblock {A quantum kinetic equation for particle production in the Schwinger
  mechanism}.
\newblock Int.J.Mod.Phys. E7 (1998) 709-722, Sep 1998.

\bibitem{hep-ph/9809233}
S.M. Schmidt, A.V. Prozorkevich, and S.A. Smolyansky.
\newblock {Creation of boson and fermion pairs in strong fields}.
\newblock 1998.
\newblock hep-ph/9809233.

\bibitem{PhysRevLett.79.1626}
D.~L. Burke, R.~C. Field, G.~Horton-Smith, J.~E. Spencer, D.~Walz, S.~C.
  Berridge, W.~M. Bugg, K.~Shmakov, A.~W. Weidemann, C.~Bula, K.~T. McDonald,
  E.~J. Prebys, C.~Bamber, S.~J. Boege, T.~Koffas, T.~Kotseroglou, A.~C.
  Melissinos, D.~D. Meyerhofer, D.~A. Reis, and W.~Ragg.
\newblock {Positron Production in Multiphoton Light-by-Light Scattering}.
\newblock {\em Phys. Rev. Lett.}, 79:1626--1629, Sep 1997.
\newblock DOI:10.1103/PhysRevLett.79.1626.

\bibitem{springerlink:10.1140/epjd/e2009-00113-x}
T.~Heinzl and A.~Ilderton.
\newblock {Exploring high-intensity QED at ELI}.
\newblock {\em The European Physical Journal D - Atomic, Molecular, Optical and
  Plasma Physics}, 55:359--364, Apr 2009.
\newblock DOI:10.1140/epjd/e2009-00113-x.

\bibitem{springerlink:10.1140/epjd/e2009-00169-6}
M.~Marklund and J.~Lundin.
\newblock {Quantum vacuum experiments using high intensity lasers}.
\newblock {\em The European Physical Journal D - Atomic, Molecular, Optical and
  Plasma Physics}, 55:319--326, Apr 2009.
\newblock DOI:10.1140/epjd/e2009-00169-6.

\bibitem{1104.0468}
R.~Soldati.
\newblock {Pairs Emission in a Uniform Background Field: an Algebraic
  Approach}.
\newblock J. Phys. A: Math. Theor. 44 (2011) 305401 (25 pp), Jun 2011.
\newblock DOI:10.1088/1751-8113/44/30/305401.

\bibitem{1110.4684}
Sang~Pyo Kim.
\newblock {Schwinger Pair Production in Solitonic Gauge Fields}.
\newblock Oct 2011.
\newblock arXiv: 1110.4684.

\bibitem{PhysRevD.84.125028}
S.~P. Kim and C.~Schubert.
\newblock {Nonadiabatic quantum Vlasov equation for Schwinger pair production}.
\newblock {\em Phys. Rev. D}, 84:125028, Dec 2011.
\newblock DOI:10.1103/PhysRevD.84.125028.

\bibitem{PhysRevD.79.065027}
C.~K. Dumlu.
\newblock {Quantum kinetic approach and the scattering approach to vacuum pair
  production}.
\newblock {\em Phys. Rev. D}, 79:065027, Mar 2009.
\newblock DOI:10.1103/PhysRevD.79.065027.

\bibitem{Hebenstreit:2011wk}
F.~Hebenstreit, R.~Alkofer, and H.~Gies.
\newblock {Particle self-bunching in the Schwinger effect in
  spacetime-dependent electric fields}.
\newblock {\em Phys.Rev.Lett.}, 107:180403, 2011.

\bibitem{Hebenstreit:2011cr}
Florian Hebenstreit, Anton Ilderton, and Mattias Marklund.
\newblock {Pair production: the view from the lightfront}.
\newblock {\em Phys.Rev.}, D84:125022, 2011.

\bibitem{PhysRevD.72.065001}
H.~Gies and K.~Klingm\"uller.
\newblock {Pair production in inhomogeneous fields}.
\newblock {\em Phys. Rev. D}, 72:065001, Sep 2005.
\newblock DOI:10.1103/PhysRevD.72.065001.

\bibitem{PhysRevLett.87.193902}
R.~Alkofer, M.~B. Hecht, C.~D. Roberts, S.~M. Schmidt, and D.~V. Vinnik.
\newblock {Pair Creation and an X-Ray Free Electron Laser}.
\newblock {\em Phys. Rev. Lett.}, 87:193902, Oct 2001.
\newblock DOI:10.1103/PhysRevLett.87.193902.

\bibitem{PhysRevLett.96.140402}
D.~B. Blaschke, A.~V. Prozorkevich, C.~D. Roberts, S.~M. Schmidt, and S.~A.
  Smolyansky.
\newblock {Pair Production and Optical Lasers}.
\newblock {\em Phys. Rev. Lett.}, 96:140402, Apr 2006.
\newblock DOI:10.1103/PhysRevLett.96.140402.

\bibitem{PhysRevLett.89.153901}
C.~D. Roberts, S.~M. Schmidt, and D.~V. Vinnik.
\newblock {Quantum Effects with an X-Ray Free-Electron Laser}.
\newblock {\em Phys. Rev. Lett.}, 89:153901, Sep 2002.
\newblock DOI:10.1103/PhysRevLett.89.153901.

\bibitem{PhysRevD.2.1191}
E.~Brezin and C.~Itzykson.
\newblock {Pair Production in Vacuum by an Alternating Field}.
\newblock {\em Phys. Rev. D}, 2:1191--1199, Oct 1970.
\newblock DOI:10.1103/PhysRevD.2.1191.

\bibitem{HebDip}
F.~Hebenstreit.
\newblock {Diploma Thesis, University Graz, Electron-Positron Pair Creation in
  Impulse-Shaped Electric Fields}, Jun 2008.

\bibitem{OrtDip}
M.~Orthaber.
\newblock {Diploma Thesis, University Graz, Electron-Positron Pair Production
  in Multiple Time Scale Electric Fields}, Feb 2010.

\bibitem{Nikishiv}
N.~B. Narozhny and A.~I. Nikishov.
\newblock {The simplest Processes in a Pair-Producing Field}.
\newblock {\em Soviet J. Nucl. Phys. 11, 596}, 1970.

\bibitem{springerlink:10.1134/1.1788038}
A.~Nikishov.
\newblock {Scattering and pair production by a potential barrier}.
\newblock {\em Physics of Atomic Nuclei}, 67:1478--1486, Nov 2004.
\newblock DOI:10.1134/1.1788038.

\bibitem{PhysRevD.84.125023}
C.~K. Dumlu and G.~V. Dunne.
\newblock {Complex worldline instantons and quantum interference in vacuum pair
  production}.
\newblock {\em Phys. Rev. D}, 84:125023, Dec 2011.
\newblock DOI:10.1103/PhysRevD.84.125023.

\bibitem{PhysRevLett.101.130404}
R.~Sch\"utzhold, H.~Gies, and G.~V. Dunne.
\newblock {Dynamically Assisted Schwinger Mechanism}.
\newblock {\em Phys. Rev. Lett.}, 101:130404, Sep 2008.
\newblock DOI:10.1103/PhysRevLett.101.130404.

\bibitem{PhysRevD.80.111301}
G.~V. Dunne, H.~Gies, and R.~Sch\"utzhold.
\newblock {Catalysis of Schwinger vacuum pair production}.
\newblock {\em Phys. Rev. D}, 80:111301, Dec 2009.
\newblock DOI:10.1103/PhysRevD.80.111301.

\bibitem{PhysRevD.85.025004}
C.~Fey and R.~Sch\"utzhold.
\newblock {Momentum dependence in the dynamically assisted Sauter-Schwinger
  effect}.
\newblock {\em Phys. Rev. D}, 85:025004, Jan 2012.
\newblock DOI:10.1103/PhysRevD.85.025004.

\bibitem{Orthaber201180}
M.~Orthaber, F.~Hebenstreit, and R.~Alkofer.
\newblock Momentum spectra for dynamically assisted schwinger pair production.
\newblock {\em Physics Letters B}, 698(1):80 -- 85, Feb 2011.
\newblock DOI:10.1016/j.physletb.2011.02.053.

\bibitem{PhysRevLett.107.180403}
F.~Hebenstreit, R.~Alkofer, and H.~Gies.
\newblock {Particle Self-Bunching in the Schwinger Effect in
  Spacetime-Dependent Electric Fields}.
\newblock {\em Phys. Rev. Lett.}, 107:180403, Oct 2011.
\newblock DOI:10.1103/PhysRevLett.107.180403.

\bibitem{HebDiss}
F.~Hebenstreit.
\newblock {PhD Thesis, University Graz, Schwinger effect in inhomogeneous
  electric fields}, Jun 2011.

\bibitem{PhysRevLett.102.080402}
M.~Ruf, G.~R. Mocken, C.~M\"uller, K.~Z. Hatsagortsyan, and C.~H. Keitel.
\newblock {Pair Production in Laser Fields Oscillating in Space and Time}.
\newblock {\em Phys. Rev. Lett.}, 102:080402, Feb 2009.
\newblock DOI:10.1103/PhysRevLett.102.080402.

\bibitem{PhysRevD.78.036008}
T.~D. Cohen and D.~A. McGady.
\newblock {Schwinger mechanism revisited}.
\newblock {\em Phys. Rev. D}, 78:036008, Aug 2008.
\newblock DOI:10.1103/PhysRevD.78.036008.

\bibitem{PhysRevD.82.045007}
C.~K. Dumlu.
\newblock {Schwinger vacuum pair production in chirped laser pulses}.
\newblock {\em Phys. Rev. D}, 82:045007, Aug 2010.
\newblock DOI:10.1103/PhysRevD.82.045007.

\bibitem{PhysRevLett.104.250402}
C.~K. Dumlu and G.~V. Dunne.
\newblock {Stokes Phenomenon and Schwinger Vacuum Pair Production in
  Time-Dependent Laser Pulses}.
\newblock {\em Phys. Rev. Lett.}, 104:250402, Jun 2010.
\newblock DOI:10.1103/PhysRevLett.104.250402.

\bibitem{PhysRevLett.102.150404}
F.~Hebenstreit, R.~Alkofer, G.~V. Dunne, and H.~Gies.
\newblock {Momentum Signatures for Schwinger Pair Production in Short Laser
  Pulses with a Subcycle Structure}.
\newblock {\em Phys. Rev. Lett.}, 102:150404, Apr 2009.
\newblock DOI:10.1103/PhysRevLett.102.150404.

\bibitem{PhysRevLett.108.030401}
E.~Akkermans and G.~V. Dunne.
\newblock {Ramsey Fringes and Time-Domain Multiple-Slit Interference from
  Vacuum}.
\newblock {\em Phys. Rev. Lett.}, 108:030401, Jan 2012.
\newblock DOI:10.1103/PhysRevLett.108.030401.

\bibitem{PhysRevD.83.065028}
C.~K. Dumlu and G.~V. Dunne.
\newblock {Interference effects in Schwinger vacuum pair production for
  time-dependent laser pulses}.
\newblock {\em Phys. Rev. D}, 83:065028, Mar 2011.
\newblock DOI:10.1103/PhysRevD.83.065028.

\bibitem{PhysRevD.78.061701}
F.~Hebenstreit, R.~Alkofer, and H.~Gies.
\newblock {Pair production beyond the Schwinger formula in time-dependent
  electric fields}.
\newblock {\em Phys. Rev. D}, 78:061701, Sep 2008.
\newblock DOI:10.1103/PhysRevD.78.061701.

\bibitem{PhysRevD.83.025011}
A.~M. Fedotov, E.~G. Gelfer, K.~Yu. Korolev, and S.~A. Smolyansky.
\newblock {Kinetic equation approach to pair production by a time-dependent
  electric field}.
\newblock {\em Phys. Rev. D}, 83:025011, Jan 2011.
\newblock DOI:10.1103/PhysRevD.83.025011.

\bibitem{1202.1557}
Gerald~V. Dunne.
\newblock {The Heisenberg-Euler Effective Action: 75 years on}.
\newblock {\em Int.J.Mod.Phys.}, A27:1260004, Feb 2012.

\bibitem{1111.5192}
T.~Heinzl.
\newblock {Strong-Field QED and High Power Lasers}.
\newblock {\em Int.J.Mod.Phys.}, A27:1260010, Nov 2011.
\newblock 1111.5192.

\bibitem{1111.3886}
A.~Di~Piazza, C.~M\"uller, K.Z. Hatsagortsyan, and C.H. Keitel.
\newblock {Extremely high-intensity laser interactions with fundamental quantum
  systems}.
\newblock Apr 2011.
\newblock arXiv: 1111.3886.

\bibitem{Ryder1996}
L.~H. Ryder.
\newblock {\em Quantum Field Theory -}.
\newblock Cambridge University Press, Cambridge, 2. aufl. edition, 1996.

\bibitem{Itzykson2006}
C.~Itzykson and J.~B. Zuber.
\newblock {\em Quantum Field Theory -}.
\newblock Dover Publications, New York, dover ed edition, 2006.

\bibitem{Maggiore2005}
M.~Maggiore.
\newblock {\em A Modern Introduction To Quantum Field Theory -}.
\newblock Oxford University Press, New York, 2005.

\bibitem{Peskin1995}
M.~E. Peskin and D.~V. Schroeder.
\newblock {\em An Introduction To Quantum Field Theory -}.
\newblock Addison-Wesley Publishing Company, Reading, 1995.

\bibitem{Mathematics1964}
Mathematics Mathematics, M.~Abramowitz, and I.~A. Stegun.
\newblock {\em Handbook of Mathematical Functions - With Formulas, Graphs, and
  Mathematical Tables}.
\newblock Courier Dover Publications, Mineola, New York, new edition edition,
  1964.

\bibitem{Press:2007:NRE:1403886}
W.~H. Press, S.~A. Teukolsky, W.~T. Vetterling, and B.~P. Flannery.
\newblock {\em Numerical Recipes 3rd Edition: The Art of Scientific Computing}.
\newblock Cambridge University Press, New York, NY, USA, 3 edition, 2007.

\bibitem{Hairer1993}
E.~Hairer and G.~Wanner.
\newblock {\em {Solving Ordinary Differential Equations: Stiff and
  differential-algebraic problems - }}.
\newblock Springer, Berlin, Heidelberg, 2nd ed. 1996. corr. 3rd printing 2004
  edition, 1993.

\end{thebibliography}


\end{document}